\documentclass[journal=jcisd8]{achemso}
\usepackage{graphicx}
\usepackage{lipsum}
\usepackage{rotating}
\usepackage[normalem]{ulem}

\newcommand{\recAllowance}{900}

\newcommand{\cmpCinebCallRatio}{7.8}

\newcommand{\cmpCinebWallSaving}{41}

\newcommand{\cmpDimerAccepted}{17}
\newcommand{\cmpDimerCallRatio}{6.4}

\newcommand{\cmpDimerShared}{17}

\newcommand{\cmpOcinebCallRatio}{7.0}

\usepackage{amssymb,amsmath}
\usepackage{mathtools}
\usepackage{float}
\usepackage{xurl}
\usepackage{hyperref}
\hypersetup{hidelinks}
\usepackage[protrusion=true,expansion=false,final]{microtype}
\usepackage{listings}
\usepackage{booktabs}
\usepackage{amsthm}
\usepackage{algpseudocode}
\newfloat{algorithm}{tbp}{loa}
\floatname{algorithm}{Algorithm}
\newtheorem{theorem}{Theorem}

\lstdefinelanguage{cpp17}{morekeywords={auto,const,double,int,void,bool,struct,class,namespace,inline,template,typename,using,static,return,if,else,for,while,switch,case,break,continue,public,private,protected,virtual,override,noexcept,constexpr},sensitive=true,morecomment=[l]{//},morecomment=[s]{/*}{*/},morestring=[b]"}
\lstdefinelanguage{Fortran}{morekeywords={program,module,use,implicit,none,type,integer,real,character,logical,call,end,subroutine,function,intent,in,out,inout,value,parameter,public,private,contains,interface,abstract,procedure,final,class,bind,import,intrinsic,allocatable,allocate},sensitive=false,morecomment=[l]{!},morestring=[b]'}
\usepackage{needspace}
\usepackage{lea-tags}
\usepackage{pgfplots}
\pgfplotsset{compat=1.18}
\usepackage{tikz}
\usepackage[export]{adjustbox}
\usetikzlibrary{positioning,arrows.meta,fit,backgrounds,calc,shapes.geometric}
\definecolor{RuhiTeal}{HTML}{004D40}
\definecolor{RuhiCoral}{HTML}{FF655D}
\definecolor{RuhiYellow}{HTML}{F1DB4B}
\SectionNumbersOn
\mciteErrorOnUnknownfalse
\author{Rohit Goswami}
\affiliation{SURF B.V., Amsterdam, The Netherlands}
\alsoaffiliation{Science Institute, University of Iceland, VR-III, 107 Reykjav\'ik, Iceland}
\email{rgoswami@ieee.org}

\author{Maxim Masterov}
\affiliation{SURF B.V., Amsterdam, The Netherlands}

\author{Satish Kamath}
\affiliation{SURF B.V., Amsterdam, The Netherlands}

\author{Alejandro Pe\~{n}a-Torres}
\affiliation{Icelandic Research e-Infrastructure (IREI), University of Iceland, 107 Reykjav\'ik, Iceland}
\alsoaffiliation{Science Institute, University of Iceland, VR-III, 107 Reykjav\'ik, Iceland}

\author{Andreas Lynge Vishart}
\affiliation{Department of Energy Conversion and Storage, Technical University of Denmark (DTU), 2800 Kgs.\ Lyngby, Denmark}

\author{Thomas Bligaard}
\affiliation{Department of Energy Conversion and Storage, Technical University of Denmark (DTU), 2800 Kgs.\ Lyngby, Denmark}

\author{Hannes J\'onsson}
\affiliation{Faculty of Physical Sciences, University of Iceland, 107 Reykjav\'ik, Iceland}
\alsoaffiliation{Science Institute, University of Iceland, VR-III, 107 Reykjav\'ik, Iceland}

\date{\today}
\date{\today}
\title{GPRidge: Gaussian-process Accelerated Calculations of Minimum Energy Paths and First-order Saddle Points}
\hypersetup{
 pdfauthor={},
 pdftitle={GPRidge: Gaussian-process Accelerated Calculations of Minimum Energy Paths and First-order Saddle Points},
 pdfkeywords={},
 pdfsubject={},
 pdfcreator={Emacs 30.2 (Org mode 9.7.11)}, 
 pdflang={English}}
\usepackage{algorithm,algpseudocode}
\usepackage{lea-tags}
\begin{document}

\begin{abstract}
GPRidge, a Gaussian-process (GP) library for accelerating calculations of minimum energy paths and first-order saddle points with compiled-language and Python bindings, is presented.
An evaluation controller uses GP predictions to iteratively adjust the Cartesian coordinates of the atoms.
The fitted GP energy surface and measured energies and forces pass from the path calculation to local refinement of the saddle point.
New evaluations improve the surrogate and check the candidate geometry.
For the molecular calculations, acceptance requires measured-force convergence and an independent Hessian with only one negative eigenvalue.
A second negative eigenvalue triggers an escape and renewed refinement.
The controller supports a joint inverse-distance GP for molecules and local GP residuals on periodic systems.
Applications to the 24 Baker--Chan and 20 Birkholz--Schlegel molecular reactions using the Universal Model for Atoms (UMA-S-1.1), and to 72 Pt heptamer and island rearrangements on Pt(111) using a Morse potential, are presented.
On matched Baker--Chan reactions, GPRidge uses \cmpCinebCallRatio{}- and \cmpOcinebCallRatio{}-fold fewer energy-and-force evaluations than climbing-image nudged elastic band (CI-NEB) and off-path climbing-image nudged elastic band (OCI-NEB), respectively, reducing validation-inclusive CI-NEB time by \cmpCinebWallSaving{}\%.
All 20 Birkholz--Schlegel saddle points pass the independent index-one check, compared with \cmpDimerAccepted{} for the classical dimer method started from the same path maxima; on those \cmpDimerShared{} reactions, the GP dimer stage uses \cmpDimerCallRatio{}-fold fewer evaluations.
All 72 Pt(111) paths meet the OptBench force criterion, with median evaluation counts of 30 and 23 for the heptamer and island problems, against 778 and 443 for CI-NEB.
\end{abstract}
\begin{tocentry}
\includegraphics[width=\linewidth]{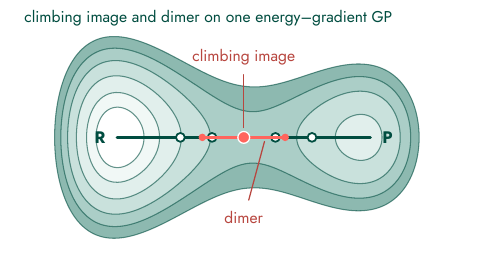}
\end{tocentry}
\section{Introduction}
\label{sec:introduction}
The identification of the mechanism and the estimation of the rate of thermally activated transitions, such as chemical reactions and diffusion events, is an important task in computational chemistry and materials science.
Since such transitions are rare events on the time scale of atomic vibrations, direct simulations of the dynamics are in most cases impractical and the calculations are instead based on transition state theory \cite{wignerTransitionStateMethod1938,vineyardFrequencyFactorsIsotope1957}.
Within the harmonic approximation, the task becomes that of finding first-order saddle points on the energy surface, the points where the atomic forces vanish and the Hessian has one and only one negative eigenvalue \cite{vineyardFrequencyFactorsIsotope1957}.
A first-order saddle point is a maximum along a minimum energy path (MEP) connecting two local minima on the energy surface, corresponding to the initial and final states of the transition.
The energy of the saddle point with respect to the initial state minimum gives the activation energy, and the vibrational frequencies at the two points give the pre-exponential factor of the rate constant.

Two types of methods are used to find first-order saddle points.
When both the initial and the final state of the transition are known, the nudged elastic band (NEB) method can be used to find the MEP \cite{millsReversibleWorkTransition1995,jonssonNudgedElasticBand1998}.
A set of replicas of the system, referred to as images, is placed between the two minima and the images are displaced iteratively until they lie on the MEP.
Each image is displaced according to the component of the atomic force perpendicular to the path, while spring forces acting along the path control the distribution of the images \cite{henkelmanImprovedTangentEstimate2000}.
In the climbing image NEB (CI-NEB), the highest energy image is in addition moved uphill in energy along the path so that it converges on the saddle point \cite{henkelmanClimbingImageNudged2000}.
The efficiency and reliability of such calculations have been improved in several ways, for example by generating the initial path from an interpolation of the interatomic distances rather than the Cartesian coordinates \cite{smidstrupImprovedInitialPath2014,schmerwitzSequentialIdpp2024,zhuGeodesicInterpolationReaction2019}, by adjusting the number of images and the effort spent on each of them during the calculation \cite{kolsbjergAutomatedNeb2016,kolsbjergAutomatedNebErratum2018,lindgrenScaledDynamicNeb2019}, by making the springs stiffer in the higher energy region of the path \cite{asgeirssonNudgedElastic2021}, and by switching the climbing image over to a search along the lowest eigenmode of the Hessian in the final stage of the calculation \cite{asgeirssonNudgedElastic2021,goswamiEnhancedClimbingImage2026}.
Recommended settings for such calculations and tools for visualizing the paths have recently been presented \cite{goswamiReproducibleNeb2026,goswamiTwodimensionalRMSDProjections2026}, and variants and assessments of the method continue to appear \cite{liangNormalModeNeb2026,manojNudgedElasticMembrane2026,wooAutonomousOptimization2026,parkHighThroughput2025,maitiBenchmarkingHierarchical2026,khandakerNebwalk2026}.

When only the initial state is known, the minimum mode following (MMF) method can be used.
The system is then displaced uphill in energy along the eigenvector corresponding to the lowest eigenvalue of the Hessian, the minimum mode, while the energy is minimized with respect to all other degrees of freedom \cite{henkelmanDimerMethodFinding1999,olsenComparisonMethodsFinding2004}.
In the dimer method, the minimum mode is estimated from the atomic forces at two closely spaced replicas of the system, so the Hessian does not need to be evaluated \cite{henkelmanDimerMethodFinding1999,heydenEfficientMethodsFinding2005,kastnerSuperlinearlyConvergingDimer2008,zengUnificationMinimumMode2014}.
This is the basis of the adaptive kinetic Monte Carlo method for simulations of long time scale dynamics, as implemented in the eOn software \cite{henkelmanLongTimeScale2001,chillEONSoftwareLong2014,goswamiEOnSoftware2026,goswamiEfficientExplorationThesis2025}, and of schemes for mapping out the transitions that lead away from a given state \cite{xuLocalTransition2026}.
An MMF search makes no use of the final state, and a search started from an arbitrary configuration can converge on a saddle point for a transition other than the one of interest.
The two types of calculations complement each other.
An NEB calculation identifies the path between the two given states, and an MMF search started from a configuration close to the maximum along that path converges efficiently on the saddle point.

Both types of calculations typically require several hundred evaluations of the energy and atomic forces.
When these are obtained from electronic structure calculations, the evaluations dominate the computational effort.
Saddle points can also be found with growing and freezing string methods \cite{petersGrowingString2004,zimmermanReliableTransition2013,zimmermanSingleEnded2015,behnFreezingString2011,marksFreezingStringInternal2025,steinmetzerPysisyphusExploringPotential2021}, starting from an interpolation based on the Hessians at the two minima \cite{tesselliHessianInterpolation2026}, and by optimization in internal coordinates as implemented, for example, in the Sella software \cite{hermesSellaOpenSource2022,wangGeometryOptimization2016}, but the number of energy and force evaluations remains the primary issue.

One way to reduce the computational effort is to replace the electronic structure calculations with a machine-learned potential energy function that has been trained in advance on a large data set \cite{woodUMAFamilyUniversal2025}.
Pretrained or system-specific potential functions have recently been used in NEB calculations \cite{wanderCatTSunami2025,jungNebscape2026,zhangSsneb2026,goncharovaCrystalNeb2025,goncharovaVibrationalCrystal2026,steffenReactionFinetune2026,ramakrishnanImplicitNeural2025} and in single-ended saddle point searches \cite{yuanAnalyticalHessian2024,cheulaFinetuneSella2026,liuNeuralSaddle2026,tangReactiveOrganometallic2025,renReactiveMachine2026,wuMachineLearnedLeftmost2026,liTransitionStateSearching2025,karasawaColabReaction2025}.
Machine-learned models have also been used to generate an estimate of the saddle point configuration that is subsequently refined \cite{marksReliableAutomated2026,haitGeodesicMachineLearned2025,darouichAdaptiveTransition2025,darouichRobustGenerative2026}, and learned policies and automated agents have been used to steer the searches \cite{kikkawaTransitionState2025,kikkawaTransferableSearch2026,meissnerEvolvingTransition2026,madhavanTsagent2026}.
Configurations in the vicinity of saddle points are, however, sparsely represented in data sets generated by sampling near the minima, so the accuracy of such potential functions there is not guaranteed and the results need to be verified or refined with electronic structure calculations.

Another approach is to construct an approximate, surrogate energy surface from the electronic structure calculations as they are carried out for the system at hand, and to use the surrogate to select the configuration for which the next calculation is carried out \cite{petersonAccelerationSaddlePoint2016,koistinenNudgedElasticBand2017}.
No data set needs to be prepared beforehand, and the surrogate is discarded once the saddle point has been found.
Gaussian process regression (GPR) is well suited for this purpose since a useful approximation of the energy surface can be obtained from a small number of data points, in particular when both the energy and the atomic forces are included.
Koistinen and co-workers showed that the number of energy and force evaluations in NEB calculations can be reduced significantly in this way \cite{koistinenNudgedElasticBand2017}, and that a covariance function based on inverse interatomic distances improves the performance for molecular systems, both in NEB calculations \cite{koistinenNudgedElasticBandInverse2019} and in MMF searches \cite{koistinenMinimumModeSaddle2020}.
Related surrogate methods have been developed for path optimization \cite{garridoTorresLowScalingAlgorithm2019,meyerNebSurrogate2019,denzelGaussianProcessRegressionMinimum2019,tengPhysicalPrior2024,zhangInstantonLineIntegral2025,fangTunnelingGaussian2024,onyangoGPRCalculator2025,fattahpourDiffusionGaussian2024,kakhandikiNnbaxNeb2025,singhTrustLimitedNeb2026} and for saddle point searches \cite{denzelGaussianProcessRegressionTransition2018,denzelHessianMatrixUpdate2020,fdezGalvanRestrictedVarianceConstrained2021,guActiveLearningSaddle2022,yuUncertaintyNebDimer2026}.

The reduction in the number of electronic structure calculations leads to a reduction in the overall computational effort only if the construction and use of the surrogate model does not take too long.
An efficient C++ implementation of the GPR accelerated dimer method was recently presented and applied to a set of 500 molecular reactions.
The median number of electronic structure calculations was reduced from 308 to 31, and the wall time was reduced in three out of four cases even though the calculations were carried out at the Hartree--Fock level \cite{goswamiEfficientImplementationGaussian2025,goswamiBayesianHierarchical2025}.
Adaptive pruning of the data set was subsequently found to increase the robustness of the calculations and to limit the overhead as the number of data points increases \cite{goswamiAdaptivePruningIncreased2026}, and the methodology has been reviewed within the framework of Bayesian optimization \cite{goswamiBayesianOptimizationGaussian2026}.
In the implementations presented so far, the calculation of the MEP and the MMF search have, however, been treated separately, with a separate surrogate model for each, and the data obtained in one type of calculation have not been used in the other.

Here, we present GPRidge, a software library for GPR accelerated CI-NEB calculations of MEPs and MMF searches for first-order saddle points in which the two types of calculations share one surrogate model.
All evaluations of the energy and atomic forces are collected in one data set, so an MMF search started from the highest energy image of a band continues with the surrogate model constructed during the NEB calculation and no additional evaluation is needed for the handoff.
Convergence is determined from the atomic forces obtained from the underlying energy surface and not from the surrogate model.
As an option, the Hessian at the final configuration is evaluated by finite differences to check that it has one and only one negative eigenvalue, and when a second negative eigenvalue is found the search is continued after a displacement along the corresponding eigenvector.
The library is written in C++ with interfaces to C, Fortran and Python, including the Atomic Simulation Environment \cite{hjorthLarsenAtomicSimulationEnvironment2017}, and it can be used with a wide range of software for evaluating the energy and atomic forces \cite{bigiMetatensorMetatomic2026}.

The performance of GPRidge is assessed in calculations of molecular reactions in the benchmark sets of Baker and Chan \cite{bakerChanLocationTransition1996} and of Birkholz and Schlegel \cite{birkholzUsingBondingGuide2015}, where the UMA-S-1.1 machine-learned potential function \cite{woodUMAFamilyUniversal2025} is used to describe the atomic interactions, and in calculations of rearrangements of Pt islands on the Pt(111) surface described with a Morse potential \cite{chillBenchmarksCharacterizationMinima2014,goswamiOptBenchReplay2026}.
The double-ended calculations are compared with CI-NEB, with the off-path climbing image NEB (OCI-NEB) \cite{goswamiEnhancedClimbingImage2026}, with the machine learning NEB of the CatLearn software \cite{garridoTorresLowScalingAlgorithm2019} and with the freezing string method \cite{marksFreezingStringInternal2025}.
The single-ended calculations are compared with the dimer method and with Sella, each started from the same configuration.
The article is organized as follows.
The methods and the software are described in Section \ref{sec:architecture}, the results of the calculations are presented in Section \ref{sec:population} and discussed in Section \ref{sec:discussion}, and conclusions are given in Section \ref{sec:conclusions}.
\section{Methods}
\label{sec:architecture}
The nudged elastic band, the dimer and the Gaussian process are set out here as they are implemented in eOn and in the earlier GPR-dimer \cite{chillEONSoftwareLong2014,goswamiEOnSoftware2026,goswamiEfficientImplementationGaussian2025}.
The library keeps one training set for the band and the dimer and exposes the search through Python, C and Fortran calls.
The molecular calculations require measured-force convergence and an independent index-one Hessian; the periodic paths use the force criterion of their benchmark.
\subsection{Climbing-image band and dimer}
\label{sec:review}
The nudged elastic band approximates the MEP by a chain of images between a known reactant and a known product.
Each image feels the component of the true force perpendicular to the path tangent and a spring force parallel to that tangent, so the chain can settle onto the path while the springs keep the images spaced \cite{henkelmanClimbingImageNudged2000,henkelmanImprovedTangentEstimate2000}.
At the highest image the parallel force is reflected, and that climbing image walks up the path to the saddle point between the chosen endpoints \cite{asgeirssonNudgedElastic2021}.
The quality of the initial path matters, because a linear interpolation in Cartesian coordinates often drives atoms through one another.
The preparation used here places each new image next to a relaxed neighbor, against an inverse-distance target: the pair potential defines the target \cite{smidstrupImprovedInitialPath2014}, sequential traversal adds one image at a time \cite{schmerwitzSequentialIdpp2024}, and inverse-distance interpolation follows the geodesic-path motivation \cite{zhuGeodesicInterpolationReaction2019}.

The dimer is a single-ended MMF search from a single geometry \cite{henkelmanDimerMethodFinding1999}.
Two replicas sit a small fixed distance apart, and a rotation that minimizes their summed energy aligns the dimer axis with the lowest curvature mode, after which a translation inverts the force along that axis and relaxes every perpendicular direction \cite{zengUnificationMinimumMode2014,heydenEfficientMethodsFinding2005}.
The rotation has been the expensive part of a dimer on an electronic-structure surface, because each trial orientation is another evaluation, and a local surrogate is useful exactly there \cite{goswamiEfficientImplementationGaussian2025,goswamiEfficientExplorationThesis2025}.
OCI-NEB uses the same reflection for the climbing image and for the dimer, and it can hand control from one to the other and back as the alignment of the mode is monitored \cite{goswamiEnhancedClimbingImage2026}.
\subsection{Gaussian process on inverse distances}
\label{sec:gpr-review}
A Gaussian process is a distribution over functions, fixed by a covariance, and here it maps Cartesian coordinates to an energy and to atomic forces \cite{koistinenMinimumModeSaddle2020,goswamiBayesianOptimizationGaussian2026}.
The covariance is a squared exponential of inverse interatomic distances, with one length scale for each pair of elements, because the inverse distance rises sharply as two atoms approach and the surrogate then resists placing them on top of one another.
Differentiating that scalar covariance supplies the energy--gradient and gradient--gradient blocks, so one evaluation, which already returns the forces, trains both the value and the slope.
The hyperparameters are fitted by maximising the marginal likelihood, and a Cholesky factor of the noisy covariance serves both the fit and the prediction \cite{goswamiAdaptivePruningIncreased2026}.
The loop of fitting the GP, relaxing on its mean, selecting a geometry and evaluating its energy and forces covers a minimization, a dimer and a band, which differ in the inner optimizer and the selection rule \cite{goswamiBayesianOptimizationGaussian2026}.
\subsection{Using GPRidge}
\label{sec:using}
A calculation needs a geometry and any calculator that returns an energy and Cartesian forces, and the Gaussian process stores each evaluation as it arrives.
The band or the dimer then moves on the posterior mean, and the calculator is called again for the geometries the acquisition selects.
A search stops when the atomic forces obtained from the calculator have dropped below the tolerance, as in the NEB and dimer methods without a surrogate.
The check of the Hessian described in Section \ref{sec:dimer} is an option of the drivers, \texttt{-{}-force-stop-certificate}, which \texttt{-{}-oracle-stencil} also requests while writing the measured Hessian to a file, and it is applied in all the molecular calculations reported here; the periodic paths of Section \ref{sec:periodic} are accepted on the force criterion of their benchmark alone.
The library separates the calculator, the Gaussian process, and the choice of a band or a dimer.
Its C++ core exposes C and Fortran interfaces, and Python calls it directly or through the Atomic Simulation Environment \cite{hjorthLarsenAtomicSimulationEnvironment2017}.
The same call reads a reactant and a product, and a stopped search resumes from the energies and forces already computed.
The compiled drivers select the potential by name when that backend is built in, among them xTB, NWChem, LAMMPS, metatomic, MOPAC and CPMD \cite{bigiMetatensorMetatomic2026}.

\begin{figure}[H]
\centering
\input{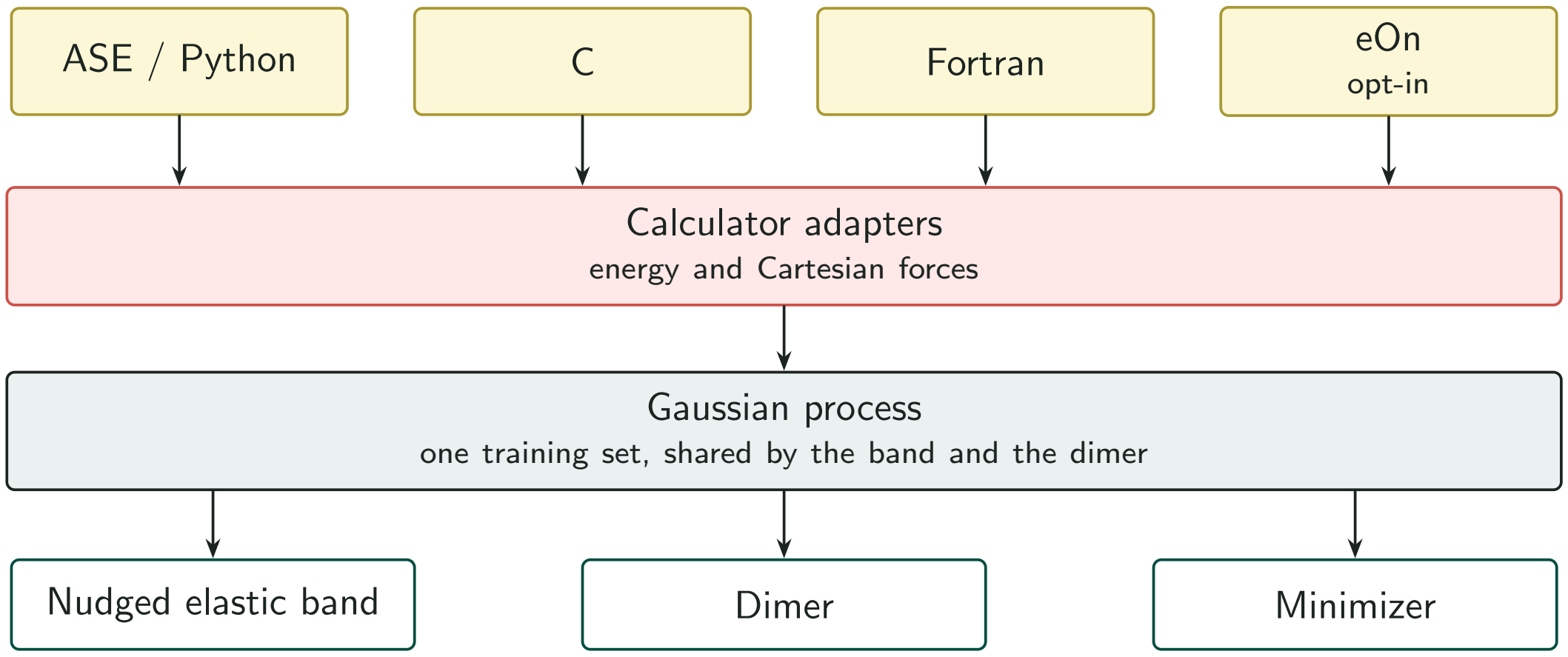}
\begin{tikzpicture}[gpr figure]
  \tikzset{entry/.style={gpr entry, minimum width=3.0cm, minimum height=0.95cm},
           wide/.style={gpr box, minimum height=1.05cm, text width=13.6cm},
           role/.style={gpr box, fill=white, draw=RuhiTeal, minimum width=3.6cm,
                        minimum height=0.8cm}}
  \node[entry] (ase) at (0.0, 0) {ASE / Python};
  \node[entry] (clang) at (3.6, 0) {C};
  \node[entry] (fort) at (7.2, 0) {Fortran};
  \node[entry] (eon) at (10.8, 0) {eOn\\[-0.1em]{\scriptsize opt-in}};
  \node[wide, gpr hot] (calc) at (5.4, -1.65)
    {Calculator adapters\\[-0.1em]{\scriptsize energy and Cartesian forces}};
  \node[wide] (gp) at (5.4, -3.3)
    {Gaussian process\\[-0.1em]{\scriptsize one training set, shared by the band and the dimer}};
  \node[role] (band) at (0.3, -4.85) {Nudged elastic band};
  \node[role] (dimer) at (5.4, -4.85) {Dimer};
  \node[role] (minim) at (10.5, -4.85) {Minimizer};
  \foreach \n in {ase, clang, fort, eon}
    \draw[gpr arrow] (\n.south) -- (calc.north -| \n.south);
  \draw[gpr arrow] (calc) -- (gp);
  \foreach \n in {band, dimer, minim}
    \draw[gpr arrow] (gp.south -| \n.north) -- (\n.north);
\end{tikzpicture}
\caption{How a calculation enters GPRidge.
A Python call supplies a geometry and takes an ASE calculator, and C and Fortran supply the geometry and the force routine, with the eOn engine available on request.
A calculator adapter returns the energy and the Cartesian forces.
One Gaussian process stores those evaluations and is shared by the nudged elastic band, the dimer and a minimization.}
\label{fig:interfaces}
\end{figure}

The following call is the Python interface, and \texttt{EMT()} stands in for any ASE calculator.
The acceptance force is \(0.001\) Ha/bohr (\(0.0514\) eV/\AA{}).
A gold atom sits in a hollow of a frozen Al(100) slab, part way from that hollow toward the bridge, which is the saddle of the hop, and the search moves the adatom.
\begin{lstlisting}[language=Python]
from ase.build import add_adsorbate, fcc100
from ase.calculators.emt import EMT
from ase.constraints import FixAtoms
from gpr_optim import GaussianProcess
from gpr_optim.dimer import GPDimer
atoms = fcc100("Al", size=(2, 2, 2), vacuum=6.0); add_adsorbate(atoms, "Au", 1.9, "hollow")
atoms.set_constraint(FixAtoms(indices=range(len(atoms) - 1))); atoms.positions[-1, 0] += 0.6 * atoms.cell[0, 0] / 2
GPDimer(atoms, oracle=EMT(), gp=GaussianProcess(uncertainty="map"), verbose=False).run(fmax=0.0514, steps=60)
\end{lstlisting}
A C call passes the calculator, the reactant, the product and the number of images, and a Fortran call passes the force routine, the atoms and the initial dimer orientation.
For a path between two minima the same \texttt{GaussianProcess} is handed to the band and then to \texttt{GPDimer} at the band maximum, so the dimer trains on energies and forces already computed.
Figure \ref{fig:interfaces} is the route those calls take, with the eOn engine available when it is requested.
The calculator returns the energy and the Cartesian forces, and one Gaussian process drives the band, the dimer and a minimization.

\subsection{What the two searches share}
\label{sec:algorithm}
The novelty is that the double-ended band and the single-ended dimer condition one posterior on one training set.
Both relax on that posterior mean between calculator calls, and every measured energy and gradient is appended to the same set, so the dimer inherits the evidence the band has gathered.
A handoff from the band to the dimer copies the set and the fitted factor, and the handoff itself does not call the potential.
In the molecular calculations, a geometry is accepted only when the calculator force and an independent Hessian meet the physical test of Section \ref{sec:dimer}.
The surrogate force is the surface on which the images move, and acceptance never reads that surrogate force.

\begin{figure}[H]
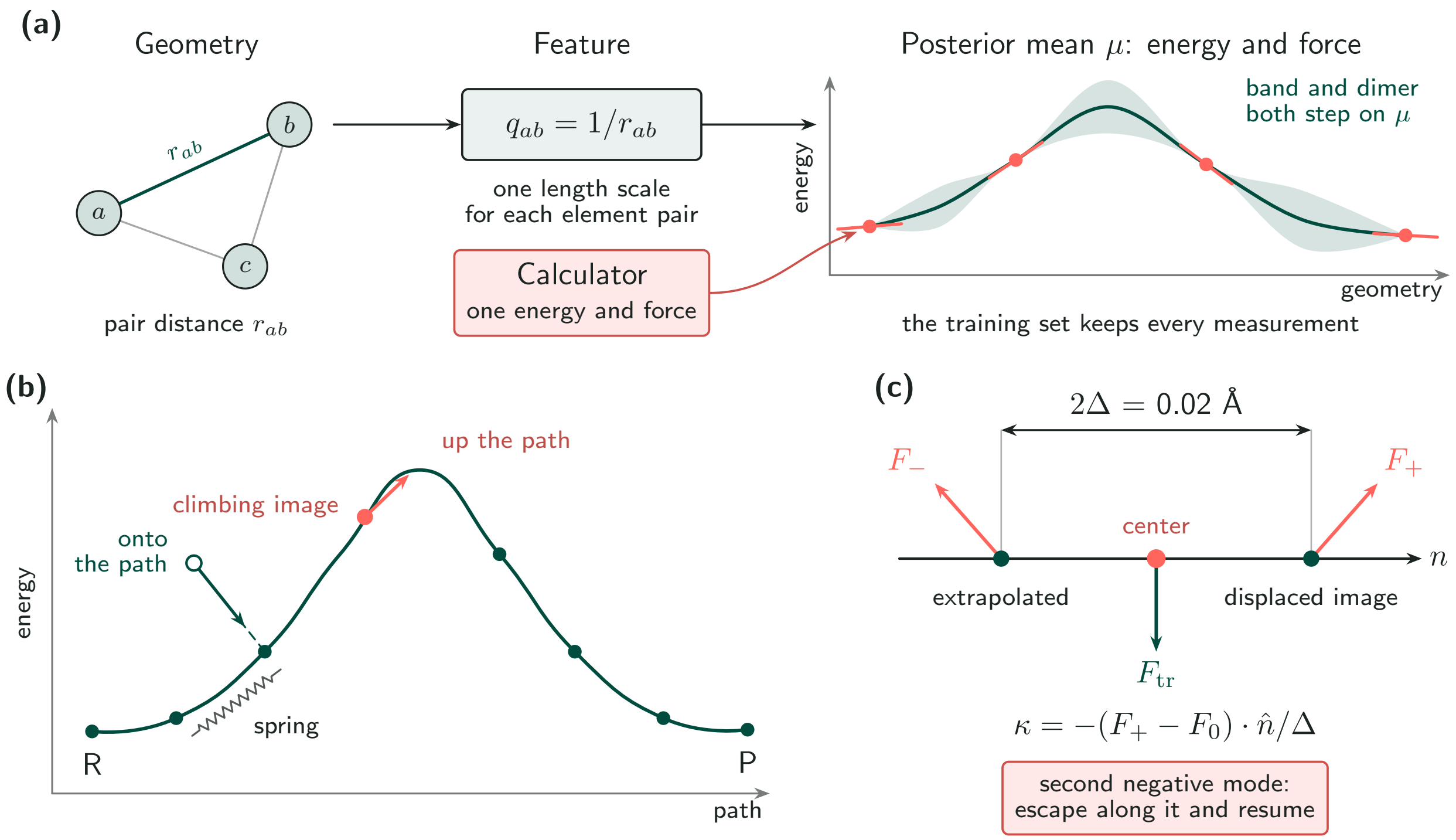

\centering
\input{figures/house_tikz}
\begin{tikzpicture}[gpr figure, baseline=(current bounding box.north)]
  \node[gpr panel] at (0,3.75) {(a)};
    \node[anchor=base] at (2.0,3.25) {Geometry};
  \coordinate (A) at (0.9,1.45);
  \coordinate (B) at (3.05,2.45);
  \coordinate (C) at (2.55,0.85);
  \draw[RuhiInk!40, line width=0.6pt] (A) -- (C) -- (B);
  \draw[RuhiTeal, line width=1.0pt] (A) -- (B);
  \foreach \p/\l in {A/a, B/b, C/c}
    \node[circle, draw=RuhiInk, line width=0.6pt, fill=RuhiTeal!18, inner sep=0pt,
          minimum size=0.5cm, font=\scriptsize\sffamily] at (\p) {$\l$};
  \node[gpr note, text=RuhiTeal, rotate=24.9, anchor=south] at ($(A)!0.5!(B)$) {$r_{ab}$};
  \node[gpr note, anchor=base] at (2.0,0.12) {pair distance $r_{ab}$};
    \node[anchor=base] at (6.35,3.25) {Feature};
  \node[gpr box, minimum width=2.7cm, minimum height=0.8cm] (feat) at (6.35,2.45) {$q_{ab} = 1/r_{ab}$};
  \node[gpr note, anchor=north] at (6.35,1.98) {one length scale\\[-0.15em]for each element pair};
  \node[gpr hot, minimum width=2.7cm, minimum height=0.8cm] (calc) at (6.35,0.55)
    {Calculator\\[-0.1em]{\scriptsize one energy and force}};
  \draw[gpr arrow] (3.55,2.45) -- (feat.west);
    \node[anchor=base] at (12.55,3.25) {Posterior mean $\mu$: energy and force};
  \begin{scope}[shift={(9.15,0.75)}]
    \draw[gpr axis] (0,0) -- (7.0,0) node[gpr note, anchor=north east, inner sep=2pt] {geometry};
    \draw[gpr axis] (0,0) -- (0,2.25);
    \node[gpr note, rotate=90, anchor=south] at (-0.05,1.1) {energy};
    \fill[RuhiTeal!16]
      plot[smooth, tension=0.6] coordinates {(0.45,0.55) (1.25,1.0) (2.1,1.3) (3.15,2.2) (4.25,1.25) (5.35,0.95) (6.5,0.45)}
      -- plot[smooth, tension=0.6] coordinates {(6.5,0.45) (5.35,0.3) (4.25,1.25) (3.15,1.6) (2.1,1.3) (1.25,0.55) (0.45,0.55)}
      -- cycle;
    \draw[RuhiTeal, line width=1.1pt]
      plot[smooth, tension=0.6] coordinates {(0.45,0.55) (1.25,0.78) (2.1,1.3) (3.15,1.9) (4.25,1.25) (5.35,0.62) (6.5,0.45)};
    \foreach \x/\y/\dx/\dy in {0.45/0.55/0.36/0.03, 2.1/1.3/0.3/0.2, 4.25/1.25/0.29/-0.22, 6.5/0.45/0.36/-0.02} {
      \draw[RuhiCoral, line width=1.0pt] (\x-\dx,\y-\dy) -- (\x+\dx,\y+\dy);
      \fill[RuhiCoral] (\x,\y) circle[radius=2.2pt];
    }
    \node[gpr note, text=RuhiTeal, anchor=west, align=left] at (4.55,1.95)
      {band and dimer\\[-0.15em]both step on $\mu$};
    \coordinate (first) at (0.45,0.55);
  \end{scope}
  \draw[gpr arrow] (feat.east) -- (9.0,2.45);
  \draw[gpr arrow, draw=RuhiCoral!80!black] (calc.east) to[out=0, in=200] ($(first)+(-0.14,-0.06)$);
  \node[gpr note, anchor=base] at (12.55,0.12)
    {the training set keeps every measurement};
\end{tikzpicture}\par\vspace{0.6em}
\input{figures/house_tikz}
\usetikzlibrary{decorations.pathmorphing}
\begin{tikzpicture}[gpr figure, baseline=(current bounding box.north)]
  \node[gpr panel] at (0,5.1) {(b)};
  \draw[gpr axis] (0.55,0.35) -- (8.65,0.35) node[gpr note, anchor=north east, inner sep=2pt] {path};
  \draw[gpr axis] (0.55,0.35) -- (0.55,4.7);
  \node[gpr note, rotate=90, anchor=south] at (0.5,2.5) {energy};
  \draw[RuhiTeal, line width=1.1pt]
    plot[smooth, tension=0.85] coordinates {
      (1.0,1.05) (1.95,1.2) (2.95,1.95) (3.8,3.05) (4.7,4.0)
      (5.6,3.05) (6.45,1.95) (7.45,1.2) (8.4,1.07)
    };
  \foreach \x/\y in {1.0/1.05, 1.95/1.2, 2.95/1.95, 5.6/3.05, 6.45/1.95, 7.45/1.2, 8.4/1.07}
    \fill[RuhiTeal] (\x,\y) circle[radius=2.2pt];
  \node[anchor=north] at (1.0,0.95) {R};
  \node[anchor=north] at (8.4,0.97) {P};
    \draw[RuhiInk!75, line width=0.6pt, decorate,
        decoration={zigzag, segment length=3.2pt, amplitude=1.6pt, pre length=3pt, post length=3pt}]
    (2.13,1.0) -- (3.13,1.75);
  \node[gpr note, anchor=north west, inner sep=2pt] at (2.75,1.3) {spring};
    \coordinate (on) at (2.95,1.95);
  \coordinate (off) at (2.15,2.95);
  \draw[RuhiTeal, densely dashed, line width=0.6pt] (on) -- (off);
  \draw[->, RuhiTeal, line width=1.0pt] (off) -- ($(off)!0.72!(on)$);
  \draw[RuhiTeal, line width=0.8pt, fill=white] (off) circle[radius=2.4pt];
  \node[gpr note, text=RuhiTeal, anchor=east, align=right] at ($(off)+(-0.15,0.1)$)
    {onto\\[-0.15em]the path};
    \coordinate (ci) at (4.08,3.47);
  \draw[->, RuhiCoral, line width=1.1pt] (ci) -- (4.58,3.95);
  \fill[RuhiCoral] (ci) circle[radius=2.6pt];
  \node[gpr note, text=RuhiCoral!75!black, anchor=east] at ($(ci)+(-0.15,0.12)$) {climbing image};
  \node[gpr note, text=RuhiCoral!75!black, anchor=south west] at (4.8,4.05) {up the path};
\end{tikzpicture}\hfill
\input{figures/house_tikz}
\begin{tikzpicture}[gpr figure, baseline=(current bounding box.north)]
  \node[gpr panel] at (0,5.1) {(c)};
  \draw[RuhiInk, ->, line width=0.8pt] (0.3,3.0) -- (6.2,3.0) node[right, inner sep=2pt] {$n$};
  \draw[RuhiInk!35, line width=0.5pt] (1.45,3.0) -- (1.45,4.45) (4.95,3.0) -- (4.95,4.45);
  \draw[<->, RuhiInk, line width=0.6pt] (1.45,4.45) -- (4.95,4.45)
    node[midway, above=1pt] {$2\Delta$ = 0.02~\AA};
  \draw[->, RuhiCoral, line width=1.1pt] (1.45,3.0) -- (0.7,3.85) node[above left, inner sep=1.5pt] {$F_{-}$};
  \draw[->, RuhiCoral, line width=1.1pt] (4.95,3.0) -- (5.7,3.85) node[above right, inner sep=1.5pt] {$F_{+}$};
  \draw[->, RuhiTeal, line width=1.1pt] (3.2,3.0) -- (3.2,1.95) node[below, inner sep=2.5pt] {$F_{\mathrm{tr}}$};
  \fill[RuhiTeal] (1.45,3.0) circle[radius=2.6pt];
  \fill[RuhiTeal] (4.95,3.0) circle[radius=2.6pt];
  \fill[RuhiCoral] (3.2,3.0) circle[radius=3.0pt];
  \node[gpr note, anchor=north] at (1.45,2.82) {extrapolated};
  \node[gpr note, anchor=north] at (4.95,2.82) {displaced image};
  \node[gpr note, text=RuhiCoral!75!black, anchor=south] at (3.2,3.14) {center};
  \node[anchor=base] at (3.3,1.02) {$\kappa = -(F_{+} - F_{0})\cdot \hat n /\Delta$};
  \node[gpr hot, anchor=north, inner sep=4pt, font=\scriptsize\sffamily]
    at (3.3,0.72) {second negative mode:\\[-0.15em]escape along it and resume};
\end{tikzpicture}
\caption{The two searches, drawn separately.
(a, b) The double-ended band.
Pair distances enter as \(q=1/r\), pairs of one element type share a length scale, and the band moves on the posterior mean of Equation \eqref{eq:posterior-mean}.
The calculator is called for the points the acquisition selects and for the tests that decide acceptance.
An image off the path is driven back onto it, and at the highest image the parallel force is reflected so that image walks up the path.
(c) The single-ended dimer.
Between calculator calls the center and one displaced image are \(0.01~\text{\AA}\) apart along the mode.
Their forces give the curvature and the translation step; the opposite force is extrapolated from them.
A second negative mode is an escape along that mode, after which refinement resumes.}
\label{fig:method-panels}
\end{figure}

\begin{figure}[H]
\centering
\input{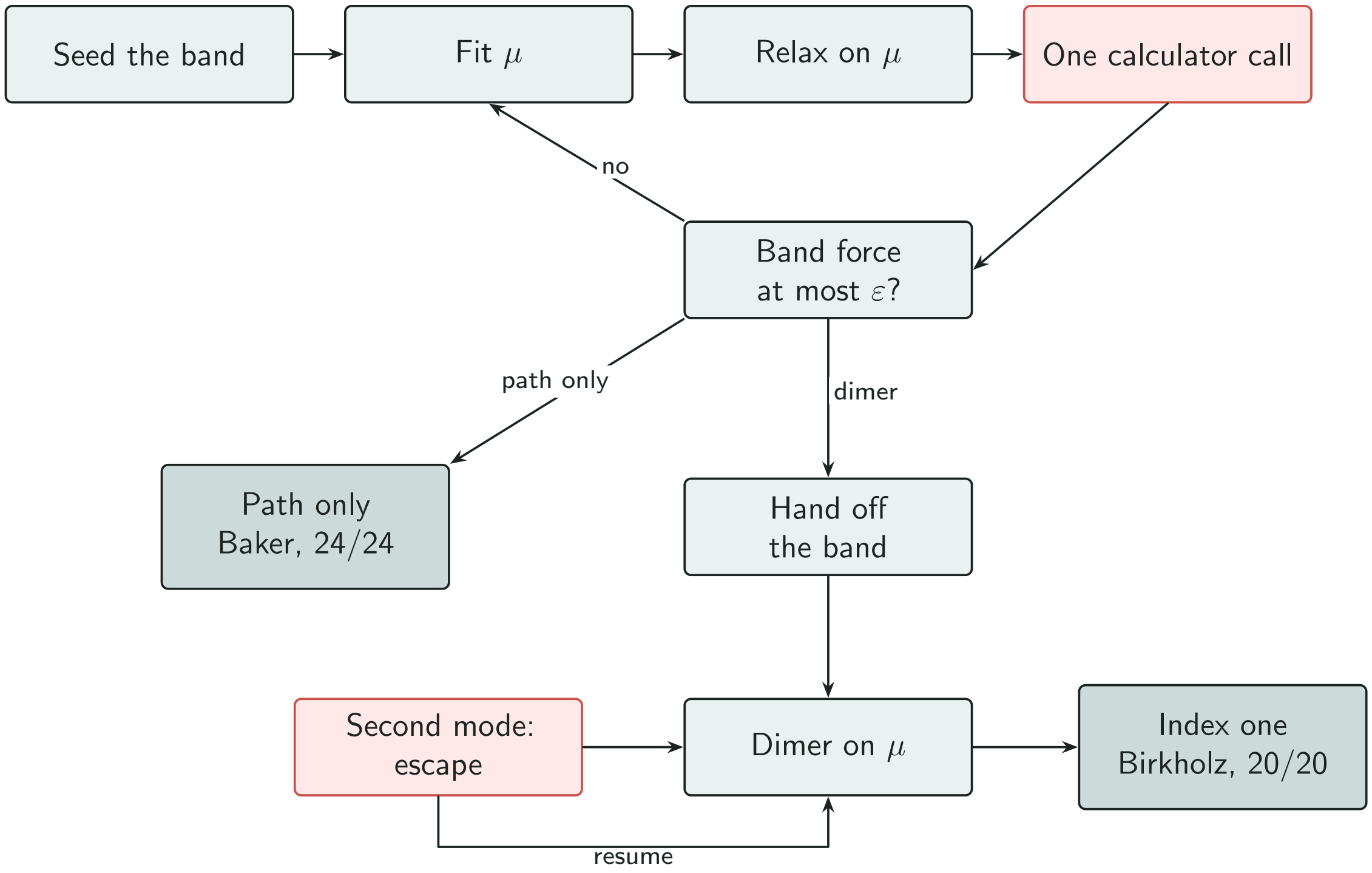}
\begin{tikzpicture}[gpr figure]
  \tikzset{box/.style={gpr box, minimum height=1.05cm, text width=2.85cm},
           hot/.style={box, gpr hot},
           exit/.style={box, gpr exit, minimum height=1.35cm},
           lab/.style={gpr note, fill=white, inner sep=1.4pt}}
  \node[box] (seed) at (0,0) {Seed the band};
  \node[box] (fit) at (3.7,0) {Fit $\mu$};
  \node[box] (relax) at (7.4,0) {Relax on $\mu$};
  \node[hot] (call) at (11.1,0) {One calculator call};
  \node[box] (force) at (7.4,-2.35) {Band force at most $\varepsilon$?};
  \node[exit] (baker) at (1.7,-5.15) {Path only\\Baker, 24/24};
  \node[box] (hand) at (7.4,-5.15) {Hand off\\the band};
  \node[box] (dimer) at (7.4,-7.55) {Dimer on $\mu$};
  \node[hot] (escape) at (3.15,-7.55) {Second mode:\\escape};
  \node[exit] (birk) at (11.7,-7.55) {Index one\\Birkholz, 20/20};
  \draw[gpr arrow] (seed) -- (fit);
  \draw[gpr arrow] (fit) -- (relax);
  \draw[gpr arrow] (relax) -- (call);
  \draw[gpr arrow] (call.south) -- (force.east);
  \draw[gpr arrow] (force.north west) -- node[lab, above, pos=0.35] {no} (fit.south);
  \draw[gpr arrow] (force.south west) -- node[lab, above, pos=0.55] {path only} (baker.north east);
  \draw[gpr arrow] (force.south) -- node[lab, right, pos=0.45] {dimer} (hand.north);
  \draw[gpr arrow] (hand) -- (dimer);
  \draw[gpr arrow] (escape) -- (dimer);
  \draw[gpr arrow] (dimer) -- (birk);
  \draw[gpr arrow] (escape.south) -- ++(0,-0.55) -| node[lab, pos=0.25, below] {resume} (dimer.south);
\end{tikzpicture}
\caption{Where the two searches return.
The double-ended Baker search, Algorithm \ref{alg:double-ended}, uses 1215 evaluations and accepts 24 of 24 reactions.
The single-ended Birkholz search, Algorithm \ref{alg:single-ended}, is started at the band maximum, uses 1846 evaluations and accepts 20 of 20 reactions, and 779 of those evaluations belong to the seed band.
A second negative mode returns the dimer center to the rotation and translation loop.}
\label{fig:flow}
\end{figure}

Let \(\mathcal D_t=\{(x_j,E_j,g_j)\}_{j=1}^{m_t}\) be the training set, with \(g_j=\nabla E(x_j)\) and with the atomic force equal to \(-g_j\).
The covariance is the squared exponential of inverse interatomic distances, Equation \eqref{eq:sexpat}, and the training target stacks each energy with its Cartesian gradient.
The noisy covariance is factored once by Cholesky, Equation \eqref{eq:training-cholesky}, and the posterior mean of the energy at a new geometry \(x_\ast\) is
\begin{equation}
\mu(x_\ast)=m(x_\ast)+k(x_\ast,\mathcal D_t)^{\mathsf T}\alpha,
\label{eq:posterior-mean}
\end{equation}
where \(m\) is the screened nuclear-repulsion prior, \(k(x_\ast,\mathcal D_t)\) is the covariance of that energy with the stacked training targets, and \(\alpha\) is the weight vector from that factor.
Analytic derivatives of the same scalar covariance supply the Cartesian forces that both searches follow.
Figure \ref{fig:method-panels} draws the double-ended surrogate and the single-ended dimer, and Figure \ref{fig:flow} shows where each search returns.

The dimer filters midpoint proposals that are near an existing observation in the earth-mover distance.
The model-dependent numerical contribution to that distance threshold is
\begin{equation}
d_{\min}=\tfrac12\,\ell_{\min}\left(\sigma_n^{2}/\sigma_f^{2}\right)^{1/2}.
\label{eq:dmin}
\end{equation}
This is a numerical admission heuristic in the fixed coordinate-unit convention; its relation to the scalar-kernel resolution example is given in Supporting Information Section \ref{sec:si-prune}.
Within the near-duplicate threshold, a calculator call is retained if the predicted force norm is lower than the smallest measured force norm, or if the GP predicts force convergence.
Otherwise the midpoint uses the GP force without another calculator call.
Measurements required for force or curvature checks take priority over this filtering.
The next calculator call is a force or a curvature the acceptance test still needs (Equations \eqref{eq:numerical-index-one} and \eqref{eq:ci-stop}), then a rejected proposal, and then an image the estimator has yet to observe.
\subsection{Double-ended search}
\label{sec:double-ended}
The double-ended search builds a band from the reactant to the product.
The initial images follow eOn's sequential traversal of an inverse-distance target along the geodesic, the preparation used for the nudged elastic band \cite{smidstrupImprovedInitialPath2014,schmerwitzSequentialIdpp2024,zhuGeodesicInterpolationReaction2019,chillEONSoftwareLong2014,goswamiEOnSoftware2026,goswamiReproducibleNeb2026}.
Image-dependent pair-potential (IDPP) interpolation uses pair-distance targets between the endpoints; sequential IDPP inserts images beside neighbors already relaxed.
The geodesic variant uses an inverse-distance metric to prepare the initial path, and the band then relaxes on the posterior mean.
An image \(i\) with tangent \(\hat\tau_i\) feels the perpendicular force and a spring,
\begin{equation}
\mathbf F_i^{\mathrm{NEB}}=\mathbf F_i-(\mathbf F_i\cdot\hat\tau_i)\hat\tau_i
+k\big(|\mathbf x_{i+1}-\mathbf x_i|-|\mathbf x_i-\mathbf x_{i-1}|\big)\hat\tau_i.
\label{eq:neb-force}
\end{equation}
At the highest image the parallel component is reflected, and that climbing image walks up the path,
\begin{equation}
\mathbf F_{\mathrm{CI}}=\mathbf F-2(\mathbf F\cdot\hat\tau)\hat\tau.
\label{eq:ci-force}
\end{equation}
Figure \ref{fig:method-panels}(a,b) is this surrogate and this climbing image.
Between calculator calls the search asks for a rejected proposal, or for one interior image chosen by the tree-structured Parzen estimator \cite{bergstraAlgorithmsHyperparameter2011}.
When the measured force on the band reaches
\begin{equation}
\max_a\lVert F_a\rVert_2\le\varepsilon_{\mathrm{band}},
\label{eq:handoff}
\end{equation}
a path-only search returns the climbing image for the independent index test.
A path-to-saddle search hands the band maximum, its tangent, the fitted Gaussian process and the training set to the dimer of Section \ref{sec:single-ended}, and Algorithm \ref{alg:double-ended} is this loop.
The Baker--Chan calculations and the Pt(111) bands are double-ended: Baker returns the climbing image, and the Pt(111) bands return the path.

\begin{algorithm}[p]
\caption{Double-ended GP climbing-image band}
\label{alg:double-ended}
\footnotesize
\begin{algorithmic}[1]
\Require Calculator $\mathcal{C}(x)=(E(x),\nabla E(x))$; reactant and product; GP $\mathcal{M}$; band threshold $\varepsilon_\mathrm{band}$
\State Place the interior images by eOn's sequential IDPP and evaluate the endpoints and the images
\State Initialize the training set $\mathcal{D}$ with those evaluations
\While{the measured band force exceeds $\varepsilon_\mathrm{band}$}
  \State Fit $\mathcal{M}$ from $\mathcal{D}$ (Equations \eqref{eq:sexpat}, \eqref{eq:training-cholesky})
  \State Relax the band on the posterior mean under Equations \eqref{eq:neb-force} and \eqref{eq:ci-force}
  \State Choose the next image: a rejected proposal, or one image chosen by the tree-structured Parzen estimator \cite{bergstraAlgorithmsHyperparameter2011}
  \State Append $\mathcal{C}(q)=(E,g)$ to $\mathcal{D}$
\EndWhile
\If{the plan continues to a dimer}
  \State Hand the band maximum, tangent, $\mathcal{M}$, and $\mathcal{D}$ to Algorithm \ref{alg:single-ended}
\Else
  \State \Return the climbing image for the independent index test
\EndIf
\end{algorithmic}
\end{algorithm}
\subsection{Single-ended search}
\label{sec:single-ended}
The single-ended search starts from one geometry, a supplied guess or the band maximum handed over by Algorithm \ref{alg:double-ended}.
The initial mode estimate \(\hat n\) is an available lowest-curvature eigenvector, the band tangent when the start is a band maximum, or the dimer orientation carried from a previous rotation.
The dimer evaluates the force \(\mathbf F_0\) at the center \(\mathbf x\) and \(\mathbf F_+\) at one displaced image \(\mathbf x+\Delta\hat n\), with \(\Delta=0.01~\text{\AA}\).
With \(\mathbf F=-\nabla E\), the finite-difference curvature is
\begin{equation}
\kappa=-\frac{(\mathbf F_+-\mathbf F_0)\cdot\hat n}{\Delta}.
\label{eq:dimer-curvature}
\end{equation}
The opposite force is extrapolated as \(\mathbf F_-=2\mathbf F_0-\mathbf F_+\).
Rotation follows the torque of the two image forces until the axis lies along the softest mode, after which translation steps the center along that mode on the posterior mean, and Figure \ref{fig:method-panels}(c) is this dimer.
The search stops when the measured force at the center satisfies Equation \eqref{eq:ci-stop} and the curvature test of Equation \eqref{eq:numerical-index-one} resolves exactly one negative mode.
A weak second negative mode is refined by partitioned rational-function optimization on the measured Hessian.
For another resolved second negative mode, the search first attempts an order-one GP escape from the measured curvature; if that fails, it walks downhill along the measured mode.
Refinement then resumes, and the candidate returns for the independent Hessian, which is the loop in Algorithm \ref{alg:single-ended}.
The Birkholz--Schlegel calculations are single-ended dimer searches started at the band maximum.

\begin{algorithm}[p]
\caption{Single-ended GP dimer}
\label{alg:single-ended}
\footnotesize
\begin{algorithmic}[1]
\Require Calculator $\mathcal{C}$; one start geometry and a mode estimate $\hat n$; GP $\mathcal{M}$; threshold $\varepsilon_\mathrm{saddle}$
\State Initialize the training set $\mathcal{D}$ with the start, or receive $\mathcal{D}$ from Algorithm \ref{alg:double-ended}
\While{the center awaits acceptance as an index-one saddle}
  \State Fit or update $\mathcal{M}$ from $\mathcal{D}$ (Equations \eqref{eq:sexpat}, \eqref{eq:training-cholesky})
  \State Rotate $\hat n$ under the torque and translate the center; curvature from Equation \eqref{eq:dimer-curvature}
  \State Stop the inner relaxation at $\varepsilon_\mathrm{saddle}$
  \State Append the calculator evaluation of the center to $\mathcal{D}$
  \If{the measured force satisfies Equation \eqref{eq:ci-stop}}
    \State Resolve the curvature (Equations \eqref{eq:numerical-index-one}, \eqref{eq:ci-stop})
    \If{a second negative mode is weak}
      \State Refine it by partitioned rational-function optimization on the measured Hessian, and include the final geometry
      \State Resume the dimer from that geometry
    \ElsIf{a second negative mode is resolved}
      \State Run the order-one GP escape from the measured curvature
      \If{that escape fails} \State Descend along the measured mode \EndIf
      \State Resume the dimer from the escaped geometry
    \ElsIf{the index is unresolved}
      \State \Return an unresolved status
    \Else
      \State \Return the center for the independent Hessian
    \EndIf
  \EndIf
\EndWhile
\end{algorithmic}
\end{algorithm}
\subsection{Computational details}
\label{sec:levers}
The molecular calculations use the UMA-S-1.1 potential on the omol task and a joint inverse-distance Gaussian process \cite{woodUMAFamilyUniversal2025}.
Baker--Chan is the double-ended set: a climbing-image band between the two minima, compared with CI-NEB, OCI-NEB, CatLearn ML-NEB and the freezing string followed by an ASE dimer.
Birkholz--Schlegel is the single-ended set: a band is built from the published freezing-string geometries, and the Gaussian-process dimer, the eOn dimer and Sella all start from the highest image of that band.
Pt(111) is a second double-ended set, a 343-atom slab treated with a pair-additive spline and a local Gaussian-process correction, and compared with CI-NEB in the same program and with the ASE implementation.
All 72 paths on that slab are scored with the OptBench force criteria, \(0.01\) eV/\AA{} on the heptamer and \(0.001\) eV/\AA{} on the island.
The cost of the search is kept separate from the cost of the independent Hessian.

Search costs count potential evaluations during initialization, the band and the dimer, and the independent Hessian and the reference-geometry energies are reported on their own.
A potential evaluation returns the potential energy and the Cartesian force at one geometry.
Both endpoints and all 18 interior images of each initial molecular band are evaluated.
Across the 24 Baker reactions, the tree-structured Parzen estimator (TPE) selects 2 of the 970 subsequent acquisition decisions; the required-measurement rules select the rest.
It selects none of the 526 Birkholz band acquisitions.
A normal acquisition round adds one geometry, but Birkholz preparation can evaluate several images in one batch when their geometries lie outside the current kernel support.
The acceptance threshold is \(0.0514\) eV/\AA{}.
Baker--Chan uses this value for the band and the saddle, and Birkholz--Schlegel uses \(0.1\) eV/\AA{} for the seed band and \(0.0514\) eV/\AA{} for the dimer.
The remaining settings are listed in the Supporting Information.
Three complete passes of GPRidge return the same per-reaction count, and each comparator time is one run.

\begin{table}[H]
\caption{\label{tab:parameters}Settings of the reported searches.}
\centering
\small
\begin{tabular}{lll}
Symbol & Meaning & Value\\
\hline
\(\varepsilon_{\mathrm{band}}\) & Double-ended Baker return & \(0.0514\) eV/\AA{}\\
\(\varepsilon_{\mathrm{band}}\) & Seed band for the single-ended Birkholz dimer & \(0.1\) eV/\AA{}\\
\(\varepsilon_{\mathrm{saddle}}\) & Measured force for acceptance & \(0.0514\) eV/\AA{}\\
& Baker and Birkholz interior images & 18\\
& Heptamer interior images & 8\\
& Spring constant, Pt(111) & \(5\) eV/\AA{}$^{2}$\\
\(\Delta\) & Dimer image separation & \(0.01\) \AA{}\\
& Active-row ceiling & 128\\
\end{tabular}
\end{table}

\subsection{Physical stationarity and saddle index}
\label{sec:dimer}
A first-order saddle is stationary on the physical potential and has exactly one negative curvature in the subspace left after rigid translation and rotation are removed.
At each returned geometry the independent check forms a centered finite-difference Hessian, projects out those rigid modes, and resolves the physical eigenvalues.
In the molecular calculations reported here, acceptance requires the measured force threshold and a resolved index of one.
Modes inside \([-\tau,\tau]\) stay unresolved, where \(\tau\) measures the antisymmetric stencil error, and finite-difference truncation is assessed on its own.
The numerical index criterion is
\begin{equation}
\widehat{\lambda}_1< -\tau,\qquad
\widehat{\lambda}_j>\tau\quad(2\leq j\leq d).
\label{eq:numerical-index-one}
\end{equation}
The force criterion is
\begin{equation}
\max_a\lVert F_a(x)\rVert_2\le\varepsilon_\mathrm{saddle},\qquad
\varepsilon_\mathrm{saddle}=0.0514~\mathrm{eV}\,\text{\AA}^{-1}.
\label{eq:ci-stop}
\end{equation}
The Supporting Information section on physical curvature and transported directions gives the mass weighting, projection, and resolution procedure.
The measured downhill walk runs for sulfolene, and the measured-stencil branch for Grignard and oxirane.
\subsection{GP representations and physical refinement}
\label{sec:derivatives}
Each molecular potential evaluation supplies an energy \(V(\mathbf{x})\) and a force \(F=-\nabla V\), and the Gaussian process trains jointly on the energies and the Cartesian gradients.
Its kernel is the squared exponential on inverse-distance features \(q_p=1/r_p\) \cite{koistinenMinimumModeSaddle2020},
\begin{equation}
k(\theta)_{nm}=\sigma_f^{2}\exp\left(-\sum_p (\Delta q_p)^2/\ell_{t(p)}^2\right).
\label{eq:sexpat}
\end{equation}
Here \(t(p)\) is the type of pair \(p\), its element pair, or on Birkholz its pair of fixed environment labels, so all pairs of one type share a length scale.
Analytic derivatives of this scalar covariance form the energy--gradient and gradient--gradient blocks, and pair-local assembly uses sparse pair Jacobians.
The mean is a screened nuclear-repulsion term with a constant offset.
The noisy training covariance is factored once by Cholesky for fitting and prediction,
\begin{equation}
C=K+N+\delta I=LL^{\mathsf T},\qquad Lb=y-\mathsf{Obs}\,m,\qquad L^{\mathsf T}\alpha=b.
\label{eq:training-cholesky}
\end{equation}
Here \(\mathsf{Obs}\) stacks the value and Cartesian derivatives of the prior mean at the training geometries; \(\mathcal D_t\) denotes the training set.
L-BFGS fits the hyperparameters from analytic likelihood derivatives and the selected priors, and the reported searches use this Cholesky factor.
The Supporting Information derives the optional dual Hilbert--Schmidt factorization and gives the acquisition schedule, the covariance derivatives, the refit conditions and the mode-resolution procedure.
The two-dimensional RMSD projection used to draw a path is a visualization, and the search itself uses the full inverse-distance model \cite{goswamiTwodimensionalRMSDProjections2026}.

The periodic Pt(111) calculations use a pair-additive spline baseline and a local Gaussian-process residual, in place of the molecular joint inverse-distance process.
The baseline describes the full slab, and the residual follows the atoms that move along the path.
Minimum-image distances and a finite cutoff define the pairs.
With cubic B-splines \(B_k\) on knots spaced by \(h\), the baseline is
\begin{equation}
V_0(x)=w_0+\sum_{i<j}\varphi(r_{ij}),\qquad
\varphi(r)=\sum_k w_k B_k(r).
\label{eq:pair-baseline}
\end{equation}
Energy and force observations determine the spline weights, and the local Gaussian-process residual describes the displacement-dependent correction.
The Supporting Information reports the knot spacing, the active region, the fitting schedule, the trust region and the batching settings.
\section{Results}
\label{sec:population}
The Baker--Chan panel contains 24 double-ended reactions, and the Birkholz--Schlegel panel contains 20 single-ended searches seeded by a band.
Both use UMA-S-1.1 on the omol task and the same measured force threshold, and acceptance requires that force together with one resolved negative physical curvature and positive curvature on every remaining physical mode (Section \ref{sec:dimer}, Equation \eqref{eq:numerical-index-one}).
The Baker endpoints are the two minima reached on UMA-S-1.1 from each Baker reference saddle, by a step along the imaginary mode of a finite-difference Hessian, damped mass-weighted steepest descent, and L-BFGS to the force threshold, and the end closer to the original Baker reactant is the reactant.
The Birkholz starts are the published freezing-string geometries \cite{behnFreezingString2011,marksFreezingStringInternal2025}.
Reference energies and geometries describe the chemical channel.
\subsection{Measured force and saddle index}
\label{sec:measured-force}
All 24 Baker and all 20 Birkholz searches meet the measured force threshold and the independent index-one test in each of three repetitions (Table \ref{tab:native-validation}).
The per-reaction search counts are the same on those three passes, which is the rank-ordered reduction of Section \ref{sec:algorithm}, and the per-reaction times, with the median and the range over the three passes, are in the Supporting Information.
Baker uses the climbing image throughout and accepts the band on the force criterion, with the independent index test applied to the returned geometry.
Birkholz uses the curvature-recovery procedure in Section \ref{sec:dimer}.

\begin{table}[htbp]
\caption{\label{tab:native-validation}Search outcomes on UMA-S-1.1 omol, one row per workflow. Baker--Chan is the double-ended climbing-image band. Birkholz--Schlegel is the single-ended dimer, and its time allowance is shared across the seed band, the dimer, and validation. Acceptance requires both force convergence and a resolved index of one.}
\centering
\small
\begin{tabular}{p{0.36\linewidth}p{0.30\linewidth}rr}
Workflow & Search & \shortstack{Force\\converged} & \shortstack{Index-one\\check}\\
\hline
Double-ended, Baker--Chan & GP-accelerated CI-NEB & 24 / 24 & 24 / 24\\
Single-ended, Birkholz--Schlegel & GP dimer from the band maximum & 20 / 20 & 20 / 20\\
\end{tabular}
\end{table}

Table \ref{tab:release-totals} separates the search evaluations from the independent checks.
The median number of search evaluations is 37.5 on Baker, with range 22--196 and mean 50.62, and 46 on Birkholz, with range 26--585 and mean 92.3.
The Baker searches use 1215 energy and force evaluations in total, and the Birkholz searches use 1846, of which 779 are the seed bands and 1067 are the dimer.
The independent Hessian and the reference geometries add 1126 and 1686 evaluations, so the combined totals are 2341 and 3532.
On every Birkholz reaction the geometry that meets the force tolerance is one already evaluated during the search.

\begin{table}[H]
\caption{\label{tab:release-totals}Evaluation counts on UMA-S-1.1, with the double-ended and single-ended searches in separate columns. Counts are summed over all reactions unless marked per reaction. Independent checks comprise the finite-difference Hessian and reference-geometry energies. Elapsed times are in the Supporting Information.}
\centering
\small
\begin{tabular}{p{0.46\linewidth}cc}
Quantity & \shortstack{Double-ended\\Baker--Chan} & \shortstack{Single-ended\\Birkholz--Schlegel}\\
\hline
Cases & 24 & 20\\
Index-one check passed & 24 & 20\\
Band evaluations, sum & 1215 & 779\\
Dimer evaluations, sum & 0 & 1067\\
Combined search evaluations, sum & 1215 & 1846\\
Search evaluations per reaction, median (range) & 37.5 (22--196) & 46 (26--585)\\
Search evaluations per reaction, mean & 50.62 & 92.3\\
Hessian and reference evaluations, sum & 1126 & 1686\\
Search and validation evaluations, sum & 2341 & 3532\\
\end{tabular}
\end{table}

The Supporting Information gives the per-reaction evaluation breakdowns and the elapsed times over the three repetitions, and the tables here carry the panel totals and the ranges.
It also shows the HCN surrogate band profile and a projected energy landscape.
The landscape projects the measured observations and the final band for visualization using the coordinates of reference \cite{goswamiTwodimensionalRMSDProjections2026}, and the search uses the full inverse-distance model.
\subsection{Double-ended comparisons}
\label{sec:classical-comparison}
The double-ended comparisons on molecules are the Baker--Chan climbing-image bands, and the Pt(111) bands are reported in Section \ref{sec:periodic}.
Table \ref{tab:protocol} lists the band settings, and Table \ref{tab:head-to-head} the outcomes on UMA-S-1.1.
Every search starts from the Baker endpoints and is scored by the force and index test of Section \ref{sec:dimer}.
The comparators are eOn CI-NEB, OCI-NEB, CatLearn ML-NEB, and the freezing string followed by an ASE dimer.
eOn stops on its projected-force criterion, and the Cartesian force is scored on the returned geometry.
The comparison with CI-NEB is the 18 reactions on which both methods meet the force and index test, and on that matched set the sums are 5964 and 767 evaluations.
OCI-NEB meets the test on 21 of 24 reactions, with sums 6672 and 952, CatLearn on 18 of 24, with sums 860 and 799, and the freezing string on 19 of 24, with sums 5520 and 1029.
Six of the 24 CI-NEB searches miss the acceptance test.
HCCH, H\(_2\)CO, HF abstraction, acrolein rotation and HCNH\(_2\) finish above the Cartesian force threshold, and ethane H\(_2\) abstraction returns two negative modes.
The 18 of 24 score is this force and index test on a band of eight interior images, against 18 interior images for GPRidge, and the Supporting Information gives every reaction.

\begin{table}[htbp]
\caption{\label{tab:protocol}Run arguments for the molecular comparisons. The GPRidge spring, image count, force, and thread counts are the driver arguments. eOn uses its published band settings and is then scored on the Cartesian force.}
\centering
\footnotesize
\begin{tabular}{p{0.28\linewidth}p{0.32\linewidth}p{0.32\linewidth}}
Quantity & GPRidge & eOn CI-NEB\\
\hline
Interior images & 18 & 8\\
Spring & \(5\) eV/\AA{}$^{2}$ & 1--10 eV/\AA{}$^{2}$, energy-weighted\\
Convergence & Cartesian maximum atomic force, \(0.0514\) eV/\AA{} & projected force, then the Cartesian test\\
Wall limit & 14400 s per Baker case & iteration cap 1000\\
Ranks and threads & 4 MPI ranks, 2 BLAS threads, 1 kernel thread & 1 process, 8 threads\\
\end{tabular}
\end{table}

\begin{table}[htbp]
\caption{\label{tab:head-to-head}Double-ended searches on Baker--Chan, UMA-S-1.1. Counts are energy-and-force evaluations summed over the reactions both methods accept. GPRidge is 24/24.}
\centering
\footnotesize
\begin{tabular}{lrrr}
Method & Accepted & Calls & GPRidge\\
\hline
CI-NEB \cite{henkelmanClimbingImageNudged2000,chillEONSoftwareLong2014} & 18/24 & 5964 & 767\\
OCI-NEB \cite{goswamiEnhancedClimbingImage2026} & 21/24 & 6672 & 952\\
CatLearn ML-NEB \cite{garridoTorresLowScalingAlgorithm2019} & 18/24 & 860 & 799\\
Freezing string, then ASE dimer \cite{marksFreezingStringInternal2025} & 19/24 & 5520 & 1029\\
\end{tabular}
\end{table}

On the 18 reactions CatLearn also certifies, the sums are 860 and 799 evaluations, the means are 47.8 and 44.4, and GPRidge uses fewer evaluations on 13 of the 18, with mean search times of 347 s and 3.9 s.
Figure \ref{fig:method-savings}(a,b) plots the four double-ended comparisons.
The multiplier, comparator divided by GPRidge, is 7.78 and 7.01 for the evaluation means of CI-NEB and OCI-NEB, 1.08 for CatLearn, and 5.36 for the freezing string, and the search-time multipliers are 1.95, 1.82, 89.5 and 2.81.
Those time multipliers use search time, with the independent Hessian kept separate.
Including independent validation, the matched CI-NEB and GPRidge times sum to \(215.601\) and \(126.923\) s, respectively, a \(41\%\) reduction.
On the 18 Baker reactions both climbing-image NEB and GPRidge certify, the means are 331 evaluations and 8.5 s against 43 evaluations and 4.3 s.

At the same Cartesian force threshold, an ASE climbing-image band with 18 interior images and the improved tangent uses 24186 evaluations across the Baker set \cite{henkelmanImprovedTangentEstimate2000,hjorthLarsenAtomicSimulationEnvironment2017}.
The ASE band is relaxed with FIRE and stops at the first iteration where the largest atomic force on the climbing image is under the threshold, which all 24 bands reach.
These ASE endpoints are not independently index-checked, so this comparison reports force convergence and evaluation counts.
The initial images come from an image-dependent pair-potential interpolation, and GPRidge uses 1215 evaluations, fewer on every one of the 24 reactions \cite{smidstrupImprovedInitialPath2014}.
The ASE bands are the \texttt{ase\_cineb\_all} target of the reproduction workflow.
The GPRidge bands in this comparison ran on Intel Xeon Gold 6248R processors, with four MPI ranks, two BLAS threads and one kernel thread, and they are the following driver.
\begin{verbatim}
gprn --reactant R.con --product P.con --potential uma --model CASE.pt2 \
  --task-name omol --uma-charge 0 --uma-spin 1 --device cpu \
  --nebmode gp_oie --use-mmf false --nimages 18 \
  --converged-force 0.0514221 --max-iterations 1000 \
  --climbing-image true --climbing-image-converged-only true \
  --ci-index-certificate false --band-refinements 1 \
  --idpp true --sidpp true --sidpp-geodesic true --seed-full-chain true \
  --spring-constant 5.0 --geometric-spring false \
  --gp-model-profile gprd_zbl --gp-zbl-prior-mean true \
  --gp-nonperiodic true \
  --gp-active-posterior true --gp-environment-residuals false \
  --gp-prior-mean-statistic max --gp-learn-noise false --ls-prior true \
  --uncertainty-mode map --hyperparameter-mode rgmin:lbfgs \
  --neb-band-engine rg-lbfgs --gp-adaptive-prune true \
  --gp-adaptive-prune-max 128 --gp-hssvd false --gp-drift-budget true \
  --kernel-support-freezes-band false --dimer-kappa-beta 5.0 \
  --linalg-mode library --linalg-threads 2 --kernel-threads 1 --seed 42 \
  --gp-observation-archive auto \
  --oracle-stencil saddle/oracle-stencil.json --oracle-stencil-step 0.01
\end{verbatim}
\subsection{Single-ended comparisons}
\label{sec:single-comparison}
The GPRidge Birkholz--Schlegel calculations share a 900 s case deadline across seed-band preparation, the dimer and independent validation.
Table \ref{tab:single-ended} reports them against the eOn dimer and against Sella, both started at the same band maximum.
The matched set for the eOn dimer is the 17 reactions both methods accept, and the matched set for Sella is the 18 reactions both accept.
On those sets the sums are 816 against 127 for the dimer stage, and 201 against 1371 for Sella, where the Sella column for GPRidge includes the seed band.
The partitioned rational-function step is the escape inside the GPRidge dimer when a second negative mode is weak.
GPRidge returns a resolved first-order saddle for oxirane and for sulfolene, where Sella reports an additional negative mode.
The eOn dimer stops on its projected-force criterion, and the Cartesian force is scored on the returned geometry.
These searches ran on Intel Xeon Gold 6248R processors, with four MPI ranks, two BLAS threads and one kernel thread.
The seed band is the Baker driver above with \texttt{--converged-force 0.1}, \texttt{--max-iterations 200}, \texttt{--band-refinements 0}, \texttt{--gp-drift-budget false}, \texttt{--kernel-support-freezes-band true}, \texttt{--gp-environment-types types.dat} and \texttt{--theta-shift-skip-auto true}.
Its annealing budget and cadence are zero, and it omits the final-geometry \texttt{--oracle-stencil} and \texttt{--oracle-stencil-step} flags used for the Baker return. The full commands are in Supporting Information Section \ref{sec:si-publication-commands}.
The dimer, started from the band maximum and seeded with that band's training set, is the following driver.
\begin{verbatim}
gprd --pos pos.con --orient direction.dat --config params.bin \
  --potential uma --model CASE.pt2 --task-name omol --uma-charge 0 \
  --uma-spin 1 --device cpu --gp-environment-types types.dat \
  --gp-model-profile gprd_zbl --gp-zbl-prior-mean true \
  --gp-nonperiodic true \
  --gp-active-posterior true --gp-environment-residuals false \
  --theta-shift-skip-auto true --gp-prior-mean-statistic max \
  --uncertainty-mode map --hyperparameter-mode rgmin:lbfgs \
  --gp-variance-subset auto --gp-anneal-stage true --gp-anneal-budget 0 \
  --gp-anneal-every 0 --gp-hssvd false --gp-no-output-norm true \
  --dimer-kappa-beta 5.0 --dimer-index-recovery true \
  --gp-seed-archive band/obs-archive --gp-seed-noise archive \
  --linalg-mode library --linalg-threads 2 --kernel-threads 1 \
  --gp-observation-archive auto \
  --oracle-stencil saddle/oracle-stencil.json --oracle-stencil-step 0.01
\end{verbatim}

\begin{table}[htbp]
\caption{\label{tab:single-ended}Single-ended searches on Birkholz--Schlegel, UMA-S-1.1. Counts are energy-and-force evaluations summed over the reactions both methods accept. GPRidge is 20/20. The Sella column for GPRidge includes the seed band; the eOn-dimer column is the dimer stage.}
\centering
\footnotesize
\begin{tabular}{lrrr}
Method & Accepted & Calls & GPRidge\\
\hline
eOn dimer \cite{henkelmanDimerMethodFinding1999} & 17/20 & 816 & 127\\
Sella \cite{hermesSellaOpenSource2022} & 18/20 & 201 & 1371\\
\end{tabular}
\end{table}

\begin{figure}[tbp]
\centering
\includegraphics[width=0.98\textwidth]{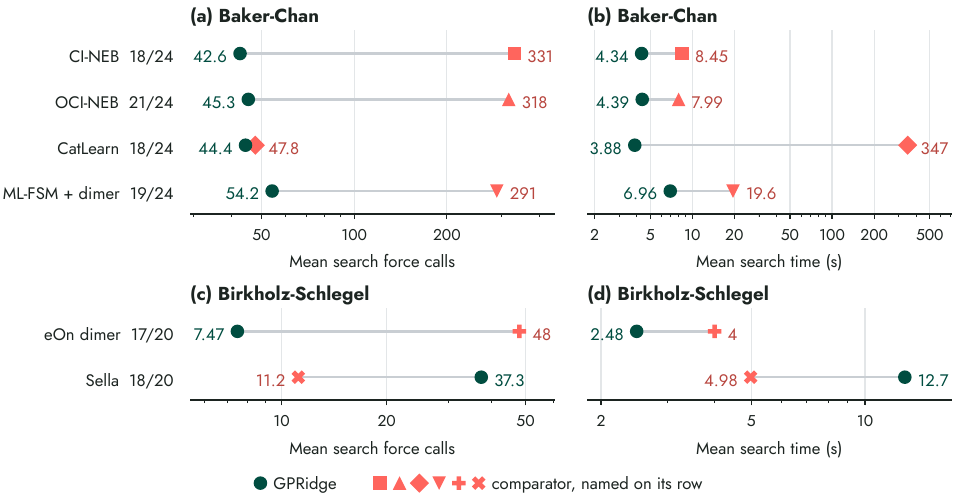}
\caption{\label{fig:method-savings}Paired search costs, labelled by workflow.
Panels (a,b) are the double-ended Baker--Chan bands, and panels (c,d) are the single-ended Birkholz--Schlegel dimer searches.
Circles mark GPRidge and each comparator has its own marker.
Printed values are means on the reactions both methods accept, and the multiplier, comparator divided by GPRidge, is given in the text.
Counts include endpoint evaluations and the construction of the freezing-string path.
Search time is the search itself, with independent validation kept separate.
The Birkholz dimer and Sella counts in the figure are the stage that starts at the band maximum, and Table \ref{tab:single-ended} includes that preparation in the GPRidge total for Sella.}
\end{figure}

Figure \ref{fig:method-savings}(c,d) plots these two single-ended comparisons.
Using only the stage that starts at the band maximum, the evaluation multiplier, comparator divided by GPRidge, is \(816/127=6.43\) for the eOn dimer and \(201/672=0.299\) for Sella.
Including the seed band in the GPRidge denominator gives \(201/1371=0.147\) for the Sella comparison in Table \ref{tab:single-ended}.
On the 17 reactions the eOn dimer also certifies, the dimer-stage means are 48 and 7.5 evaluations.
On the 18 reactions Sella certifies, the sum favours Sella: 201 evaluations from the band maximum, against 672 for the GPRidge dimer stage.
The GPRidge total from the endpoints, band included, is 1371. 

\subsection{Chemical character of the returned structures}
\label{sec:bonding}

\begin{figure}[tbp]
\centering
\includegraphics[width=0.92\textwidth]{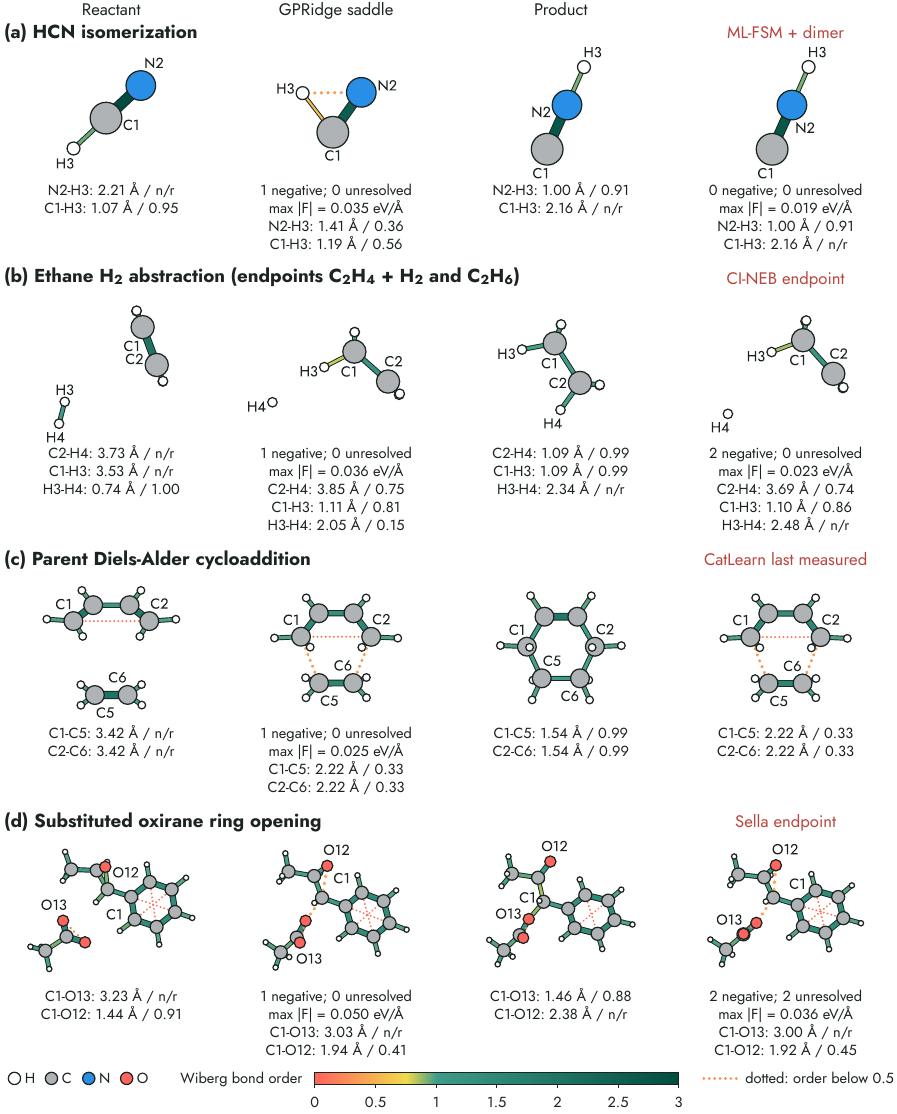}
\caption{\label{fig:reaction-chemistry}Bonding at the calculated geometries. Panels (a--c) are double-ended Baker reactions: two with failed comparator endpoints, and the Diels--Alder cycloaddition that CatLearn left short of the force threshold. Panel (d) is the single-ended Birkholz oxirane ring opening. Columns show the reactant, GPRidge saddle, product, and comparator endpoint. Bond color encodes the GFN2-xTB Wiberg bond order. n/r marks a pair the bond-order output leaves unmarked. The Supporting Information gives the electronic calculation settings, the density surfaces, two further reactions, and every saddle of both panels.}
\end{figure}

We compare the structures in Figure \ref{fig:reaction-chemistry} with GFN2-xTB Wiberg bond orders \cite{bannwarthGFN2xTB2019}.
These single-point calculations describe the bonding at the benchmark geometries, and the forces and the saddle indices remain those of the benchmark potential.
For HCN isomerisation the GPRidge saddle shares the migrating hydrogen between carbon and nitrogen, with C--H and N--H bond orders of 0.56 and 0.36.
The freezing-string endpoint followed by a dimer has the product N--H distance and bond order, 1.00 \AA{} and 0.91, and the curvature at that geometry is positive, so the low force is a minimum.

On ethane H\(_2\) abstraction the GPRidge saddle lies where the H\(_2\) unit has dissociated: H3 is bonded to C1 at 1.11 \AA{} with bond order 0.81, and H4 lies 2.05 \AA{} from H3 and 3.85 \AA{} from C2.
The CI-NEB endpoint keeps the C1--H3 bond at 1.10 \AA{}, with H4 at 2.48 \AA{} from H3, and the curvature check reports two negative modes there against one at the GPRidge saddle.
CatLearn leaves the Diels--Alder cycloaddition short of the force threshold, with the two forming C--C bonds at 2.221 and 2.224 \AA{} and a maximum atomic force of 0.235 eV/\AA{}.
The GPRidge saddle has the same synchronous structure, both forming bonds at 2.22 \AA{} with bond order 0.33.
At the oxirane saddle the ring C1--O12 bond is elongated from 1.44 to 1.94 \AA{} and the bond order goes from 0.91 to 0.41, while Sella returns 1.92 \AA{} and 0.45 together with two negative modes and two unresolved modes.
The Supporting Information gives the electronic settings, the density surfaces and every saddle of both panels.

\subsection{A periodic example: Pt(111) heptamer and island}
\label{sec:periodic}
The two periodic problems are the Pt(111) problems of OptBench in its modernized form \cite{chillBenchmarksCharacterizationMinima2014,goswamiOptBenchReplay2026}: both hold a seven-atom Pt island on a Pt(111) slab, 343 atoms in all, of which 175 are free.
The heptamer problem has 13 reactant--product pairs, and the island problem has one reactant and 59 products, each reached by moving a few island atoms.
Every atom interacts through the energy-shifted Morse potential of that collection with a \(9.5\) \AA{} cutoff.
The heptamer band carries eight interior images and converges at \(0.01\) eV \AA{}\(^{-1}\); the island band carries three interior images and converges at \(0.001\) eV \AA{}\(^{-1}\).
For each image the force norm is \(\lVert F\rVert_2\) over all \(3N_{\mathrm{free}}\) components, with \(N_{\mathrm{free}}=175\), rather than the maximum per-atom norm used for molecules.

\begin{figure}[tbp]
\centering
\includegraphics[width=0.98\textwidth]{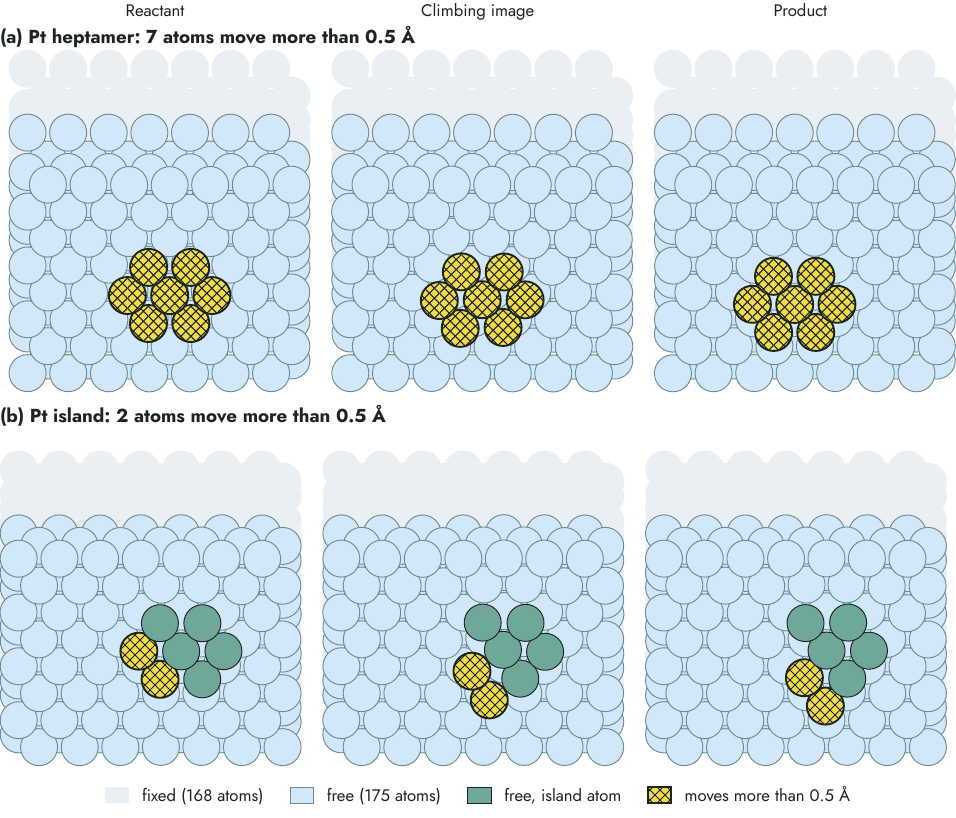}
\caption{\label{fig:pt-structures}Reactant, climbing image and product of the matched Pt(111) starts. Hatching marks atoms displaced by more than \(0.5\) \AA{} between the endpoints: seven heptamer atoms and two island atoms. The 168 fixed atoms of the bottom layers are drawn as plain discs, and the 175 free atoms are outlined. The view is tilted \(40^{\circ}\) from the surface normal and the slab is wrapped so that the moving atoms sit in the middle of the cell.}
\end{figure}

The control is CI-NEB on the measured forces in the same program, with the same linear initial band, spring constant, step limit and criterion, relaxed by L-BFGS, and both searches count every evaluation, including the two endpoints.
The heptamer searches stop at 10000 evaluations.
Summing the campaign search times of the converged starts gives 24.93 s for the heptamer band against 20.68 s for CI-NEB, and 85.83 s against 76.80 s on the island.
Both bands ran on Intel Xeon Gold 6248R processors, one thread, through \texttt{gprn} with a linear initial band and a spring constant of 5 eV/\AA{}$^{2}$.
The surrogate band sets \texttt{--nebmode gp\_local}, and the control uses the standard band mode, relaxed by L-BFGS on the measured forces.
Figure \ref{fig:pt-structures} shows the reactant, the climbing image and the product of the two matched starts.

\begin{figure}[tbp]
\centering
\includegraphics[width=0.98\textwidth]{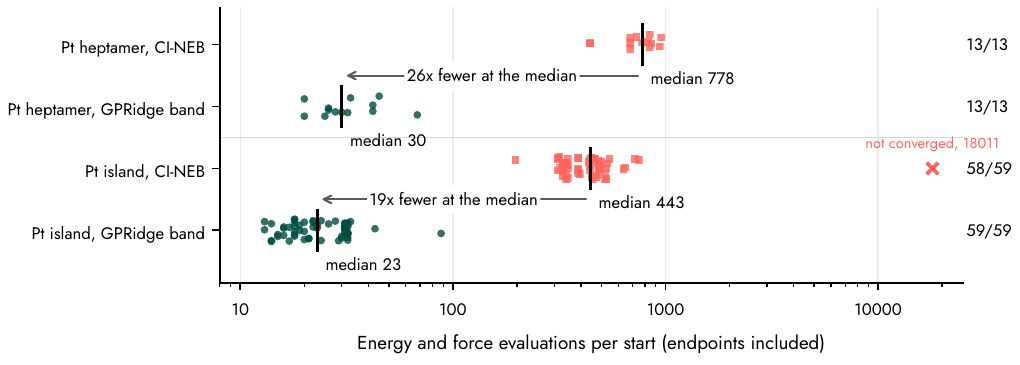}
\caption{\label{fig:pt-calls}Double-ended Pt(111) searches.
Energy-and-force evaluations of every start on the energy-shifted Morse potential, endpoints included.
Each marker is one start, the vertical bars and printed values are medians over the starts each search converges, and the right-hand counts give the converged starts.
The cross marks the island start on which CI-NEB reaches 18011 evaluations with the force still past the criterion.
Convergence is the OptBench force criterion, and these paths are accepted on that criterion.}
\end{figure}

Table \ref{tab:pt-cells} gives the outcomes, and Figure \ref{fig:pt-calls} shows the evaluations of every start.
The local pair-baseline band converges every start of both problems, 13 of 13 and 59 of 59.
The control converges every heptamer start and 58 of the 59 island starts, and on the remaining island start it reaches 18011 evaluations with the force still past the criterion.
On every start converged by both, the band uses fewer evaluations.
Over the starts each search converges, the medians are 30 against 778 on the heptamer and 23 against 443 on the island, a 26-fold and a 19-fold drop.
The matched heptamer start, pair 1, takes 20 evaluations against 442, and the matched island start, product 0, takes 17 against 347.
The climbing-image energies of those two starts differ by \(3.07\times10^{-5}\) eV on the heptamer and by \(1.51\times10^{-7}\) eV on the island.
The Supporting Information pairs every start of both panels by evaluations and by time.
Acceptance on these paths is the OptBench force criterion.

\begin{table}[H]
\caption{\label{tab:pt-cells}Double-ended Pt(111) bands on the energy-shifted Morse potential. Evaluations count every energy-and-force call, endpoints included; medians and ranges are over the starts converged by each search. The matched starts are heptamer pair 1 and island product 0.}
\centering
\small
\begin{tabular}{p{0.30\linewidth}p{0.12\linewidth}p{0.12\linewidth}p{0.12\linewidth}p{0.12\linewidth}}
Quantity & Heptamer band & Heptamer CI-NEB & Island band & Island CI-NEB\\
\hline
Starts converged & 13 / 13 & 13 / 13 & 59 / 59 & 58 / 59\\
Evaluations, median & 30 & 778 & 23 & 443\\
Evaluations, range & 20--68 & 442--954 & 13--88 & 197--749\\
Matched start, evaluations & 20 & 442 & 17 & 347\\
ASE 3.29 CI-NEB, evaluations, median (IQR) &  & 586 (554--610) &  &  \\
Matched start, climbing-image energy (eV) & $-1775.190$ & $-1775.190$ & $-1774.806$ & $-1774.806$\\
\end{tabular}
\end{table}
ASE 3.29 climbing-image NEB on these thirteen pairs, with the energy-shifted Morse potential, eight interior images and the same \(L_2\) force criterion of \(0.01\) eV/\AA{}, has a median of 586 evaluations.
The interquartile range is 554--610, against a median of 30 for the surrogate band on the same starts.
The Supporting Information tabulates the per-start results.

\section{Discussion}
\label{sec:discussion}
The Birkholz--Schlegel comparisons are single-ended searches, and the plain dimer, Sella and the Gaussian-process dimer all start from the highest image of the same band.
On the 18 reactions Sella and GPRidge both accept, Sella uses 201 evaluations from that image, against 672 for the GPRidge dimer stage.
Grignard accounts for most of the GPRidge evaluations on that matched set, and Sella works in internal coordinates that follow the wide range of stiffness in these molecules.
The GPRidge total from the endpoints on those 18 reactions, seed band included, is 1371, while the seed bands of all 20 reactions use 779 evaluations.
GPRidge returns a resolved first-order saddle point for oxirane and for sulfolene, the two reactions on which Sella reports an additional negative mode.
The similar oxirane bond lengths in Figure \ref{fig:reaction-chemistry}(d) therefore require the curvature check to distinguish the returned stationary points.
On the 17 reactions where the dimer method without a surrogate also reaches a first-order saddle point, it uses a mean of 48 evaluations from the band maximum, against 7.5 for the GPRidge dimer stage.

On Pt(111) the median number of evaluations falls from 778 to 30 on the heptamer and from 443 to 23 on the island.
The summed campaign search times are 24.93 s against 20.68 s on the heptamer and 85.83 s against 76.80 s on the island, so the climbing-image band is faster in wall time on both panels while the surrogate band uses fewer evaluations.

In the earlier study of the GPR accelerated dimer method on 500 molecular reactions, the median number of Hartree--Fock evaluations fell from 308 to 31.
The wall time fell in three cases out of four, even though each Hartree--Fock evaluation was already inexpensive \cite{goswamiEfficientImplementationGaussian2025,goswamiBayesianHierarchical2025}.
The surrogate overhead was a minority of that total, so a higher-level electronic-structure evaluation makes the wall time follow the count of evaluations.
That comparison is why a shared on-the-fly surrogate is worth carrying into a single-ended search as well.

A small force alone does not distinguish a minimum from a saddle, as the HCN comparison in Figure \ref{fig:reaction-chemistry}(a) shows.
The HCN band profile and projected observations in Supporting Information Figures \ref{fig:hcn-band} and \ref{fig:hcn-landscape} show the path and the accepted GPRidge saddle between the supplied endpoints.
For ethane abstraction the independent curvature check instead distinguishes a first-order saddle from the second-order CI-NEB endpoint.
Five further CI-NEB searches, HCCH, H\(_2\)CO, HF abstraction, acrolein rotation and HCNH\(_2\), finish above the Cartesian force threshold.
The molecular comparison therefore scores both measured force and saddle index.
The band moves on the GP mean between calculator calls, which supply its initial observations, correct unsupported or rejected proposals, and check candidate convergence.

The matched Pt(111) climbing-image energies agree to \(3.07\times10^{-5}\) eV and \(1.51\times10^{-7}\) eV.
This agreement measures consistency of the periodic path energies; those paths are accepted on the OptBench force criterion.
\section{Conclusions}
\label{sec:conclusions}
A double-ended search for the minimum-energy path and a single-ended minimum-mode search can share one Gaussian-process surrogate.
The molecular calculations accept a geometry on its measured force and an independent Hessian with a single negative eigenvalue.
On UMA-S-1.1, with that acceptance at \(0.0514\) eV/\AA{}, the band reaches a first-order saddle point for all 24 Baker--Chan reactions and the dimer for all 20 Birkholz--Schlegel reactions.

Against the climbing-image band on the 18 Baker--Chan reactions both methods accept, the search uses 7.78-fold fewer evaluations and 1.95-fold less search time, with the independent Hessian kept separate.
On the 18 Birkholz--Schlegel reactions Sella also accepts, Sella uses 201 evaluations from the band maximum, against 672 for the GPRidge dimer stage, and the GPRidge total from the endpoints on those 18 reactions is 1371.
On the Pt(111) heptamer the median falls from 778 to 30 evaluations, and the summed search time is 24.93 s for the surrogate band against 20.68 s for the climbing-image band.
The same search is a Python call that takes an ASE calculator, or a C or Fortran call.

For a larger number of movable atoms the training set can be pruned to the configurations near the saddle point \cite{goswamiAdaptivePruningIncreased2026}.
The molecular searches restrict the factor to at most 128 training geometries, and the local GP residual on Pt(111) follows the atoms that move along the path.
\section{Reproduction details}
\label{sec:reproduction}
The molecular searches and the Pt(111) timings ran on Intel Xeon Gold 6248R processors under Linux on the x86-64 architecture.
Each molecular search reserved four MPI ranks, two BLAS threads and one kernel thread.
The calculations use the archived GPRidge 0.5.0 source snapshot \texttt{3ddd83b774f49b80031f1e940c9c5f1787e695d6}, built through the accompanying pixi environment, with ScaLAPACK factorizations.
Parallel sums are gathered to one rank, accumulated in rank order in extended precision and broadcast.
Three complete passes return the same per-reaction search count.
\section{Data and software availability}
\label{sec:data-availability}
GPRidge is available under the MIT license.
The Baker--Chan and Birkholz--Schlegel reactions are those of Baker and Chan and of Birkholz and Schlegel, and the Pt(111) problems are those of OptBench \cite{bakerChanLocationTransition1996,birkholzUsingBondingGuide2015,chillBenchmarksCharacterizationMinima2014,goswamiOptBenchReplay2026}.
The potential on the molecular reactions is UMA-S-1.1 \cite{woodUMAFamilyUniversal2025}.
The geometries, GPRidge 0.5.0, and the inputs for the 24 Baker--Chan and 20 Birkholz--Schlegel index-one calculations, including DACP2 and Grignard, are deposited at \href{https://doi.org/10.5281/zenodo.23776354}{doi:10.5281/zenodo.23776354}.
The Baker driver takes the reactant and the product, the UMA potential on the omol task, 18 images, a force threshold of \(0.0514\) eV/\AA{} and the climbing image.
CatLearn ML-NEB is version 0.6.2 at revision \texttt{6a307f32dbef18da9cd848f632d082b8e4f117b1}; Sella is version 2.6.0.
eOn uses tag \texttt{oci-neb-correction-2026.10.29}, revision \href{https://github.com/TheochemUI/eOn/commit/8356aa8a23658be7904b1a600e228aadad917114}{\texttt{8356aa8a2365}} \cite{goswamiEOnSoftware2026}.
The eOn documentation is at \url{https://eondocs.org}.
The freezing-string geometries are those published by Marks and Gomes \cite{marksFreezingStringInternal2025}.
The bond orders use the GFN2-xTB single points, and the Python calls use ASE.
\section{Supporting Information}
\label{sec:org1294b81}
Publication protocol and per-reaction results, force and curvature validation, model and search derivations, scaling and arithmetic measurements, and the software interfaces, with a table of contents.
The Baker driver arguments and the comparator sources are restated there.
\section{Acknowledgments}
\label{sec:acknowledgments}
The calculations used the Icelandic Research e-Infrastructure at the University of Iceland, with resources and user support from that project, supported by the Icelandic Infrastructure Fund.
Rohit Goswami acknowledges support from his family, particularly Ruhila Goswami, Prof.\ D.\ Goswami, S.\ Goswami, A.\ Goswami and M.\ Sallermann, and from his pets and plants.
\appendix
\renewcommand{\thesection}{S\arabic{section}}
\section*{Supporting Information}
The following appendices are the journal Supporting Information, included here for the preprint.

\noindent
This supplement gives the publication protocol and settings, the
per-reaction force and index assessments with search
and validation costs, the pair-local joint energy--force covariance
tiles, the tied environment-type evidence rule, the one-image
restore-versus-continue bandit and Brent inner-cycle detection, and the
active-residual model used for Birkholz--Schlegel. The contents list
groups the calculations, model derivations, numerical measurements,
and software documentation.

The reported calculations factorize the noisy Gram by Cholesky. Birkholz
uses fixed environment labels generated offline from endpoint
k-nearest-neighbour radial and SOAP descriptors. SOAP assigns those labels;
the covariance is the squared exponential on inverse pair distances. The
descriptor comparison, the dual Hilbert--Schmidt factorization, and the internal chart are
derived here. The reported searches use the Cholesky factor. Analytical identities, symbolic checks and
Lean declarations document their respective assumptions. The manuscript proof library states
the chart and tied-parameter identities over real matrices and
differentiable maps, with their rank and curvature assumptions explicit.

\setcounter{tocdepth}{1}
\makeatletter
\renewcommand{\@pnumwidth}{2.4em}
\renewcommand{\l@section}[2]{\@dottedtocline{1}{0em}{3em}{\bfseries#1}{\bfseries#2}}
\makeatother
\tableofcontents

\section{Data and code availability}
\label{sec:si-data}
GPRidge 0.5.0, the geometries, and the inputs and outputs of the calculations and figures in this supplement are deposited at \href{https://doi.org/10.5281/zenodo.23776354}{doi.org/10.5281/zenodo.23776354}. From that deposit, \texttt{cd gpridge\_repro \&\& pixi install -e uma} prepares the potential, and \texttt{cd umaRuns \&\& uv run --with snakemake --with pyyaml snakemake --configfile config/<campaign>.yaml --config mint\_python=\$(pixi run -e uma which python) --cores 8 --keep-going} repeats a campaign. The molecular searches use the \texttt{gprn} and \texttt{gprd} settings in Section~\ref{sec:si-publication-commands}. Snakemake can schedule one calculation or a reaction campaign.

\section{Publication protocol}
\label{sec:si-publication-protocol}
The primary reaction calculations and the scaling study use GPRidge 0.5.0. The tabulated pass is one of three complete repetitions. Both use UMA-S-1.1 on the omol task. These are energy and force evaluations of the machine-learned potential.
\subsection{Effective method settings}
\label{sec:org94990a9}

Each panel uses four MPI ranks and two CPU slots per rank, one local
kernel worker and two BLAS threads, with ScaLAPACK and library arithmetic. The campaign fixes the thread counts; OpenBLAS selects its CPU kernel. The potential adapter runs the compiled model with one OpenMP thread. Each system's UMA-S-1.1 model is exported with \texttt{torch.export} (non-strict \texttt{make\_fx} tracing) into an AOTInductor package, with the roll gauge of the edge rotations fixed; at its export geometry, each of the 44 packages reproduces the FAIRChem ASE calculator to \(2.0\times10^{-6}\) eV in energy and \(1.4\times10^{-5}\) eV \AA{}\(^{-1}\) in force. Both use \texttt{gprd\_zbl}, a switched ZBL
mean, inverse-distance covariance, \texttt{rgmin:lbfgs} hyperparameter fitting,
MAP uncertainty, a length-scale prior, and Cholesky factorization.
The calculations set \texttt{GPR\_GRADIENT\_DENSE=1} for dense
likelihood-gradient contraction, with a fixed contraction order and thread count.
The reported fits use Cholesky. SOAP and environment residual kernels, manifold charts, output normalization
in GPRD, and the dual HS-SVD factorization are analysed separately in this supplement. The Cartesian force criterion is
\begin{equation}
 f_{\max}(x)=\max_{a=1,\ldots,N}
 \sqrt{F_{ax}^2+F_{ay}^2+F_{az}^2}\leq
 0.0514221\;\mathrm{eV}\,\text{\AA}^{-1}.
\end{equation}
It is neither the largest Cartesian component nor the norm over all
\(3N\) components. A climbing-force reflection preserves the latter norm,
but need not preserve the maximum atomic norm. The final check therefore
uses the raw measured atomic forces.

\begin{table}[htbp]
\caption{Settings that distinguish the two fixed publication protocols. The native argument files, not a repository default, determine each run.}
\centering
\small
\begin{tabular}{p{0.30\linewidth}p{0.28\linewidth}p{0.31\linewidth}}
Control & Baker--Chan & Birkholz--Schlegel\\
\hline
Search & GP-accelerated CI-NEB & GP-NEB preparation, then GP-dimer\\
Intermediate band images & 18 & 18\\
Initial observations & both endpoints and all interior images & both endpoints and all interior images\\
Acquisition & required measurements and tree-structured Parzen estimator (TPE) & same priorities and TPE\\
Band optimizer & rgmin L-BFGS & rgmin L-BFGS\\
Band force threshold (eV/Angstrom) & 0.0514221 & 0.1\\
Nominal band outer limit & 1000 & 200\\
Kernel-support batch & off & on\\
Drift budget & on & off\\
Band index certificate & off & off\\
Environment pair labels & off & on\\
Active posterior refits & enabled & enabled\\
Dimer seed observations/noise & no dimer handoff & imported from the band\\
Dimer variance subset & not applicable & search footprint (auto)\\
Annealing stage & default off & first full fit and measured contradiction\\
Force reporting & final accepted geometry & final accepted geometry\\
Wall limit & 14400 s per case & shared 900 s\\
\end{tabular}
\end{table}

The Birkholz dimer uses L-BFGS rotation and translation, separation
0.01 Angstrom, and the basin-constrained kappa option with beta 5.
The nominal limits are 500 outer phases, 200 inner steps,
and 10 rotation iterations. A shared wall deadline bounds the complete
workflow. The band chooses its peak and a normalized centered tangent;
its measured energies, gradients, pair labels, fitted kernel and physical
noise initialize the dimer. Importing them adds no energy and force evaluations.
The band search and its initial endpoint evaluations remain charged to
the Birkholz total. A band is a seed here, not a second certificate.
\subsection{Publication commands}
\label{sec:si-publication-commands}
The reported calculations run the drivers below under \texttt{srun} with four MPI
tasks and two CPUs per task. The Baker band
returns the climbing image directly:

\begin{verbatim}
gprn --reactant R.con --product P.con --potential uma --model CASE.pt2 \
  --task-name omol --uma-charge 0 --uma-spin 1 --device cpu \
  --nebmode gp_oie --use-mmf false --nimages 18 \
  --converged-force 0.0514221 --max-iterations 1000 \
  --climbing-image true --climbing-image-converged-only true \
  --ci-index-certificate false --band-refinements 1 \
  --idpp true --sidpp true --sidpp-geodesic true --seed-full-chain true \
  --spring-constant 5.0 --geometric-spring false \
  --gp-model-profile gprd_zbl --gp-zbl-prior-mean true \
  --gp-nonperiodic true \
  --gp-active-posterior true --gp-environment-residuals false \
  --gp-prior-mean-statistic max --gp-learn-noise false --ls-prior true \
  --uncertainty-mode map --hyperparameter-mode rgmin:lbfgs \
  --neb-band-engine rg-lbfgs --gp-adaptive-prune true \
  --gp-adaptive-prune-max 128 --gp-hssvd false --gp-drift-budget true \
  --kernel-support-freezes-band false --dimer-kappa-beta 5.0 \
  --linalg-mode library --linalg-threads 2 --kernel-threads 1 --seed 42 \
  --gp-observation-archive auto \
  --oracle-stencil saddle/oracle-stencil.json --oracle-stencil-step 0.01
\end{verbatim}

The Birkholz band uses the same model settings with a 0.1 eV \AA{}\(^{-1}\) seed
threshold, 200 outer iterations, no band refinement, the fixed environment
labels, and a frozen band once the kernel support settles. It omits the final-geometry stencil flags of the Baker command:

\begin{verbatim}
gprn ... --converged-force 0.1 \
  --max-iterations 200 --band-refinements 0 \
  --gp-environment-types types.dat --theta-shift-skip-auto true \
  --gp-anneal-stage true --gp-anneal-budget 0 --gp-anneal-every 0 \
  --gp-drift-budget false --kernel-support-freezes-band true
\end{verbatim}

The dimer starts from the band maximum and its direction and seeds its GP
from the band's training set:

\begin{verbatim}
gprd --pos pos.con --orient direction.dat --config params.bin \
  --potential uma --model CASE.pt2 --task-name omol --uma-charge 0 \
  --uma-spin 1 --device cpu --gp-environment-types types.dat \
  --gp-model-profile gprd_zbl --gp-zbl-prior-mean true \
  --gp-nonperiodic true \
  --gp-active-posterior true --gp-environment-residuals false \
  --theta-shift-skip-auto true --gp-prior-mean-statistic max \
  --uncertainty-mode map --hyperparameter-mode rgmin:lbfgs \
  --gp-variance-subset auto --gp-anneal-stage true --gp-anneal-budget 0 \
  --gp-anneal-every 0 --gp-hssvd false --gp-no-output-norm true \
  --dimer-kappa-beta 5.0 --dimer-index-recovery true \
  --gp-seed-archive band/obs-archive --gp-seed-noise archive \
  --linalg-mode library --linalg-threads 2 --kernel-threads 1 \
  --gp-observation-archive auto \
  --oracle-stencil saddle/oracle-stencil.json --oracle-stencil-step 0.01
\end{verbatim}

Both drivers write the validation stencil of the final structure with
central differences of 0.01 \AA{} (\texttt{-{}-{}oracle-stencil-step 0.01}).
\subsection{Controls exercised by the publication calculations}
\label{sec:org3bbdd7b}

The searches exercise six controls: the measured-column dimer certificate, the measured descent escape, replacement in the dense factor, the best-force continuation reward, an active-set cap of 128, and the kernel hold for a resolved prediction. The certificate runs for 1 of 20 Birkholz reactions, factor replacement for 17 of 20, the continuation reward for 15 of 24 Baker reactions and 16 of 20 Birkholz reactions, and the kernel hold for 12 of 24 and 16 of 20. The descent escape is the sulfolene search.
The calculations also use the geodesic S-IDPP seed, dimer index recovery with the
complete-stencil restart, the bounded first trial and the line-search zoom
safeguard of the hyperparameter optimizer, the kept measured step radius of
the climbing image, and a force-readiness ceiling that includes the numerical
jitter.
Birkholz--Schlegel uses fixed atom-environment labels; Baker--Chan uses element-pair types. The preparer reads the reactant and product CONs with \texttt{gp\_environment\_types.py} using \texttt{knn+soap}, a \(4\)~\AA{} cutoff and four neighbors per species, and writes each \texttt{types.dat} label file. The search reads that file through \texttt{-{}-{}gp-environment-types}. These labels are the pair-length-scale groups, tied to element and reaction role. SOAP assigns the labels; the covariance is the squared exponential on inverse pair distances. The supplement analyses SOAP residuals, manifold charts, and the dual HS-SVD factorization separately.

Each calculation keeps its \texttt{gprn} argument file and, when a dimer is used, its \texttt{gprd} argument file,
together with the binary parameter file, the run log, the fitted kernel and the observation database.
With the annealing budget set to zero, each fit still uses
\(\max(n_{\mathrm{previous\ fit}},d_\theta+1)\) evaluations, where
\(d_\theta\) is the number of independent parameters.
\subsection{Scalar energy, derivative observations, and the prior}
\label{sec:org860106c}

For positions \(x\), pair \(p=(a,b)\) has descriptor \(q_p(x)=1/r_{ab}\).
With positive element/pair-type length scales, the signal covariance is
\begin{equation}
 k(x,x')=\sigma_f^2\exp\!\left[-\sum_p
 \frac{(q_p(x)-q_p(x'))^2}{\ell_{t(p)}^2}\right].
\end{equation}
The same atom mapping is used at both arguments. Pair distances enforce
translation and rotation invariance, but tying element parameters alone
does not symmetrize the kernel over independent permutations of atoms.
For \(d=q(x)-q(x')\), \(\Lambda=2\operatorname{diag}(\ell_{t(p)}^{-2})\),
and \(J=\partial q/\partial x\), the mixed derivative is
\begin{equation}
 \partial_x\partial_{x'}^\top k
 =J(x)^\top k\,[\Lambda-\Lambda dd^\top\Lambda]J(x').
\end{equation}
Energy--gradient blocks use first derivatives, and forces have the
opposite sign to the gradients. All force and Hessian predictions
are derivatives of the same scalar posterior, including the prior mean.
Pair-local tiles evaluate these contractions without storing a dense
Jacobian for every pair and Cartesian coordinate.

The repulsive mean is
\begin{equation}
 m_{\mathrm{ZBL}}(x)=\sum_{a<b}s(r_{ab})\,
 \frac{14.3996454784255\,Z_aZ_b}{r_{ab}}
 \sum_{j=1}^4 a_j\exp(-b_j r_{ab}/a_{ab}),
\end{equation}
with energies in eV, distances in Angstrom,
\(a_{ab}=0.88534(0.529177210903)/(Z_a^{0.23}+Z_b^{0.23})\),
\(a=(0.1818,0.5099,0.2802,0.02817)\) and
\(b=(3.2,0.9423,0.4029,0.2016)\).
Let \(R\) be the sum of the Cordero covalent radii. The switch is one below
\(0.55R\), zero above \(0.75R\), and
\(1-10u^3+15u^4-6u^5\) between them, with
\(u=(r-0.55R)/(0.20R)\). Thus the prior acts on strong compression and is
not an approximate bonding potential. These switch values are fixed
model settings, not a learned chemical feasibility theorem. Residual
observations subtract both its energy and its gradient. Prediction adds
both back; an energy offset does not contribute to the gradient.

For joint residual targets \(r=y-\mathsf{Obs}\,m\) and noisy covariance \(C\),
the fitted negative log posterior is
\begin{equation}
 \mathcal L(\theta)=\tfrac12r^\top C^{-1}r+
 \tfrac12\log\det C+\tfrac{n}{2}\log(2\pi)-\log p(\theta).
\end{equation}
The parameters are optimized in log coordinates. A Gaussian density on
a physical length must therefore include its change-of-variable term;
a Gaussian defined directly on log length must not include it twice.
The enabled two-sided log-length penalty uses descriptor-derived centers
and a log-distance width when those statistics are nondegenerate. Tied
lengths count one prior term per independent group. The selected
\texttt{gprd\_zbl} profile also uses its signal-scale prior. The complete
profile, seed statistics, bounds and fallback cases are implemented in
\texttt{SexpatCF} and \texttt{GaussianProcessRegression}, and the fitted parameters
make these choices inspectable. MAP uncertainty is conditional on the
fitted hyperparameters; it is not an integral over their uncertainty.
\subsection{Observation units, conditioning, and stopping}
\label{sec:org4e78c2e}

One energy and \(3N\) Cartesian derivatives observe one latent energy.
The implementation stores \(G=\nabla E\). The physical force is \(F=-G\).
The two observations differ by a length. Their counts do not set their
relative accuracy. For every positive \(\varepsilon\),
\(E_\varepsilon(x)=\varepsilon\sin(x/\varepsilon^2)\) satisfies
\(|E_\varepsilon|\le\varepsilon\) while \(F_\varepsilon(0)=-1/\varepsilon\).
Sampled energies can therefore agree while every force is large. A
constant energy offset changes the energies and none of the forces.

The published stopping test is the measured maximum atomic force norm,
\begin{equation}
 f_{\max}(x)=\max_a\lVert F_a(x)\rVert_2
 \le 0.0514221\;\mathrm{eV}\,\text{\AA}^{-1}.
\end{equation}
It is not the root-mean-square over \(3N\) components: one nonzero
component of size \(A\) has squared mean \(A^2/(3N)\). In the calculations, Baker--Chan has 24 of 24 measured-force
successes and 24 of 24 index-check passes. Birkholz--Schlegel has 20 of
20 force-threshold passes and 20 of 20 index-check passes.
The Birkholz workflow sets
\texttt{band\_\allowbreak{}hssvd=false}, \texttt{dimer\_\allowbreak{}hssvd=false}, \texttt{hyperparameter\_\allowbreak{}mode=rgmin:lbfgs}, \texttt{gp\_\allowbreak{}drift\_\allowbreak{}budget=false}, \texttt{kernel\_\allowbreak{}support\_\allowbreak{}freezes\_\allowbreak{}band=true}, \texttt{sidpp\_\allowbreak{}geodesic=true}, \texttt{dimer\_\allowbreak{}index\_\allowbreak{}recovery=true}, \texttt{linalg\_\allowbreak{}threads=2}, \texttt{kernel\_\allowbreak{}threads=1}, \texttt{force\_\allowbreak{}norm=max\_\allowbreak{}force\_\allowbreak{}on\_\allowbreak{}atom}, and \texttt{prior\_\allowbreak{}mean\_\allowbreak{}statistic=max}.

The reported molecular likelihood uses one fitted scalar noise parameter. The noise block \(N\) assigns its value to the energy and gradient rows in the numerical observation convention, and \(\delta I\) is the numerical jitter: \(C=K+N+\delta I\).
The scalar identity-noise formulas in optional derivations specialize this general form; \(N_g\) denotes its gradient-row noise block and is not a separately fitted parameter. No row count, including one energy row
against \(3N\) gradient rows, calibrates them. The gradient-difference
variance uses \(2N_g\) because two gradient observations carry that shared
variance, not because the force noise has been identified from the
component count.

An invertible observation transformation \(W\), fixed during a solve,
acts on residuals, the covariance, and every cross-covariance by
\(y'=Wy\), \(C'=WCW^{\top}\), and \(k'=Wk\). Energy and derivative operators
take the scale of their own row. Noise and jitter are included in \(C\)
before the congruence. The physical posterior mean and covariance
reduction are unchanged, and \(F=-G\) is unchanged after the rows are
mapped back. The Gaussian negative log marginal likelihood gains
\(\log|\det W|\). Omitting that Jacobian changes the reported objective.
A scalar jitter added after the transformation is not the original
jitter: its physical value is \(\delta W^{-1}W^{-\top}\).

Covariance equilibration is a transformation of this form. For a positive diagonal entry,
\(W_{ii}=2^{-\lfloor\mathrm{ilogb}(C_{ii})/2\rfloor}\), so each scaled
diagonal entry lies in \([1,4)\). A factor \(L'\) of \(WCW^{\top}\) is
returned to physical units by \(L=W^{-1}L'\) before the solve, the log
determinant, and cached factor use. The transformation changes row
units and conditioning. It does not change the relative physical
energy and force noise. Exact SymPy checks cover the transport, the
Jacobian, operator scaling, noise scores, and the energy--force
counterexample. The Lean development checks the general factor and
precision transport.
\subsection{Notation and optional methods}
\label{sec:si-notation}
The reaction tables refer to the archived GPRidge 0.5.0 snapshot.
The band-stepper experiments reproduce 0.4.0 controls, and the full-search layout and prediction-arithmetic experiments identify their 0.6.0 sources in their own sections.
The optional method derivations describe capabilities that the headline reaction calculations leave disabled.

\begin{table}[H]
\caption{Notation used across the supplementary derivations. A section that uses a symbol for a local scalar defines that scalar at its equation.}
\centering\small
\begin{tabular}{p{0.23\linewidth}p{0.68\linewidth}}
Symbol & Meaning\\\hline
$\mathcal D_t$; $\mathsf{Obs}$ & Training observations; operator stacking the energy and Cartesian derivatives\\
$n$; $D$; $N$ & Observation or matrix order as stated locally; derivative dimensions; image or atom count with its local definition\\
$K$; $N_g$; $C$ & Signal covariance; gradient-noise block; complete noisy covariance including jitter\\
$\alpha$; $\mu$ & Posterior weights; posterior mean. Optimizer step sizes carry an iteration subscript\\
$q$; $J$ & Inverse-distance descriptor; its Cartesian Jacobian\\
$A$; $\Phi$; $\Psi$; $S$ & Feature design; Mercer features; orthonormal basis; diagonal singular values in the dual derivation\\
$Q$; $P$ & Measured Hessian subspace basis; projector in the section that defines it\\
$\lambda$; $\kappa$; $\tau$ & Eigenvalue, curvature, and sign-resolution threshold in the curvature section; local condition-number targets and trust parameters are defined separately\\
$\varepsilon_F$; $\varepsilon_{\rm mach}$ & A force tolerance; floating-point unit roundoff. Unsubscripted tolerances retain the units stated at their equations\\
\end{tabular}
\end{table}

One-image-evaluated (OIE) and all-image-evaluated (AIE) specify the number of images acquired in an ordinary outer round.
The Ziegler--Biersack--Littmark (ZBL) screened-repulsion prior, integrated mean squared prediction error (IMSPE), and partitioned rational-function optimization (P-RFO) are named where used.
The implementation name Sexpat denotes the squared-exponential atomic inverse-distance covariance.
The local optimizer library is rgmin; Software for Linear Algebra Targeting Exascale (SLATE) is an optional distributed linear algebra backend.
Scheme II denotes the residue-based reproducible matrix-product method defined in Section \ref{sec:si-predict-arithmetic}.

\subsection{Frozen background and minimum-mode active refits}
\label{sec:org9ba6d57}

The active construction is enabled in the Birkholz run and occurs in its
native fit logs. Let \((m_B,C_B)\) be the fitted background posterior and
\(K_\theta\) an independent residual kernel. Zero residual constraints at
anchor channels \(B\), with physical noise \(N_B\), give
\begin{equation}
 \widetilde K_\theta(X,Y)=K_\theta(X,Y)
 -K_\theta(X,B)[K_\theta(B,B)+N_B]^{-1}K_\theta(B,Y).
\end{equation}
Conditioning \(C_B+\widetilde K_\theta\) on new measured channels uses their
residual \(y_A-m_B(A)\). The background hyperparameters remain fixed; its
uncertainty remains in the combined covariance. This is not equivalent
to refitting a single full GP, nor to adding the background mean without
its covariance. The article gives the posterior,
conditional likelihood, derivative through the anchor solve and PSD
argument. Soft anchor noise does not enforce exact interpolation.

For a normalized minimum mode \(n\), group \(t\) has directional score
\begin{equation}
 w_t=\sum_{p:t(p)=t}
 \left[\ell_t^{-1}J_p(x)n\right]^2.
\end{equation}
The implementation sorts these scores and selects the smallest
prefix covering 99 percent of their sum. It requests a full fit when
force uncertainty or measured force error reaches the convergence
threshold, when the mode cannot be resolved, when successive modes have
absolute overlap below 0.9, or after seven restricted refits. These are
explicit algorithmic settings, not consequences of a convergence proof.
Every pair remains in the background covariance; the selection restricts
which residual length scales are fitted. A contradicted combined model
triggers a full reassessment rather than indefinitely adjusting only the
residual. Optional SOAP and Cartesian residual candidates are described
as library capabilities, not as the source of the headline coverage.
\subsection{Measured-trial model management}
\label{sec:orgc9208c7}

A saddle-search step need not decrease energy or force norm. The radius
controller therefore compares measured and predicted changes in the
Cartesian gradient. At accepted center \(x_0\) and trial \(x_1\), predictions
must come from the same proposing posterior:
\begin{align}
 \widehat g_c(x_1)&=g(x_0)+\widehat g(x_1)-\widehat g(x_0),\\
 d&=g(x_1)-g(x_0),\qquad
 \widehat d=\widehat g(x_1)-\widehat g(x_0),\\
 \rho_g&=1-\frac{\lVert d-\widehat d\rVert_2}
                       {\max(\lVert d\rVert_2,\lVert\widehat d\rVert_2)}.
\end{align}
The zero-change case receives neutral agreement 0.5. Otherwise the ratio
lies in \([-1,1]\) by the triangle inequality; it is not an energy-reduction
ratio. Radius growth and contraction thresholds are 0.75 and 0.25, with
factors 2 and 0.5 and a numerical radius floor of \(10^{-6}\) Angstrom.
Growth requires a boundary trial (at least 0.9 of the radius). A measured
force increase with prediction error larger than the accepted force
scale or tolerance triggers a refit and restores the measured center;
the trial observation is kept. Accurate uphill motion remains
admissible during escape from a minimum.

For the gradient difference, uncertainty includes cross covariance:
\begin{equation}
 \operatorname{Var}(G_1-G_0)=C_{11}+C_{00}-C_{10}-C_{01}+2N_g.
\end{equation}
Independent likelihood noise supplies \(2N_g\). A discrepancy within the
reported spread is not used to justify contraction; a boundary trial
can extend the radius. The next burst also scales its force variance
by \(\max(1,e^2/v)\), using the preceding measured prediction error \(e\)
and the proposing variance \(v\). This calibration is a local correction,
not a guarantee of posterior coverage. Replacing the cross covariance by
a sum of independent variances would change this decision rule.

Surrogate relaxation ends at uncertainty or measured-trust boundaries,
a resolved local residual condition, an iteration limit, or a detected
cycle; it then requests an energy and force evaluation. The local analysis covers residual descent
at a negative-curvature root of one surrogate relaxation. Force convergence is established only by a measured gradient.
No monotone global dimer or universal chemical-convergence theorem is
claimed.
\subsection{Per-system costs}
\label{sec:org6d61854}

\begin{table}[htbp]
\caption{\label{tab:fixed-baker}Baker--Chan: energy and force evaluations and elapsed time per reaction. Search counts the evaluations during band preparation and the search, including the dimer when the reaction runs one. Validation counts the evaluations of the independent Hessian check and of the reference-geometry energies. The Wall column is the elapsed time including setup, band preparation, the search, and validation, in seconds. Force is the maximum atomic force of the final point in eV/Angstrom. The imaginary frequency is that of the independent Hessian at the final point. The barrier is the measured climbing-image energy less the reactant energy.}
\centering
\scriptsize
\begin{tabular}{lrrrrrrrl}
Case & Search & Validation & Total & Wall (s) & Force & Barrier (eV) & \(\nu^\ddagger\) (cm\(^{-1}\)) & Certified\\
\hline
HCN & 25 & 20 & 45 & 4.100 & 0.035161 & 2.034 & 1234i & yes\\
HCCH & 32 & 26 & 58 & 4.336 & 0.035151 & 0.080 & 956i & yes\\
H2CO & 80 & 26 & 106 & 7.357 & 0.025323 & 3.515 & 1871i & yes\\
CH3O & 31 & 32 & 63 & 4.789 & 0.015965 & 1.573 & 2093i & yes\\
cyclopropyl & 53 & 50 & 103 & 7.440 & 0.046194 & 1.098 & 1110i & yes\\
bicyclobutane & 46 & 62 & 108 & 7.505 & 0.046222 & 2.173 & 343i & yes\\
formyloxyethyl & 48 & 62 & 110 & 7.232 & 0.035699 & 0.834 & 1172i & yes\\
Diels-Alder & 46 & 98 & 144 & 13.524 & 0.024815 & 0.870 & 628i & yes\\
tetrazine & 28 & 50 & 78 & 5.109 & 0.034403 & 2.314 & 887i & yes\\
trans-butadiene & 22 & 62 & 84 & 4.937 & 0.034382 & 0.003 & 165i & yes\\
Ethane H\(_2\) abstraction & 196 & 54 & 250 & 56.256 & 0.035560 & 2.973 & 111i & yes\\
HF abstraction & 33 & 50 & 83 & 5.235 & 0.041316 & 2.508 & 1696i & yes\\
vinyl alcohol & 39 & 44 & 83 & 5.144 & 0.037527 & 2.963 & 2171i & yes\\
HOCl & 30 & 26 & 56 & 4.381 & 0.012004 & 1.703 & 1249i & yes\\
H2PO4 anion & 34 & 44 & 78 & 5.115 & 0.039292 & 1.425 & 1227i & yes\\
Claisen & 44 & 86 & 130 & 9.904 & 0.031645 & 1.341 & 610i & yes\\
silylene insertion & 41 & 68 & 109 & 7.064 & 0.032683 & 0.955 & 1217i & yes\\
HNCCS & 134 & 32 & 166 & 19.380 & 0.046812 & 0.721 & 460i & yes\\
HCONH3 cation & 36 & 44 & 80 & 5.295 & 0.042515 & 2.064 & 736i & yes\\
acrolein rotation & 35 & 50 & 85 & 5.752 & 0.048528 & 0.324 & 205i & yes\\
HCONHOH & 32 & 44 & 76 & 5.129 & 0.042188 & 1.989 & 1885i & yes\\
HCN + H2 & 45 & 32 & 77 & 6.266 & 0.033641 & 2.710 & 1753i & yes\\
H2CNH & 33 & 32 & 65 & 4.611 & 0.039359 & 2.179 & 1904i & yes\\
HCNH2 & 72 & 32 & 104 & 8.197 & 0.032500 & 1.277 & 1246i & yes\\
\end{tabular}
\end{table}

\begin{table}[htbp]
\caption{\label{tab:fixed-birkholz}Birkholz--Schlegel: energy and force evaluations and elapsed time per reaction. Search counts the evaluations during band preparation and the search, including the dimer when the reaction runs one. Validation counts the evaluations of the independent Hessian check and of the reference-geometry energies. The Wall column is the elapsed time including setup, band preparation, the search, and validation, in seconds. Force is the maximum atomic force of the final point in eV/Angstrom. The imaginary frequency is that of the independent Hessian at the final point. 900 s is the configured timeout.}
\centering
\scriptsize
\begin{tabular}{lrrrrrrl}
Case & Search & Validation & Total & Wall (s) & Force & \(\nu^\ddagger\) (cm\(^{-1}\)) & Certified\\
\hline
C2NO2 & 38 & 56 & 94 & 7.301 & 0.049205 & 573i & yes\\
C5HT & 40 & 80 & 120 & 10.179 & 0.034405 & 1646i & yes\\
HCN & 26 & 20 & 46 & 5.920 & 0.034949 & 1233i & yes\\
Cope & 51 & 98 & 149 & 15.581 & 0.045801 & 697i & yes\\
CPHT & 35 & 68 & 103 & 8.060 & 0.036786 & 1172i & yes\\
CYCBUT & 39 & 62 & 101 & 8.033 & 0.036228 & 770i & yes\\
DACP2 & 68 & 134 & 202 & 34.576 & 0.050075 & 556i & yes\\
DACP-eth & 39 & 104 & 143 & 14.786 & 0.040683 & 614i & yes\\
DFCP & 73 & 56 & 129 & 12.102 & 0.022467 & 723i & yes\\
Ene & 40 & 92 & 132 & 12.275 & 0.030184 & 1311i & yes\\
Grignard & 585 & 236 & 821 & 263.173 & 0.037993 & 243i & yes\\
H2CO & 75 & 26 & 101 & 7.766 & 0.032958 & 1874i & yes\\
HF-eth & 36 & 50 & 86 & 7.331 & 0.026658 & 1698i & yes\\
Hydro & 57 & 126 & 183 & 24.970 & 0.043529 & 1472i & yes\\
MeOH & 45 & 38 & 83 & 6.917 & 0.015895 & 2072i & yes\\
Oxirane & 327 & 168 & 495 & 84.373 & 0.050208 & 515i & yes\\
Oxycope & 49 & 86 & 135 & 12.566 & 0.034251 & 608i & yes\\
Silane & 47 & 68 & 115 & 9.163 & 0.043853 & 1214i & yes\\
SN2 & 28 & 38 & 66 & 5.875 & 0.019300 & 483i & yes\\
Sulfolene & 148 & 80 & 228 & 13.059 & 0.046520 & 412i & yes\\
\end{tabular}
\end{table}

\begin{figure}[htbp]
\centering
\includegraphics[width=\textwidth]{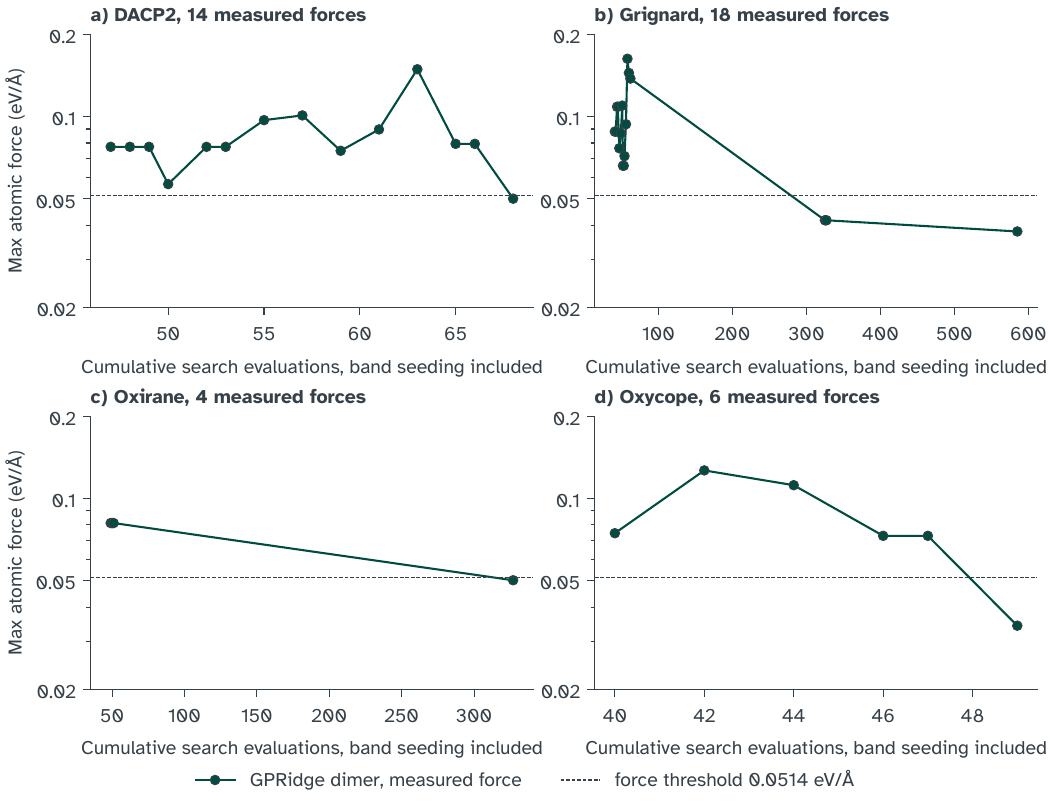}
\caption{Measured atomic force versus cumulative search evaluations for
four Birkholz cases (DACP2, Grignard, Oxirane and Oxycope).
The curves include band-seeding cost and terminate at the measured center
of the complete Hessian stencil, with the full search-call count. Markers
are the measurements, joined by straight lines. The dashed line is the force threshold. Forces are
the measured force norms.
Both traces follow the publication protocol.}
\label{fig:measured-dimer-force}
\end{figure}

The Baker searches use 1215 calls and 2341 including validation and
reference energies. The Birkholz searches use 779 band and 1067 dimer calls, giving 1846 search calls and 3532
including validation and reference energies. Elapsed times include setup, band preparation, the search, and validation. The database audit
matches the position and force arrays of the geometry that first reaches the
force tolerance, for all 20 Birkholz cases, exactly to earlier potential evaluations
without new evaluations. Figure \ref{fig:measured-dimer-force} shows the
measured atomic force along the search for four Birkholz cases.

\subsection{Repetitions and primary timing components}
\label{sec:orgdfc1bf7}

All three repetitions accept all 24 Baker and 20 Birkholz reactions, with identical search counts for each reaction. The primary results use one of the three passes. Table \ref{tab:si-release-repetitions} reports its elapsed time alongside the three-run mean and sample standard deviation for the slowest Baker reaction, Ethane H2 abstraction, and the two longest Birkholz calculations. These times include setup, band preparation, search and independent validation. The deposited per-reaction summary gives means, sample standard deviations and ranges for all 44 reactions; this table prints the three longest calculations.

\begin{table}[H]
\caption{\label{tab:si-release-repetitions}Complete-calculation timings from three repetitions of GPRidge \texttt{v0.5.0}. All listed reactions pass the measured force and independent saddle-index checks in every repetition. Search counts include band preparation and are identical across repetitions; uncertainty is the sample standard deviation of elapsed time.}
\centering
\small
\begin{tabular}{lrrr}
Reaction & Search evaluations & Primary time (s) & Mean \(\pm\) SD (s)\\
\hline
Ethane H2 abstraction & 196 & 56.256 & \(56.771 \pm 0.446\)\\
Grignard & 585 & 263.173 & \(266.614 \pm 3.122\)\\
Oxirane & 327 & 84.373 & \(84.327 \pm 0.407\)\\
\end{tabular}
\end{table}

Table \ref{tab:si-primary-components} gives the disjoint timing regions of the primary Grignard and oxirane calculations. Potential evaluations, model updates, surrogate relaxation, independent validation and process start-up each appear once. Each dimer calculation contains two complete Cartesian stencils: the first reports unresolved weak modes, and the next certifies the refined saddle. The earlier stencil is search work and the final stencil is independent validation. GP mode extraction is timed on its own. The remainder of the dimer interval is the dimer search time minus the oracle, objective, conditioning, prediction, retry, and GP mode-extraction times. That remainder, and the pipeline time outside the named timers, is the time outside those components.

\begin{table}[H]
\caption{\label{tab:si-primary-components}Components of the primary elapsed time, in seconds, at four MPI ranks and two CPU slots per rank. Each column describes one of the three passes; the uncertainty across complete repetitions is reported separately in Table \ref{tab:si-release-repetitions}. Values are rounded to three decimal places.}
\centering
\small
\begin{tabular}{lrr}
Component & Grignard & Oxirane\\
\hline
Potential evaluations & 143.585 & 8.612\\
Model updates & 22.734 & 8.544\\
Surrogate relaxation & 7.607 & 3.209\\
Independent validation & 33.198 & 13.068\\
Process start-up & 5.565 & 3.461\\
Repeated Cartesian stencils & 27.103 & 10.135\\
GP mode extraction & 0.363 & 0.023\\
Dimer phase remainder & 12.623 & 29.747\\
Unassigned & 10.396 & 7.576\\
Total & 263.173 & 84.373\\
\end{tabular}
\end{table}

\begin{figure}[H]
\centering
\includegraphics[width=\textwidth]{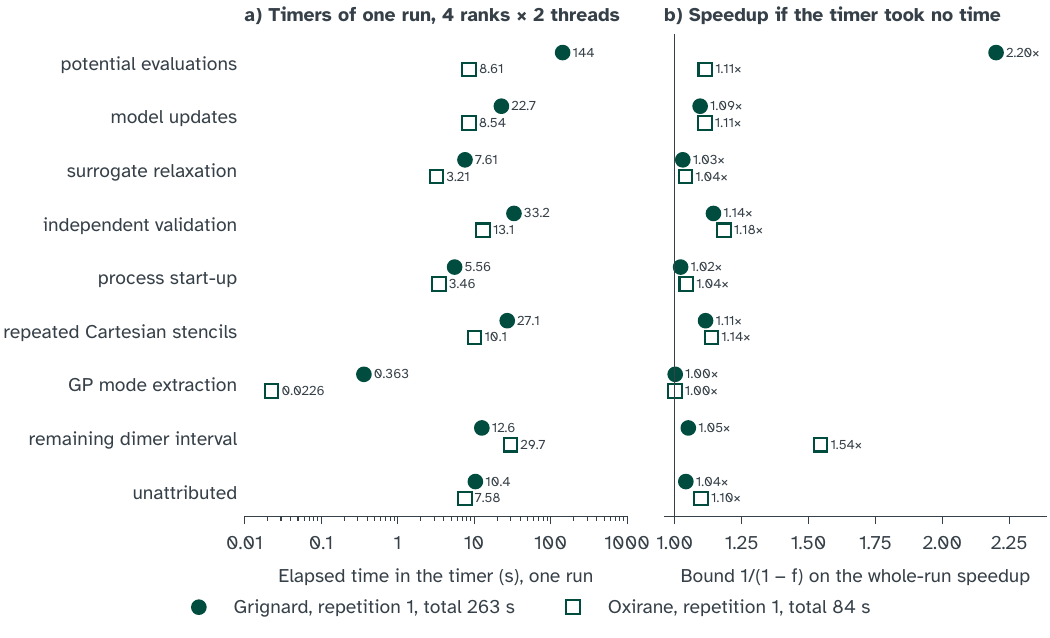}
\caption{\textbf{Where the elapsed time goes at $4 \times 2$.}
Left: nonoverlapping measured regions of the primary Grignard and oxirane
calculations from one of the three passes: potential evaluations, model updates,
surrogate relaxation, independent validation, process start-up, repeated
Cartesian stencils, GP mode extraction, and the remaining dimer interval. The unattributed region is pipeline time
outside these components. The main
elapsed-time figure and the Supporting Information quantify
variation across three complete repetitions; these component regions describe
one of the three passes. Right: the idealized speedup $1/(1-f)$ if an elapsed-time share
$f$ took no time. This is a mathematical bound, not a measured optimization; the
remaining residual identifies no kernel or removable code region.}
\label{fig:where-time}
\end{figure}

The primary Grignard and oxirane timing components are shown in Figure \ref{fig:where-time}. Grey marks time outside the named phase timers.

\subsection{Per-reaction search evaluations and elapsed time}
\label{sec:org1f1a2dd}

These plots show the reaction-level search and validation costs behind the main-text totals. The evaluation bars separate search calls from the independent checks. The timing bars include setup, preparation, search, and validation; error bars summarize three complete repetitions.

\begin{figure}[p]
\centering
\includegraphics[width=\textwidth]{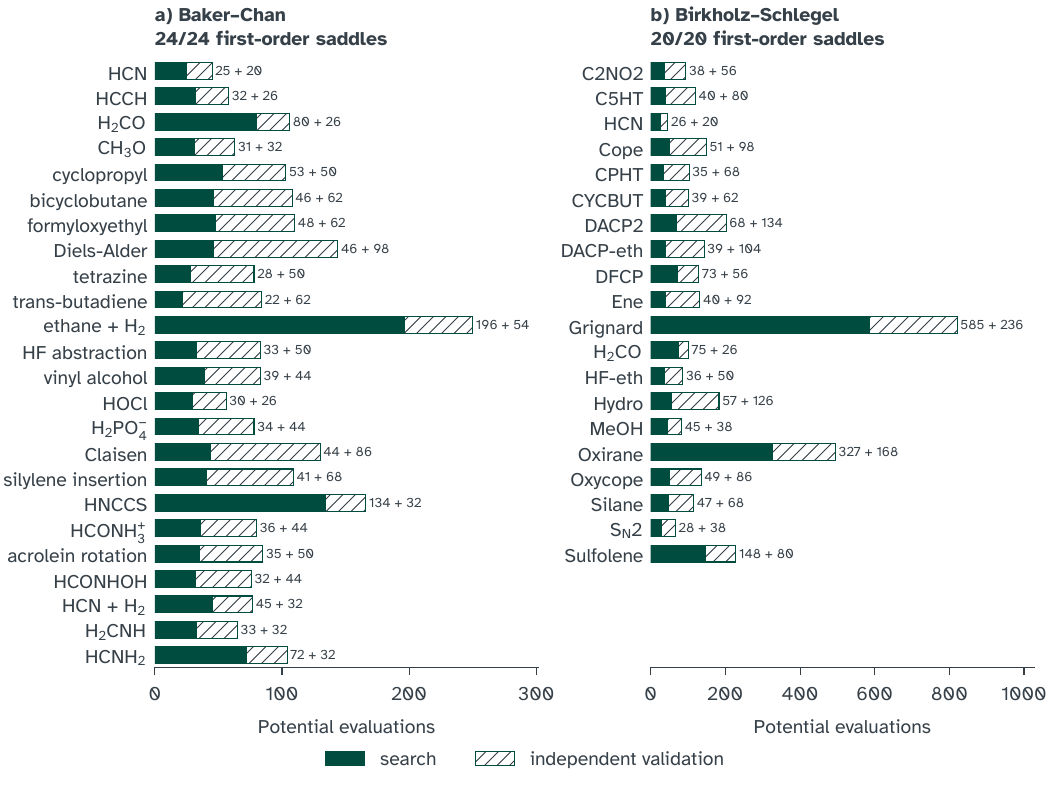}
\caption{\textbf{Potential evaluations by reaction.}
Each bar represents one reaction in one of the three passes and separates search
evaluations from independent checks. Baker search counts cover CI-NEB;
Birkholz counts cover band preparation and the dimer. The checks comprise
a finite-difference Hessian and reference-geometry energies. Every
evaluation uses UMA-S-1.1.}
\label{fig:campaign-calls}
\end{figure}

\begin{figure}[p]
\centering
\includegraphics[width=\textwidth]{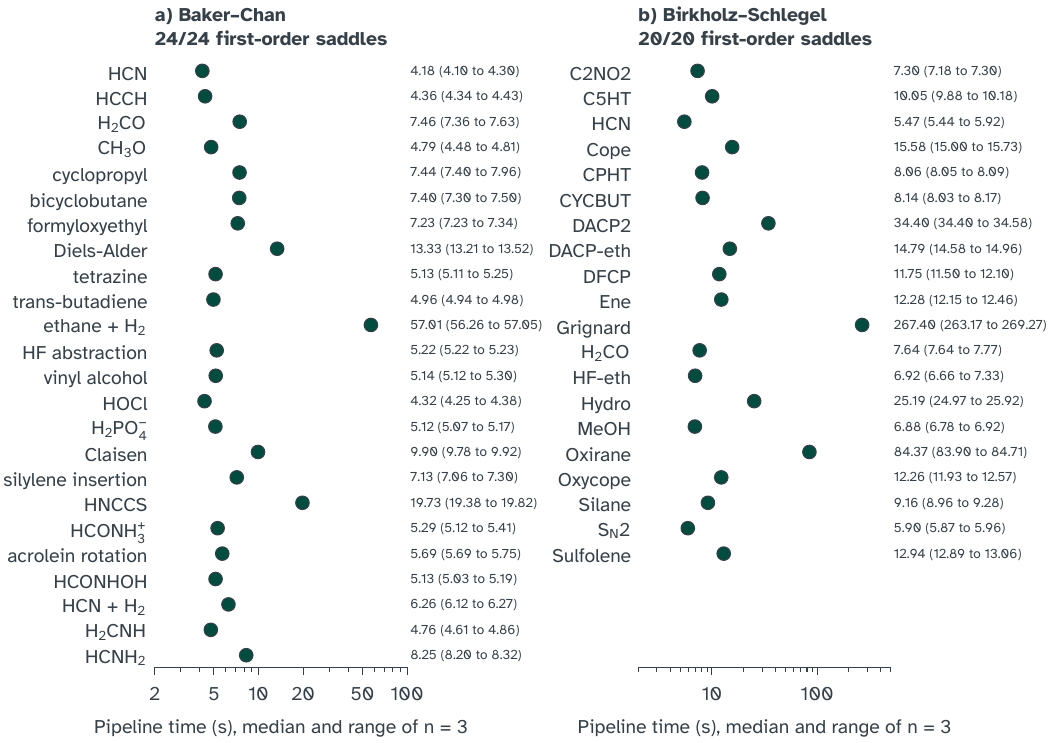}
\caption{\textbf{Complete elapsed time by reaction.}
Markers show medians across three exclusive-node repetitions; bars
span the minimum and maximum. The accompanying table gives the means
and sample standard deviations. Times include setup, band preparation,
saddle search, a finite-difference Hessian, and reference-energy checks.
The axis is logarithmic. The Birkholz limit of \recAllowance{} s
applies to the complete calculation.}
\label{fig:campaign-wall}
\end{figure}
The geometry that first reaches the force tolerance matches an earlier potential evaluation in every component.
The force and index audits of these panels stand apart from the library
test suite. Scalar Lean identities and sampled numerical checks establish
their explicit assumptions.
\subsection{Controlled removal of model and index controls}
\label{sec:si-method-ablations}
The separately measured ablation study comprises 128 calculations: two Baker variants
and four Birkholz variants. Each reaction reserves an exclusive Intel
node with four MPI ranks and two CPU slots per rank. All variants in a
set use identical inputs, potential packages, force criteria, validation
and resource settings. Birkholz uses the 900 s case deadline.
Each effective campaign differs from its control in exactly one field:
active posterior refits, environment pair labels, or dimer index recovery.
Removing active refits or environment labels affects both band preparation
and dimer search; their dimer starting geometries can therefore differ.

\begin{table}[H]
\centering
\small
\setlength{\tabcolsep}{4pt}
\begin{tabular}{llrrrrrr}
\toprule
Set & Removed control & Strict & Band & Dimer & Total & Sum / s & Max. / s\\
\midrule
Baker & none & 24/24 & 1215 & 0 & 1215 & 217.2 & 56.9\\
Baker & active refits & 24/24 & 1080 & 0 & 1080 & 159.1 & 15.8\\
Birkholz & none & 20/20 & 779 & 1067 & 1846 & 569.5 & 269.5\\
Birkholz & active refits & 19/20 & 1000 & 1138 & 2138$^{a}$ & 1436.8 & 900.1\\
Birkholz & environment labels & 20/20 & 793 & 1774 & 2567 & 664.9 & 283.6\\
Birkholz & index recovery & 17/20 & 779 & 194 & 973 & 369.4 & 108.4\\
\bottomrule
\end{tabular}
\caption{\textbf{Controlled method ablations.}
Band and dimer columns count energy and force evaluations during search;
their sum is the total. Refused curvature checks are included in the search count.
Independent validation and reference evaluations are reported separately. Elapsed times include setup, band preparation,
search and independent checks. Both sums and maxima include unsuccessful
cases. Each reaction and variant has one timing measurement.
The control runs the release revision and settings; its search counts
equal those of the calculations. Its elapsed times are a separate
measurement: 217.2 and 569.5 s summed, against 214.1 and 564.0 s in
one of the three passes.
$^{a}$A lower bound: DACP2 reaches the 900 s case deadline after 278
evaluations.}
\label{tab:method-ablations}
\end{table}

Disabling dimer index recovery leaves Grignard and oxirane with one
negative mode but five and three unresolved vibrational modes,
respectively. Sulfolene has two negative modes and no unresolved mode.
All three meet the force criterion: their maximum atomic forces are
0.04117970, 0.04043602 and 0.02475407 eV \AA{}\(^{-1}\).
The control passes all 20 reactions. The lower cost without recovery
therefore accompanies lower strict acceptance.

Removing active refits keeps all 24 Baker reactions and lowers the Baker
total from 1215 to 1080 evaluations and the longest reaction from 56.9 to
15.8 s. On Birkholz the same removal loses DACP2 to the 900 s deadline and
raises the evaluation count of the completed reactions. Removing
environment labels keeps 20/20 Birkholz reactions but raises the total
from 1846 to 2567 evaluations. The table reports single timing
observations.

\subsection{Comparator settings and computational resources}
\label{sec:si-comparator-settings}
GPRidge relaxes the Baker surrogate band with rgmin L-BFGS. The comparators use the algorithms listed in Table \ref{tab:si-comparator-settings}; the measured thread layout is stated below.
Each comparison reserves eight CPU cores on an exclusive node and uses the same compiled UMA-S-1.1 potential, charge, spin and independent validator. The execution layout follows the implementation: GPRidge uses four MPI processes and two BLAS threads per process, while eOn and the Python comparators use one process with eight threads available to the potential and numerical libraries. The latter set \texttt{OMP\_NUM\_THREADS}, \texttt{MKL\_NUM\_THREADS} and \texttt{OPENBLAS\_NUM\_THREADS} to eight. GPRidge costs use one of the three passes of release \texttt{v0.5.0}; each comparator contributes one measurement per reaction.

\begin{table}[tbp]
\caption{\label{tab:si-comparator-settings}Method-specific choices in the measured comparisons. Physical inputs and the force threshold are matched; each method uses its own optimization algorithm. These are fixed, reproducible configurations. The CatLearn thread control is the only comparator resource sweep reported here.}
\centering
\small
\begin{tabular}{p{0.18\linewidth}p{0.70\linewidth}}
Method & Effective settings\\
\hline
GPRidge & Publication argument files; inverse distances, Cholesky fitting, L-BFGS hyperparameter optimization, compensated posterior products; four MPI processes\\
eOn NEB, CI-NEB, OCI-NEB & 8 intermediate images, SIDPP initialization, L-BFGS, 0.1 \AA{} maximum move, energy-weighted springs of 1 to 10 eV \AA{}\(^{-2}\), maximum iteration count 1000\\
CatLearn ML-NEB & 20 images, IDPP interpolation, improved tangent, default acq\textsubscript{5} acquisition and MDMin relaxation; 1000 acquisition iterations and 400 surrogate steps allowed\\
ML-FSM + dimer & Redundant-coordinate interpolation, 18 minimum nodes, 100 interpolation points; the L-BFGS-B recipe, then ASE dimer refinement\\
Sella & Internal coordinates, first-order saddle search, package default initial mode; the Birkholz band maximum\\
eOn dimer & The Birkholz band maximum and tangent; L-BFGS, with shared preparation charged separately\\
\end{tabular}
\end{table}

CatLearn supports an MPI wrapper that fits on rank zero and broadcasts the model. The measured CatLearn comparison uses one process, so its fit and factorization run in that process. Threaded BLAS and the potential remain available in the measured one-process configuration. The four-reaction one-thread control in the article checks the sensitivity to the thread setting. Thus, the elapsed-time comparison includes the parallel capabilities supplied by each implementation. The results describe these pinned implementations and their documented configurations.

\begin{table}[htbp]
\caption{\label{tab:ase-cineb-baker}ASE climbing-image band on Baker--Chan, run by the reproduction workflow on the compiled potential packages of the other comparator arms. FIRE relaxes 18 interior images with the improved tangent, a spring of 5 eV \AA{}\(^{-2}\), and an image-dependent pair-potential initial path. The band stops at the first iteration where the largest atomic force on the climbing image is at or below 0.0514 eV \AA{}\(^{-1}\); the counts include one evaluation of each endpoint. GPRidge counts are the search evaluations in Table \ref{tab:fixed-baker}.}
\centering
\small
\begin{tabular}{lrr}
Case & ASE evaluations & GPRidge evaluations\\
\hline
HCN & 740 & 25\\
HCCH & 974 & 32\\
H2CO & 668 & 80\\
CH3O & 704 & 31\\
cyclopropyl & 920 & 53\\
bicyclobutane & 1154 & 46\\
formyloxyethyl & 938 & 48\\
Diels-Alder & 776 & 46\\
tetrazine & 992 & 28\\
trans-butadiene & 38 & 22\\
Ethane H\(_2\) abstraction & 3332 & 196\\
HF abstraction & 1010 & 33\\
vinyl alcohol & 740 & 39\\
HOCl & 758 & 30\\
H2PO4 anion & 1118 & 34\\
Claisen & 1712 & 44\\
silylene insertion & 740 & 41\\
HNCCS & 1046 & 134\\
HCONH3 cation & 956 & 36\\
acrolein rotation & 938 & 35\\
HCONHOH & 1028 & 32\\
HCN + H2 & 992 & 45\\
H2CNH & 830 & 33\\
HCNH2 & 1082 & 72\\
\end{tabular}
\end{table}

\section{Periodic example: Pt(111) heptamer and island}
\label{sec:si-periodic}
The two problems are \texttt{neb/Pt-heptamer} and \texttt{double-ended-saddle-search/pt-island} of OptBench \cite{chillBenchmarksCharacterizationMinima2014} in its modernized form \cite{goswamiOptBenchReplay2026}, taken from the archived \texttt{con} geometries of that repository. The slab has 343 Pt atoms in a cell of \(19.21\times19.01\) \AA{} in the surface plane; the 168 atoms of the bottom layers are fixed and 175 are free. The interaction is the energy-shifted Morse potential of the collection, \(D_e=0.7102\) eV, \(a=1.6047\) \AA{}\(^{-1}\), \(r_e=2.897\) \AA{}, cut at \(9.5\) \AA{}. The heptamer problem supplies 13 reactant--product pairs, a band of eight interior images and the criterion \(\lVert F\rVert_2\le0.01\) eV \AA{}\(^{-1}\) over the free atoms, with a cap of 10000 evaluations. The island problem supplies one reactant with 59 products, a band of three interior images and \(\lVert F\rVert_2\le0.001\) eV \AA{}\(^{-1}\). One island start reaches 18011 evaluations without meeting the force criterion. The evaluation counts are those of the program's own counter and include the two endpoints.
\subsection{Settings}
\label{sec:org979a4b7}

Both searches run through the \texttt{gprn} driver with a linear initial band, a spring constant of 5 eV \AA{}\(^{-2}\), a step limit of \(0.2\) \AA{}, the climbing image active from the first iteration, and one thread. The control selects the standard band mode, which relaxes the band by L-BFGS on the measured forces and evaluates every image at every step. The local pair-baseline band selects \path{--nebmode gp_local} with its defaults: local atoms by a tenth of the largest endpoint displacement, a shell radius of \(3\) \AA{}, a residual window of 16 measurements with noise \(10^{-3}\), a local trust radius of \(0.3\) \AA{} and a substrate trust radius of \(0.5\) \AA{}, at most 20 band steps per round with the round tolerance at half the convergence force, a refit cadence of four outer iterations, a pair cutoff of \(9.5\) \AA{} with the knot spacing rule \(h=0.05\,\text{\AA}\,(\varepsilon/0.01\,\mathrm{eV}\,\text{\AA}^{-1})^{1/4}\), which is \(0.050\) \AA{} on the heptamer and \(0.028\) \AA{} on the island, a tail split at \(7\) \AA{}, the local phase on, the band memory kept until a refit, and the step safeguard on. One image is measured per round until the eighth outer iteration, after which the batch grows by one every eight iterations to at most three. Section \ref{sec:si-local-band-options} lists the options.
\subsection{Outcomes on every start}
\label{sec:orgbbee0a4}

\begin{figure}[tbp]
\centering
\includegraphics[width=0.98\textwidth]{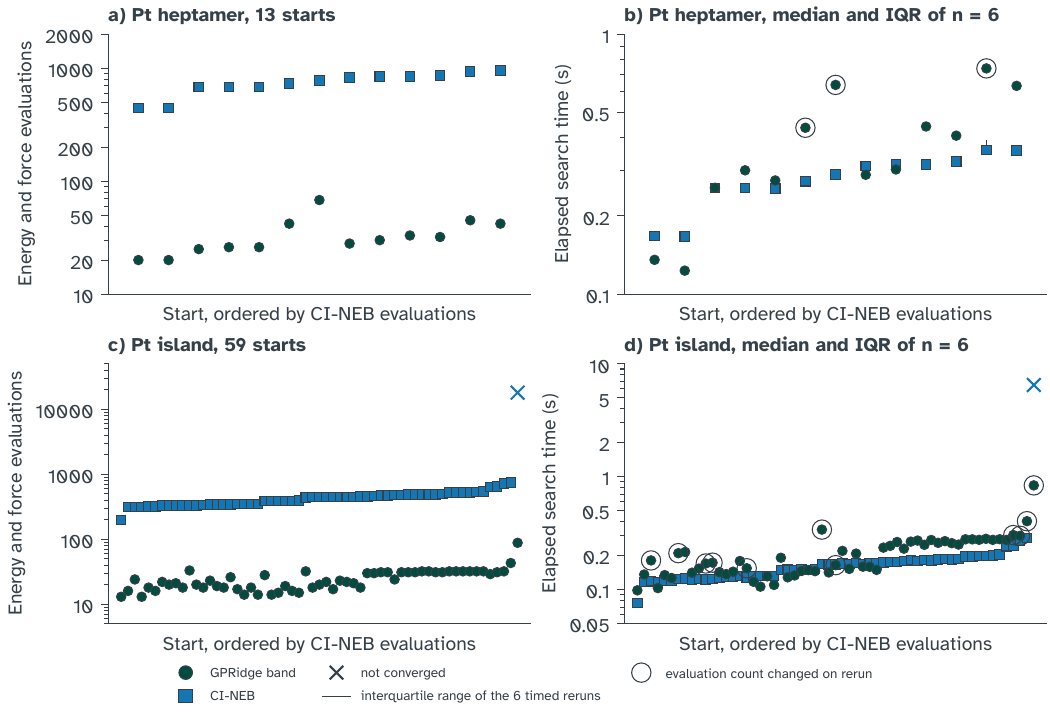}
\caption{\label{fig:pt-cells}Every start of the two Pt(111) panels, ordered by the CI-NEB evaluations. Left: campaign energy-and-force evaluations. Right: median time inside the search from six interleaved fresh-process reruns on one pinned core, after one warm-up pass; bars give the interquartile range. Circles are the local pair-baseline band, squares CI-NEB; the cross marks the island start at which CI-NEB reaches 18011 evaluations. Rings mark starts whose rerun count differs from the campaign count.}
\end{figure}

Figure \ref{fig:pt-cells} pairs the starts of both panels. The band converges all 13 heptamer and all 59 island starts; the control converges 13 and 58, and one island start stops at 18011 evaluations, short of the force criterion. The band uses fewer evaluations on every start converged by both. The evaluation counts, and the campaign search times summed below, are the release campaign on Intel Xeon Gold 6248R nodes. Summing those campaign search times of the converged starts gives 24.93 s for the heptamer band against 20.68 s for CI-NEB, and 85.83 s against 76.80 s on the island. The climbing-image energies of the matched starts, heptamer pair 1 and island product 0, differ by \(3.07\times10^{-5}\) eV and \(1.51\times10^{-7}\) eV. The timing study in Figure \ref{fig:pt-cells} uses six interleaved repetitions per start after a warm-up and reports time inside the search. It is separate from the campaign timing totals and the matched-start evaluation counts.

\subsection{Multi-fidelity scale}
\label{sec:si-multifidelity}
On the 13 Pt(111) heptamer starts, and with a budget of 36 evaluations, a Kennedy--O'Hagan scaled Lennard-Jones prior \cite{kennedyPredictingOutput2000} lowers the best force on 11 of the 13 starts.
The median of that force is 0.40 eV/\AA{} against 1.25 eV/\AA{} for fixed images, and the energy lies within a median 0.44 eV of the \texttt{gp\_local} barrier.
That \texttt{gp\_local} search certifies its barrier in a median 30 evaluations, which the scaled prior does not.
Nakajima's \(\Delta\)-HGPR, a difference potential for dynamics rather than a saddle search, is also written as a radial term plus a many-body Gaussian-process correction and uses the ensemble force standard deviation to choose the next reference evaluation \cite{nakajimaDeltaHGPR2026}.
\subsection{Options of the local pair-baseline band}
\label{sec:si-local-band-options}
\begin{itemize}
\item \path{--nebmode gp_local} selects the mode; \path{--gp-local-fraction} (0.1) and \path{--gp-local-radius} (0) choose the local atoms by endpoint displacement and by a neighbor radius; \path{--gp-local-grow-radius} (3 \AA{}) is the shell that moves with them in the local phase, widened by the same amount after three stalled rounds, twice at most.
\item \path{--gp-local-window} (16) and \path{--gp-local-noise} (\(10^{-3}\)) fix the residual model; \path{--gp-local-refit-every} (4) is the outer cadence at which the baseline is refitted, if it has new data, and the residual rebuilt; \path{--gp-local-residual} (true) keeps the residual on top of the baseline.
\item \path{--gp-local-pair-cutoff} (9.5 \AA{}) and \path{--gp-local-pair-spacing} (0, which applies the fourth-root rule) define the spline; \path{--gp-local-pair-split} (7 \AA{}) is the distance beyond which pairs are summed once per round and held as a linear model, with \path{--gp-local-pair-tail-refresh} (0) an optional in-round resum after that displacement.
\item \path{--gp-local-inner-max} (20) and \path{--gp-local-inner-tol} (0.5) bound a round; \path{--gp-local-trust} (0.3 \AA{}) and \path{--gp-local-far-trust} (0.5 \AA{}) are the trust spheres of the local atoms and of the substrate about the last measurement; \path{--gp-local-far-refresh} (0.05 \AA{}) refreshes an image's own substrate factor once it moves that far.
\item \path{--gp-local-phase} (true) opens each round in the local phase; \path{--gp-local-keep-memory} (true) keeps the band quasi-Newton memory until a refit; \path{--gp-local-safeguard} (true) halves a step that raises the largest surrogate band force by more than half.
\item \path{--gp-local-batch} (1), \path{--gp-local-batch-growth} (8) and \path{--gp-local-batch-max} (3) set how many unmeasured images are evaluated per round and how that number grows with the outer iteration; \path{--gp-local-initial-sweep} (true) evaluates every image of the initial band before the first fit.
\end{itemize}

\section{Search control and other saddle software}
\label{sec:si-control-and-software}
At each outer iteration, a priority planner selects measurements for model support, rejected-step correction, band-drift checks, or force and curvature acceptance. Required measurements have priority; posterior acquisition can replace only requests marked as eligible. Each calculator result joins the training set before the GP is updated. At band-to-dimer handoff, the training set and fitted GP move with the band maximum and tangent, so refinement continues from the measured path state. Posterior proposals guide relaxation; measured forces and an independent Hessian establish physical convergence and saddle index. A second negative mode triggers an order-one escape before refinement resumes.

pysisyphus includes NEB and growing-string paths followed by transition-state optimization \cite{steinmetzerPysisyphusExploringPotential2021,petersGrowingString2004,zimmermanReliableTransition2013,zimmermanSingleEnded2015}. PyDMF, dmf-g16, autodE, and KinBot are separate double-ended and automated stationary-point searches \cite{kodaLocatingTransition2024,kodaPyDMF2026,kodaDmfG162026,youngAutodeReaction2021,vandevijverKinBot2020}. Kendall and Z{\'a}dor benchmark UMA on gas-phase chemical kinetics \cite{kendallBenchmarkingUMA2026}. Our method comparisons configure named algorithms on shared inputs and acceptance criteria, enabling method-level comparisons across packages. Transition1x provides molecular structures for training reactive machine-learning potentials, and React-OT generates transition-state structures from paired reactants and products \cite{schreinerTransition1xDatasetBuilding2022,grambowReactantsProducts2020,duanOptimalTransportGenerating2025}. Their reported outputs are structures; a force-call comparison requires iterative calculator-driven searches under the same protocol \cite{wanderCatTSunami2025,zhaoHarnessingMachine2025,liTransitionStateSearching2025,yuanAnalyticalHessian2024,karasawaColabReaction2025,kurylaEfficientExploration2025,kovacsMACEOFF2025,batatiaFoundationModel2025,batatiaMACE2022,zhangReactiveMachine2024}.

\section{Search controls and supplementary comparisons}
\label{sec:si-moved}
\subsection{Baker driver and comparator sources}
\label{sec:si-driver-sources}
The Baker climbing-image driver takes the reactant and the product, selects the UMA potential on the omol task, uses 18 images, sets the converged force to \(0.0514221\) eV/\AA{} and turns the climbing image on.
CatLearn ML-NEB is version 0.6.2 at revision \texttt{6a307f32dbef18da9cd848f632d082b8e4f117b1}; Sella is version 2.6.0.
The eOn comparison uses tag \texttt{oci-neb-correction-2026.10.29}, revision \href{https://github.com/TheochemUI/eOn/commit/8356aa8a23658be7904b1a600e228aadad917114}{\texttt{8356aa8a2365}} \cite{goswamiEOnSoftware2026}.
The eOn documentation is at \url{https://eondocs.org}.
The freezing-string geometries are the Marks and Gomes deposit.

\subsection{Tree-structured Parzen estimator}
Acquisition uses a tree-structured Parzen estimator \cite{bergstraAlgorithmsHyperparameter2011}.
It selects only unobserved interior images whose posterior variance exceeds the noise floor.
Besides planner-approved replacements, it can select an image when the force is still above the tolerance and the planner is idle or has selected a rejected proposal.
Its force-accuracy utility is
\begin{equation}
u_i=z(f_i)+\beta_t z(v_i^F),\qquad
\beta_t=\sqrt{\max\!\left(0,2\log\frac{m t^3\pi^2}{6}\right)},
\label{eq:main-acquisition-utility}
\end{equation}
The implementation sets \(\delta_t=1/t\), so \(t^2/\delta_t=t^3\).
This is a growing exploration weight inspired by GP-UCB, not a fixed-confidence regret guarantee for the TPE search.
Here \(f_i\) is the predicted NEB-force norm, \(v_i^F\) is posterior force variance, \(m\) is the number of interior images, and \(t\) is the acquisition round \cite{srinivasInformationTheoreticRegret2012}.
The utility standardizes the force and force-variance channels over the band; once the saddle energy is known to within its confidence threshold, standardized energy also contributes.
The estimator divides eligible images into upper and lower utility groups and selects the largest log-density ratio.
TPE selected 2 of 970 Baker acquisitions; the planner chose the rest.
On the Birkholz bands it selected none of 526 acquisitions.
The Supporting Information gives the eligibility test.
Support batching can add several images when the band extends beyond the kernel support.

\subsection{Rows that enter the Gram}
\begin{figure}[tbp]
\centering
\providecolor{RuhiTeal}{HTML}{004D40}
\providecolor{RuhiCoral}{HTML}{FF655D}
\providecolor{RuhiInk}{HTML}{1C2421}
\begin{tikzpicture}
\begin{axis}[
  width=0.96\linewidth,
  height=5.3cm,
  tick label style={font=\small\sffamily},
  label style={font=\small\sffamily},
  axis line style={RuhiInk},
  tick style={RuhiInk},
  axis lines=left,
  clip=false,
  xlabel={Separation $s/\ell$},
  ylabel={Relative conditional variance},
  xmin=0, xmax=3, ymin=0, ymax=1.08,
  every axis plot/.append style={line width=1.1pt}
]
\addplot[RuhiTeal, domain=0:3, samples=160] {1-exp(-2*x^2)};
\addplot[RuhiCoral, dashed, domain=0:3, samples=2] {0.04000000000000001};
\draw[RuhiCoral, line width=0.95pt] (axis cs:0.1,0) -- (axis cs:0.1,0.72);
\node[font=\footnotesize\sffamily, text=RuhiCoral, anchor=west]
  at (axis cs:0.18,0.62) {$d_{\min}$, $\sigma_{\mathrm{rel}}=0.2$};
\node[font=\footnotesize\sffamily, text=RuhiCoral, anchor=west]
  at (axis cs:1.55,0.11000000000000001) {$\sigma_{\mathrm{rel}}^{2}$};
\node[font=\footnotesize\sffamily, anchor=north west] at (rel axis cs:0,1) {(a)};
\end{axis}
\end{tikzpicture}
\par\vspace{0.9em}
\begin{tikzpicture}
\begin{axis}[
  width=0.68\linewidth,
  height=4.4cm,
  scale only axis,
  tick label style={font=\small\sffamily},
  label style={font=\small\sffamily},
  axis line style={RuhiInk},
  tick style={RuhiInk},
  axis lines=left,
  xbar,
  bar width=11pt,
  xmin=0, xmax=175,
  ymin=-0.55, ymax=1.55,
  ytick={0,1},
  yticklabels={Birkholz--Schlegel, Baker--Chan},
  xlabel={Peak observation archive (geometries)},
  nodes near coords,
  point meta=x,
  every node near coord/.append style={font=\footnotesize\sffamily, anchor=west, /pgf/number format/fixed},
  clip=false
]
\addplot[fill=RuhiTeal, draw=none] coordinates {(70,0) (133,1)};
\draw[RuhiInk, dashed, line width=0.8pt]
  (axis cs:128,-0.45) -- (axis cs:128,1.35)
  node[pos=1, anchor=south, font=\footnotesize\sffamily, inner sep=1pt] {128};
\node[font=\footnotesize\sffamily, anchor=south west] at (rel axis cs:0,1) {(b)};
\end{axis}
\end{tikzpicture}
\caption{Rows that enter the Gram, and the training-set ceiling.
(a) Relative conditional variance of a second squared-exponential observation.
A candidate closer than \(d_{\min}\) stays in the training set and out of the factorization; the mark is that distance at a relative noise of \(0.2\).
(b) Largest training set on each panel.
The dashed line is the cap of 128 geometries.
The complete observation archive reaches 133 geometries on Baker, in the ethane \(\mathrm{H}_{2}\) abstraction, where 120 of the 136 cap events fall.
The ceiling applies to the geometries selected for the factor; observations outside that active subset remain in the archive.
Birkholz reaches 70, below the cap.}
\label{fig:active}
\end{figure}
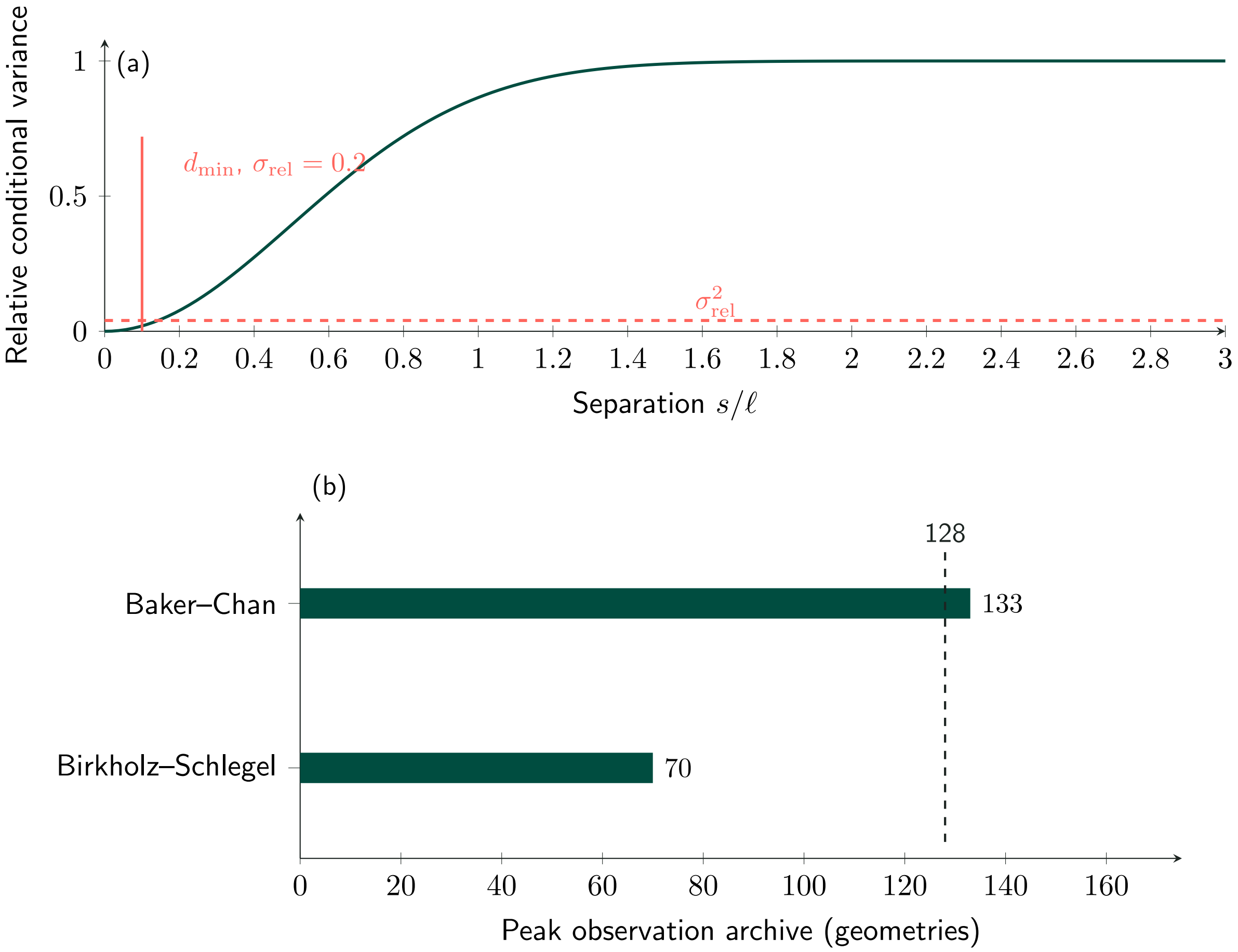

\subsection{Evidence-driven search transitions}
\begin{figure}[tbp]
\centering
\resizebox{0.98\textwidth}{!}{\begin{tikzpicture}[
  font=\small\sffamily,
  box/.style={draw=RuhiTeal, fill=RuhiTeal!7, rounded corners=3pt,
              minimum width=3.15cm, minimum height=0.78cm, align=center,
              line width=0.8pt},
  state/.style={box, fill=white},
  gate/.style={draw=RuhiTeal, fill=RuhiYellow!30, diamond, aspect=2.5,
               minimum width=3.4cm, minimum height=1.0cm, align=center,
               line width=0.8pt},
  accept/.style={box, fill=RuhiTeal!18},
  recover/.style={box, fill=RuhiCoral!18, draw=RuhiCoral},
  arrow/.style={-{Stealth[length=5pt]}, RuhiTeal, line width=0.8pt},
  label/.style={font=\scriptsize\sffamily, fill=white, inner sep=1pt},
  node distance=0.72cm and 1.05cm
]
\node[state] (archive) {Training set $\mathcal D$\\measured energies and gradients};
\node[box, below=of archive] (fit) {Fit selected GP\\representation};
\node[box, below=of fit] (relax) {Relax active band or dimer\\on posterior mean};
\node[gate, below=of relax] (due) {Required physical or\\coverage check due?};
\node[box, below left=of due] (required) {Select highest-priority\\required geometry};
\node[box, below right=of due] (acquire) {Select by the active\\stage acquisition rule};
\node[box, below=1.0cm of due] (oracle) {Calculator $\mathcal C(x)=(E,g)$};
\node[state, below=of oracle] (append) {Append measured $(x,E,g)$\\to $\mathcal D$};
\node[gate, below=of append] (force) {Measured force\\within tolerance?};
\node[accept, below left=of force] (path) {Path-only plan:\\accept band maximum};
\node[box, below right=of force] (curv) {Band-to-dimer handoff\\maximum, tangent, GP, $\mathcal D$};
\node[gate, below=of curv] (index) {Measured curvature\\has index one?};
\node[recover, below left=of index] (escape) {Extra negative mode:\\escape and resume};
\node[accept, below right=of index] (check) {Independent Hessian\\and accepted saddle};
\node[state, right=1.4cm of index] (unresolved) {Weak mode remains?\\refine or return unresolved};

\draw[arrow] (archive) -- (fit);
\draw[arrow] (fit) -- (relax);
\draw[arrow] (relax) -- (due);
\draw[arrow] (due) -- node[label, above left] {yes} (required);
\draw[arrow] (due) -- node[label, above right] {no} (acquire);
\draw[arrow] (required.south) |- (oracle.west);
\draw[arrow] (acquire.south) |- (oracle.east);
\draw[arrow] (oracle) -- (append);
\draw[arrow] (append) -- (force);
\draw[arrow] (append.west) -| (archive.west);
\draw[arrow] (force.west) -- node[label, above] {no} ++(-1.05,0) |- (archive.west);
\draw[arrow] (force) -- node[label, above left] {yes: band-only} (path);
\draw[arrow] (force) -- node[label, above right] {yes: coupled plan} (curv);
\draw[arrow] (curv) -- (index);
\draw[arrow] (index) -- node[label, above left] {extra mode} (escape);
\draw[arrow] (escape.west) -| (relax.west);
\draw[arrow] (index) -- node[label, above right] {index one} (check);
\draw[arrow] (index) -- node[label, above] {weak / unresolved} (unresolved);
\end{tikzpicture}}
\caption{\textbf{Evidence-driven search control.} The calculator supplies measured energies and gradients; a selectable GP representation guides inner relaxation. The controller spends evaluations on required coverage, force, and curvature checks before optional acquisition. A band-to-dimer handoff carries the maximum, tangent, fitted state, and full training set. Higher-index recovery returns to local search. Benchmark acceptance requires measured force and an independently resolved index-one Hessian.}
\label{fig:controller}
\end{figure}
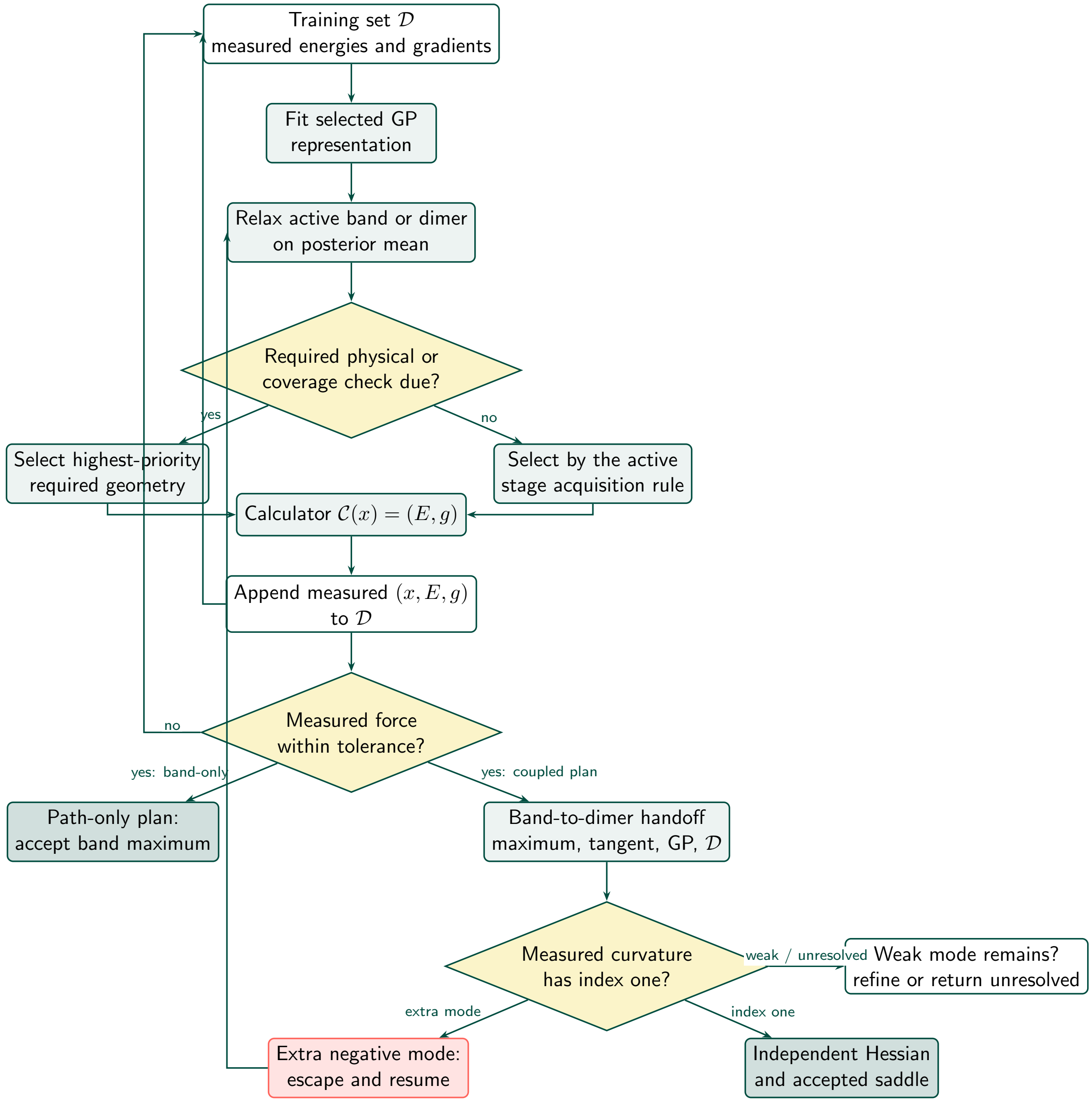

Figure \ref{fig:controller} summarizes these transitions.

\subsection{Calculation settings}
Search costs are the potential evaluations during initialization, band relaxation, and dimer refinement. The independent Hessian and the reference-geometry energies are reported on their own. Reported elapsed times include setup, search, and validation for each reaction. The primary calculations use four MPI ranks, two BLAS threads and one kernel thread, with ScaLAPACK, an inverse-distance covariance, a screened nuclear-repulsion mean, a log-length-scale prior, Cholesky fitting, and compensated binary64 posterior products. The Birkholz calculations use a 900 s limit.

\subsection{Software and workflow integration}
\label{sec:software}
GPRidge is an MIT-licensed C++ library with C, Fortran, Python, and ASE interfaces. Its \texttt{gprn} and \texttt{gprd} drivers read standard CON structures. A calculator adapter returns the potential energy and Cartesian gradient for each requested geometry; the controller uses this contract for electronic-structure, classical, and learned potentials. The band and dimer commands, including the driver flags, are in the Supporting Information.
The stage plan either returns a force-converged path or passes the fitted Gaussian process and the measured energies and forces to the dimer.
Structures, energies, and forces can restart a calculation.
Snakemake can schedule the same command for a set of reactions. The molecular calculations use a joint inverse-distance GP; periodic Pt(111) uses a pair-additive spline baseline with a local GP residual. chemparseplot and rgpycrumbs analyze the calculations \cite{goswamiChemparseplot2026,goswamiRgpycrumbs2026,goswamiTwodimensionalRMSDProjections2026}.

\subsection{Stable-basis posterior}
The training factorization in the article is
\begin{equation}
C=K+N+\delta I=LL^{\mathsf T},\qquad Lb=y-\mathsf{Obs}\,m,\qquad L^{\mathsf T}\alpha=b.
\label{si-eq:training-cholesky}
\end{equation}

The library also provides an optional stable-basis posterior for this joint energy--gradient system. For the noise-free training kernel, write \(K=\Psi\operatorname{diag}(s_i^2)\Psi^{\mathsf T}\) with orthonormal columns in \(\Psi\). For residual targets \(y\) and isotropic noise \(\sigma^2 I\), the noisy solve separates the included kernel modes from the orthogonal residual:
\begin{equation}
\alpha=\Psi\operatorname{diag}\!\left((s_i^2+\sigma^2)^{-1}\right)\Psi^{\mathsf T}y+\sigma^{-2}(I-\Psi\Psi^{\mathsf T})y.
\label{eq:stable-dual-solve}
\end{equation}
This identity is exact when \(\Psi\) spans the nonzero range of \(K\); dropping positive eigenvalues instead inverts the corresponding truncated covariance. The dual Hilbert--Schmidt SVD (HS-SVD) obtains this basis from the assembled training kernel in the training dimension, avoiding an explicit tensor-product Mercer feature matrix \cite{fasshauerStableEvaluationGaussian2012,fornbergStableComputationsGaussian2011}. With this option enabled, the basis supplies posterior predictions and analytic dimer Hessian-vector products; hyperparameter fitting uses a separate Cholesky reduction for the noisy log determinant and quadratic form. The basis is orthonormal, while the noisy inverse remains sensitive to small eigenvalues; the implementation's rank and noise policies are detailed in the Supporting Information. The molecular release calculations reported here disable dual HS-SVD and use the Cholesky path in Eq. \ref{si-eq:training-cholesky}. The reported call and time figures belong to these pinned Cholesky configurations.

\subsection{CatLearn work between evaluations}
The implementations perform different work between potential evaluations. The pinned CatLearn implementation obtains surrogate forces by finite differences of predicted energies, requiring \(6N\) displaced predictions per image with \(N\) free atoms, plus its energy prediction. GPRidge evaluates analytic posterior derivatives and distributes GP calculations. CatLearn also fits its GP on rank zero. The Supporting Information documents these implementation differences; the aggregate timings combine their contributions. The one-thread CatLearn control preserves the large elapsed-time difference.

\subsection{Per-reaction ratios and performance profiles}
\begin{figure}[tbp]
\centering
\includegraphics[width=0.98\textwidth]{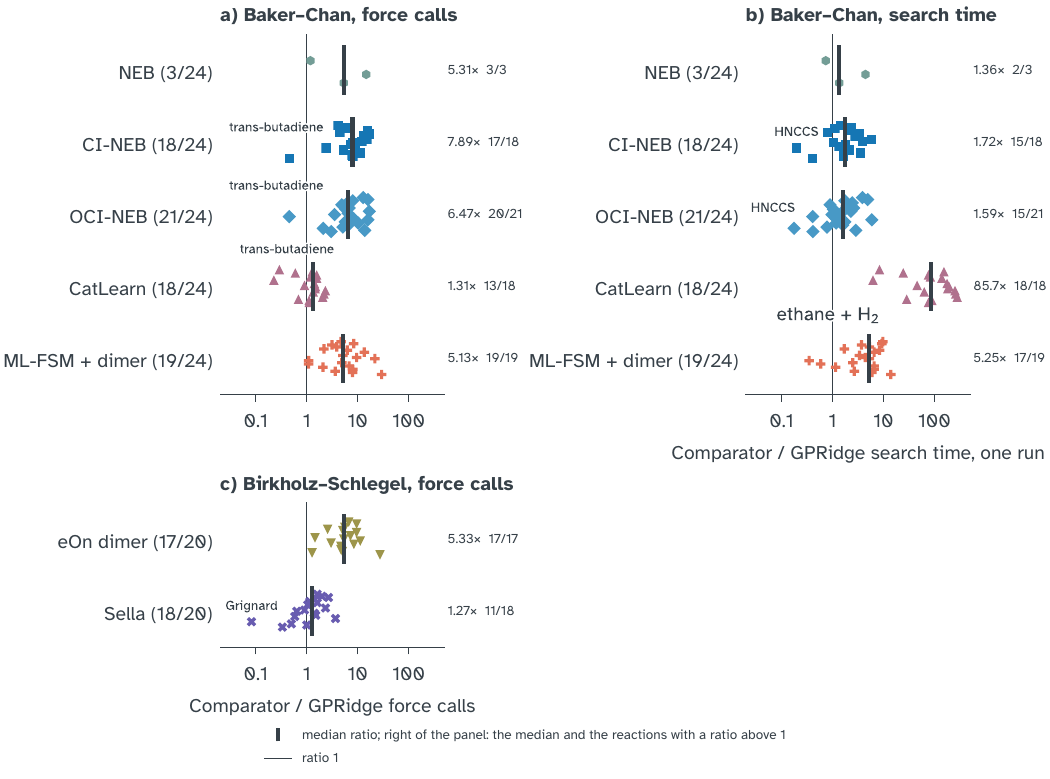}
\caption{\label{fig:paired-ratios}Per-reaction cost ratios, labelled by workflow. The upper panels are the double-ended Baker--Chan bands, in force calls (a) and search time (b); the lower panel (c) is the single-ended Birkholz--Schlegel dimer searches, in force calls. This profile omits Birkholz elapsed times; the single-ended stage times are compared in Figure \ref{fig:method-savings}(d) using the implementation layouts in Table \ref{tab:si-comparator-settings}. Each point is one reaction at comparator/GPRidge cost on a shared logarithmic axis; points right of 1 favour GPRidge. Vertical bars mark the median, and the right-hand columns give the median and the number of reactions on which GPRidge is cheaper. Labels name the reaction with the lowest ratio when it is below 0.5. Method labels give full-panel acceptance. Costs, accepted sets and exclusions are those of Figure \ref{fig:method-savings}. The Supporting Information shows the GPRidge saddle of every labelled reaction, with its bonding changes.}
\end{figure}

The Dolan--Moré profiles \cite{dolanBenchmarkingOptimization2002} include unsuccessful outcomes in the main comparison. Figure \ref{fig:paired-ratios} shows the same pairs reaction by reaction. In evaluations, GPRidge is cheaper on 17 of 18 CI-NEB, 20 of 21 OCI-NEB, 19 of 19 ML-FSM and 17 of 17 eOn dimer reactions, with medians of 7.9, 6.5, 5.1 and 5.3. Against Sella, the median evaluation ratio is 1.3; the summed totals favour Sella because Grignard alone accounts for most GPRidge evaluations. The Supporting Information tabulates every per-reaction ratio (\emph{Freezing-string and Sella comparisons}). Per-reaction costs and cumulative acceptance against absolute budgets are in the Supporting Information.

\begin{figure}[tbp]
\centering
\includegraphics[width=0.98\textwidth]{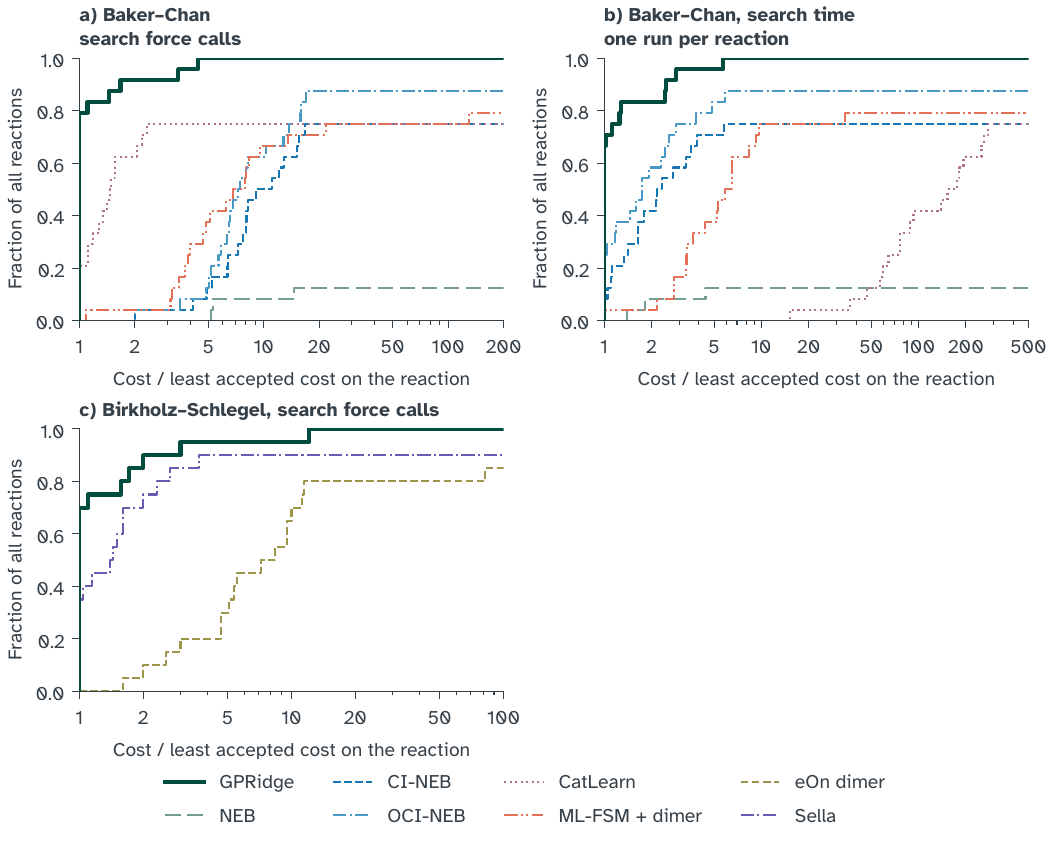}
\caption{\label{fig:performance-profiles}Dolan--Moré performance profiles, labelled by workflow. Panels (a,b) are the double-ended Baker--Chan bands. Panel (c) is the single-ended Birkholz--Schlegel dimer searches. (a,c) are search force calls and (b) is elapsed search time. This profile omits Birkholz elapsed times; the single-ended stage times are compared in Figure \ref{fig:method-savings}(d) using the implementation layouts in Table \ref{tab:si-comparator-settings}. The horizontal coordinate is cost divided by the least accepted cost on the same reaction. Failed checks are assigned infinite cost and remain in the denominator. GPRidge attains the lowest force-call cost on 19/24 double-ended reactions and 14/20 single-ended reactions, including ties. Counts include initialization; time excludes independent validation and the shared seed band of the single-ended searches. Comparator costs use one measurement per reaction; GPRidge uses one of the three passes.}
\end{figure}

Figure \ref{fig:performance-profiles} compares each search with the least costly accepted result for the same reaction. The left edge gives the fraction of reactions on which a method is least costly, including ties; the final height gives the fraction accepted. Every failed reaction remains in the denominator. GPRidge has the lowest evaluation count on 19/24 Baker reactions and 14/20 Birkholz reactions. The Baker time profile shows a wider separation from CatLearn than the evaluation-count profile. The Supporting Information gives the Pareto comparison on common successes and the absolute-budget curves.

\subsection{Pt(111) wall-clock times}
The search times in the article are the campaign times of the converged starts, summed to 24.93 s for the heptamer band against 20.68 s for CI-NEB, and to 85.83 s against 76.80 s on the island.
The host is an Intel Xeon Gold 6248R.
These totals cover every converged start. The separate six-repetition study in Figure \ref{fig:pt-cells} compares individual starts and records search time and whole-process time separately; its medians are not panel totals.
The climbing-image energies of those starts differ by \(3.07\times10^{-5}\) eV and \(1.51\times10^{-7}\) eV.

\subsection{Where the calculation time goes}
\label{sec:scaling}
The primary Grignard and Oxirane runs take 263.173 s and 84.373 s at four MPI ranks with two CPU slots per rank. Potential evaluations account for 143.6 s and 8.6 s. Each dimer calculation has an earlier full Cartesian stencil that reports unresolved weak modes, followed by the final certified stencil. The earlier stencil takes 27.1 s for Grignard and 10.1 s for Oxirane. That stencil is search work, and the final stencil is independent validation. GP mode extraction takes 0.363 s for Grignard and 0.023 s for Oxirane. The remaining dimer intervals are 12.6 and 29.7 s.

One fixed 5,376-row Grignard fit falls from 197.12 s on one core to 12.91 s on 96 cores across two nodes (15.27-fold speedup; 15.9\% efficiency). Its strong- and dense weak-scaling measurements, including MPI diagnostics and repeated runs, are reported in the Supporting Information.

\begin{figure}[H]
\centering
\includegraphics[width=\textwidth]{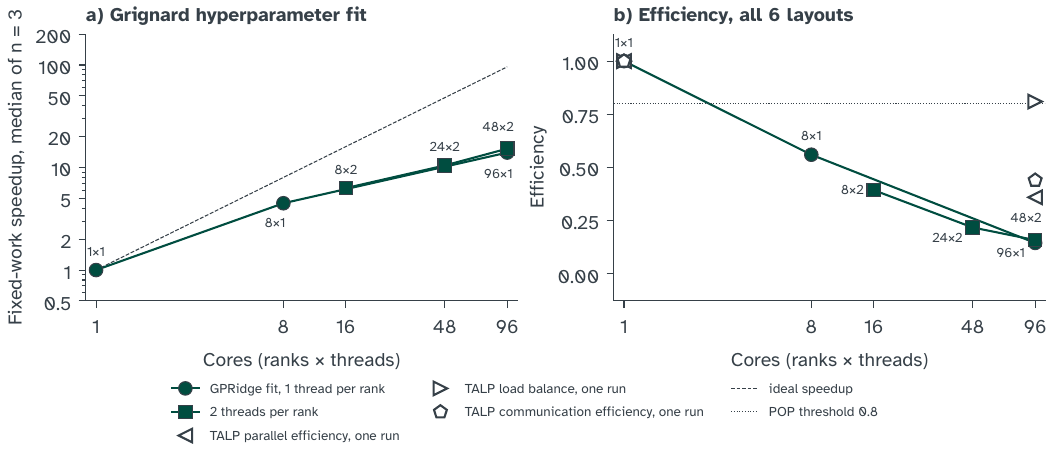}
\caption{\textbf{Parallel efficiency of a GP fit.}
A complete hyperparameter fit of 48 Grignard observations with GPRidge 0.5.0,
a 5376-row covariance and reproducible arithmetic. Every layout performs
47 objective evaluations and 94 gradient evaluations for the two covariance
components, with no potential evaluations.
Left: speedup from the median of three unprofiled repetitions against the
single core; layouts with the same core count are shown separately.
Right: speedup divided by core count, together with the MPI parallel
efficiency, load balance and communication efficiency from separate TALP
(Tracking Application Live Performance) invocations, covering MPI
initialization to finalization of the native replay. TALP reports the
Performance Optimisation and Productivity (POP) metrics; see
\cite{lopezTALPLightweightTool2021}. TALP factors are measured only for the
one-core and 96-process layouts. The dotted line marks 0.8. The two 96-core
layouts span two nodes.}
\label{fig:pop-factors}
\end{figure}

Figure \ref{fig:pop-factors} shows the repeated timings and parallel efficiency for this fit. It measures the GP fitting stage without potential evaluations; the full-search timings above include preparation, search and independent validation.

\subsection{Handoff from the band to the dimer}
A path-only plan returns a climbing image only after measuring its force at the current geometry; independent Hessian analysis then checks its index. A path-to-saddle plan transfers four items to the dimer: the band maximum, neighboring-image tangent, fitted GP state, and complete training set. The dimer measures physical force and curvature at its candidate, then adds probes until weak modes are resolved. A second negative mode triggers an order-one escape. The search first attempts a GP-based escape initialized with measured curvature rows; if that fails, it follows a measured downhill mode. When the full stencil identifies a weak second mode, partitioned RFO steps start from the measured Hessian and use each measured gradient to update a local Bofill Hessian; only the final geometry enters the training set \cite{banerjeeSearchStationary1985,bakerAlgorithmLocationTransition1986,bofillUpdatedHessian1994}. The candidate returns to local refinement and receives new physical measurements. Acceptance requires a measured maximum atomic force below tolerance and an independent Hessian with exactly one negative mode and positive curvature for every remaining physical mode. The Supporting Information gives these transitions.

\subsection{Index recovery on the Birkholz search}
The Baker search uses CI-NEB alone; the Birkholz search uses a GP-NEB band followed by dimer refinement. Their climbing-image and dimer updates use the reflected-force formulation published with OCI-NEB \cite{goswamiEnhancedClimbingImage2026}. The contribution here is the saddle-stage integration of measured mode resolution with recovery and the shared training set. During dimer refinement, measured curvature probes extend the mode span when signs remain unresolved. A second resolved negative mode triggers an order-one escape. The search first runs an order-one partitioned rational-function (RFO) search on the GP refit with the measured rows along the offending mode, and measures only the returned point; if that search is unavailable, it walks downhill along the mode on the measured surface, doubling the step while the energy falls. A second negative mode is weak when the refinement resolves it as soft or its coarse curvature lies within twice the stencil resolution. For a weak mode, partitioned RFO steps on the potential itself use the measured Hessian and gradient differences to update a local Bofill Hessian; only the final geometry enters the training set \cite{banerjeeSearchStationary1985,bakerAlgorithmLocationTransition1986,bofillUpdatedHessian1994}. The measured downhill walk runs for sulfolene and the measured-stencil branch for Grignard and oxirane. An unresolved mode prevents acceptance. When the search has formed a full Hessian at the returned geometry, the validator reuses that stencil. Search-triggered stencils and refinements are included in the potential-evaluation count.

\subsection{How the comparisons were run}
The comparisons use the supported algorithms of each package with the same potential and physical inputs, reserving eight CPU cores per calculation. GPRidge uses four MPI processes with two BLAS threads per process; eOn and the Python comparators use one process with eight threads available to the potential and numerical libraries. Method-specific choices follow each package: Sella uses internal coordinates, ML-FSM uses its redundant-coordinate string construction, and CatLearn uses its acquisition and surrogate-relaxation routines. The Supporting Information lists the effective settings and thread-count controls. Elapsed times measure the resulting implementations, including their numerical backends and available parallelism. GPRidge distributes GP work across MPI processes and reuses measured observations between the band and dimer stages; these capabilities contribute to its practical performance.

\subsection{Bonding at three further stationary points}
Ethane H\(_2\) abstraction (row b; the endpoints run from ethene and H\(_2\) to ethane) shows a second-order point near the saddle. The GPRidge saddle lies where the H\(_2\) unit has dissociated: H3 is bonded to C1 at 1.11 \AA{} with bond order 0.81, and H4 has not yet bonded to C2, lying 2.05 \AA{} from H3 and 3.85 \AA{} from C2. The eOn CI-NEB endpoint keeps the same C1-H3 bond, 1.10 \AA{}, with H4 farther out at 2.48 \AA{} from H3, and the benchmark-potential curvature check reports two negative modes there against one at the GPRidge saddle. The GFN2-xTB HOMO-LUMO gap is 0.56 eV at the GPRidge saddle and 0.20 eV at the CI-NEB endpoint, against 5.6 eV at the separated reactants. At these nearly degenerate geometries the closed-shell density assigns a C2-H4 bond order of 0.75 across 3.85 \AA{}; the figure draws no bond between atoms more than 2.5 covalent-radius sums apart, and the value remains in the annotation.

The substituted oxirane illustrates a subtler distinction. At the GPRidge saddle, the ring C1-O12 bond is elongated from 1.44 to 1.94 \AA{}, with its bond order reduced from 0.91 to 0.41. Sella returns similar local bonding, 1.92 \AA{} and 0.45, but its curvature check reports two negative modes and two unresolved modes. One negative mode exceeds 50 cm\(^{-1}\) in magnitude; the unresolved modes prevent a definite saddle-order assignment. GPRidge has one negative mode and no unresolved modes. Bonding pictures explain the structural change, while the measured curvatures distinguish the returned stationary points.

CatLearn's Diels--Alder timeout (row c) has a different origin: the search has not yet reached a stationary point. Its last measured geometry is the maximum of the surrogate path, with the two forming C-C bonds at 2.221 and 2.224 \AA{}. The GPRidge saddle has the same synchronous structure, both forming bonds at 2.22 \AA{} with bond order 0.33, so the two differ in force convergence rather than in bonding. Its maximum atomic force is 0.235 eV/\AA{}, 4.6 times the target, although the reported path energy uncertainty is below its 0.05 eV threshold. The predicted barrier varies by only 0.034 eV over the final ten iterations. The measured force remains above the target despite the nearly steady barrier and plausible bond formation. Formyloxyethyl also ends on its predicted path maximum, at a measured force of 2.35 eV/\AA{}; the last acquired points of the other four timeouts differ from their predicted path maxima. The Supporting Information separates these unfinished searches from the minima, extra negative modes and unresolved curvatures returned by other methods.

\subsection{Learned initializers}
Wu, Yuan, Hegazy, Blau, and Head-Gordon learn the leftmost Hessian eigenvector and use it in a transition-state optimization \cite{wuMachineLearnedLeftmost2026}; Yang and Zhong's MARC-TS learns a continuous reaction path and can initialize a nudged elastic band from it \cite{yangLearningContinuous2026}; Li, Sun, Yu, Yan, Gerstein, and Pham's Tr\`efle generates a transition-state candidate in one step \cite{liTrefle2026}; and Zhao, Zhang, Qin, Xu, and Xiao's ASPIRE proposes a set of saddle candidates from one initial geometry and refines them with a dimer \cite{zhaoAspire2026}.

\subsection{Climbing-image band on the heptamer}
The heptamer comparison in the article is the in-program climbing-image band on the same linear path, spring and force criterion as the surrogate band.
That control converges all 13 starts, with a median of 778 evaluations against 30 for the surrogate band.
The island control converges 58 of 59 starts, with a median of 443 evaluations against 23.

\subsection{Ethane, Diels--Alder, and oxirane}
Ethane H\(_2\) abstraction shows a second-order point near the saddle: the curvature check reports two negative modes at the CI-NEB endpoint and one at the GPRidge saddle.
CatLearn's Diels--Alder timeout has not yet reached a stationary point.
The GPRidge saddle has the same synchronous structure, both forming bonds at 2.22 \AA{} with bond order 0.33, so the two differ in force convergence rather than in bonding, and the CatLearn force is 0.235 eV/\AA{}, 4.6 times the target.
At the substituted oxirane, Sella returns similar local bonding, but its curvature check reports two negative modes and two unresolved modes, while GPRidge has one negative mode and no unresolved modes.

\section{Software stack}
\label{sec:si-stack}
A saddle search also depends on a structure format and its reader, an
append-only store of the calculator evaluations, an ahead-of-time
compiled potential, and the workflow that orders the three. This
section says what each one does and gives the commands that run a
campaign end to end.
\subsection{Maintained implementation source configuration}
\label{sec:si-maintained-sources}
This separate implementation comparison pairs source \texttt{ae8849da1} with \texttt{00106b604}; the headline reaction tables use the archived GPRidge 0.5.0 snapshot \texttt{3ddd83b77}.
The \texttt{rgpot} configuration includes the UMA batch-capability and Eigen fallback patches supplied with GPRidge.

Within that paired implementation experiment, coordinates, force and index checks, and search-call counts agree across all 44 cases. Its search-time comparison passes on 30 of 44 cases, so the overall comparison reports a failed timing criterion.
Its 2769 validation calls exclude 44 reference evaluations and belong to this source pair. The publication snapshot instead has 2768 validation calls and 44 reference calls, totaling 2812 independent-check evaluations.

\begin{table}[htbp]
\caption{\label{tab:maintained-validation}Strict endpoint acceptance and measured costs for the paired implementation check. Calls are totals over 44 cases; times are per-case maxima. The full pipeline includes oracle validation. These are measurements of the stated source pair, separate from the publication snapshot.}
\centering
\small
\begin{tabular}{lrrrrr}
Source & Endpoints & Search calls & Validation calls & Search (s) & Pipeline (s)\\
\hline
\texttt{ae8849da1} & 44/44 & 3061 & 2769 & 215.628 & 251.769\\
\texttt{00106b604} & 44/44 & 3061 & 2769 & 216.441 & 252.671\\
\end{tabular}
\end{table}
\subsection{The \texttt{con} format and \texttt{readcon}}
\label{sec:si-readcon}
Geometries enter and leave the drivers as eOn \texttt{con} files. The reader
is \texttt{readcon-core}, a Rust library exposed through a C ABI carrying an
explicit stamp with major, minor, and layout-revision fields. A host
binary linked against one revision refuses a mismatched library
instead of reading a shifted struct.

A frame is built rather than parsed from text:
\texttt{rkr\_frame\_builder} takes atoms, an energy
(\texttt{rkr\_frame\_builder\_set\_energy}), a frame index and a metadata JSON
blob, and \texttt{rkr\_frame\_builder\_build} produces the frame. Reading back
gives the same fields (\texttt{rkr\_frame\_energy}, \texttt{rkr\_frame\_frame\_index},
\texttt{rkr\_frame\_spec\_version}). Forces and positions are reachable as
zero-copy views over the frame's own storage, and a band passed to the
potential is a view of that storage.

CON version 3 requires explicit, valid length and energy units. The
reported molecular campaigns use \texttt{angstrom} and \texttt{eV}. The maintained
reader and writer also preserve reduced-unit labels such as \texttt{sigma} and
\texttt{epsilon} without assigning them physical scales. Missing or invalid
version-3 unit metadata is rejected. Every \texttt{con} file consumed here is
written by \texttt{readcon} itself, through the \texttt{readcon-chemfiles} Python
bindings, so the writer and the reader use the same implementation of
the format.
\subsection{Saved evaluations}
\label{sec:si-readcondb}
Every energy and force the calculator returns is appended to the training set and written with the trajectory. A restarted search rebuilds the training set from those evaluations, so it pays for each observation once. The reported evaluation counts are the number of those evaluations.

Taking the frame handle directly, instead of serialising a frame to
text and re-parsing it, removes two filesystem round trips per
observation from the dimer path.
\subsection{Ahead-of-time compiled potentials}
\label{sec:si-aoti}
A search calls the ahead-of-time compiled potential. It is
exported once with \texttt{torch.export} and compiled by AOTInductor into a
\texttt{.pt2} package, and the drivers load that package. This removes the
interpreter from the inner loop, where a GP-accelerated search makes
tens to hundreds of potential evaluations and each call would otherwise cross
into Python.

The export traces the model at one geometry, and the trace fixes the
neighbor graph. A package is therefore valid over the geometries that
keep that graph and invalid outside it, which is a statement about
coverage rather than a technicality. Traced at the default setting the
graph is captured under a neighbor cap, and the first search step that
changes the neighbor list aborts inside the compiled kernel with an
index past the atom count. A complete neighbor graph, with the cutoff at \(0.4L\) and
no neighbor cap, is exact for a
molecule isolated in a box of side \(L\). One package then stays valid as
the geometry moves.

Each package here is traced at the reference saddle rather than at an
endpoint. A search that starts from the endpoints and climbs spends its
last and most delicate iterations near the barrier top, and tracing
there keeps the whole path inside the graph the package was built for.
The box used for tracing and the box the endpoints are written in must
agree, since the box sets the cutoff. The larger Birkholz systems use larger boxes than the Baker set.
\subsection{Package consumption in \texttt{rgpot}}
\label{sec:si-rgpot}
\texttt{rgpot} presents one potential interface to the drivers and hides which
backend answers. Its UMA backend loads the AOTInductor \texttt{.pt2} and
supplies neighbor lists from \texttt{vesin} rather than from the model, which
is what lets the exported module run with \texttt{otf\_graph=False} and the
graph fixed at export time.

Evaluation is batched over the band: one AOTI call serves a chunk of up
to the package's \texttt{batch\_max} same-composition systems. A NEB band is
exactly such a chunk, so an all-image-evaluated iteration costs one call
into the compiled module rather than one per image.
The reported runs factor the noisy training matrix by Cholesky and leave a near-duplicate geometry out of that factor. The sections that follow keep the derivations, including constructions the reported runs do not enable, as reference material.

\subsection{Classical path and saddle comparisons}
\label{sec:si-classical-comparison}
The comparison uses the same potential, the same endpoints, and the same acceptance test. All searches use the same compiled
UMA-S-1.1 omol potential files and oracle shared library, with identical
charge and spin. An audit reads both endpoint files through readcon and
checks equal coordinates, element order, cell and angles. The validator
source is identical, including its adaptive refinement of weak modes.
Eight CPU slots are reserved per calculation: one process with eight
oracle threads for eOn, and four MPI ranks with two slots each for
GPRidge.

The eOn band calculations use 8 intermediate images, SIDPP
initialization, L-BFGS, a maximum move of 0.1 \AA{}, and at most 1000
iterations. Energy-weighted spring constants span 1 to
10 eV \AA{}\(^{-2}\). The relative climbing-image trigger is 0.8 and the
dimer rotation tolerance is 10 degrees. OCI-NEB uses a relative MMF
trigger of 0.31 and an alignment tolerance of 0.85, following the
published settings \cite{goswamiEnhancedClimbingImage2026}. The potential
remains UMA-S-1.1. The projected-force threshold is
0.0514221 eV \AA{}\(^{-1}\), using the maximum norm over atoms.
The selected eOn stopping condition is unchanged. The OCI-NEB handoff
compares raw climbing-image forces using this same norm, restores the
saved image at positive curvature, and redistributes the band after a
helpful walk. Molecular coordinates remain in a continuous chart during
path interpolation and redistribution.
The separate acceptance test applies this threshold to the Cartesian
force and requires one resolved negative curvature and no unresolved
physical modes.

\begin{table}[H]
\caption{\label{tab:si-classical-totals}Classical and GP searches on the publication inputs. Evaluations include the initial endpoints or dimer origin, and exclude independent validation. Birkholz dimer counts exclude the shared band preparation, which costs 779 evaluations over 20 reactions. Optimizer convergence and strict saddle acceptance are separate columns.}
\centering
\small
\begin{tabular}{llrrr}
Set & Method & Optimizer stops & Accepted saddles & Mean evaluations\\
\hline
Baker & Plain NEB & 24/24 & 3/24 & 518.67\\
Baker & CI-NEB & 24/24 & 18/24 & 336.33\\
Baker & OCI-NEB & 24/24 & 21/24 & 356.08\\
Baker & GPRidge & 24/24 & 24/24 & 50.62\\
Birkholz & eOn dimer & 20/20 & 17/20 & 165.55\\
Birkholz & GP-dimer & 20/20 & 20/20 & 53.35\\
\end{tabular}
\end{table}

Every classical calculation writes a final evaluation count.
Plain NEB minimizes the force perpendicular to the path, so its
stopping criterion permits a nonstationary highest-energy image.
The saddle acceptance rate therefore measures a different condition
from its projected-force stop. All 24 CI-NEB and all 24 OCI-NEB
calculations meet their selected optimizer stopping condition.
CI-NEB fails the Cartesian force test on HCCH, H\(_2\)CO, HF abstraction,
acrolein rotation and \(\mathrm{HCNH_2}\).
Ethane H\(_2\) abstraction has two negative modes.
OCI-NEB fails the Cartesian force test on \(\mathrm{CH_3O}\) and returns two negative modes on ethane H\(_2\)
abstraction and one unresolved mode on \(\mathrm{HCONH_3^+}\).

Each classical dimer starts at the publication band peak,
along its tangent. The GP-dimer imports the measured band
observations. Both workflows are charged the same band preparation,
separately from the dimer stage. Grignard and oxirane have 12 and
three unresolved modes, respectively, in the classical dimer results;
sulfolene has two negative modes. All 20 classical dimer runs meet their optimizer's stopping
condition.

\begin{table}[H]
\caption{\label{tab:si-classical-paired}Costs on reactions accepted by both methods. Calls include initial potential evaluations. Search time excludes validation and, for the dimer, band preparation. Each entry is the sum over the shared successful reactions. Failures remain in the denominators of Table \ref{tab:si-classical-totals}.}
\centering
\small
\begin{tabular}{lrrrrr}
eOn method & Shared cases & eOn calls & GP calls & eOn search (s) & GP search (s)\\
\hline
CI-NEB & 18 & 5964 & 767 & 152.182 & 78.180\\
OCI-NEB & 21 & 6672 & 952 & 167.874 & 92.107\\
Dimer & 17 & 816 & 127 & 67.934 & 42.239\\
\end{tabular}
\end{table}

Each comparator reaction has one timing measurement; GPRidge uses
one of the three passes. 
Complete Baker calculation
times, including setup and validation, are 215.601 versus 126.923 s
on the CI-NEB intersection and 239.214 versus 147.716 s on the OCI-NEB
intersection.
\subsection{CatLearn ML-NEB comparison}
\label{sec:si-catlearn-comparison}
CatLearn ML-NEB \cite{garridoTorresLowScalingAlgorithm2019} uses the same 24 Baker endpoint pairs, compiled UMA potentials,
charge and spin as the publication searches. Each calculation reserves
eight CPUs on an exclusive node. CatLearn uses one process with eight
threads and GPRidge uses four MPI ranks with two threads each.
The CatLearn paths contain 20 images, including the two fixed endpoints.
The search limit is 1200 s; validation has a separate 540 s limit.
The independent force and curvature criteria are identical to those
applied to GPRidge. The adapter includes measured endpoint values in ASE
single-point calculators and counts each evaluation through the shared
native oracle.

CatLearn accepts 18/24 cases and GPRidge accepts 24/24. The six CatLearn
failures reach the search limit. Their evaluation counts are
lower bounds on the cost of convergence. Among the 18 common successes,
CatLearn and GPRidge use 860 and 799 potential evaluations in the search, and
1922 and 1519 with the true-surface confirmation included.
Search times sum to 6247.632 and 69.841 s. Times including independent
validation sum to 6276.963 and 114.340 s, a ratio of 54.90.
All 24 reactions remain in Table \ref{tab:si-catlearn-cases}; failed searches
are excluded only from the comparison of costs on common successes.
Comparator times are single measurements; GPRidge uses one of the three passes.

\begin{table}[H]
\caption{\label{tab:si-catlearn-cases}CatLearn ML-NEB and GPRidge on the Baker reactions. Both call columns are the search plus the true-surface confirmation. CatLearn's search is the two endpoints, one interpolated image and one acquired geometry on each iteration, and its library stop is the last acquired force together with the interior surrogate uncertainty; the confirmation remeasures the returned path and evaluates the crest stencil. The GPRidge column adds the same certificate calls to the band search. A superscript \(+\) is a CatLearn search that stopped short of acceptance, with no confirmation. Times include that confirmation. CatLearn outcomes are accepted (A) or timeout (T).}
\centering
\small
\begin{tabular}{llrrrr}
Reaction & Outcome & CatLearn calls & GP calls & CatLearn (s) & GP (s)\\
\hline
HCN & A & 97 & 45 & 472.54 & 4.10\\
HCCH & A & 83 & 58 & 174.58 & 4.34\\
H2CO & A & 100 & 106 & 146.12 & 7.36\\
CH3O & A & 79 & 63 & 104.76 & 4.79\\
Cyclopropyl & A & 137 & 103 & 750.71 & 7.44\\
Bicyclobutane & T & 81\textsuperscript{+} & 108 & 1200.47 & 7.50\\
Formyloxyethyl & T & 66\textsuperscript{+} & 110 & 1200.49 & 7.23\\
Diels--Alder & T & 57\textsuperscript{+} & 144 & 1200.52 & 13.52\\
Tetrazine & A & 130 & 78 & 650.89 & 5.11\\
trans-Butadiene & A & 86 & 84 & 16.07 & 4.94\\
Ethane H2 abstraction & T & 86\textsuperscript{+} & 250 & 1200.47 & 56.26\\
HF abstraction & A & 106 & 83 & 203.29 & 5.23\\
Vinyl alcohol & A & 120 & 83 & 476.89 & 5.14\\
HOCl & A & 92 & 56 & 156.40 & 4.38\\
H2PO4 anion & A & 133 & 78 & 745.48 & 5.12\\
Claisen & T & 80\textsuperscript{+} & 130 & 1200.54 & 9.90\\
Silylene insertion & A & 145 & 109 & 736.62 & 7.06\\
HNCCS & A & 90 & 166 & 141.78 & 19.38\\
HCONH3 cation & T & 68\textsuperscript{+} & 80 & 1200.48 & 5.29\\
Acrolein rotation & A & 116 & 85 & 199.89 & 5.75\\
HCONHOH & A & 113 & 76 & 375.23 & 5.13\\
HCN + H2 & A & 101 & 77 & 349.72 & 6.27\\
H2CNH & A & 100 & 65 & 431.78 & 4.61\\
HCNH2 & A & 94 & 104 & 144.22 & 8.20\\
\end{tabular}
\end{table}

A one-thread control repeats HCN, HCCH, HF abstraction and acrolein
rotation with the same CatLearn settings. All four pass, taking 392.2,
176.3, 198.9 and 182.7 s including validation. The eight-thread times
are 472.5, 174.6, 203.3 and 199.9 s in the set. These controls show that the elapsed
time difference on these cases is not explained by the eight-thread
setting alone.

\subsubsection{CatLearn timeout diagnostics}
\label{sec:org83be09f}

All 24 printed force histories are checked against the measured trajectories. CatLearn reprints its full progress table at every iteration with the current timestamp; the analysis uses each step's first occurrence. Printed force values agree with the maximum atomic force to the six-decimal output precision. The time between entries includes fitting, surrogate relaxation, acquisition, potential evaluation and logging. For the six timeouts, its median rises from 1--2 s over the first ten intervals to 36--54 s over the last ten. These intervals measure the complete iteration.

\begin{figure}[p]
\centering
\includegraphics[width=0.98\textwidth]{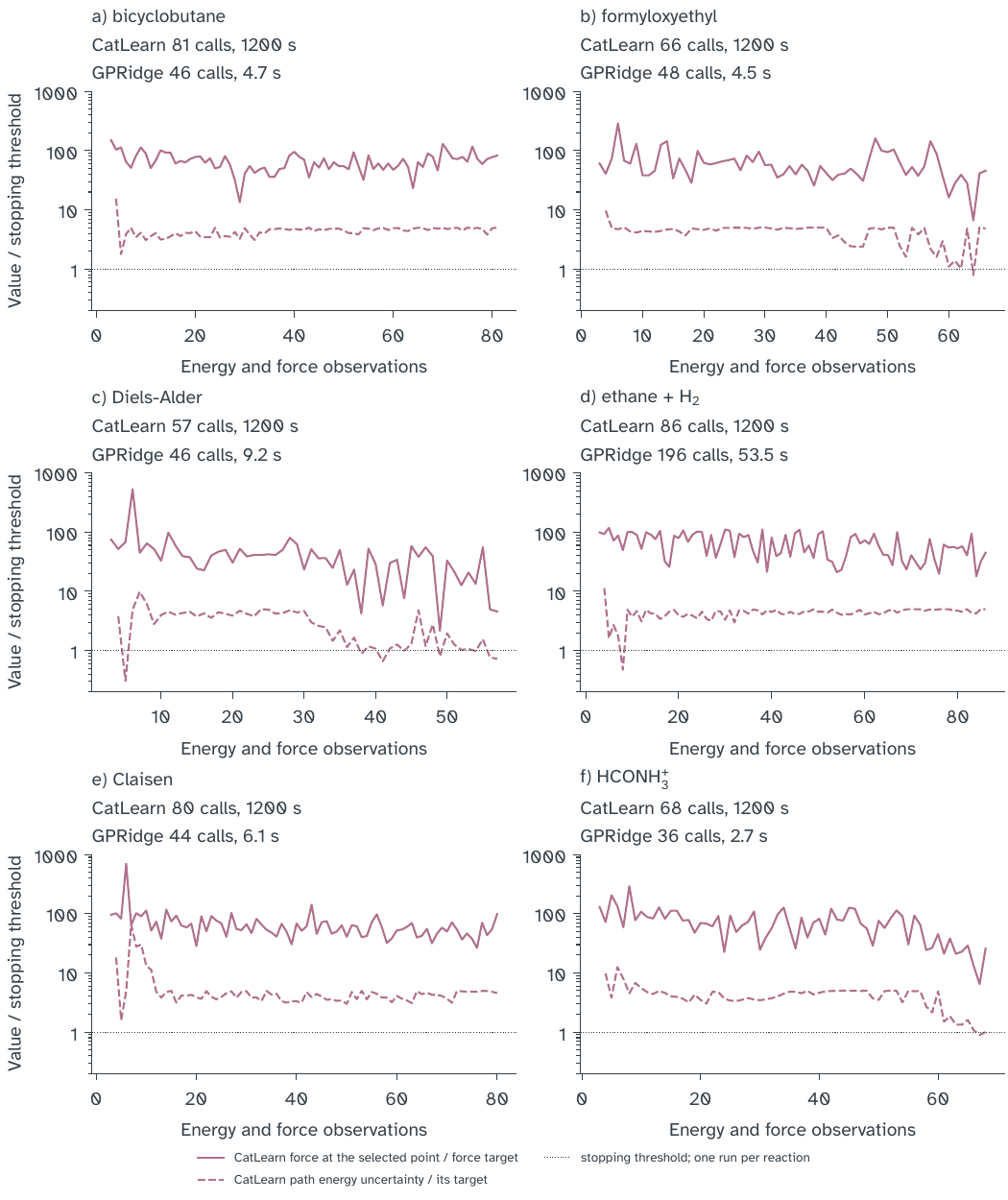}
\caption{\label{fig:catlearn-failures}Progress of the six timed-out CatLearn searches. Solid curves show the measured maximum atomic force at each selected geometry divided by the force target; dashed curves show the preceding path's energy uncertainty divided by its own target. The dotted line marks one in these separate ratios. Endpoints are excluded from the curves but included in the observation count. Neither a low energy uncertainty nor a stable predicted barrier certifies a saddle. The histories include all 24 reactions.}
\end{figure}

Table \ref{tab:catlearn-timeout-diagnostics} and Figure \ref{fig:catlearn-failures} distinguish the true force at the selected acquisition geometry from the energy uncertainty along the surrogate path. The force target is 0.0514221 eV/\AA{} and the path uncertainty target is 0.05 eV. CatLearn's reported uncertainty is twice the predicted standard deviation including regularization; it is not a force-error bound. Its surrogate relaxation stops when this quantity reaches 0.25 eV. Bicyclobutane and ethane H2 abstraction have values of at least 0.225 eV in eight of the final ten iterations; Claisen does so in nine. Acquisition remains active in these searches.

\begin{table}[tbp]
\caption{\label{tab:catlearn-timeout-diagnostics}The six CatLearn searches unfinished at 1200 s. CatLearn counts are partial. \(F\) is the true maximum atomic force at the last sampled geometry, and \(U\) is the maximum energy uncertainty on the preceding surrogate path. GPRidge columns give completed search costs for the same reaction; validation is excluded.}
\centering
\small
\begin{tabular}{lrrrrr}
Reaction & CatLearn calls & \(F\) (eV/\AA{}) & \(U\) (eV) & GP calls & GP time (s)\\
\hline
Bicyclobutane & 81\textsuperscript{+} & 4.245 & 0.2492 & 46 & 4.72\\
Formyloxyethyl & 66\textsuperscript{+} & 2.348 & 0.2411 & 48 & 4.48\\
Diels--Alder & 57\textsuperscript{+} & 0.235 & 0.0361 & 46 & 9.19\\
Ethane H2 abstraction & 86\textsuperscript{+} & 2.312 & 0.2497 & 196 & 53.55\\
Claisen & 80\textsuperscript{+} & 5.040 & 0.2278 & 44 & 6.15\\
HCONH3 cation & 68\textsuperscript{+} & 1.320 & 0.0500 & 36 & 2.68\\
\end{tabular}
\end{table}

Diels--Alder satisfies the path-energy uncertainty threshold but has not met the force threshold. Its last selected geometry exactly matches the highest-energy internal image of the predicted path. Both forming C-C distances are near 2.22 \AA{}, between endpoint values of 3.42 and 1.54 \AA{}. The final ten predicted barriers span 0.03384 eV while the selected-point forces range from 0.110 to 2.865 eV/\AA{}. The formyloxyethyl search also ends on its predicted path maximum, at a measured force of 2.348 eV/\AA{} with a path uncertainty of 0.241 eV, near the 0.25 eV surrogate-relaxation stop and above the 0.05 eV target. The other four final acquisition points differ from their respective predicted path maxima by maximum atomic displacements of 0.374--1.644 \AA{}. Those four geometries are not the predicted path maxima.

The structures resolve the chemical coordinates involved. Bicyclobutane opens two C-C contacts, formyloxyethyl cleaves C-O, and Diels--Alder forms two C-C bonds. The ethane case changes H-H and C-H contacts; Claisen combines C-O cleavage with C-C formation; HCONH3 cation changes C-N, N-H and C-H contacts. At CatLearn's bicyclobutane path maximum, the two opening C-C distances are 1.537 and 2.225 \AA{}, whereas the last sampled point has 2.517 and 2.445 \AA{}. The sampled structure and the predicted maximum therefore represent different stages of bond opening along the unfinished path.

At timeout the full Cartesian derivative matrices contain 1496--3440 rows. In the pinned implementation, \texttt{ASECalc.calculate} estimates surrogate forces from two displaced predicted energies per free coordinate. Its \texttt{train\_gp\_model} fits a single Cartesian length scale on rank zero. The production GPRidge path uses analytic derivatives and MPI distribution. The measured clocks include all of these operations; no percentage of the speedup is assigned to an individual operation.

\subsection{Acceptance as a function of search cost}
\label{sec:si-method-cost-curves}
\begin{figure}[tbp]
\centering
\includegraphics[width=0.98\textwidth]{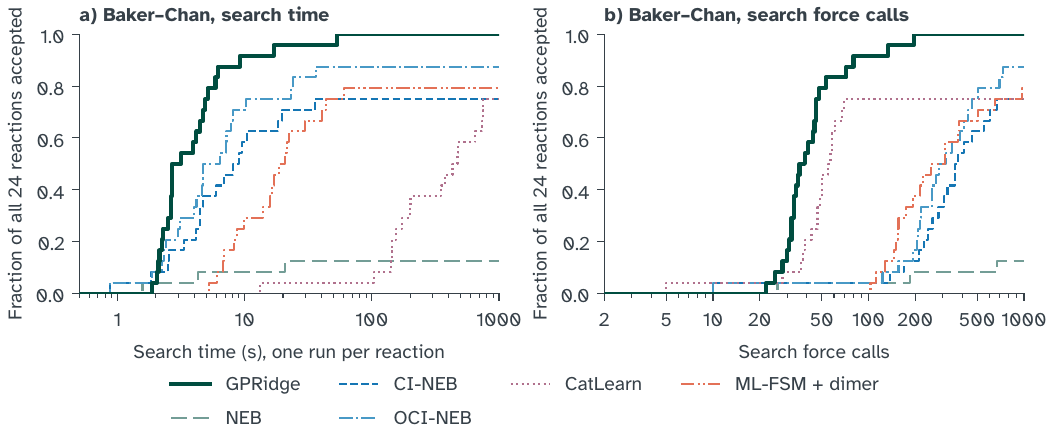}
\caption{\label{fig:method-acceptance-baker}Baker--Chan acceptance versus (a) search time and (b) energy and force evaluations. The denominator is all 24 reactions for every method. Counts include endpoint evaluations and ML-FSM string construction. Independent validation is excluded from both horizontal coordinates. The final heights report acceptance under the same force and saddle-index criteria.}
\end{figure}

\begin{figure}[tbp]
\centering
\includegraphics[width=0.98\textwidth]{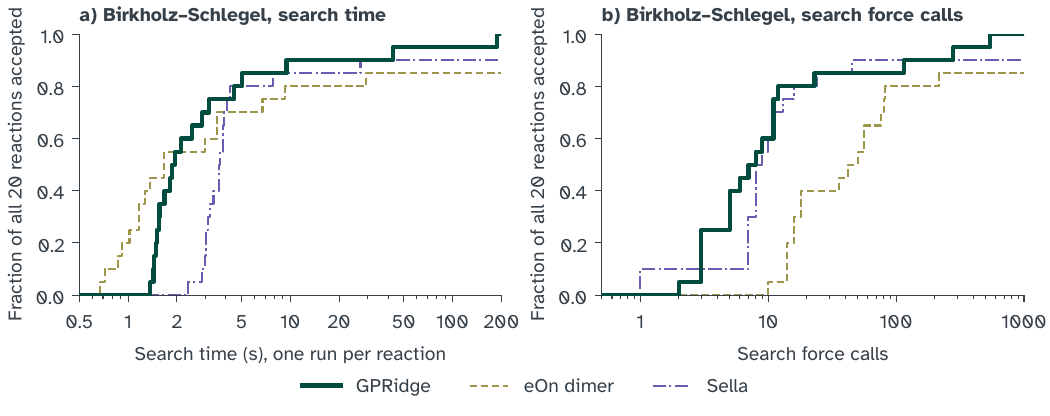}
\caption{\label{fig:method-acceptance-birkholz}Birkholz--Schlegel acceptance versus (a) search time and (b) energy and force evaluations for all 20 starts. The methods use the band maxima; GPRidge also reuses band observations. Band preparation and final independent validation are excluded from both horizontal coordinates. Grignard and unsuccessful searches remain represented in the full-panel comparison.}
\end{figure}

Figures \ref{fig:method-acceptance-baker} and \ref{fig:method-acceptance-birkholz} use the same cases as the method comparisons. A step occurs only when a search returns an accepted saddle. Each denominator contains the entire benchmark set, so a failed force, index or execution check leaves the curve below one. Timeouts also remain in that denominator. The horizontal coordinate is the measured cost of the completed search. Each empirical curve uses one run per method and reaction, with one of the three passes for GPRidge. This cumulative view is the absolute-budget counterpart of a cactus plot.

For the Pareto comparison in this supplement, a method is dominated if another measured method has no larger mean cost in either coordinate and a strictly smaller cost in at least one. Plain NEB contributes to the full-panel acceptance comparison; its maximum-energy image has no saddle-refinement stage.

\subsection{Relative performance profiles}
\label{sec:si-performance-profiles}
The performance profiles in the main manuscript follow Dolan and Moré \cite{dolanBenchmarkingOptimization2002}. For reaction \(p\) and method \(s\), let \(c_{ps}\) be its measured search cost when its returned saddle passes the common checks, and infinity otherwise. The ratio \(r_{ps}=c_{ps}/\min_j c_{pj}\) compares each accepted result with the least costly accepted result for that reaction. The profile is
\begin{equation}
\rho_s(\alpha)=\frac{1}{N}\sum_{p=1}^{N}\mathbf{1}[r_{ps}\leq\alpha],\qquad \alpha\geq1.
\end{equation}
A reaction with no accepted result contributes zero to every method. Thus, \(\rho_s(1)\) gives the fraction of reactions on which a method attains the lowest cost, including ties, while the final height gives its full-panel acceptance. Force calls and time have separate profiles. These relative profiles complement the absolute-budget curves and the paired means; their denominator is all 24 Baker or 20 Birkholz reactions.
\subsection{Common-case cost fronts}
\label{sec:si-method-cost-frontiers}
\begin{figure}[tbp]
\centering
\includegraphics[width=0.98\textwidth]{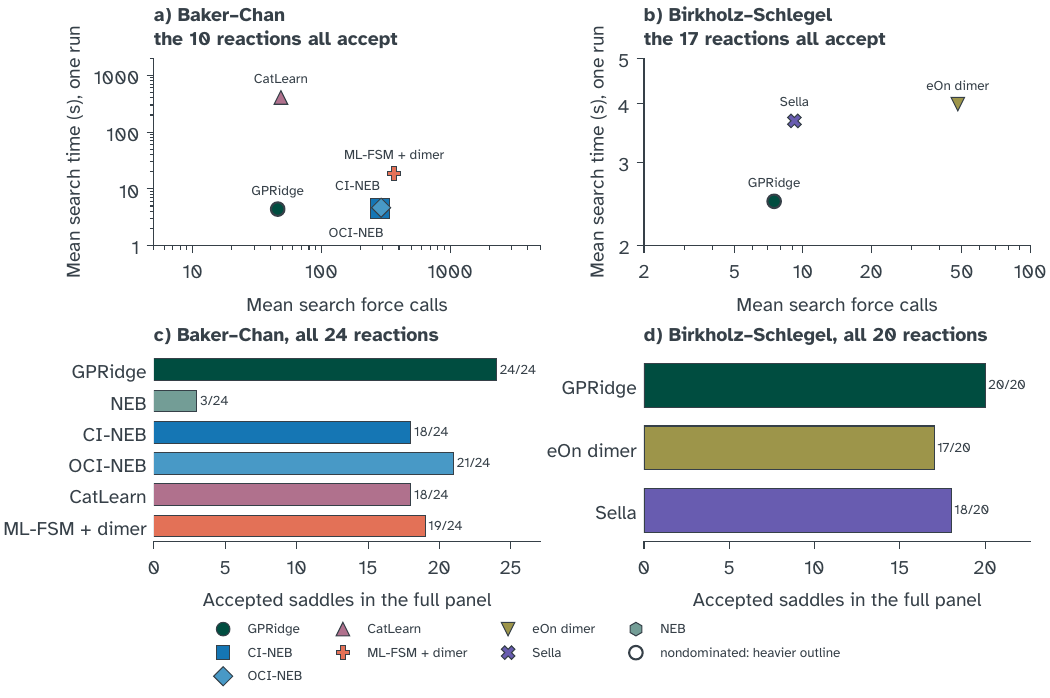}
\caption{\label{fig:method-cost-frontiers}Search cost and acceptance. (a,b) Mean force calls and search time on reactions accepted by every plotted saddle-search method. Heavier outlines mark nondominated measured methods; a dashed line joins them when there are two or more. A covered marker is drawn larger so its edge remains visible. (c,d) Acceptance on the full panels, including plain NEB. Search costs exclude independent validation and, for Birkholz, band preparation. Each point uses one run per reaction, with one of the three passes for GPRidge. These fronts describe the measured methods and settings.}
\end{figure}

Figure \ref{fig:method-cost-frontiers} compares both costs on identical sets of accepted reactions and reports success on the full panels. GPRidge is the only nondominated measured method on the 10 common Baker reactions. On the 17 common Birkholz reactions, GPRidge is the only nondominated method in evaluations: it uses fewer evaluations than either comparator. The common cases exclude Grignard because the classical dimer fails its saddle check, and oxirane and sulfolene because both comparators fail their checks. The cumulative cost curves include every unsuccessful attempt in the full-panel denominator.

\subsection{Per-reaction cost comparisons}
\label{sec:si-method-cost-matrices}
\begin{figure}[p]
\centering
\includegraphics[width=0.98\textwidth]{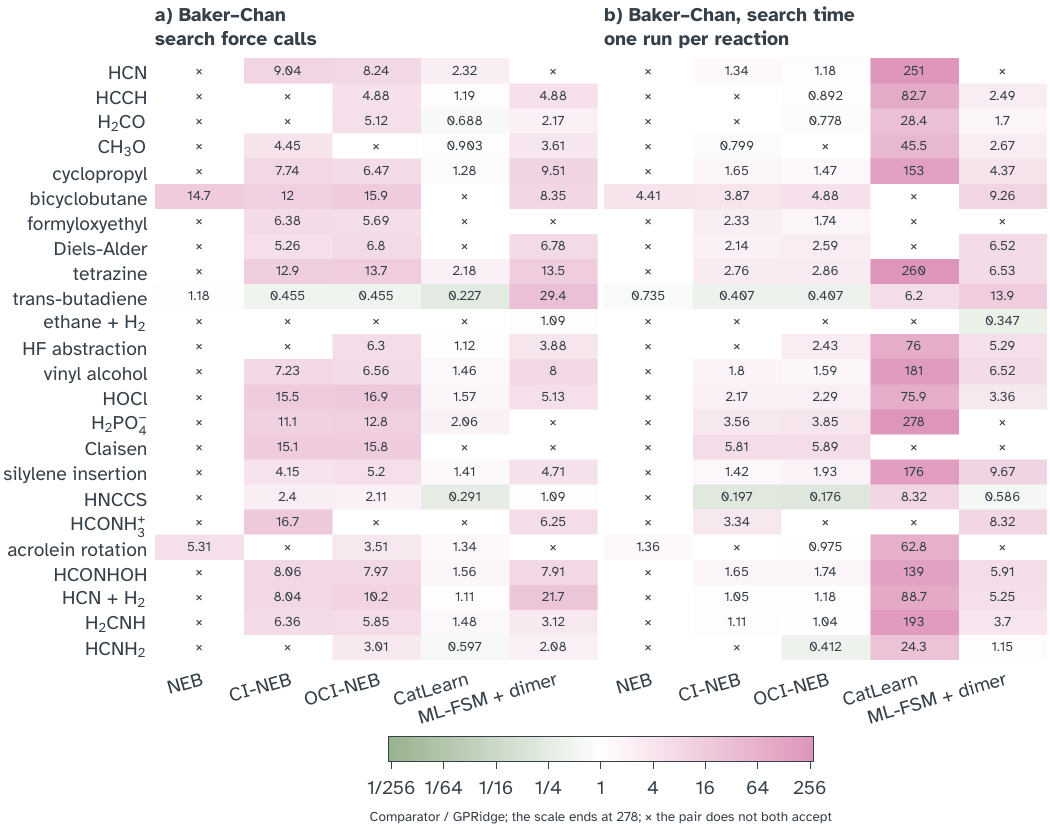}
\caption{\label{fig:reaction-costs-baker}Baker--Chan costs by reaction. Values are comparator/GPRidge search costs: (a) energy and force evaluations and (b) elapsed search time. Values above one favour GPRidge; crosses denote outcomes that fail the common checks. All 24 GPRidge reactions pass. Validation is excluded, while endpoint evaluations and ML-FSM string construction are included. Plain NEB has no saddle-refinement stage. The color scale is symmetric in the logarithm of the ratio and extends to a factor of 278 in either direction, the largest ratio or inverse ratio on either panel. No cell saturates; printed values give the measured ratios.}
\end{figure}

\begin{figure}[p]
\centering
\includegraphics[width=0.98\textwidth]{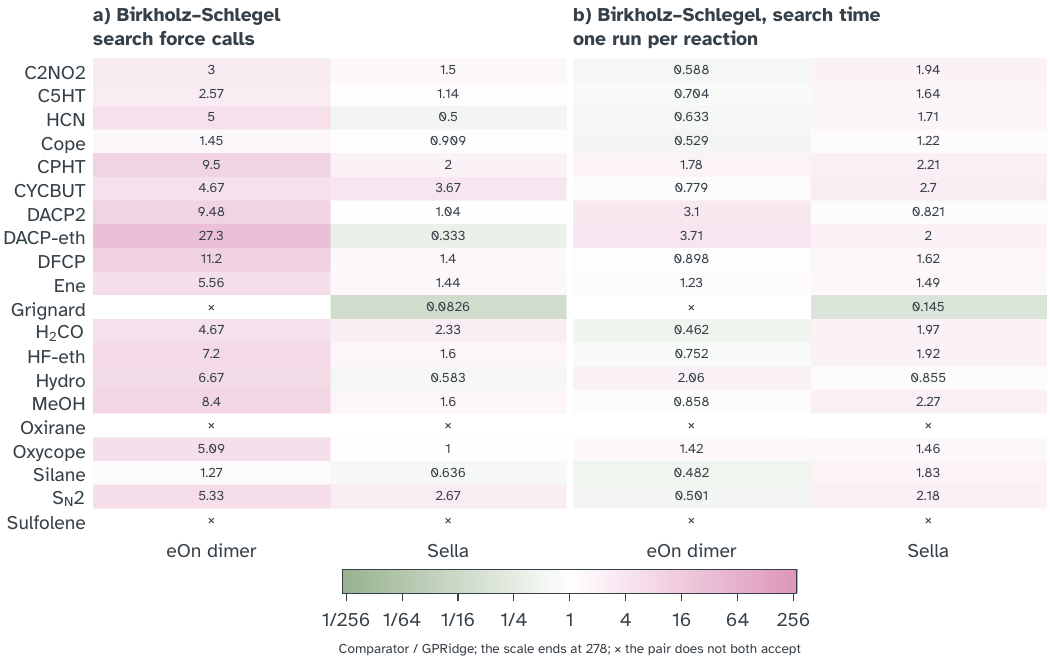}
\caption{\label{fig:reaction-costs-birkholz}Birkholz--Schlegel costs by reaction, using the same ratio and cross conventions as Figure \ref{fig:reaction-costs-baker}. All 20 GPRidge reactions pass. The comparison excludes shared band preparation and independent validation. Each method starts from the band maximum, and GPRidge also reuses the band's measured observations. Sella uses fewer evaluations on six common successes; the crosses show its unsuccessful oxirane and sulfolene outcomes.}
\end{figure}

Figures \ref{fig:reaction-costs-baker} and \ref{fig:reaction-costs-birkholz} show where the paired mean costs arise. Every cell compares a method with GPRidge on the same reaction. Crosses identify outcomes that fail a force, index or execution check; the tables give the reason. Thus, a timeout, an oracle error and an unresolved curvature are distinct outcomes despite sharing a plotting symbol. The matrices include all 24 Baker and 20 Birkholz reactions.

The shared subset is specific to each comparator, whereas the Pareto figure uses the intersection accepted by every plotted method.

\subsection{Freezing-string and Sella comparisons}
\label{sec:si-reference-comparison}
The comparison uses ML-FSM 1.0.1 and Sella 2.6.0
\cite{marksFreezingStringInternal2025,hermesSellaOpenSource2022} with the publication
potential packages, oracle library, physical inputs and independent
validator. Each calculation reserves eight CPU cores on an exclusive
node. Comparator costs use one timing measurement per case; GPRidge uses one of the three passes.

ML-FSM uses redundant internal-coordinate interpolation and Cartesian
L-BFGS-B optimization, with 18 minimum nodes, 100 interpolation points,
two optimizer iterations, three line searches and a maximum move of
0.3 \AA{}. ASE dimer refines the highest-energy image using a Gaussian
initial displacement with seed 706253457. This follows the
freezing-string benchmark recipe; the dimer is a separate refinement
stage. The string and dimer evaluation counts are reported separately.

Sella uses internal coordinates, saddle order one and its default
initial mode. Each starting geometry is the exact publication band peak,
without an added displacement. The band's 779 evaluations over
20 reactions are shared preparation and are excluded from both methods'
saddle-stage costs. Sella receives the starting geometry; the GP-dimer
also includes the band's measured observations.

\begin{table}[H]
\caption{\label{tab:si-reference-paired}Reference-method comparisons on the common accepted reactions. Evaluations include string construction for ML-FSM and exclude shared band preparation for Sella. Search time excludes the independent validation. Call and time pairs list the reference method first and GPRidge second. All failures remain in the acceptance denominators.}
\centering
\small
\begin{tabular}{lrrrr}
Reference & Accepted & Shared & Calls & Search time (s)\\
\hline
ML-FSM + dimer & 19/24 & 19 & 5520 / 1029 & 371.567 / 132.272\\
Sella & 18/20 & 18 & 201 / 672 & 89.604 / 229.373\\
\end{tabular}
\end{table}

Tables \ref{tab:si-paired-ratios-baker} and \ref{tab:si-paired-ratios-birkholz} give every comparator/GPRidge ratio behind the paired-ratio figure of the main text, with the medians and win counts it quotes.

\begin{table}[H]
\caption{\label{tab:si-paired-ratios-baker}Comparator/GPRidge cost per Baker--Chan reaction, as search evaluations / search time, on the reactions both methods accept; -- marks a reaction either method fails. Values above one favour GPRidge. Costs and exclusions are those of the paired comparison tables. The last rows give the median ratio and the number of reactions on which GPRidge is cheaper.}
\centering
\footnotesize\setlength{\tabcolsep}{4pt}
\begin{tabular}{llllll}
Reaction & NEB & CI-NEB & OCI-NEB & CatLearn ML-NEB & ML-FSM\\
\hline
HCN & -- & 9.04 / 1.34 & 8.24 / 1.18 & 2.32 / 251.08 & --\\
HCCH & -- & -- & 4.88 / 0.89 & 1.19 / 82.74 & 4.88 / 2.49\\
H2CO & -- & -- & 5.12 / 0.78 & 0.69 / 28.42 & 2.17 / 1.70\\
CH3O & -- & 4.45 / 0.80 & -- & 0.90 / 45.50 & 3.61 / 2.67\\
cyclopropyl & -- & 7.74 / 1.65 & 6.47 / 1.47 & 1.28 / 152.66 & 9.51 / 4.37\\
bicyclobutane & 14.65 / 4.41 & 12.04 / 3.87 & 15.87 / 4.88 & -- & 8.35 / 9.26\\
formyloxyethyl & -- & 6.38 / 2.33 & 5.69 / 1.74 & -- & --\\
Diels-Alder & -- & 5.26 / 2.14 & 6.80 / 2.59 & -- & 6.78 / 6.52\\
tetrazine & -- & 12.93 / 2.76 & 13.75 / 2.86 & 2.18 / 259.97 & 13.46 / 6.53\\
trans-butadiene & 1.18 / 0.73 & 0.45 / 0.41 & 0.45 / 0.41 & 0.23 / 6.20 & 29.41 / 13.86\\
Ethane H\(_2\) abstraction & -- & -- & -- & -- & 1.09 / 0.35\\
HF abstraction & -- & -- & 6.30 / 2.43 & 1.12 / 76.01 & 3.88 / 5.29\\
vinyl alcohol & -- & 7.23 / 1.80 & 6.56 / 1.59 & 1.46 / 180.73 & 8.00 / 6.52\\
HOCl & -- & 15.53 / 2.17 & 16.87 / 2.29 & 1.57 / 75.93 & 5.13 / 3.36\\
H2PO4 anion & -- & 11.12 / 3.56 & 12.79 / 3.85 & 2.06 / 277.80 & --\\
Claisen & -- & 15.14 / 5.81 & 15.82 / 5.89 & -- & --\\
silylene insertion & -- & 4.15 / 1.42 & 5.20 / 1.93 & 1.41 / 175.87 & 4.71 / 9.67\\
HNCCS & -- & 2.40 / 0.20 & 2.11 / 0.18 & 0.29 / 8.32 & 1.09 / 0.59\\
HCONH3 cation & -- & 16.72 / 3.34 & -- & -- & 6.25 / 8.32\\
acrolein rotation & 5.31 / 1.36 & -- & 3.51 / 0.98 & 1.34 / 62.77 & --\\
HCONHOH & -- & 8.06 / 1.65 & 7.97 / 1.74 & 1.56 / 139.31 & 7.91 / 5.91\\
HCN + H2 & -- & 8.04 / 1.05 & 10.24 / 1.18 & 1.11 / 88.68 & 21.69 / 5.25\\
H2CNH & -- & 6.36 / 1.11 & 5.85 / 1.04 & 1.48 / 192.98 & 3.12 / 3.70\\
HCNH2 & -- & -- & 3.01 / 0.41 & 0.60 / 24.30 & 2.08 / 1.15\\
\hline
Median & 5.31 / 1.36 & 7.89 / 1.72 & 6.47 / 1.59 & 1.31 / 85.71 & 5.13 / 5.25\\
GPRidge cheaper & 3/3 / 2/3 & 17/18 / 15/18 & 20/21 / 15/21 & 13/18 / 18/18 & 19/19 / 17/19\\
\end{tabular}
\end{table}

\begin{table}[H]
\caption{\label{tab:si-paired-ratios-birkholz}Comparator/GPRidge cost per Birkholz--Schlegel reaction, as search evaluations, on the reactions both methods accept; -- marks a reaction either method fails. Values above one favour GPRidge. Costs and exclusions are those of the paired comparison tables. The last rows give the median ratio and the number of reactions on which GPRidge is cheaper.}
\centering
\small
\begin{tabular}{lll}
Reaction & eOn dimer & Sella\\
\hline
C2NO2 & 3.00 & 1.50\\
C5HT & 2.57 & 1.14\\
HCN & 5.00 & 0.50\\
Cope & 1.45 & 0.91\\
CPHT & 9.50 & 2.00\\
CYCBUT & 4.67 & 3.67\\
DACP2 & 9.48 & 1.04\\
DACP-eth & 27.33 & 0.33\\
DFCP & 11.20 & 1.40\\
Ene & 5.56 & 1.44\\
Grignard & -- & 0.08\\
H2CO & 4.67 & 2.33\\
HF-eth & 7.20 & 1.60\\
Hydro & 6.67 & 0.58\\
MeOH & 8.40 & 1.60\\
Oxirane & -- & --\\
Oxycope & 5.09 & 1.00\\
Silane & 1.27 & 0.64\\
SN2 & 5.33 & 2.67\\
Sulfolene & -- & --\\
\hline
Median & 5.33 & 1.27\\
GPRidge cheaper & 17/17 & 11/18\\
\end{tabular}
\end{table}

ML-FSM plus dimer ends at a minimum for HCN. Formyloxyethyl has an
unresolved mode; H2PO4 anion has two negative and two unresolved modes;
Claisen has two negative modes. Acrolein stops when the compiled potential
returns an evaluation error, after 681 calls. The acrolein outcome
is an oracle execution failure. For the 19 shared successes, elapsed
times including validation sum to 407.193 and 181.953 s, a ratio of 2.24.

Sella fails the index criterion for two reactions. Oxirane has two negative and two unresolved modes; sulfolene has two negative modes and no unresolved mode. Both meet the force threshold, with maximum atomic forces of 0.0365 and 0.0435 eV/\AA{}, respectively. GPRidge returns a resolved first-order saddle for each.

Sella's sulfolene structure has two resolved imaginary frequencies, 423\(i\) and 65\(i\) cm\(^{-1}\). ML-FSM plus dimer similarly returns two resolved negative modes for Claisen, 249\(i\) and 67\(i\) cm\(^{-1}\), at 0.0477 eV/\AA{}. These are extra unstable directions at low-force structures. Unresolved modes in oxirane, formyloxyethyl and H2PO4 leave the complete saddle index undetermined. The HCN endpoint is a minimum, and the acrolein oracle error supplies no final stationary point. These outcomes are distinct.

On the 18 shared successes, Sella uses 0.30 times the dimer-stage evaluations of GPRidge, with shared seed-band preparation excluded. Sella uses fewer evaluations on six reactions. Grignard dominates these sums: the GP-dimer spends 545 evaluations on it, against Sella's 45. On the other shared reactions, Sella uses 1.23 times the evaluations of GPRidge. Both failures remain in the full-panel performance profiles. The call pairs exclude the shared band preparation.

\subsection{Electronic analysis of the structures}
\label{sec:si-reaction-chemistry}
\begin{figure}[tbp]
\centering
\includegraphics[width=0.98\textwidth]{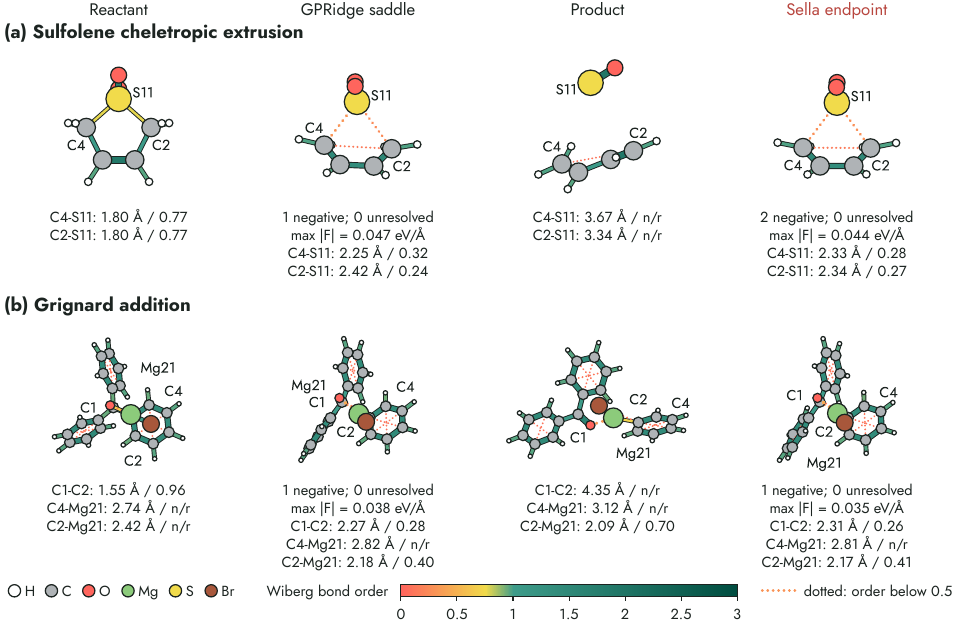}
\caption{\label{fig:reaction-hard}Bonding at the geometries for two Birkholz reactions: (a) sulfolene SO\(_2\) extrusion, where Sella returns the synchronous second-order point, and (b) the Grignard addition, where Sella and GPRidge return the same saddle. Conventions follow the main-text bonding figure.}
\end{figure}

\begin{figure}[p]
\centering
\includegraphics[width=0.98\textwidth]{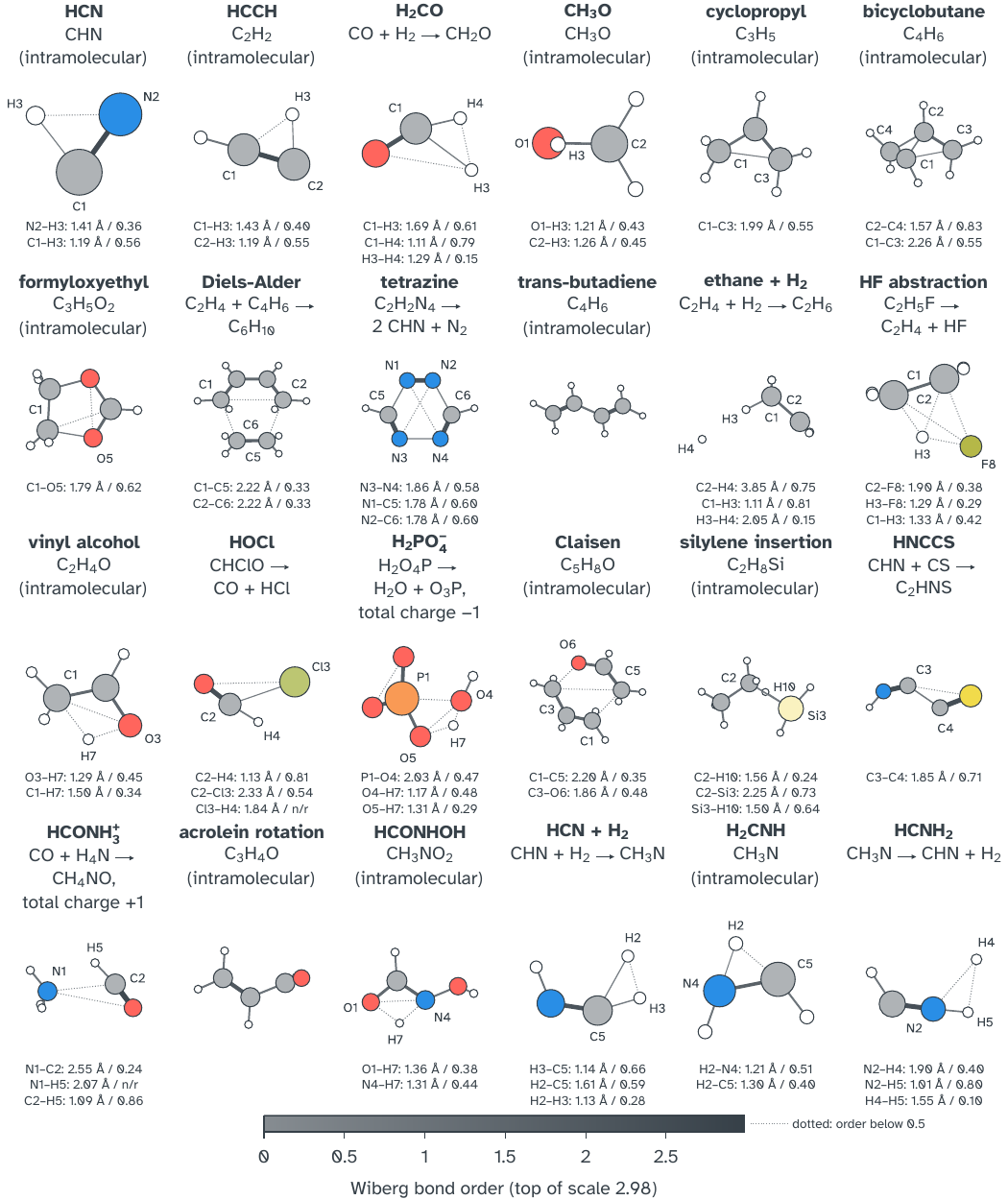}
\caption{\label{fig:reaction-gallery-baker}GPRidge saddle of every Baker reaction, titled with the label the comparison figures use. The line below each title gives the fragments of the reactant and product, by covalent-radius connectivity, and the total charge when it is not zero. Bonds and annotations follow the main-text bonding figure; the annotated pairs are those whose distance changes most between the endpoints, so the conformational rotations of trans-butadiene and acrolein carry no annotation.}
\end{figure}

\begin{figure}[p]
\centering
\includegraphics[width=0.98\textwidth]{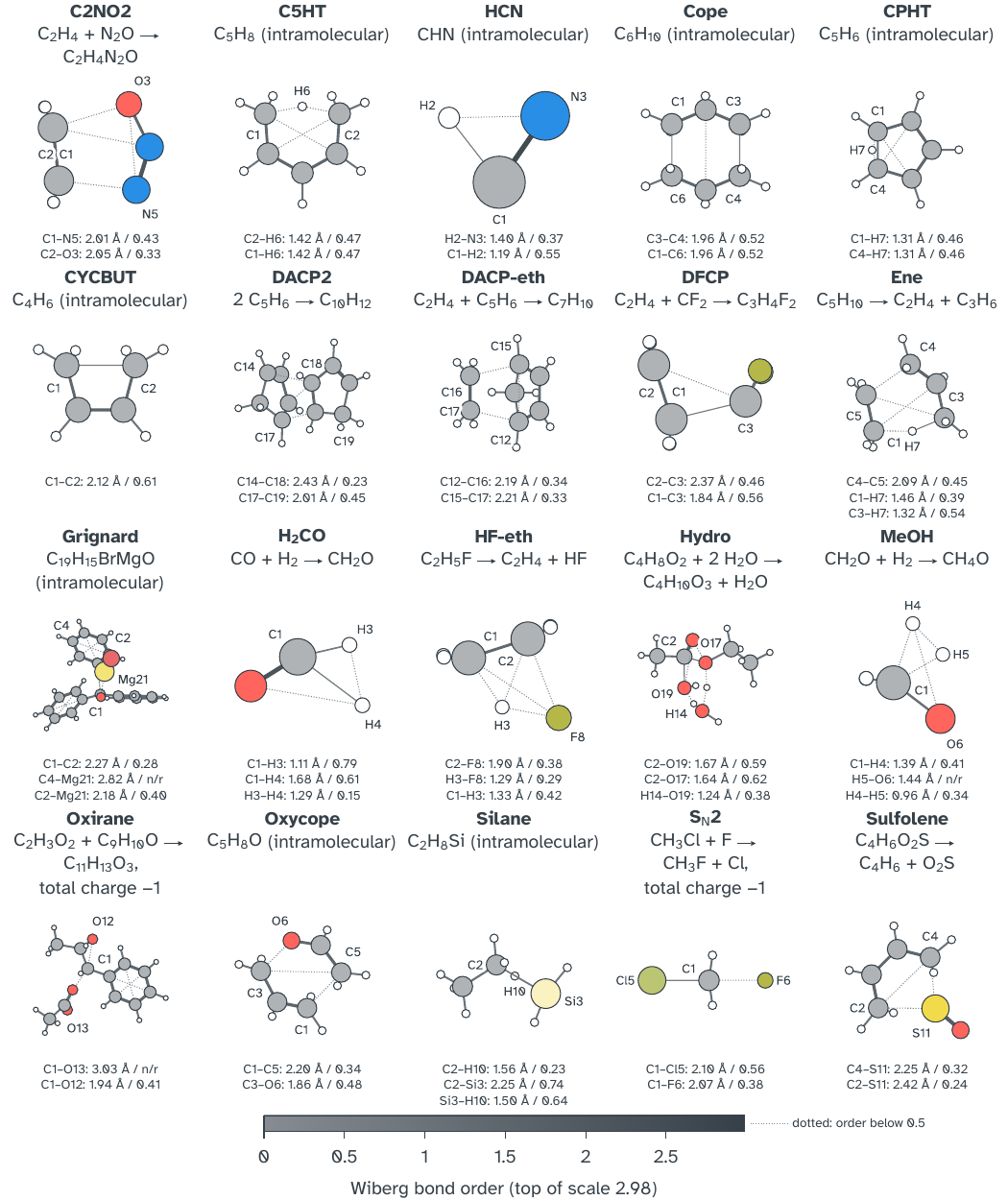}
\caption{\label{fig:reaction-gallery-birkholz}GPRidge saddle of every Birkholz reaction, with the conventions of Figure \ref{fig:reaction-gallery-baker}.}
\end{figure}

GFN2-xTB 6.7.1 single-point calculations provide the Wiberg bond orders and electron-density cubes \cite{bannwarthGFN2xTB2019}. Each comparison uses four geometries: the reactant, GPRidge saddle, product, and comparator endpoint. The main text shows HCN, ethane H\(_2\) abstraction, the parent Diels--Alder cycloaddition and the substituted oxirane; Figure \ref{fig:reaction-hard} adds sulfolene and Grignard. The galleries in Figures \ref{fig:reaction-gallery-baker} and \ref{fig:reaction-gallery-birkholz} evaluate only the GPRidge saddle of every reaction. Each calculation uses the charge and multiplicity of the benchmark cell: the oxirane, SN2 and dihydrogen phosphate systems have charge \(1i\) and the HCONH\(_3^+\) cation \(+1\); CH\(_3\)O, cyclopropyl and formyloxyethyl are neutral doublets, and the other systems are singlets. Structures are rigidly aligned to a common molecular orientation, with atom order and every interatomic distance preserved. The alignment treats the molecules as nonperiodic.

\begin{figure}[tbp]
\centering
\includegraphics[height=0.75\textheight,keepaspectratio]{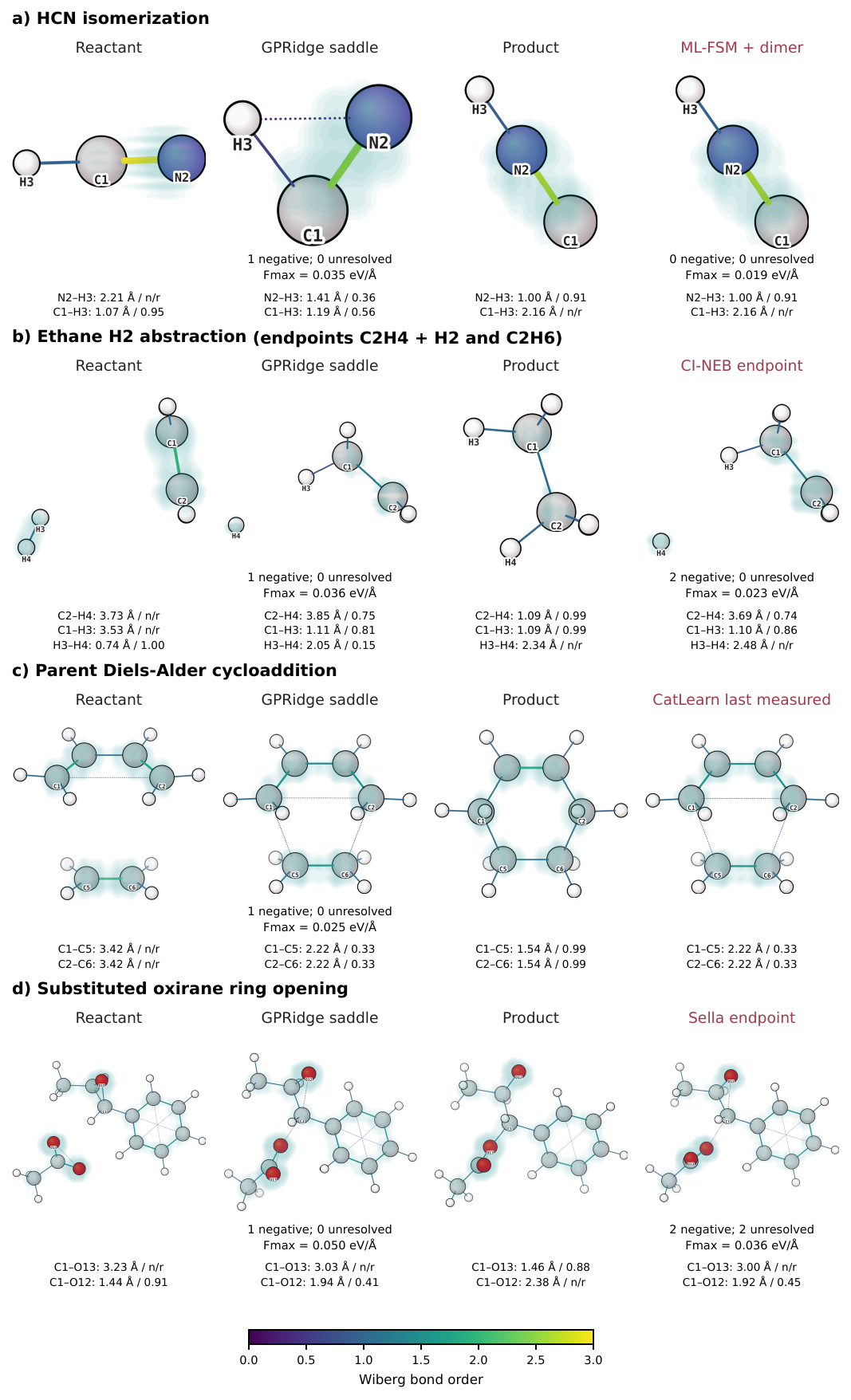}
\caption{\label{fig:reaction-density}GFN2-xTB electron-density isosurfaces at the same structures as the main-text bonding comparison. Teal surfaces use a common isovalue of 0.02 \(e\,\mathrm{bohr}^{-3}\). Distances, bond orders, and benchmark force and mode counts follow the main-text conventions. Electronic analysis uses fixed geometries and does not contribute to the search evaluation counts or timings.}
\end{figure}

The calculations use the default SCC mixing parameter of 0.4 and set an iteration limit of 2000. All 68 single points terminate normally. Density cubes use a 0.3 bohr grid and no density-matrix screening. Figure \ref{fig:reaction-density} shows the 0.02 \(e\,\mathrm{bohr}^{-3}\) isosurface with opacity 0.3. These approximate densities illustrate the electronic structure at the supplied coordinates; the UMA potential supplies the benchmark forces and Hessians.

chemparseplot draws the bonding structures as vector diagrams from the coordinates, and xyzrender 0.3.8 draws the density surfaces \cite{goodfellowGraphInternalCoordinates2026}. Connectivity, color and bond width use the computed bond orders. Values from 0.1 to below 0.5 use dotted lines, and values below 0.1 are omitted. No bond is drawn between atoms more than 2.5 times their covalent-radius sum apart; that distance keeps the forming or breaking contacts at every saddle except in ethane H\(_2\) abstraction, where H4 has separated from both H3 and C2, and the values of omitted pairs remain in the annotations. A missing pair in the sparse output is left unreported, rather than assigned zero. The numbered pairs are selected by their change in endpoint distance. Density and bond-order rendering share the same rigid orientation.

For HCN, the ML-FSM plus dimer endpoint has an N-H bond order of 0.9146, compared with 0.9145 in the product and 0.3610 at the GPRidge saddle. For oxirane, Sella's two computed negative frequencies are \(579.7i\) and \(43.3i\,\mathrm{cm}^{-1}\), and the endpoint also has two unresolved modes. GPRidge has one negative frequency, \(514.7i\,\mathrm{cm}^{-1}\), and no unresolved modes. For sulfolene, the GPRidge saddle extrudes SO\(_2\) asynchronously, with C-S distances of 2.25 and 2.42 \AA{} and bond orders 0.32 and 0.24. Sella's endpoint is the synchronous structure, 2.33 and 2.34 \AA{} with bond orders 0.28 and 0.27, and its curvature check reports two negative modes. For Grignard, whose endpoints run from the magnesium alkoxide to the reagent--ketone complex, so that the C1-C2 bond breaks from 1.55 \AA{}, both methods return essentially the same saddle: C1-C2 is 2.27 \AA{} (GPRidge) and 2.31 \AA{} (Sella), Mg-C2 2.18 and 2.17 \AA{}, each with one negative mode. Sella reaches it in 45 evaluations and the GPRidge dimer in 545. These diagnostics characterize local bonding and curvature; endpoint connectivity requires separate path-following calculations.

The cost and comparison figures use rgpycrumbs 1.13.0 and chemparseplot 1.13.2. The bonding figures use a later revision of chemparseplot, which is recorded with the figure sources.

\begin{figure}[tbp]
\centering
\includegraphics[width=0.98\textwidth]{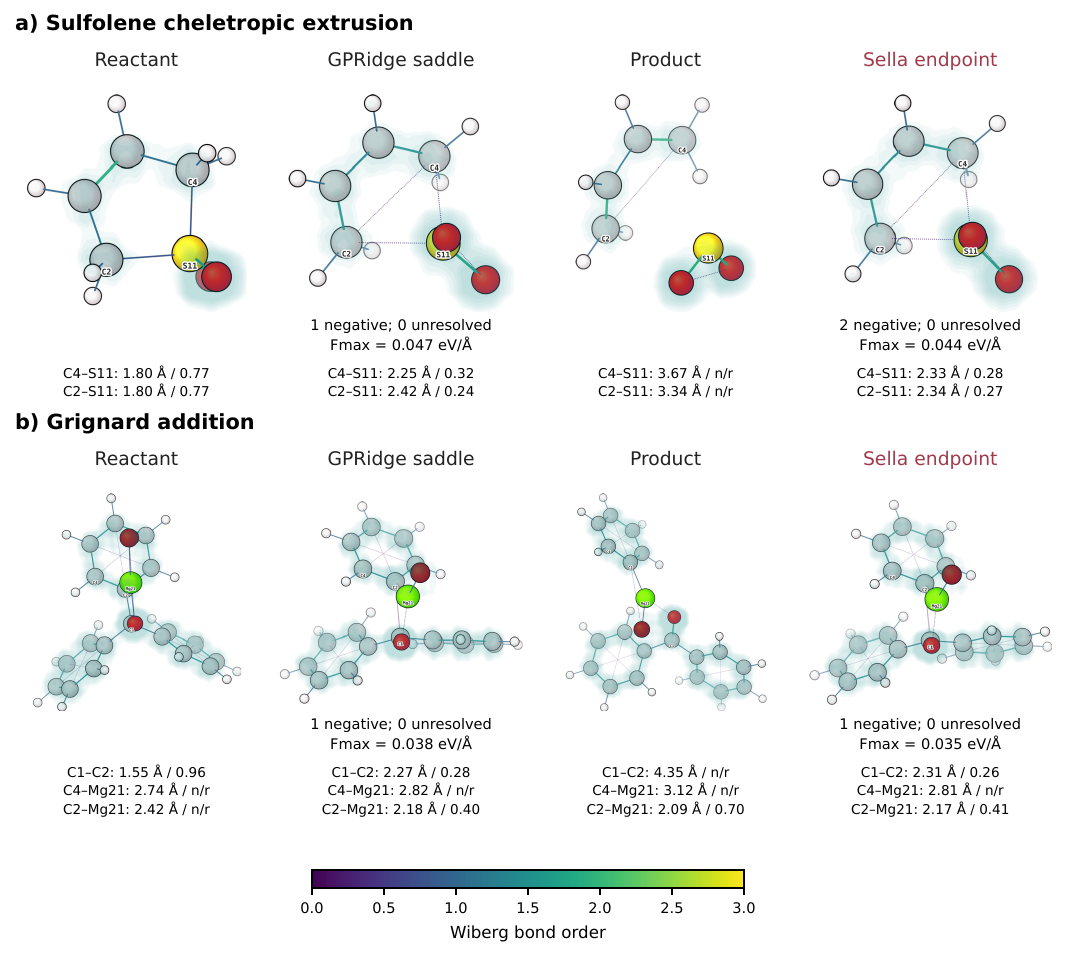}
\caption{\label{fig:reaction-hard-density}GFN2-xTB electron-density isosurfaces for the sulfolene and Grignard structures of Figure \ref{fig:reaction-hard}, with the conventions of Figure \ref{fig:reaction-density}.}
\end{figure}

\subsection{Full searches across CPU layouts}
\label{sec:org8fd1668}

Two repetitions of ethane H2 abstraction,
Grignard and oxirane compare one, four and eight MPI ranks, each with two
CPU slots per rank, at GPRidge 0.6.0. This full-search scaling experiment is separate from the GPRidge 0.5.0 retained-fit scaling and reaction results. Each reaction has an exclusive allocation and uses the
publication inputs, library arithmetic and force/index acceptance rules.
The table gives elapsed-time means and sample standard deviations over the two
repetitions. Stage calls are band calls for ethane H2 abstraction and dimer
calls for Grignard and oxirane; dimer counts exclude band seeding.

\begin{table}[H]
\caption{Full-search cost across CPU layouts. Every reaction finishes inside the configured 900 s allowance.}
\centering
\small
\begin{tabular}{lrrrr}
Case & Ranks & Passes & Stage calls & Elapsed time (s)\\
\hline
Grignard & 1 & 2/2 & 556 & 534.355 \(\pm\) 1.573\\
Grignard & 4 & 2/2 & 558 & 303.599 \(\pm\) 13.613\\
Grignard & 8 & 2/2 & 605 & 328.777 \(\pm\) 0.940\\
Ethane H2 abstraction & 1 & 2/2 & 82 & 10.331 \(\pm\) 0.018\\
Ethane H2 abstraction & 4 & 2/2 & 92 & 8.593 \(\pm\) 0.126\\
Ethane H2 abstraction & 8 & 2/2 & 76 & 12.065 \(\pm\) 1.490\\
Oxirane & 1 & 2/2 & 531 & 321.382 \(\pm\) 114.356\\
Oxirane & 4 & 2/2 & 520 & 125.317 \(\pm\) 1.070\\
Oxirane & 8 & 2/2 & 518 & 136.194 \(\pm\) 0.122\\
\end{tabular}
\end{table}

The call counts agree between repetitions of each layout and change with
rank count. The Grignard dimer search grows from 556 calls at one rank to
605 at eight, the oxirane search shrinks from 531 to 518, and the ethane
band takes 82, 92 and 76 calls. These timings measure full-search performance, combining parallel
execution with changes in search length. Four ranks give the shortest
elapsed time for every case; the one-rank oxirane repetitions differ by
162 s at identical call counts, the only pair whose times disagree by more
than 20 s. The four-rank layout remains the recommended configuration
for the complete panels.

Six additional trials set \texttt{OPENBLAS\_THREAD\_TIMEOUT=4} at four ranks,
shortening the pthread idle spin interval. All six pass with unchanged
call counts. Their mean elapsed times are 302.972 s for Grignard, 8.849 s
for ethane H2 abstraction and 131.574 s for oxirane, against 303.599,
8.593 and 125.317 s with the compiled idle-wait setting. The recommended
configuration uses that compiled setting. The complete study has 24
passes in 24 trials.
\subsection{Full searches with reproducible arithmetic}
\label{sec:org8f8f53a}

Nine exclusive reactions repeat the three-case layout
comparison with \texttt{-{}-{}linalg-mode reproducible} in both drivers, two CPU slots per rank and the
unchanged 900 s Birkholz allowance. For each case, the search-call count,
terminal energy, maximum atomic force and complete vibrational frequency
array agree exactly across one, four and eight ranks. Grignard and oxirane
pass with one imaginary mode and no unresolved modes. The ethane H2
abstraction band ends, at every rank count, at a stationary point with two
imaginary frequencies, \(101.7i\) and \(47.7i\) cm\(^{-1}\); the second lies
below the 50 cm\(^{-1}\) frequency tolerance, but the acceptance rule counts
every imaginary mode, so the cell fails the index criterion three times
with identical outputs.

\begin{table}[H]
\caption{Reproducible-mode full searches at fixed search-call counts. Each time is one measurement of the complete calculation, including independent validation; no repetition uncertainty is estimated. The ethane rows fail the one-imaginary-mode criterion at every rank count.}
\centering
\small
\begin{tabular}{lrrrr}
Case & Ranks & Search calls & Dimer calls & Elapsed time (s)\\
\hline
Grignard & 1 & 639 & 595 & 489.465\\
Grignard & 4 & 639 & 595 & 308.923\\
Grignard & 8 & 639 & 595 & 291.551\\
Ethane H2 abstraction & 1 & 74 & 0 & 11.026\\
Ethane H2 abstraction & 4 & 74 & 0 & 8.548\\
Ethane H2 abstraction & 8 & 74 & 0 & 9.033\\
Oxirane & 1 & 369 & 325 & 176.836\\
Oxirane & 4 & 369 & 325 & 106.057\\
Oxirane & 8 & 369 & 325 & 101.642\\
\end{tabular}
\end{table}

Eight ranks give elapsed-time speedups of 1.68 for Grignard and 1.74 for
oxirane relative to one rank, parallel efficiencies of 0.210 and 0.217. The fixed search-call
counts permit this comparison without a change in search length.

At four ranks, reproducible arithmetic increases Grignard from 558 to 595
dimer calls and its elapsed time from 303.6 to 308.9 s, and decreases
oxirane from 520 to 325 dimer calls and from 125.3 to 106.1 s. Ethane H2
abstraction takes 74 search calls against 92 but does not pass. Rank
invariance therefore costs little in search length on this release, and
the complete-panel recommendation uses library arithmetic because of
the ethane index failure. Both arithmetic modes finish the one-rank
Grignard search within the allowance, in 534.4 s with library arithmetic
and 489.5 s with reproducible arithmetic.

With library arithmetic the search length depends on the layout: ethane
H2 abstraction takes 82, 92 and 76 band calls, Grignard 556, 558 and 605
dimer calls and oxirane 531, 520 and 518 at one, four and eight ranks.
The controlled arithmetic change removes that layout dependence in all
three cases.
\subsection{SLATE verification}
\label{sec:org323b52f}

The SLATE backend passes the three-case comparison
at one, four and eight MPI ranks, with two CPU slots per rank and
\texttt{-{}-{}linalg-mode library}. All nine reactions have one imaginary mode and
no unresolved modes; all 32 configured suites pass, with 1785 test
invocations.

\begin{table}[H]
\caption{SLATE full searches with library arithmetic. Each elapsed time is one measurement including band preparation and independent validation. Search calls include band seeding; dimer calls exclude it.}
\centering
\small
\begin{tabular}{lrrrr}
Case & Ranks & Search calls & Dimer calls & Elapsed time (s)\\
\hline
Grignard & 1 & 604 & 561 & 508.639\\
Grignard & 4 & 633 & 595 & 357.653\\
Grignard & 8 & 557 & 513 & 270.356\\
Ethane H2 abstraction & 1 & 94 & 0 & 18.592\\
Ethane H2 abstraction & 4 & 103 & 0 & 17.853\\
Ethane H2 abstraction & 8 & 96 & 0 & 14.978\\
Oxirane & 1 & 318 & 274 & 168.172\\
Oxirane & 4 & 317 & 273 & 96.563\\
Oxirane & 8 & 315 & 271 & 81.930\\
\end{tabular}
\end{table}

The call counts vary across layouts, so elapsed time also reflects
changes in search length. The four-rank ScaLAPACK configuration remains
the recommendation for the complete panels.

An isolated
350-row test of \(I+vv^T\) checks the inverse
\(I-vv^T/(1+v^Tv)\) over 5000 repetitions at eight ranks and two workers
per rank. Its maximum entry error is \(6.7\times10^{-16}\).
The backend test repeats this analytic comparison in the
supported test layouts.

\section{Campaign option list}
\label{sec:option-manual}
The settings used in the calculations are in the article.
This section lists the drivers and acquisition schedules used in the calculations.
\subsection{Model profile and search protocol}
\label{sec:org3cb58e1}

A model profile supplies a group of GP defaults. The names \texttt{chemgp}
and \texttt{gprd\_zbl} select the same classic joint energy-gradient model;
\texttt{extended} exposes the independently configured alternatives. The two
classic identifiers preserve compatibility with reference inputs. They
do not identify the Baker and Birkholz datasets.

In the tagged source, the classic default optimizer is SCG, and an
explicit \path{--hyperparameter-mode rgmin:lbfgs} selects rgmin L-BFGS. The
optimizer, the factorization and the search driver are separate choices.
The profile name alone does not enable the switched ZBL mean.
\subsection{Run parameters that fix the arithmetic}
\label{sec:si-run-parameters}
The following settings determine the arithmetic of a run and are
given with each case's arguments and runtime environment.
\begin{itemize}
\item \texttt{-{}-{}linalg-threads N} (default 1) sets the dense linear-algebra
thread count for Eigen and, when linked, OpenBLAS.
\texttt{-{}-{}kernel-threads N} (default 1) sets independent kernel-block
workers per rank; parallel kernel blocks require
\texttt{-{}-{}linalg-threads 1}.
\item \texttt{-{}-{}gp-variance-convention latent|predictive} (default
\texttt{latent}) names the posterior variance the loop reads. The
Cholesky backend takes its prior from the training
covariance, which carries the likelihood nugget (the classic
\texttt{get\_variance.m} definition); the stable dual basis takes it
from the kernel alone. \texttt{latent} removes the nugget on the
first and \texttt{predictive} adds it on the second, so both report
one number and an acquisition that compares uncertainty
against the nugget behaves the same on either backend.
\item The dense-algebra library selects its kernels by the host
processor. Two nodes of one cluster then round the same
factorization differently from identical arguments.
\texttt{OPENBLAS\_CORETYPE} selects the kernel family. The thread count
is the separate \texttt{-{}-{}linalg-threads} setting. The oracle adapter
pins the compiled model to one OpenMP thread.
\item The oracle package is exported with a fixed roll in every
edge frame and evaluates a canonically ordered neighbor
list. Without both, the universal potential's output
depends on the frame and the neighbor order, and the
acquisition sequence of a cell can diverge between runs. These
choices fix the frame and neighbor order. Bitwise agreement of
the complete search also depends on the GP backend, processor
kernels, MPI layout and thread counts.
\end{itemize}
\subsection{Drivers and rg sessions}
\label{sec:si-drivers}
\texttt{ActiveGP} answers as a potential: from the GP when the
posterior is resolved, from the oracle otherwise. rg
sessions drive that potential (\texttt{SellaMin}, \texttt{SellaSaddle},
\texttt{MinMode}, \texttt{RgsaddleBandEngine}). \texttt{RgsaddleClient}
loads \texttt{librgsaddle} with \texttt{dlopen} from a path or
\texttt{RGSADDLE\_LIBRARY} and requires C ABI major 1.
\texttt{SellaSaddle} is the order-1 saddle check.

The reported searches relax the band by L-BFGS. A FIRE2 band optimizer is also available, because the NEB force is not a potential gradient. The drivers read the reactant and product geometries and write the relaxed chain.
\subsection{One-image outer schedule}
\label{sec:si-oie-schedule}
The reported executable defaults to \texttt{-{}-{}acquisition tpe}, and the
publication launcher does not override it. It evaluates the initial
18-image band and its endpoints through \texttt{-{}-{}seed-full-chain true}.
Subsequent outer rounds generally acquire one image; support batching
adds unsupported images in the Birkholz preparation stage.

The measurement priorities include geometries needed to release a
kernel-support boundary, activate or confirm the climbing image, assess
a rejected proposal, and resolve the force-readiness tests. A required
measurement keeps its priority over an ordinary posterior acquisition.
The climbing image receives a potential evaluation at its current
geometry before force convergence is accepted. A geometry already
present in the earlier evaluations is not acquired as a new row merely
because its image index has moved.

For replaceable acquisitions, the tree-structured Parzen estimator \cite{bergstraAlgorithmsHyperparameter2011} considers
unobserved interior images whose variance exceeds the likelihood floor.
Its force-accuracy utility is
\begin{equation}
 u_i=z(f_i)+\beta_t z(v_i^F),\qquad
 \beta_t=\sqrt{\max\!\left(0,2\log\frac{m t^3\pi^2}{6}\right)},
\end{equation}
where \(f_i\) is the predicted NEB-force norm, \(v_i^F\) the posterior
gradient variance, \(m\) the number of interior images and \(t\geq1\) the
acquisition round. The code uses \(\delta_t=1/t\) in \(2\log(mt^2\pi^2/(6\delta_t))\), producing the displayed \(t^3\). This schedules exploration; a fixed-confidence GP-UCB regret bound is not asserted for this standardized TPE utility. Each \(z\) subtracts that channel's mean over the
interior images and divides by its population standard deviation;
a constant channel contributes zero. This standardization acts only
on acquisition scores. When the band maximum is confident, \(z(E_i)\)
is added to the utility, and an eligible interior maximum above the
endpoints can be selected directly.

The remaining eligible image indices are split by descending utility.
The upper group contains
\(\min(\lceil0.1m_{\mathrm{eligible}}\rceil,25)\) indices.
Gaussian Parzen densities are constructed on the upper and lower groups;
upper-group weights are proportional to \(\exp(u_i-u_{\max})\).
The policy chooses the eligible integer index maximizing
\(\log p_{\mathrm{upper}}(i)-\log p_{\mathrm{lower}}(i)\).
The source fixes the bandwidth clipping and prior component in
\path{gpr/neb/oie/AcquisitionPolicy.cpp}.

A separate utility resolves saddle identity when the climbing image is
active, its identity is uncertain and the force-readiness criterion is
met. It weights each challenger's posterior energy standard deviation
by the marginal Gaussian comparison
\begin{equation}
 \pi_i=\Phi\!\left(\frac{\mu_i-\mu_c}{\sqrt{v_i^E+v_c^E}}\right),
\end{equation}
where \(c\) is the candidate. This comparison uses the sum of marginal
variances. The candidate receives the sum of the comparisons times its
own standard deviation; endpoint comparisons contribute to that sum too.
This utility supplies the same tree-structured Parzen estimator ranking \cite{bergstraAlgorithmsHyperparameter2011}.

Force acceptance uses
\(\max_a\lVert g_a(x_{i^\star}^{(k)})\rVert_2<\varepsilon\) at a
geometry measured in that iteration. The independent Hessian assessment
is separate. The publication Baker band disables its internal
index probe; Birkholz passes the band maximum and observations to
the separate dimer search.

Expected improvement is an alternative selected by \texttt{-{}-{}acquisition ei}.
For energy maximization its score is
\cite{jonesEfficientGlobalOptimization1998,rasmussenGaussianProcessesMachine2006}
\begin{equation}
 \mathrm{EI}(x_i)=(\mu_i-\mu_{\mathrm{best}})\Phi(z_i)
                 +\sigma_i\phi(z_i),\qquad
 z_i=(\mu_i-\mu_{\mathrm{best}})/\sigma_i.
\label{eq:si-jones-ei}
\end{equation}
The calculations use the tree-structured Parzen estimator above \cite{bergstraAlgorithmsHyperparameter2011}, not this expected-improvement score.
The optional OCI-NEB strategy adds dimer bursts to a band search
\cite{goswamiEnhancedClimbingImage2026}; the reported Baker and Birkholz
protocols use the two sequences stated in the main text.
\subsection{Energies that seed the dimer}
\label{sec:si-archive}
Every calculated energy and force is kept. Adaptive pruning decides
which geometries the covariance carries. A pruned geometry can still be
evaluated again.

Before the dimer's first fit, each band energy is shifted by the
midpoint energy, as every other training energy is. The dimer is seeded
with those shifted energies. The unshifted band energies are not a
valid training set for the dimer.
Both drivers keep the likelihood noise at the profile's
fixed nugget: the oracle is deterministic, and a learned
noise absorbs the surrogate's misfit as a floor under every
variance gate. On the band it floors the pruning and
acquisition criteria; on the dimer it floors the
posterior force uncertainty above the trust hand-off's own
scale. The reported calculations keep that noise fixed.
\subsection{Scoring on a path}
\label{sec:si-scoring}
A pair of minima can be joined by several minimum energy
paths. Correctness is the index-1 test of the main text,
not agreement with an earlier geometry. A barrier subtracts
an endpoint that each code relaxes separately, so
disagreement in one direction and agreement in the other
localises the discrepancy to an endpoint. Structural
comparison against published saddles is a description of
which path a run found.

\section{Physical curvature and transported directions}
\label{sec:si-physical-curvature}
Let \(R\) contain the Cartesian rigid translations and, for an isolated
molecule, its independent rigid rotations. Let \(M\) contain the positive
atomic masses, repeated for each Cartesian coordinate, and let \(H\) be
the Cartesian Hessian. For a Cartesian displacement \(x\),
\((M^{1/2}R)^\top(M^{1/2}x)=R^\top Mx\), so \(\ker(R^\top M)\) is the
set of displacements orthogonal to rigid motion in the mass-weighted
product. Removing rigid motion from the mass-weighted Hessian
\(M^{-1/2}HM^{-1/2}\) selects Cartesian displacements in that kernel.
For a Cartesian orthonormal basis \(Q\) of this space, define
\(W=Q^\top M Q\). In the frame \(M^{1/2}QW^{-1/2}\), the physical block
of \(M^{-1/2}HM^{-1/2}\) is
\begin{equation}
 W^{-1/2}(Q^\top H Q)W^{-1/2}.
\end{equation}
That matrix is congruent to \(Q^\top H Q\). Since \(W\) is positive definite,
this matrix and \(Q^\top H Q\) have the same
inertia. Their eigenvalues need not coincide. This construction permits
Cartesian force differences while including the input isotope masses.
Periodic systems remove translations; fixed atoms define a constrained
coordinate space. A linear molecule contributes only two independent
rigid rotations.

At a nonstationary geometry, rigid rotations need not be null vectors of
the Cartesian Hessian. Unit-mass and mass-weighted projection can therefore
report different indices even when the force lies below the convergence
threshold. The exact three-atom witness in
\nolinkurl{scripts/derive_mass_physical_space.py} distinguishes these two
indices. Native and independent validation use the same physical mass
convention; the force criterion remains the maximum Euclidean atomic force.

A measured negative subspace can supply directions for a nearby geometry. If \(U\) contains those directions and \(P=QQ^\top\), the transported proposal is an orthonormal basis of \(PU\), provided its rank is kept, and its curvature is measured again on the true surface. No curvature sign is transported. The exact example in \nolinkurl{scripts/derive_transported_curvature.py} shows that projection can change the sign of the compressed Hessian.

Two freshly resolved negative Ritz values reject an index-one candidate, and residuals on that subspace alone leave the complement open. Every real complement curvature \(c\) is compatible with the observed products, including \(c<0\), and the sign is measured before it counts. Refining that pair evaluates the potential gradient at both signs of the separation and of half the separation, eight energy and force evaluations for the two directions. Here \(A_h\) is the central-difference Hessian action at separation \(h\), and the combination \((4A_{h/2}-A_h)/3\) cancels the leading \(O(h^2)\) error in the Hessian action on those axes. The stencil difference is the empirical resolution of the pair. A negative upper bound is an exit along the measured descent, and the search must still satisfy the force tolerance and independent index check.

Along a measured direction the local model is \(fq+cq^{2}/2+bq^{3}/6\), and the stationary curvature is \(\mathrm{sign}(c)\sqrt{c^{2}-2bf}\) when \(c^{2}-2bf>0\). The coefficient \(b\) uses the even gradient residuals of the measured axes. Two orthonormal axes determine the symmetric cubic in their plane, and a coefficient on three distinct axes is absent from the sum. While a mode's measured upper bound stays nonnegative, the same model can lower the force tolerance through the dual norm of the mode, and the floor \(\sqrt{2\max(bf,0)}\) is reported beside the spectrum. The climbing-image index counts posterior curvatures on its complete measured complement, and the dimer index counts the Ritz values of its measured span once the Lanczos residuals lie inside the stencil error. Independent force positions are distributed across MPI ranks, with every potential evaluation included in the call ledger.
\section{Pair-local tiles, tied types, and one-image control}
\label{sec:si-pair-local-ties-control}
This section gives the production covariance tiles, the evidence
rule that releases an environment pair type from its base type, and
the one-image outer/inner controls: a restore-versus-continue bandit,
Brent cycle detection and the kernel-support freeze.
\subsection{Pair-local covariance tiles}
\label{sec:si-pair-local-tiles}
The Sexpat kernel on inverse pair distances is
\begin{equation}
  K(x,x')=m\exp(-S),\qquad
  S=\sum_f w_f\bigl(q_f(x)-q_f(x')\bigr)^2,\qquad
  w_f=1/\ell_{t(f)}^2,
\end{equation}
with $q_f=1/r_f$. Differentiating with respect to the Cartesian
coordinates of $x$ and $x'$ gives the joint energy--gradient tile
\begin{equation}
\begin{gathered}
K\begin{pmatrix}
1 & h^{\mathsf T} \\
-g & -g h^{\mathsf T}+2 S_J
\end{pmatrix},\\
g=2\sum_f w_f d_f J_f(x),\qquad
h=2\sum_f w_f d_f J_f(x'),\\
S_J=\sum_f w_f J_f(x)J_f(x')^{\mathsf T}.
\end{gathered}
\label{eq:si-pair-local-tile}
\end{equation}
where $d_f=q_f(x)-q_f(x')$ and $J_f$ is the Cartesian gradient of
$q_f$. On the diagonal $x=x'$ one has $d=0$, so $g=h=0$ and the
tile is $K[1,0;0,2 S_J]$. The derivative with respect to
$\log\ell_t$ is
\begin{equation}
  2 S_t\cdot(\text{tile})
  +
  K
  \begin{pmatrix}
    0 & -2 h_t^{\mathsf T}\\
    2 g_t & 2 g_t h^{\mathsf T}+2 g h_t^{\mathsf T}-4 S_{J,t}
  \end{pmatrix},
\end{equation}
with $S_t=\sum_{f\in t} w_f d_f^2$ and $g_t$, $h_t$, $S_{J,t}$ the
type-$t$ parts of $g$, $h$ and $S_J$. The derivative with respect
to $\log m$ is the tile itself.

\nolinkurl{scripts/derive_pair_local_tiles.py} derives both closed
forms by differentiating the kernel on a two-atom, two-type model
that carries the sparse pair Jacobian and checks every entry against
the expressions \texttt{PairLocalKernel} evaluates. The tests
\nolinkurl{PairLocalTilesAssembleTheBlockGeometryCovariance} and
\nolinkurl{PairLocalTilesReproduceTheHyperparameterGradient} compare the
tiled assembly to the block-geometry covariance and the
hyperparameter gradient. Tiles fill the training covariance, the
hyperparameter-gradient contractions, the quadratic forms, the
ScaLAPACK factor, and the cross covariances between training rows
and a query chain.

The chain-variance probe is memoised on the exact band geometry,
training rows, and factor epoch (\texttt{ChainVarianceMemo}), so the restricted-variance target and
the proposal readiness of one inner step share one triangular solve.
A measured dimer endpoint that coincides with the axis probe at
roundoff is reused; bitwise equality is not required.

Dense linear algebra can raise the BLAS thread count for the
posterior triangular solves, the inverse diagonal, and the cached
inverse, and it restores that count afterwards. The thread setting is restored around each backend operation. The reported publication fits use the two BLAS threads per rank specified in the run arguments.
\subsection{Periodic pair features and fixed substrates}
\label{sec:si-periodic-pairs}
The covariance expressions above also apply to compact radial features.
For a cutoff radius \(R_c\) and switching width \(w\), define
\begin{equation}
q(r)=\frac{s(r)}{r},\qquad
s(r)=
\begin{cases}
1, & r\leq R_c-w,\\
10u^3-15u^4+6u^5, & R_c-w<r<R_c,\\
0, & r\geq R_c,
\end{cases}
\qquad u=\frac{R_c-r}{w}.
\end{equation}
The switching polynomial has zero first and second derivatives at both
ends. Thus \(q\), its Cartesian gradient and its Cartesian Hessian join
continuously at the switching radii. With minimum-image displacement
\(\delta_f\) from the partner atom to the first atom,
\begin{equation}
J_{f,a}=\left(\frac{s'(r_f)}{r_f}-\frac{s(r_f)}{r_f^2}\right)
\frac{\delta_f}{r_f},\qquad J_{f,b}=-J_{f,a}.
\end{equation}
A fixed partner contributes to \(q_f\) and \(J_{f,a}\), but has no independent
Cartesian channel. The unconstrained molecular feature is recovered by
\(s(r)=1\).

The implementation stores each radial value and its three-component pair
gradient once per configuration. Every Cartesian covariance tile and
hyperparameter contraction reuses this geometry. Pair indices use the
same ordering across configurations; a pair outside the cutoff contributes
an exact zero. This common ordering is necessary when a pair crosses the
switching shell between a training configuration and a query. The cutoff defines compact pair features; the assembled covariance remains
dense.

Native tests compare all energy and gradient covariance blocks with the
independent radial-feature implementation for periodic configurations,
fixed atoms, three length-scale types and pairs on both sides of the
switching radii. The absolute covariance tolerance is \(10^{-12}\).
Central differences check the log-length derivatives at a tolerance of
\(2\times10^{-8}\). These tests establish agreement of the two implementations
and their derivatives; search performance is measured separately.
\subsection{Environment types tied to the base pair types}
\label{sec:si-tied-types}
The base pair types are the element-and-reaction-role split of the
bond-graph rule
\cite{goswamiAdaptivePruningIncreased2026}.
\texttt{scripts/gp\_environment\_types.py} assigns SOAP/knn
environment sub-types by a largest-gap split within each base type.
The types file carries the base type as a fourth column. Each
sub-type starts tied to its base: the optimizer works on group
parameters $\theta$ with fine log-lengths $w=J\theta$. Each column
of $J$ identifies one nonempty, disjoint group, so $J^{\mathsf T}J$
is invertible and
\begin{equation}
  \frac{\mathrm{d}F}{\mathrm{d}\theta}=J^{\mathsf T}\frac{\mathrm{d}F}{\mathrm{d}w}.
\end{equation}
A stationary point of the tied problem is $J^{\mathsf T}\nabla F=0$.
Let $g$ be the derivative of the objective with respect to the
length-scale weights. The production code keeps the fine storage: the
group leader is the optimizer's parameter and carries $J^{\mathsf T}g$;
members are held at the leader's value. The averaging projector
$P=J(J^{\mathsf T}J)^{-1}J^{\mathsf T}$ has the same stationary
points, but it replaces each group sum by the group mean, the sum
divided by the group size, which is not the problem the optimizer solves.

The seed and the prior anchor of a group are the base type's motion
scale, the root of the summed feature variances of the union of its
pairs. The covariance function's two-sided length-scale prior
counts each group once (a prior weight of one on a leader and zero
on a member). A model with every fine type tied is therefore the
base-type model. The tests
\nolinkurl{TiedFitReproducesTheBaseTypeFit} and
\nolinkurl{TiedObjectiveGradientAndCurvatureMatchDifferences} check
that identity and the wrapped objective's gradient and curvature
against finite differences.

Releasing a child adds one parameter. With $g$ the fine gradient of
the negative log marginal likelihood at the tied value and $h$ its
curvature there, a positive $h$ gives a minimizing step $-g/h$
and a predicted decrease $g^2/(2h)$ in the quadratic model. Laplace's approximation of the evidence charges an
Occam factor whose $n$-dependence is $(\log n)/2$, the BIC term.
The code uses that BIC form as a screen, not as the decision:
\begin{equation}
  \text{screen }c\text{ iff }\frac{g_c^2}{2 h_c}>\tfrac12\log n.
\end{equation}
Curvatures $h_c$ come from Hutchinson's diagonal estimator
\cite{bekasEstimatorDiagonalMatrix2007}.
Gradient differences run along random sign vectors over the candidate
coordinates, one probe at a time. Probing stops at one standard error,
or when the still-ambiguous coordinates number no more than the probes
already spent. Those coordinates are then probed exactly. The screened set is
released in a trial refit. The realized change of the negative log
marginal likelihood is attributed to the candidates in proportion
to their predicted gains. Each share, minus the Occam cost, feeds
the type's cumulative sum
\cite{pageContinuousInspectionSchemes1954},
floored at zero. A type is released when its sum reaches the Occam
cost; the others return to their groups. Releases are monotone. A
tie group of one keeps its motion scale bit for bit.

\texttt{scripts/derive\_tied\_length\_scales.py} checks the
group-sum identity, the averaging projector, the quadratic gain,
the Laplace Occam term, the Hutchinson diagonal identity, and the
pooled motion scale in exact SymPy arithmetic.

The manuscript file \nolinkurl{lean/GprOptim/TiedLengthScales.lean}
formalizes differentiation through the fixed linear expansion,
projector/group-sum stationarity under the inverse Gram relation,
and the completed-square identity and minimum for positive curvature.
The gain is a property of the local quadratic model; the nonlinear
refit is assessed by its realized objective change.
\subsection{Restore-versus-continue bandit, Brent detection, and support freeze}
\label{sec:si-continuation-brent}
Two controls in the one-image outer and inner loops, a
restore-versus-continue bandit and Brent cycle detection, are the
production controls of the searches.

\paragraph{Restore as a fixed point.}
A lower-barrier rule restores the best climb whenever the
relaxed band has a lower barrier. A restore that returns to the same
snapshot every outer is a fixed point of the outer iteration, not a
search.

\paragraph{Two-armed bandit.}
Whether to restore the best climb or continue from the relaxed
band is a two-armed bandit with unbounded real rewards
\cite{gaglioloAlgorithmPortfolioSelection2011}.
Each time the barrier rule proposes a restore, one arm is pulled.
The reward is the relative improvement, at the next outer, of the
best measured climbing force. The best climb holds that best
force, so a restore returns to it by the restore itself and earns
the improvement only when a later measurement lowers the best force.
Each arm's rewards are treated as normal with
unknown mean $\mu$ and variance $s^2$ under the Jeffreys prior
$p(\mu,s^2)\propto 1/s^2$. After $n$ rewards with mean
$\bar x$ and sample variance $v$, the marginal posterior of
$\mu$ is Student's $t$ with $n-1$ degrees of freedom, location
$\bar x$ and scale $\sqrt{v/n}$. Thompson sampling
\cite{thompsonLikelihoodUnknownProbability1933}
draws one $\mu$ per arm and pulls the larger draw. Until each arm
has two rewards, the arm with fewer pulls is chosen. A restore that
only reproduces the previous rejection earns nothing and stops
being chosen; a restore that lowers the force keeps being chosen.

\paragraph{Brent cycle detection and doubling-window progress.}
The inner loop uses no fixed iteration cap as its liveness bound. Brent's cycle detection
\cite{brentImprovedMonteCarloFactorization1980}
moves the anchor to the current accepted band iterate at every
power of two accepted steps. An iterate that coincides with the
anchor, at the resolution the acquisition uses for one point,
closes a limit cycle of the non-conservative band force and ends
the inner loop. With tail length $\mu$ and cycle length
$\lambda$ the cycle is found within the first power of two at or
above $\max(\mu,\lambda)$ plus $\lambda$ steps. A quasi-cycle that
never returns exactly is caught on the same doubling windows: the
relaxation force's running minimum must improve within each
window, judged only once the window is at least the band L-BFGS
memory, the horizon over which the stepper's quasi-Newton model is
built. No new constants. A configured inner limit remains an
explicit user bound; zero is unbounded.

\paragraph{Kernel-support freeze.}
An image farther than one length scale from every training row, in
the kernel metric, is held. With the freeze on, every such image
joins that outer's oracle batch: the surrogate has no prediction
for it, and waiting one round per image stalls the band.

\nolinkurl{scripts/derive\_continuation\_bandit.py} derives the
$t$ marginal by integrating $s^2$ out of the normal--Jeffreys
model, checks Brent's detection bound on an explicit orbit
(tail 3, cycle 5), and states the doubling-window progress rule.
\section{Conditioning an active residual on training observations}
\label{sec:si-active-residual}
Let the background be a Gaussian process already conditioned on its
training energy and Cartesian-gradient channels. Write \(m_B\) and \(C_B\)
for that posterior mean and covariance, and write \(B\) for those
channels. An active refit leaves \((m_B,C_B)\) fixed, background
hyperparameters included, and adds an independent residual
\(f_\theta\sim\mathrm{GP}(0,K_\theta)\). The model on the new channels is
the sum
\begin{equation}
 f=f_B+f_\theta .
 \label{eq:si-residual-sum}
\end{equation}
The new channels \(A\) are disjoint from \(B\). The residual kernel may be
the inverse-distance kernel, a species-resolved smooth overlap of atomic
positions, a Cartesian chart kernel, or a sum of those kernels.

On a pair of index sets the sum has covariance \(C_B+K_\theta\).
Conditioning that sum on the new observations alone, with likelihood
noise \(N_A\), uses \(Q=C_B+K_\theta\) and \(r=y_A-m_B(A)\):
\begin{equation}
 m_{\mathrm{free}}(X)
 =m_B(X)+Q(X,A)\,[Q(A,A)+N_A]^{-1}r .
 \label{eq:si-free-residual-mean}
\end{equation}
The block \(Q(B,A)\) need not vanish, so \(m_{\mathrm{free}}(B)\) need not
reproduce the background mean.

One training channel and one new channel, in that order, make the move
explicit. Take \(N_A=0\) and \(m_B=0\),
\begin{equation}
 C_B=\begin{pmatrix}0&0\\0&3/4\end{pmatrix},\qquad
 K_\theta=\frac14\begin{pmatrix}1&1\\1&1\end{pmatrix}.
\end{equation}
Adding the blocks gives
\begin{equation}
 Q=C_B+K_\theta=\begin{pmatrix}1/4&1/4\\1/4&1\end{pmatrix}.
\end{equation}
The new observation \(y_A=1\) then has conditional mean
\begin{equation}
 m_{\mathrm{free}}(B)
 =Q_{BA}Q_{AA}^{-1}y_A
 =\frac14\cdot 1^{-1}\cdot 1
 =\frac14 .
 \label{eq:si-free-residual-counterexample}
\end{equation}
The noiseless background mean was \(0\). The free residual moved it
by \(1/4\).

The anchored model constrains the residual before it sees \(A\).
Pseudo-observations of zero residual at the training channels,
\begin{equation}
 f_\theta(B)+\varepsilon_B=0,
 \qquad
 \varepsilon_B\sim\mathcal N(0,N_B),
 \label{eq:si-anchor-observation}
\end{equation}
use a covariance \(N_B\) equal to the training observations' physical
likelihood noise plus numerical jitter. That covariance is fixed while
\(\theta\) is fitted. The pseudo-observations are a prior constraint on
the residual. They are not a second measurement of the physical surface,
and the training physical observations are not appended to \(A\).

The joint covariance of \(\bigl(f_\theta(X),\,f_\theta(B)+\varepsilon_B\bigr)\)
has Schur complement
\begin{equation}
 \widetilde K_\theta(X,Y)=K_\theta(X,Y)
 -K_\theta(X,B)[K_\theta(B,B)+N_B]^{-1}K_\theta(B,Y).
 \label{eq:si-anchored-residual}
\end{equation}
On the anchors themselves, with \(M=K_{BB}+N_B\),
\begin{align}
 \widetilde K_\theta(B,B)
 &=K_{BB}-K_{BB}M^{-1}K_{BB}
 =K_{BB}M^{-1}N_B \nonumber\\
 &=N_B-N_B M^{-1}N_B .
 \label{eq:si-anchor-variance}
\end{align}
The reductions substitute \(K_{BB}=M-N_B\). The difference of the anchor noise and
this variance is the Gram matrix \(N_B M^{-1}N_B\), so
\(\widetilde K_\theta(B,B)\) does not exceed \(N_B\) in the semidefinite
order. If \(N_B=0\) and \(K_{BB}\) is invertible, the cross row vanishes as
well:
\(\widetilde K_\theta(B,Y)=K_\theta(B,Y)-K_{BB}K_{BB}^{-1}K_\theta(B,Y)=0\).

In the two-channel example, \(N_B=0\) leaves \(K_{BB}=1/4\) invertible and
\(W=M^{-1}K_{BA}=1\). Then
\(\widetilde K_{AA}=\tfrac14-\tfrac14\cdot 1=0\) and
\(\widetilde K_{BA}=\tfrac14-\tfrac14\cdot 1=0\). The background cross
block \(C_B(B,A)\) is already zero, so the anchored update leaves the
background mean at \(0\). This \(K_\theta\) has rank one, and the noiseless
anchor removes it on \(A\) as well as on \(B\). A kernel with a component
orthogonal to \(B\) keeps residual variance on \(A\); the figure
is that comparison on one coordinate, where both refits reach the new
observation and only the free residual moves the background gradient.

The posterior conditions \(C_B+\widetilde K_\theta\) on \(A\). The
observable is \(y_A=f_B(A)+f_\theta(A)+\varepsilon_A\) with
\(\varepsilon_A\sim\mathcal N(0,N_A)\) independent of the anchor. Here
\(N_A\) is the new channels' physical variance plus jitter, the same
construction as \(N_B\). The observable has mean \(m_B(A)\) and covariance
\begin{equation}
 S=C_B(A,A)+\widetilde K_\theta(A,A)+N_A .
 \label{eq:si-active-S}
\end{equation}
Its cross-covariance with a query \(X\) is
\(C_B(X,A)+\widetilde K_\theta(X,A)\). With \(r=y_A-m_B(A)\), the conditional
Gaussian is
\begin{align}
 m(X)&=m_B(X)+[C_B(X,A)+\widetilde K_\theta(X,A)]S^{-1}r,\\
 C(X,Y)&=C_B(X,Y)+\widetilde K_\theta(X,Y) \nonumber\\
 &\quad-[C_B(X,A)+\widetilde K_\theta(X,A)]S^{-1}
        [C_B(A,Y)+\widetilde K_\theta(A,Y)].
 \label{eq:si-active-posterior}
\end{align}
Every matrix uses the joint energy-gradient ordering and physical
units. Conditioning the unmodified sum of the two priors on the whole
data set is a different model: the anchor constrains the residual, and
the background observations are not reused as independent evidence.
Kernel candidates are compared through the conditional likelihood and
the whole-geometry leave-one-out density on \(A\).

The parameter derivative runs through the anchor solve. Partition the
joint residual kernel with \(B\) before \(A\), and write
\(W=M^{-1}K_{BA}\). The matrix \(M\) is symmetric, so
\(K_{AB}M^{-1}=W^{\mathsf T}\). For fixed \(N_B\),
\(\partial_\theta M=\partial_\theta K_{BB}\) and
\(\partial_\theta(M^{-1})=-M^{-1}(\partial_\theta K_{BB})M^{-1}\).
Differentiating \(\widetilde K_{AA}=K_{AA}-K_{AB}W\) therefore gives
\begin{align}
 \partial_\theta\widetilde K_{AA}
 &=\partial_\theta K_{AA}
  -(\partial_\theta K_{AB})W
  -K_{AB}M^{-1}(\partial_\theta K_{BA}) \nonumber\\
 &\quad+K_{AB}M^{-1}(\partial_\theta K_{BB})W \nonumber\\
 &=\partial_\theta K_{AA}
  -(\partial_\theta K_{AB})W
  -W^{\mathsf T}(\partial_\theta K_{BA})
  +W^{\mathsf T}(\partial_\theta K_{BB})W .
 \label{eq:si-anchored-residual-derivative}
\end{align}
Set \(T=[-W^{\mathsf T},\,I]\). The joint derivative times the column
factor of \(T^{\mathsf T}\) is
\begin{equation}
 (\partial_\theta K)\begin{pmatrix}-W\\ I\end{pmatrix}
 =\begin{pmatrix}
    \partial_\theta K_{BA}-(\partial_\theta K_{BB})W\\
    \partial_\theta K_{AA}-(\partial_\theta K_{AB})W
  \end{pmatrix}.
\end{equation}
Left multiplication by \([-W^{\mathsf T},\,I]\) is
Eq. \ref{eq:si-anchored-residual-derivative}, so
\begin{equation}
 \partial_\theta\widetilde K_{AA}=T(\partial_\theta K)T^{\mathsf T}.
 \label{eq:si-anchored-derivative-lift}
\end{equation}

Positive semidefiniteness is the same factor. Let
\(R=K+\operatorname{diag}(N_B,0)\), with the zero block on \(A\). Then
\((K_{BB}+N_B)W=K_{BA}\), and
\begin{equation}
 R\begin{pmatrix}-W\\ I\end{pmatrix}
 =\begin{pmatrix}
    -(K_{BB}+N_B)W+K_{BA}\\
    -K_{AB}W+K_{AA}
  \end{pmatrix}
 =\begin{pmatrix}0\\ \widetilde K_{AA}\end{pmatrix}.
\end{equation}
Left multiplication by \(T\) leaves the lower block, hence the congruence
\begin{equation}
 \widetilde K_{AA}
 =T\left[K+\operatorname{diag}(N_B,0)\right]T^{\mathsf T}.
 \label{eq:si-anchored-congruence}
\end{equation}
A valid joint kernel is positive semidefinite, and a nonnegative
diagonal \(N_B\) keeps \(R\) positive semidefinite. Congruence preserves
that property, so \(\widetilde K_{AA}\) is positive semidefinite. The
conditional likelihood uses Eq. \ref{eq:si-anchored-derivative-lift}
in the trace and quadratic form below. A sum of residual kernels is
conditioned as a sum, so its components share one anchor solve.

Exact SymPy arithmetic checks
the rational counterexample, its zero-noise correction, and the derivative
and covariance identities. Native tests check
anchor variance, parameter derivatives, and the Hessian action against
finite differences. These checks establish the stated finite-dimensional
identities.
\subsection{Kernel products for posterior predictions}
\label{sec:org31736b6}

The mean in the active posterior requires products of joint kernels with
fitted energy and gradient coefficients. These products admit a direct
feature-space contraction. Let \(q\) and \(t\) denote query and training
features, \(\delta=q-t\), and
\(k(q,t)=s^2\exp(-\delta^{\mathsf T}\Lambda\delta/2)\).
The nonnegative diagonal metric \(\Lambda\) contains inverse squared length
scales. For one training geometry, write its energy coefficient as \(c\)
and its Cartesian-gradient coefficients as \(d\). With the Jacobians kept
as Cartesian rows and feature columns, define \(v=J_t^{\mathsf T}d\).
The contribution to the scalar mean and its Cartesian gradient is
\begin{align}
 m_{qt}&=k(q,t)\left(c+\delta^{\mathsf T}\Lambda v\right),\\
 \nabla_x m_{qt}
 &=k(q,t)J_q\left[\Lambda v-\Lambda\delta
          \left(c+\delta^{\mathsf T}\Lambda v\right)\right].
 \label{eq:si-direct-kernel-mean}
\end{align}
The species-environment kernel sums these contributions over matching
central elements. Its metric is \(\Lambda=\ell_Z^{-2}I\) for element \(Z\).
Sexpat uses its sparse pair-feature contraction with
\(\Lambda=2\operatorname{diag}(\ell_p^{-2})\), including the derivatives
of cutoff functions and moving--frozen pairs.

The implementation projects the training gradient coefficients once per
application and shares these feature vectors across query atoms. Each
worker owns one atom contribution; the final reduction follows a fixed
atom order. The same projected coefficients enter the kernel derivatives
used in dimer Hessian actions. Sum kernels distribute the product over
their components. For the anchored residual,
\begin{equation}
 \widetilde K(X,A)w=K(X,A)w-K(X,B)
 [K(B,B)+N_B]^{-1}\bigl(K(B,A)w\bigr),
\end{equation}
so the factor at the training channels is also shared by the composed
operator. The explicit covariance and its product define the same model.
The scalar mean of a kernel applied to a coefficient vector is linear
in those coefficients, and so are its Cartesian gradient and its
Hessian-vector product. For coefficients \(w\) on \(A\), the display above
is that mean with coefficients \(w\) on the new rows and coefficients
\(-[K(B,B)+N_B]^{-1}\bigl(K(B,A)w\bigr)\) on the anchor rows. The
prediction and the Hessian action of the anchored residual are this
expansion of the underlying kernel.

For a fitted mean, the feature centers, energy coefficients and projected
gradient coefficients \(J_t^{\mathsf T}d\) form an immutable representation.
Training Jacobians are discarded from that representation. Every query
still evaluates its descriptor and Cartesian Jacobian. Sum kernels own
the fitted component functions; the anchored kernel also owns the fixed
coefficient vector \([K(B,B)+N_B]^{-1}K(B,A)w\). The representation is rebuilt
when the fitted posterior changes, so no query result or geometry lookup
is kept. Exact SymPy arithmetic checks the nonlinear
feature and composition identities, and the scalar mean and its feature derivatives
for isotropic and diagonal anisotropic metrics. Cartesian
Jacobians compose these identities by the chain rule.
\subsection{Adjoint evaluation of conditional likelihood gradients}
\label{sec:orgeed33c6}

Write \(\alpha=S^{-1}r\) and
\(G=(S^{-1}-\alpha\alpha^{\mathsf T})/2\). The derivative of the conditional
negative log likelihood is the Frobenius product
\(\langle G,T(\partial_\theta K)T^{\mathsf T}\rangle_F\). Cyclicity of the
trace gives
\begin{equation}
 \partial_\theta\mathcal L
 =\langle T^{\mathsf T}GT,\partial_\theta K\rangle_F.
 \label{eq:si-conditional-gradient-adjoint}
\end{equation}
The implementation forms \(T^{\mathsf T}GT\) once per objective evaluation.
Each parameter then requires a contraction with its joint covariance
derivative. Pair, Cartesian and species-environment residuals use this
operator. Length-scale derivatives are assembled in batches bounded by
the worker count, so their joint-matrix storage does not grow with the
number of optimized parameters. The optimizer uses its convergence
criteria.

The exact SymPy check includes the adjoint identity. Native regressions
compare forward and adjoint contractions in channel-major layouts and
check the scalar objective by finite differences.
\subsection{Square-root evaluation of geometry leave-one-out evidence}
\label{sec:org426b35d}

For the positive definite observation covariance \(S=LL^{\mathsf T}\),
let \(E_g\) select every energy and gradient channel belonging to geometry
\(g\). Define \(z=L^{-1}r\) and the thin QR factorization
\(L^{-1}E_g=Q_gR_g\). The precision block and its residual satisfy
\begin{equation}
 P_g=E_g^{\mathsf T}S^{-1}E_g=R_g^{\mathsf T}R_g,\qquad
 \alpha_g=E_g^{\mathsf T}S^{-1}r=R_g^{\mathsf T}Q_g^{\mathsf T}z.
\end{equation}
The sum of negative log predictive densities is therefore
\begin{equation}
 \mathcal L_{\mathrm{LOO}}
 =\sum_g\left[
 \frac12\lVert Q_g^{\mathsf T}z\rVert_2^2
 -\sum_{i=1}^{d_g}\log\lvert(R_g)_{ii}\rvert
 +\frac{d_g}{2}\log(2\pi)\right],
\end{equation}
where \(d_g\) is the number of channels withheld for geometry \(g\).
This evaluates the density through triangular solves and orthogonal
transformations. Forming \(P_g\) explicitly squares the condition number
of its square root and can lose positive definiteness in floating-point
arithmetic. With only one geometry, withholding all channels gives the
joint negative log likelihood, which is evaluated directly from \(L\).
Exact SymPy arithmetic verifies the
precision, quadratic-form, determinant, and single-geometry identities.
Native regressions cover ill-conditioned
independent geometry groups and direct Gaussian conditioning.

\section{Illustrations and elapsed-time figures}
\label{sec:si-illustrations}
The figures in this section illustrate identities used in the article. Labels are unchanged. A drawing of a few coordinates illustrates an identity under the values printed on it.  Where a construction is switched off in the reported calculations, the subsection says so.
\subsection{Band acquisition in the two calculation sequences}
\label{sec:org1e5bcea}

The schematic shows which band images supply measured observations. The complete sequence, including the separate Hessian check, is given in the main-text calculation table.

\begin{figure}[H]
\centering
\begin{tikzpicture}[>=Stealth, font=\sffamily\scriptsize,
                    line cap=round, line join=round,
                    panel/.style={draw=RuhiTeal!70, rounded corners=2pt,
                                  inner sep=6pt, align=left}]
  \node[font=\sffamily\small\bfseries, text=RuhiTeal, anchor=west] at (0,3.55) {(a) Baker--Chan};
  \node[font=\sffamily\small\bfseries, text=RuhiTeal, anchor=west] at (7.55,3.55) {(b) Birkholz--Schlegel};

  \begin{scope}[shift={(0.35,1.70)}]
    \draw[black!45] (0,0) .. controls (1.2,0.7) and (2.4,0.7) .. (3.6,0)
                  .. controls (4.6,-0.45) and (5.4,-0.35) .. (6.3,-0.15);
    \foreach \x/\y in {0/0, 1.05/0.42, 2.15/0.58, 4.35/-0.22, 5.35/-0.28, 6.3/-0.15} {
      \draw[RuhiTeal, fill=white, line width=0.6pt] (\x,\y) circle[radius=2.2pt];
    }
    \fill[RuhiCoral] (3.25,0.22) circle[radius=2.6pt];
    \node[anchor=south, text=RuhiCoral] at (3.25,0.40) {climb};
    \node[anchor=north, align=center] at (3.15,-0.85)
      {one measured image\\no dimer};
  \end{scope}

  \begin{scope}[shift={(7.90,1.70)}]
    \draw[black!45] (0,0) .. controls (1.2,0.7) and (2.4,0.7) .. (3.6,0)
                  .. controls (4.6,-0.45) and (5.4,-0.35) .. (6.3,-0.15);
    \foreach \x/\y in {0/0, 2.15/0.58, 4.35/-0.22, 6.3/-0.15} {
      \draw[RuhiTeal, fill=white, line width=0.6pt] (\x,\y) circle[radius=2.2pt];
    }
    \fill[RuhiCoral] (1.05,0.42) circle[radius=2.3pt];
    \fill[RuhiCoral] (5.35,-0.28) circle[radius=2.3pt];
    \fill[RuhiTeal] (3.25,0.22) circle[radius=2.6pt];
    \draw[RuhiCoral, line width=0.8pt] (2.55,0.72) -- (3.95,-0.28);
    \fill[RuhiCoral] (2.55,0.72) circle[radius=1.5pt];
    \fill[RuhiCoral] (3.95,-0.28) circle[radius=1.5pt];
    \node[anchor=south, text=RuhiCoral] at (3.15,0.95) {dimer axis};
    \node[anchor=south, text=RuhiCoral] at (1.05,0.58) {support};
    \node[anchor=north, text=RuhiCoral] at (5.35,-0.42) {support};
    \node[anchor=north] at (3.15,-1.05) {imported observations};
  \end{scope}
\end{tikzpicture}
\caption{\textbf{Two fixed protocols.}
Both panels use an inverse-distance kernel, a ZBL mean, Cholesky
factorization, rgmin L-BFGS, the one-image schedule, $18$ interior
images ($N=20$), and the uncapped condition-driven inner loop. Dual HS-SVD,
SOAP residuals, and manifold charts are off.
(a) Baker--Chan. One measured climbing image, and kernel-support
freezing is off. The panel runs no dimer.
(b) Birkholz--Schlegel. Images outside kernel support join the seed.
The dimer driver imports the band observations and continues with dimer index
recovery at $\beta=5$.}
\label{fig:protocols}
\end{figure}
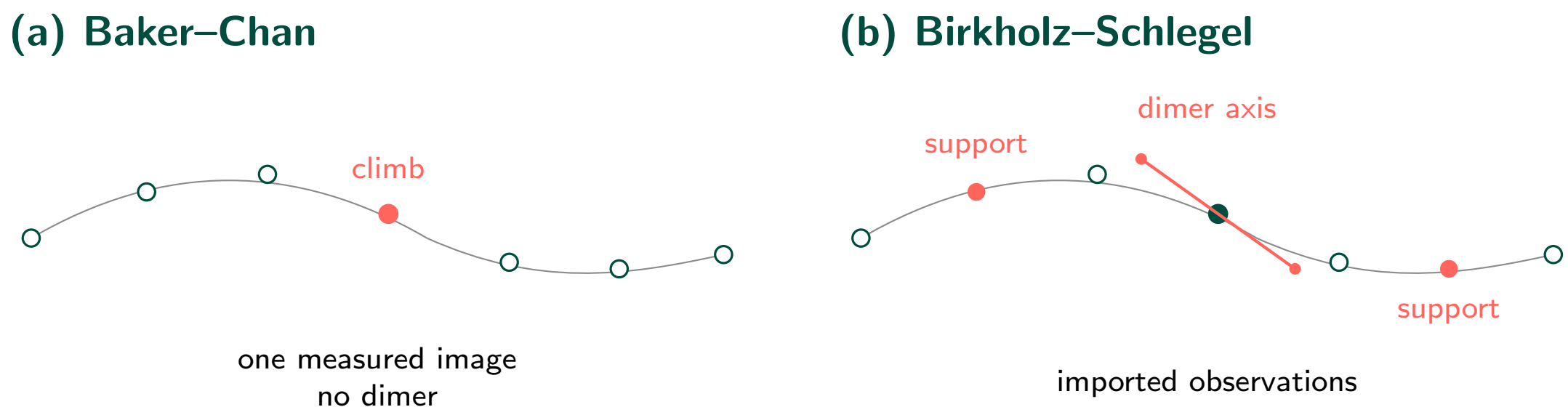
\subsection{GP-NEB acquisition cycle}
\label{sec:orgcff9cee}

The reported search relaxes the band on the GP posterior mean, then evaluates one selected image on the supplied potential. Cycle and progress checks return unproductive surrogate relaxation to acquisition. The measured force determines convergence; a separate Hessian determines the index.

\begin{figure}[H]
\centering
\includegraphics[width=0.52\textwidth]{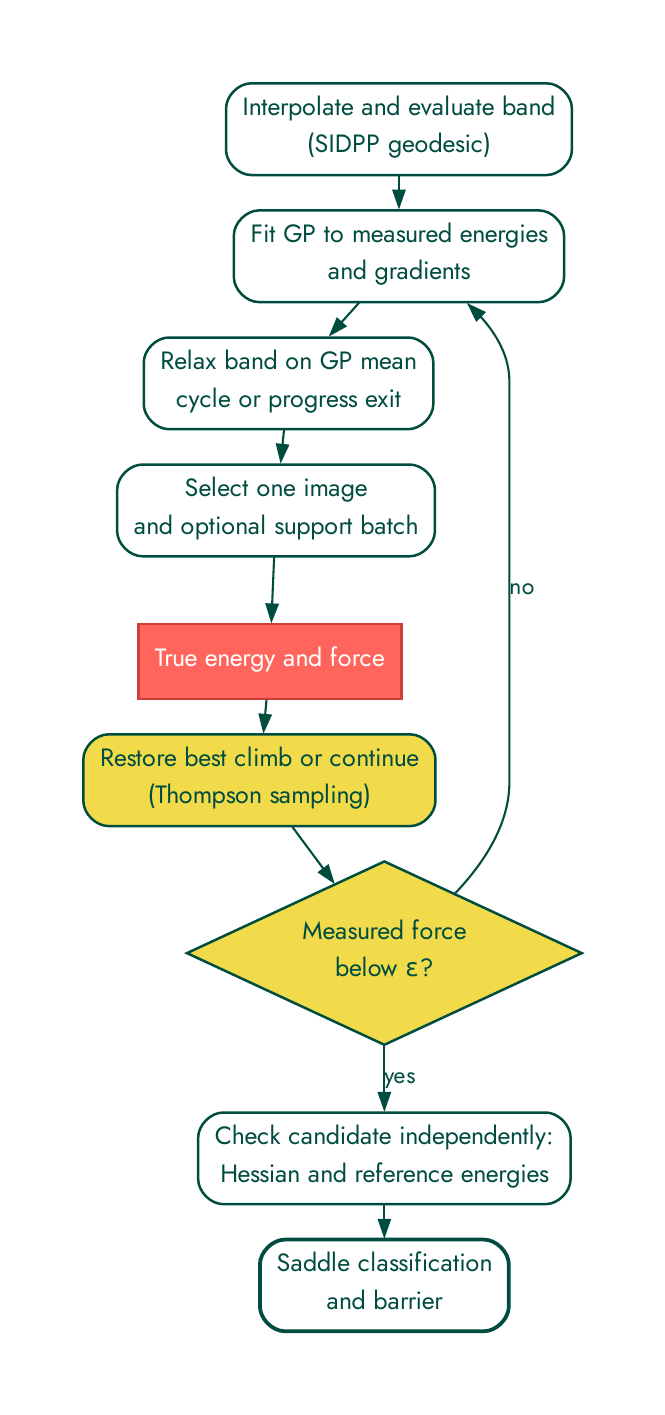}
\caption{\textbf{One-image GP-NEB.}
The band alternates between relaxation on the GP mean and a potential
evaluation at a selected image. Cycle and progress checks terminate
unproductive surrogate relaxation. A barrier-decrease proposal selects
between restoring the best measured climb and continuing. Birkholz
preparation also evaluates images outside kernel support. The measured
force determines convergence; a separate Hessian determines the index.}
\label{fig:gp-oie-loop}
\end{figure}
\subsection{HCN path and projected observations}
\label{sec:org79f343d}

The HCN profile joins the final-image GP means by straight segments and shows measured energies where images coincide with training rows. The second figure projects the final band and the potential evaluations into a plane of reaction progress and orthogonal deviation. It displays the energies without fitting an additional surface; the search uses the full inverse-distance model.

\begin{figure}[H]
\centering
\includegraphics[width=\textwidth]{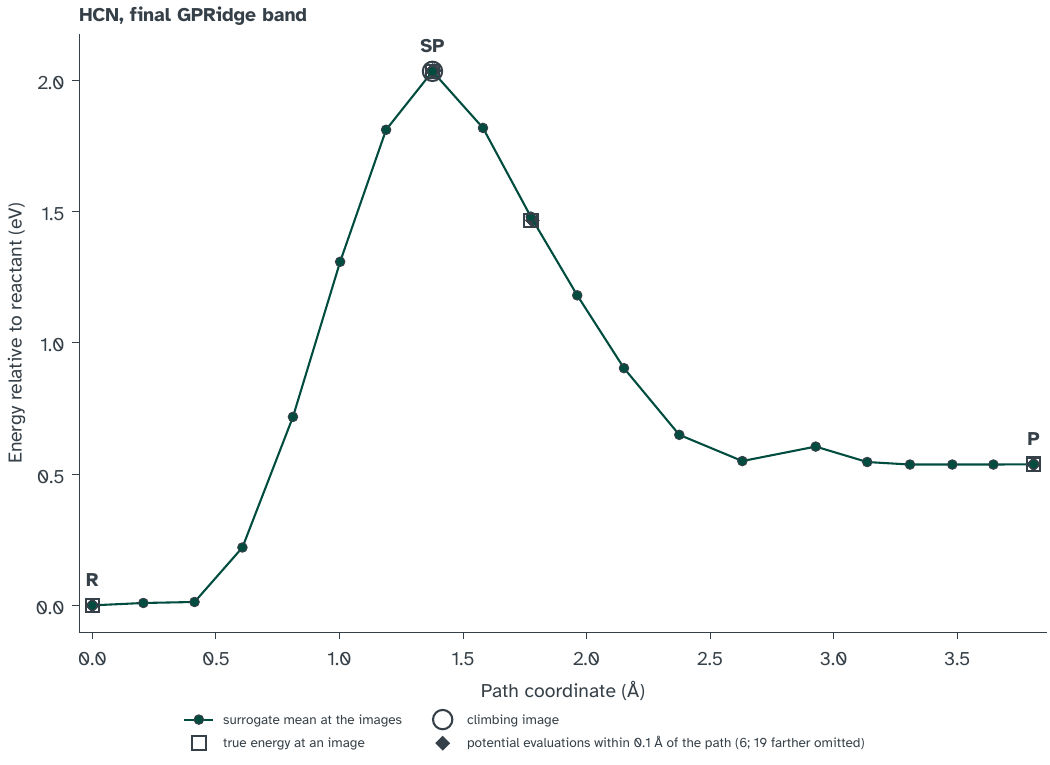}
\caption{\textbf{HCN isomerization: surrogate energy profile.}
Straight segments join the GP surrogate means at the final images
along the path coordinate, in angstrom. Squares are potential energies at
images that coincide with training rows; the ringed point is the climbing
image, $2.034$~eV above the reactant.
Energies are relative to the reactant. Diamonds are potential evaluations
projected onto the final path and drawn when they lie within $0.1$~\AA{} of
it (6 of the 25 training rows shown; the other 19 are omitted).}
\label{fig:hcn-band}
\end{figure}

\begin{figure}[H]
\centering
\includegraphics[width=\textwidth]{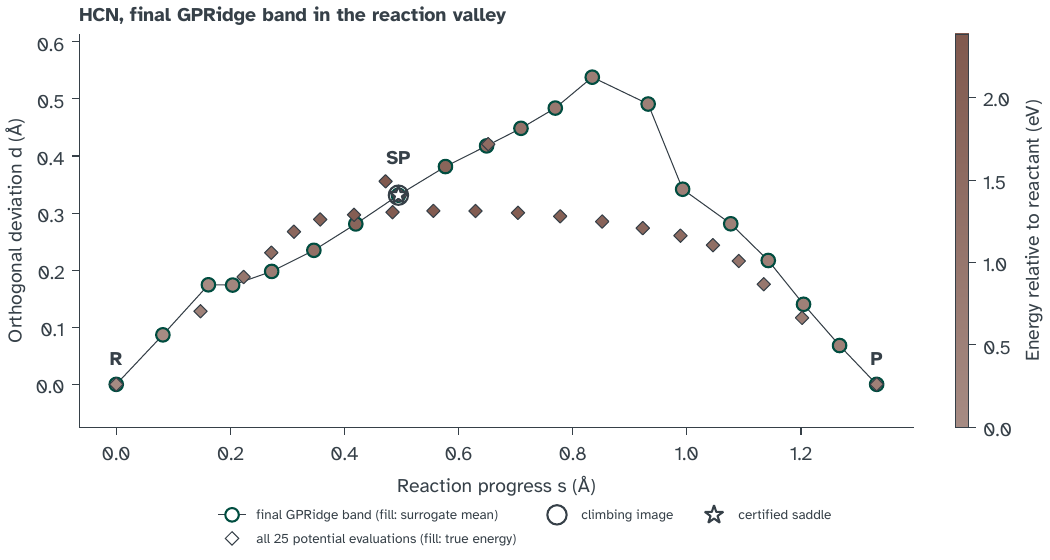}
\caption{\textbf{HCN isomerization: projected observations.}
Plane of reaction progress and orthogonal deviation, both RMSD-based and
given in \AA. Circles are the final band images, filled by their
surrogate mean; diamonds are all potential evaluations, filled by their
measured energy. Both use the colorbar scale relative to the reactant.
The ring marks the climbing image and the star the saddle that passes the
numerical index check. The figure displays the observations and
final band without an additional fitted surface or variance contours.}
\label{fig:hcn-landscape}
\end{figure}
\subsection{Inverse distances}
\label{sec:org2bf8d9f}

\begin{figure}[H]
\centering
\begin{tikzpicture}[>=Stealth, font=\sffamily\scriptsize,
                    line cap=round, line join=round]
  \def\s{1.15}
  \coordinate (A) at (0,0);
  \coordinate (B) at (3*\s,0);
  \coordinate (C) at (0,4*\s);
  \draw[RuhiTeal, line width=0.9pt] (A) -- (B) -- (C) -- cycle;
  \foreach \p/\name/\anc in {A/A/north east, B/B/north west, C/C/south east} {
    \fill[RuhiTeal] (\p) circle[radius=2.1pt];
    \node[anchor=\anc, inner sep=2pt] at (\p) {$\name$};
  }
  \node[anchor=north, inner sep=2pt] at ($(A)!0.5!(B)+(0,-0.08)$)
    {$r=3$, $q=1/3$};
  \node[anchor=east, inner sep=3pt] at ($(A)!0.5!(C)$)
    {$r=4$, $q=1/4$};
  \node[anchor=south west, inner sep=1pt] at ($(B)!0.55!(C)+(0.06,0.02)$)
    {$r=5$, $q=1/5$};
\end{tikzpicture}
\caption{\textbf{Inverse distances.}
Three atoms sit at $(0,0)$, $(3,0)$ and $(0,4)$. The side lengths are
$3$, $4$ and $5$, and the feature on each side is the inverse distance
$q=1/r$.}
\label{fig:inverse-distances}
\end{figure}
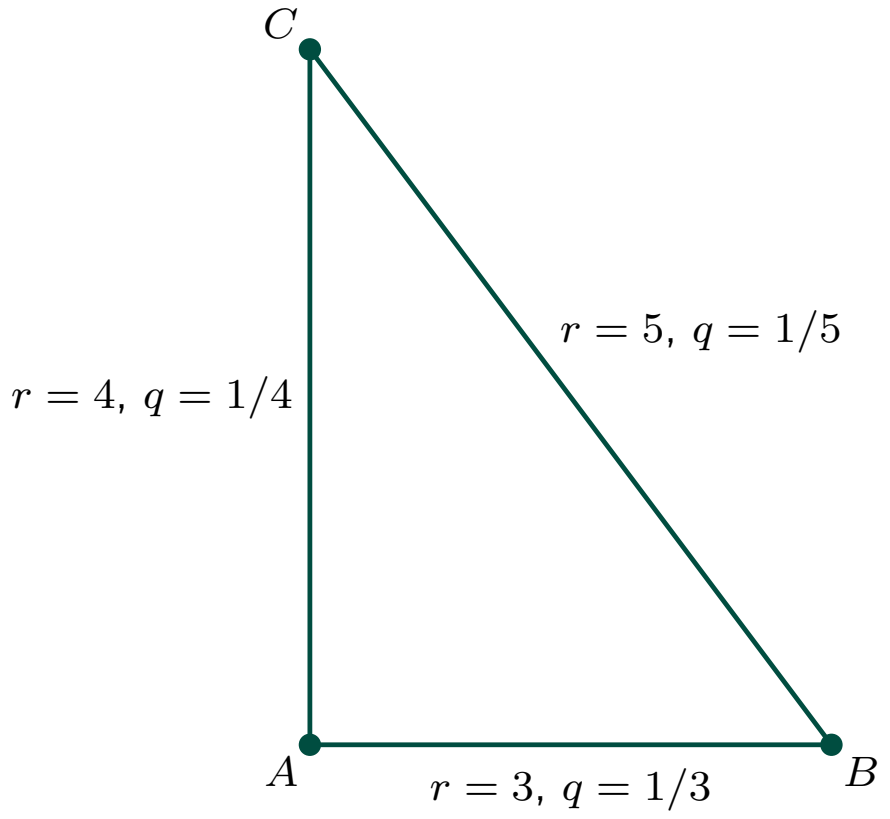

Pair distances enter the reported kernel as inverse distances. Figure \ref{fig:inverse-distances} is that feature on three atoms. The distances are \(3\), \(4\) and \(5\), so the inverse distances are \(1/3\), \(1/4\) and \(1/5\).

\subsection{One pair-local tile}
\label{sec:orgfab723b}

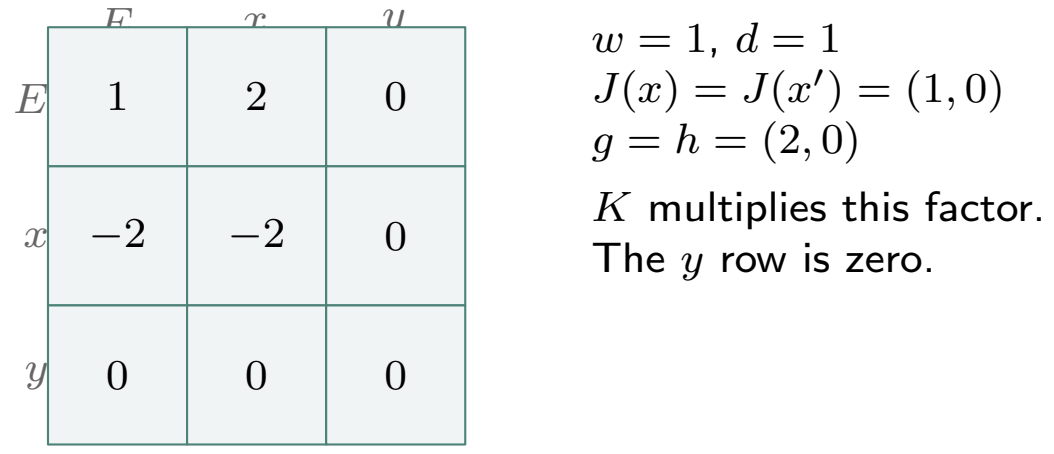
\begin{figure}[H]
\centering
\begin{tikzpicture}[>=Stealth, font=\sffamily\scriptsize,
                    line cap=round, line join=round]
  \def\cs{0.92}
  \foreach \i/\lab in {0/E, 1/x, 2/y} {
    \node[text=black!60] at ({1.15+\i*\cs}, 2.55) {$\lab$};
    \node[text=black!60, anchor=east] at (0.85, {2.05-\i*\cs}) {$\lab$};
  }
  \foreach \r/\c/\val in {
      0/0/1, 0/1/2, 0/2/0,
      1/0/-2, 1/1/-2, 1/2/0,
      2/0/0, 2/1/0, 2/2/0} {
    \draw[RuhiTeal!70, fill=RuhiTeal!6] ({0.70+\c*\cs}, {1.60-\r*\cs})
      rectangle ++(\cs,\cs);
    \node at ({0.70+\c*\cs+0.46}, {1.60-\r*\cs+0.46}) {$\val$};
  }
  \node[anchor=west, align=left] at (4.15,1.70)
    {$w=1$, $d=1$\\
     $J(x)=J(x')=(1,0)$\\
     $g=h=(2,0)$\\[0.4em]
     $K$ multiplies this factor.\\
     The $y$ row is zero.};
\end{tikzpicture}
\caption{\textbf{One pair-local tile.}
One feature, $D=2$, $w=1$, $d=1$, and $J(x)=J(x')=(1,0)$.
Then $g=h=(2,0)$, and $S_J$ is zero apart from a $1$ in the first
Cartesian entry. The drawn matrix is the factor that $K$ multiplies, with
rows $(1,\;2,\;0)$, $(-2,\;-2,\;0)$ and $(0,\;0,\;0)$. The $y$ row is
zero because the feature is independent of that component.}
\label{fig:pair-tile}
\end{figure}

Figure \ref{fig:pair-tile} is the pair-local factor in the article for one feature and \(D=2\). With \(w=1\), \(d=1\) and \(J(x)=J(x')=(1,0)\), the sums are \(g=h=(2,0)\) and \(S_J\) has a \(1\) in the first Cartesian entry. The factor that \(K\) multiplies has rows \((1,2,0)\), \((-2,-2,0)\) and \((0,0,0)\). The last row is zero because this feature does not depend on the second Cartesian component. These Jacobians are the algebra of this one feature. The reported calculations assemble the joint Gram from the same tile on the inverse-distance features of each molecule.

\subsection{Dual HS-SVD on three modes}
\label{sec:orgfea64a2}

Every reported calculation factorizes the noisy Gram by Cholesky. Dual HS-SVD is an optional inverse of the same Gram, and both panels leave it off. The identities below assume that \(\Psi\) spans the exact nonzero range of the assembled \(K\). They are algebraic statements under that hypothesis. They do not recover information lost while assembling \(K\), they do not bound the rounding error of a computed factor, and they are not a measurement of the Cholesky solver used for the 44 reactions. Section \ref{sec:si-hssvd} gives the derivation.

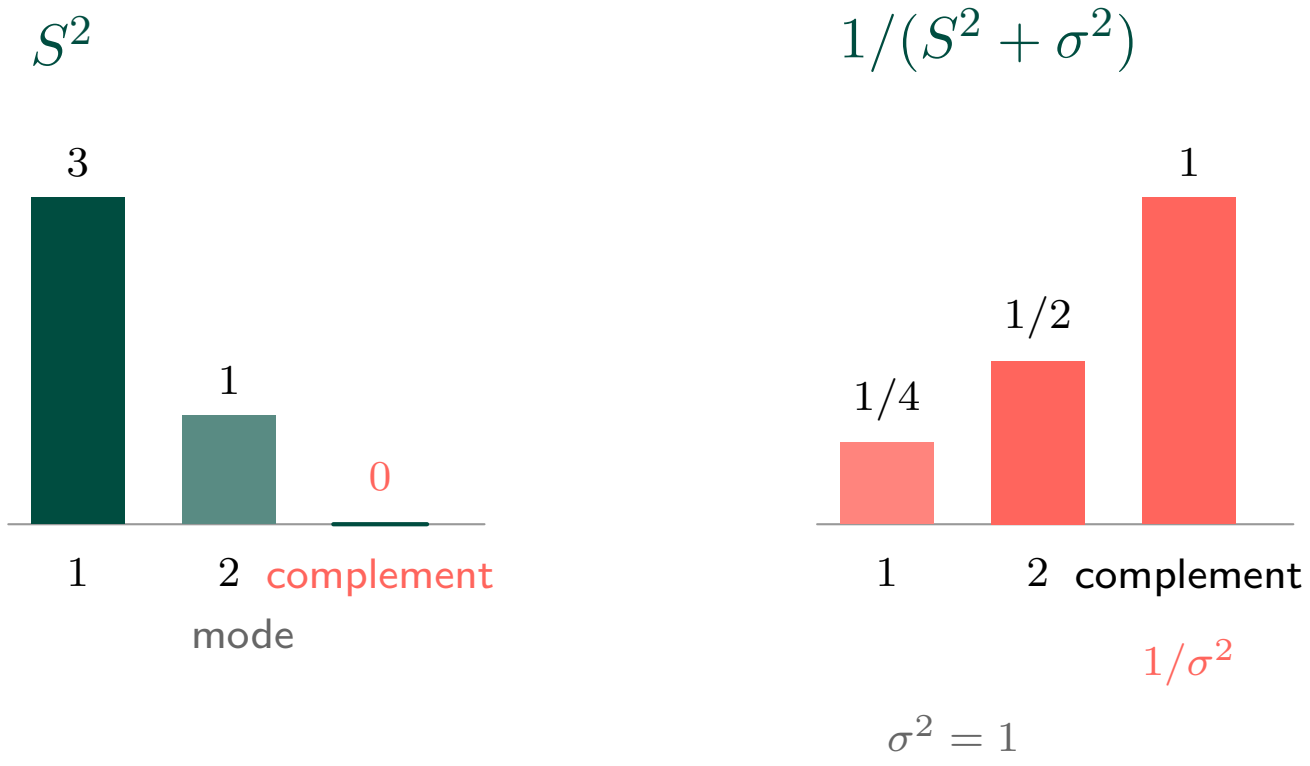
\begin{figure}[H]
\centering
\begin{tikzpicture}[>=Stealth, font=\sffamily\scriptsize,
                    line cap=round, line join=round]
  \def\bw{0.62}
  \def\gap{0.38}
    \node[anchor=west, font=\sffamily\small\bfseries, text=RuhiTeal] at (0,3.20) {$S^{2}$};
  \draw[black!40] (0,0) -- (3.15,0);
  \fill[RuhiTeal] (0.15,0) rectangle ++(\bw,2.16);
  \fill[RuhiTeal!65] (0.15+\bw+\gap,0) rectangle ++(\bw,0.72);
  \draw[RuhiTeal, line width=0.8pt] (0.15+2*\bw+2*\gap,0) -- ++(\bw,0);
  \node[anchor=south] at (0.15+0.5*\bw,2.16) {$3$};
  \node[anchor=south] at (0.15+\bw+\gap+0.5*\bw,0.72) {$1$};
  \node[anchor=south, text=RuhiCoral] at (0.15+2*\bw+2*\gap+0.5*\bw,0.08) {$0$};
  \node[anchor=north] at (0.15+0.5*\bw,-0.08) {$1$};
  \node[anchor=north] at (0.15+\bw+\gap+0.5*\bw,-0.08) {$2$};
  \node[anchor=north, text=RuhiCoral] at (0.15+2*\bw+2*\gap+0.5*\bw,-0.08) {complement};
  \node[anchor=north, text=black!60] at (1.55,-0.48) {mode};

    \begin{scope}[shift={(5.35,0)}]
    \node[anchor=west, font=\sffamily\small\bfseries, text=RuhiTeal] at (0,3.20)
      {$1/(S^{2}+\sigma^{2})$};
    \draw[black!40] (0,0) -- (3.15,0);
    \fill[RuhiCoral!80] (0.15,0) rectangle ++(\bw,0.54);
    \fill[RuhiCoral] (0.15+\bw+\gap,0) rectangle ++(\bw,1.08);
    \fill[RuhiCoral] (0.15+2*\bw+2*\gap,0) rectangle ++(\bw,2.16);
    \node[anchor=south] at (0.15+0.5*\bw,0.54) {$1/4$};
    \node[anchor=south] at (0.15+\bw+\gap+0.5*\bw,1.08) {$1/2$};
    \node[anchor=south] at (0.15+2*\bw+2*\gap+0.5*\bw,2.16) {$1$};
    \node[anchor=north] at (0.15+0.5*\bw,-0.08) {$1$};
    \node[anchor=north] at (0.15+\bw+\gap+0.5*\bw,-0.08) {$2$};
    \node[anchor=north] at (0.15+2*\bw+2*\gap+0.5*\bw,-0.08) {complement};
    \node[anchor=north, text=RuhiCoral]
      at (0.15+2*\bw+2*\gap+0.5*\bw,-0.62) {$1/\sigma^{2}$};
    \node[anchor=north, text=black!60] at (0.90,-1.12) {$\sigma^{2}=1$};
  \end{scope}
\end{tikzpicture}
\caption{\textbf{Dual HS-SVD on three modes.}
$S^{2}=3$, $1$, $0$ and $\sigma^{2}=1$. The inverse weights are $1/4$,
$1/2$ and $1$. The zero mode is the complement, and its weight is
$1/\sigma^{2}$. In this basis $\mathbf{y}=(1,0,1)$ gives
$(K+\sigma^{2}I)^{-1}\mathbf{y}=(1/4,0,1)$. This identity assumes that
$\Psi$ spans the nonzero range of $K$. The reported calculations
factorize the noisy Gram by Cholesky and leave dual HS-SVD off.
$\mathrm{cond}(\Psi)=1$ does not bound these weights.}
\label{fig:hssvd}
\end{figure}

Figure \ref{fig:hssvd} illustrates the range and complement inverse in Equation \eqref{eq:camp-noisy}.
For signal eigenvalues $3,1,0$ and noise variance $1$, the inverse weights are $1/4,1/2,1$.
The zero mode belongs to the complement. The orthogonality of the basis does not bound these inverse weights.

\subsection{Linear bend in an internal chart}
\label{sec:org023457b}

\begin{figure}[H]
\centering
\begin{tikzpicture}[>=Stealth, font=\sffamily\scriptsize,
                    line cap=round, line join=round]
  \def\bendphideg{32}
  \pgfmathsetmacro{\bendphi}{\bendphideg}
  \pgfmathsetmacro{\cx}{2.30*cos(\bendphi)}
  \pgfmathsetmacro{\cy}{2.30*sin(\bendphi)}
  \pgfmathsetmacro{\vx}{0.90*cos(\bendphi)}
  \pgfmathsetmacro{\vy}{0.90*sin(\bendphi)}

  \node[font=\sffamily\small\bfseries, text=RuhiTeal] at (-2.45,4.72) {(a)};
  \node[font=\sffamily\small\bfseries, text=RuhiTeal] at (4.55,4.72) {(b)};

    \begin{scope}[shift={(0,1.72)}]
    \draw[black!45, dashed] (-2.62,0) -- (2.62,0);
    \node[anchor=west, text=black!55] at (2.66,0.02) {axis};
    \draw[RuhiTeal, line width=0.95pt] (-2.30,0) -- (0,0);
    \draw[RuhiTeal, line width=0.95pt] (0,0) -- (\cx,\cy);
    \draw[-{Stealth[length=4.4pt]}, line width=0.65pt] (0,0) -- (-0.90,0);
    \node at (-0.58,-0.24) {$\hat u$};
    \draw[-{Stealth[length=4.4pt]}, line width=0.65pt] (0,0) -- (\vx,\vy);
    \node[anchor=south, inner sep=1pt, yshift=3pt] at (\vx,\vy) {$\hat v$};
    \draw[-{Stealth[length=4.4pt]}, RuhiTeal, line width=0.7pt] (0,0) -- (0,1.02);
    \node[text=RuhiTeal, anchor=south] at (0.16,1.02) {$n_2$};
    \draw[black!75] (180:0.50) arc[start angle=180, end angle=\bendphi, radius=0.50];
    \node at (-0.36,0.58) {$\theta$};
    \draw[RuhiCoral] (0:1.05) arc[start angle=0, end angle=\bendphi, radius=1.05];
    \node[text=RuhiCoral] at (14:1.30) {$\varphi$};
    \draw (0.48,-0.62) circle[radius=0.105];
    \fill (0.48,-0.62) circle[radius=0.032];
    \node[anchor=west] at (0.64,-0.62) {$n_1$};
    \fill[RuhiTeal] (-2.30,0) circle[radius=2.15pt];
    \fill[RuhiTeal] (0,0) circle[radius=2.15pt];
    \fill[RuhiTeal] (\cx,\cy) circle[radius=2.15pt];
    \node[anchor=north] at (-2.30,-0.12) {$a$};
    \node[anchor=north] at (0,-0.12) {$b$};
    \node[anchor=south west] at (\cx,\cy) {$c$};
    \node[align=center, anchor=north] at (0.15,-1.08)
      {$\beta_n=n\cdot(\hat u+\hat v)$\\[1pt]
       $\beta_{n_2}=\sin\varphi,\quad \beta_{n_1}=0$};
  \end{scope}

    \begin{scope}[shift={(4.85,0.22)}, x=0.0565cm, y=0.50cm]
    \draw[->, black!80] (0,0) -- (96,0);
    \draw[->, black!80] (0,0) -- (0,8.65);
    \node[anchor=south, rotate=90] at (-20,4.15) {row factor};
    \foreach \tick in {1,2,4,8}
      \draw (0,\tick) -- (-2.2,\tick) node[anchor=east] {$\tick$};
    \foreach \tick in {0,30,60,90}
      \draw (\tick,0) -- (\tick,-0.28) node[anchor=north] {$\tick$};
    \node[anchor=north] at (45,-1.45) {bend $\varphi$ (degrees)};
    \begin{scope}
      \clip (-1,-0.05) rectangle (90.3,8.08);
      \draw[RuhiCoral, line width=0.95pt, domain=6.4:90, samples=160]
        plot (\x, {1/sin(\x)});
      \draw[RuhiTeal, line width=0.95pt] (0,1) -- (90,1);
      \draw[black!40, dashed] (\bendphi,0) -- (\bendphi,{1/sin(\bendphi)});
      \fill[RuhiTeal] (\bendphi,1) circle[radius=1.25pt];
      \fill[RuhiCoral] (\bendphi,{1/sin(\bendphi)}) circle[radius=1.25pt];
    \end{scope}
    \node[text=RuhiCoral, anchor=west] at (16,5.85) {$1/\sin\theta$};
    \node[text=RuhiTeal, anchor=west] at (52,2.55) {linear-bend row};
    \node[anchor=west] at (34.2,{1/sin(\bendphi)+0.15}) {$\bendphideg^{\circ}$};
  \end{scope}
\end{tikzpicture}
\caption{\textbf{Linear bend in the internal chart.}
(a) A unit leg bent by $\varphi=32^{\circ}$ from the axis. The valence
angle is $\theta=\pi-\varphi$. The references $n_1$ and $n_2$ are
orthonormal and perpendicular to the axis, and $n_1$ is perpendicular
to the page. This in-plane bend has $\beta_{n_2}=\sin\varphi$ and
$\beta_{n_1}=0$.
(b) The angle row carries $1/\sin\theta$. On the unit leg, with $n$
perpendicular to $\hat u$, the linear-bend row has length $1$ at every
bend. The mark is the geometry in (a). Manifold charts are off in the
reported calculations. The Supporting Information
section on linear bends gives the gradient and the conditioning cut.}
\label{fig:linear-bend}
\end{figure}
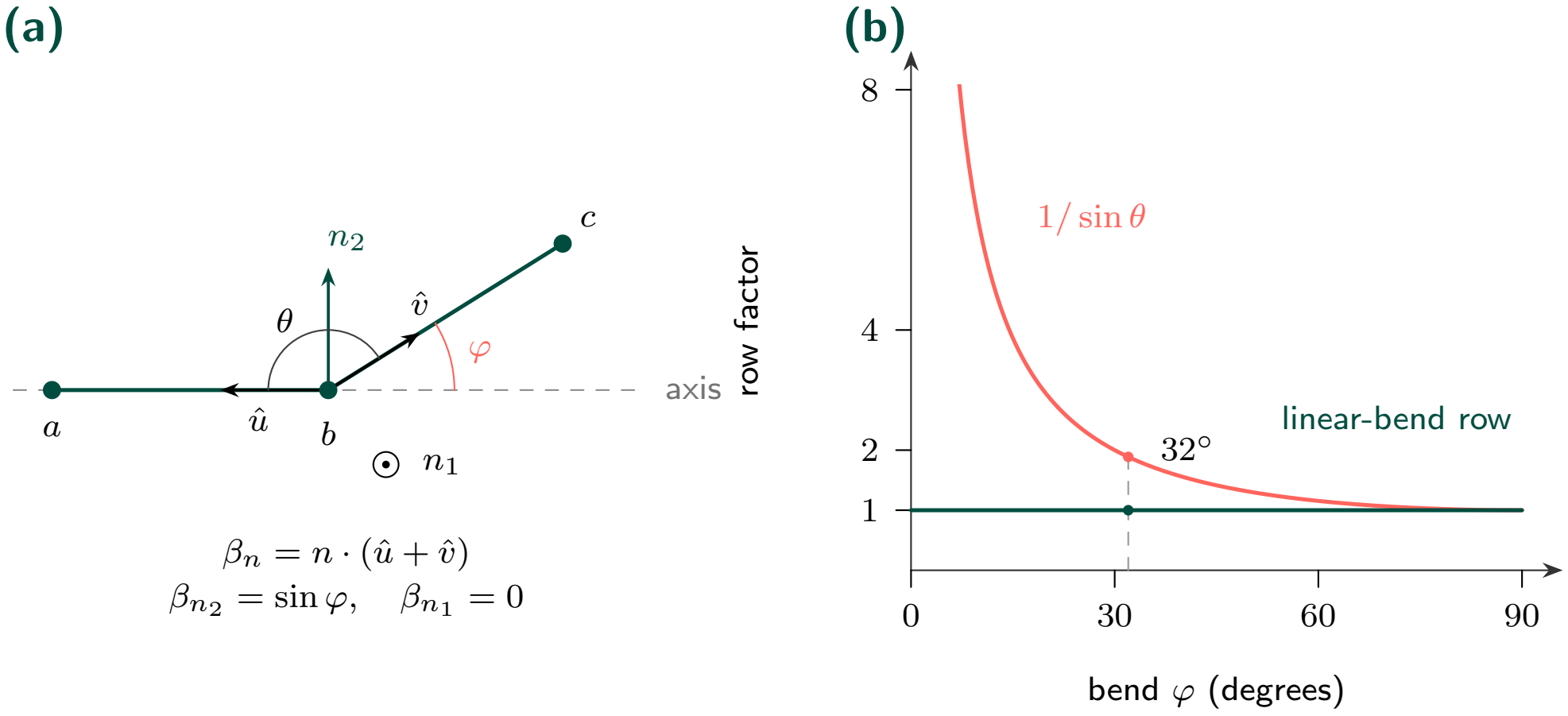

The reported panels use Cartesian inverse-distance features. The linear-bend coordinate belongs to an optional manifold chart. Section \ref{sec:si-linear-bends} gives its gradient and the conditioning cut. Figure \ref{fig:linear-bend} is the pair \(\beta_n=n\cdot(\hat u+\hat v)\) on one bend from the axis. The references \(n_1\) and \(n_2\) are orthonormal and perpendicular to the axis. On a unit leg, with \(n\) perpendicular to \(\hat u\), the linear-bend row has length \(1\). A collinear angle is replaced in the chart because its row carries \(1/\sin\theta\).

\subsection{Gabriel graph}
\label{sec:org5c753f7}

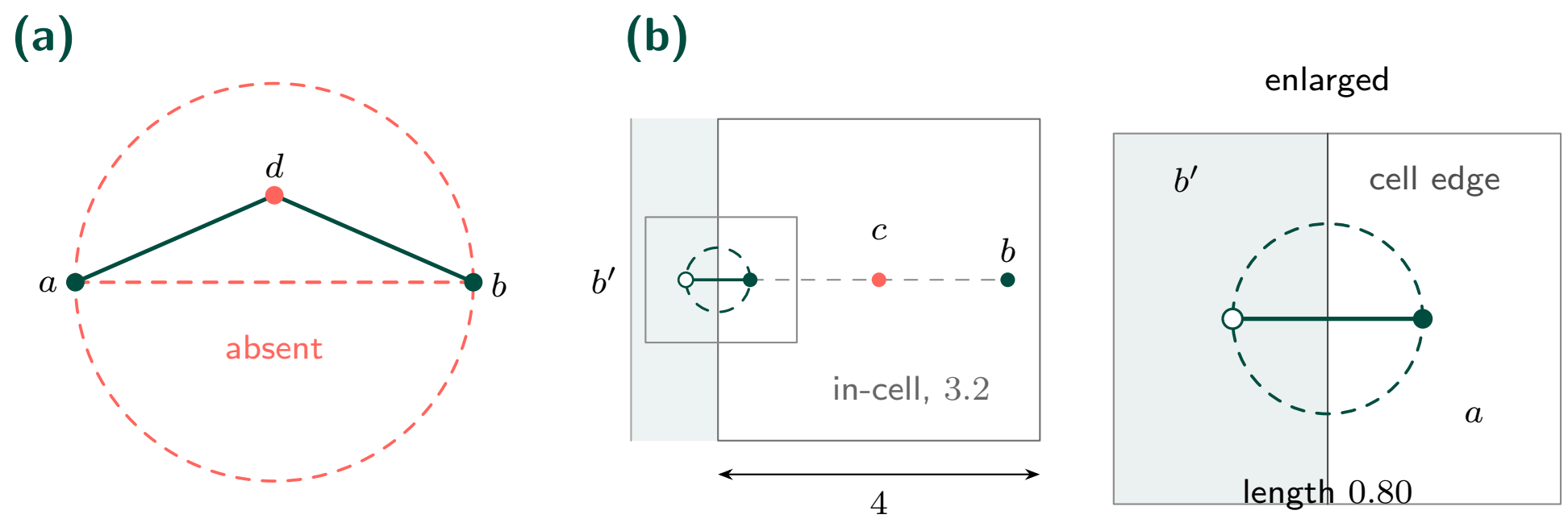
\begin{figure}[H]
\centering
\begin{tikzpicture}[>=Stealth, font=\sffamily\scriptsize,
                    line cap=round, line join=round]
  \def\GAx{0}\def\GAy{0}
  \def\GBx{3.2}\def\GBy{0}
  \def\GDx{1.6}\def\GDy{0.70}
  \pgfmathsetmacro{\GMx}{(\GAx+\GBx)/2}
  \pgfmathsetmacro{\GMy}{(\GAy+\GBy)/2}
  \pgfmathsetmacro{\GRad}{veclen(\GBx-\GAx,\GBy-\GAy)/2}

  \def\Cell{4}
  \def\PAx{0.40}\def\PAy{2}
  \def\PBx{3.60}\def\PBy{2}
  \def\PCx{2}\def\PCy{2}
  \pgfmathsetmacro{\rawx}{\PBx-\PAx}
  \pgfmathsetmacro{\rawy}{\PBy-\PAy}
  \pgfmathsetmacro{\micx}{\rawx-\Cell*round(\rawx/\Cell)}
  \pgfmathsetmacro{\micy}{\rawy-\Cell*round(\rawy/\Cell)}
  \pgfmathsetmacro{\PMx}{\PAx+0.5*\micx}
  \pgfmathsetmacro{\PMy}{\PAy+0.5*\micy}
  \pgfmathsetmacro{\PRad}{veclen(\micx,\micy)/2}
  \pgfmathsetmacro{\Bpx}{\PAx+\micx}
  \pgfmathsetmacro{\Bpy}{\PAy+\micy}
    \def\WinL{-0.90}\def\WinR{0.98}
  \def\WinB{1.22}\def\WinT{2.78}

  \node[font=\sffamily\small\bfseries, text=RuhiTeal] at (0.02,4.78) {(a)};
  \node[font=\sffamily\small\bfseries, text=RuhiTeal] at (5.05,4.78) {(b)};

    \begin{scope}[shift={(0.28,2.78)}, x=1.02cm, y=1.02cm]
    \draw[RuhiCoral, dashed, line width=0.7pt] (\GMx,\GMy) circle[radius=\GRad];
    \draw[RuhiCoral, dashed, line width=0.8pt] (\GAx,\GAy) -- (\GBx,\GBy);
    \draw[RuhiTeal, line width=0.95pt] (\GAx,\GAy) -- (\GDx,\GDy);
    \draw[RuhiTeal, line width=0.95pt] (\GBx,\GBy) -- (\GDx,\GDy);
    \node[text=RuhiCoral, anchor=north] at (\GMx,-0.28) {absent};
    \fill[RuhiTeal] (\GAx,\GAy) circle[radius=2.15pt];
    \fill[RuhiTeal] (\GBx,\GBy) circle[radius=2.15pt];
    \fill[RuhiCoral] (\GDx,\GDy) circle[radius=2.15pt];
    \node[anchor=east] at (\GAx,-0.02) {$a$};
    \node[anchor=west] at (\GBx,-0.02) {$b$};
    \node[anchor=south] at (\GDx,\GDy) {$d$};
  \end{scope}

    \begin{scope}[shift={(5.55,1.48)}, x=0.66cm, y=0.66cm]
    \fill[RuhiTeal!7] (-1.08,0) rectangle (0,\Cell);
    \draw[black!55] (0,0) rectangle (\Cell,\Cell);
    \draw[black!35] (-1.08,0) -- (-1.08,\Cell);
    \draw[black!45, dashed] (\PAx,\PAy) -- (\PBx,\PBy);
    \draw[RuhiTeal, dashed, line width=0.6pt] (\PMx,\PMy) circle[radius=\PRad];
    \draw[RuhiTeal, line width=0.7pt] (\Bpx,\Bpy) -- (\PAx,\PAy);
    \fill[RuhiTeal] (\PAx,\PAy) circle[radius=1.7pt];
    \fill[RuhiTeal] (\PBx,\PBy) circle[radius=1.7pt];
    \draw[RuhiTeal, fill=white, line width=0.5pt] (\Bpx,\Bpy) circle[radius=1.7pt];
    \fill[RuhiCoral] (\PCx,\PCy) circle[radius=1.7pt];
    \node[anchor=east] at (-1.02,\Bpy) {$b'$};
    \node[anchor=south] at (\PBx,\PBy) {$b$};
    \node[anchor=south] at (\PCx,2.28) {$c$};
    \node[anchor=north west, text=black!60] at (1.20,1.02) {in-cell, $3.2$};
    \draw[{Stealth[length=3.5pt]}-{Stealth[length=3.5pt]}]
      (0,-0.42) -- (\Cell,-0.42) node[midway, below] {$4$};
    \draw[black!45] (\WinL,\WinB) rectangle (\WinR,\WinT);
  \end{scope}

    \begin{scope}[shift={(10.55,-1.42)}, x=1.95cm, y=1.95cm]
    \fill[white] (\WinL,\WinB) rectangle (\WinR,\WinT);
    \fill[RuhiTeal!8] (\WinL,\WinB) rectangle (0,\WinT);
    \draw[black!45] (\WinL,\WinB) rectangle (\WinR,\WinT);
    \draw[black!70] (0,\WinB) -- (0,\WinT);
    \draw[RuhiTeal, dashed, line width=0.7pt] (\PMx,\PMy) circle[radius=\PRad];
    \draw[RuhiTeal, line width=1.0pt] (\Bpx,\Bpy) -- (\PAx,\PAy);
    \fill[RuhiTeal] (\PAx,\PAy) circle[radius=2.3pt];
    \draw[RuhiTeal, fill=white, line width=0.6pt] (\Bpx,\Bpy) circle[radius=2.3pt];
    \node[anchor=south east] at (-0.46,2.46) {$b'$};
    \node[anchor=north west] at (0.50,1.70) {$a$};
    \node[anchor=north] at (0,1.40) {length $0.80$};
    \node[anchor=west, text=black!70] at (0.10,2.58) {cell edge};
    \node[anchor=south] at (0,2.86) {enlarged};
  \end{scope}
\end{tikzpicture}
\caption{\textbf{Gabriel graph on the minimum image.}
(a) The open ball with diameter $ab$ contains $d$, and the edge $ab$
is absent. The balls on $ad$ and $bd$ are empty, and those edges are
joined. The coordinates are $a=(0,0)$, $b=(3.2,0)$ and
$d=(1.6,0.70)$: the radius is $1.6$, and $d$ lies $0.70$ from the
center.
(b) A square periodic cell of side $4$. The in-cell vector from $a=(0.40,2)$
to $b=(3.60,2)$ has length $3.2$. The minimum-image displacement has
length $0.80$, and the open ball of that diameter is centered on the
cell edge. The ball contains no third atom, so $a$ and $b$ are joined.
The atom $c=(2,2)$ lies on the in-cell chord and outside that ball.
The inset enlarges the same ball. This graph supplies bonds to the
optional internal chart, which is off in the reported calculations.
The section on the minimum-image proximity graph gives the blocker prefix.}
\label{fig:gabriel-graph}
\end{figure}

Figure \ref{fig:gabriel-graph} is the Gabriel test used to choose bonds of the optional internal chart \cite{gabrielNewStatisticalApproach1969}. Atoms \(a\) and \(b\) are joined when the open ball with diameter \(ab\) contains no third atom, and every vector in that test is the shortest periodic image. The displacement comes from the \texttt{minimage} wrap. An absent periodic cell keeps the Cartesian difference, as does the cubic vacuum cell of an isolated molecule. Any other periodic cell uses the shortest image. \texttt{vesin} lists the pairs inside the neighbor cutoff, and \texttt{minimage} measures their lengths. The reported Baker--Chan and Birkholz--Schlegel cases are isolated molecules, and the chart that consumes this graph is disabled.

\subsection{Bond-graph and environment length scales}
\label{sec:org1a00a63}

\begin{figure}[H]
\centering
\includegraphics[width=\textwidth,height=0.62\textheight,keepaspectratio]{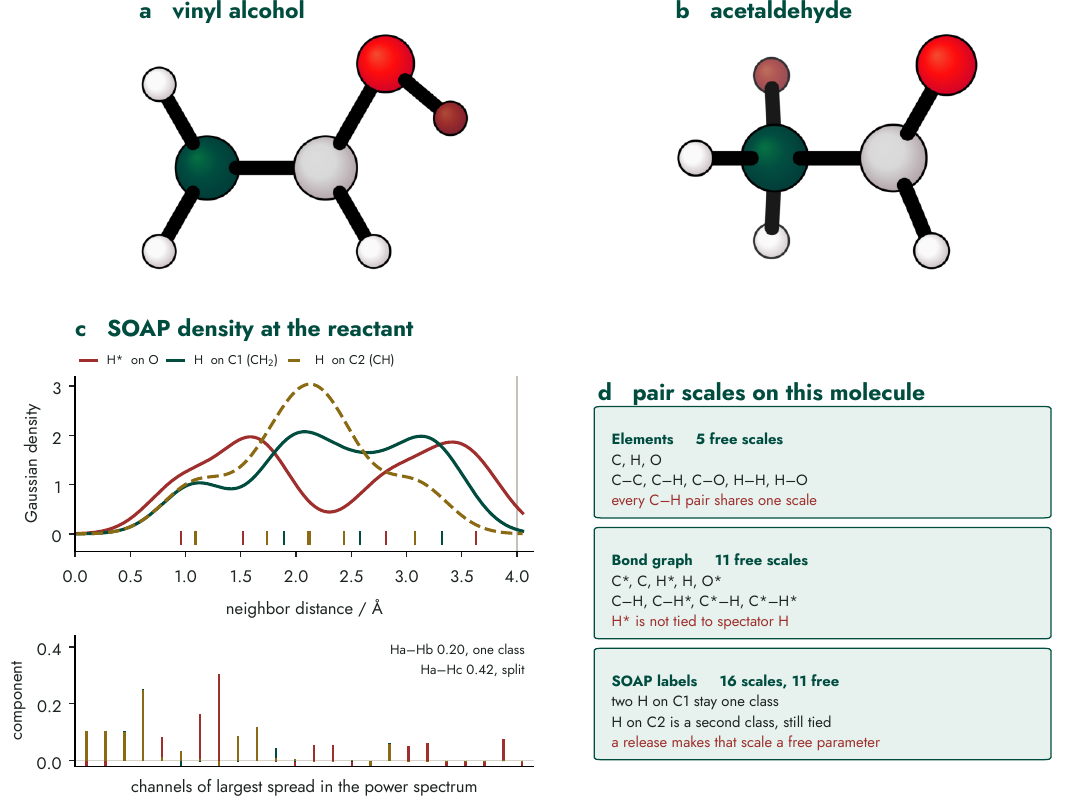}
\caption{\textbf{Length scales on the Baker--Chan vinyl alcohol case.}
Vinyl alcohol (a) and acetaldehyde (b), in one atom order, at standard
bond lengths. Teal is the carbon whose bond partners change, and
coral is the hydrogen that leaves oxygen and binds to that carbon.
(c) Reactant SOAP density for the descriptor comparison. Each neighbor inside $4$~\AA{} contributes a
Gaussian of width $0.3$~\AA{} ($n_{\max}=5$, $l_{\max}=4$). The sticks
are the power-spectrum channels of largest spread among the three
hydrogens. On the reactant and product concatenated, the two hydrogens
on the teal carbon are $0.20$ apart and one class, and the hydrogen on
the other carbon is $0.42$ away and is split off.
(d) A star is the reactive role. Element labels give five free pair
scales, so every C--H pair shares one. The bond-graph split gives the transferring pair its own length scale.
The SOAP-only illustration adds one spectator-hydrogen class and sixteen
pair labels. The reported Birkholz labels use the combined k-nearest-neighbour
and SOAP descriptor; the reported fits release no finer length scale. The
SOAP descriptors determine fixed labels offline, and the covariance uses
inverse pair distances. A released scale is a kernel parameter, not a coordinate of the environment
\cite{goswamiAdaptivePruningIncreased2026}.}
\label{fig:length-scales}
\end{figure}

Figure \ref{fig:length-scales} draws vinyl alcohol and acetaldehyde in one atom order at standard bond lengths. The bond-graph split gives reactive atoms a separate base type, so a transferring pair need not share a length scale with every spectator pair of the same elements. Panel (c) shows the SOAP density at the reactant, and panel (d) compares element, bond-graph, and SOAP-only labels for this illustration. The measured Birkholz searches use fixed types generated from combined k-nearest-neighbour and SOAP descriptors at the reactant and product. Panel (d)'s SOAP-only count is for this illustration. Each environment type starts tied to its element-and-reaction-role base. The reported fits release no finer scale. SOAP contributes to label preparation; the search covariance uses inverse pair distances.

\subsection{Active residual on one coordinate}
\label{sec:orgf2a823d}

\begin{figure}[H]
\centering
\includegraphics[width=\textwidth,height=0.42\textheight,keepaspectratio]{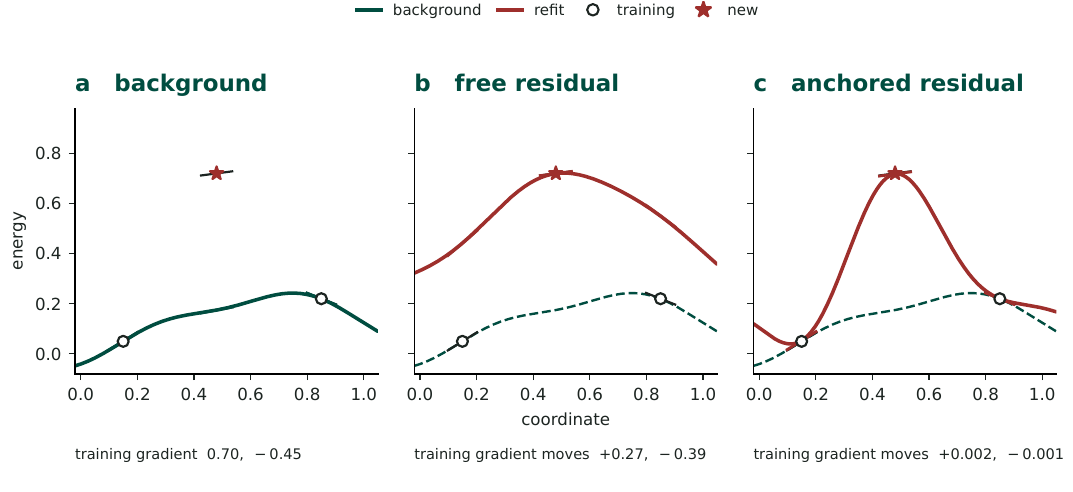}
\caption{\textbf{Anchoring an active residual.}
One coordinate, with a squared-exponential energy and gradient. The
curves illustrate the conditioning. Circles are
two training observations and the star is a new one. Short lines are
energy gradients; the force is the negative of that slope.
(a) The background matches the training gradients $0.70$ and $-0.45$ and
is then held fixed. At the new site it predicts energy $0.17$; the
observation is $0.72$.
(b) A free residual, conditioned only on the new observation, reaches
it and changes the training gradients by $+0.27$ and $-0.39$.
(c) The Schur anchor conditions that residual toward zero on the training
channels, at gradient noise $0.004$. The new observation is still
reached. The training gradients change by $+0.002$ and $-0.001$.
A two-channel case in the section on the active residual moves a background mean
from $0$ to $1/4$ when the residual is free, and leaves it at $0$ when
the anchor noise vanishes. The reported Birkholz--Schlegel residual uses
inverse pair distances. SOAP and chart residuals are off.}
\label{fig:active-residual}
\end{figure}

The Birkholz--Schlegel calculations enable a conditioned active residual on inverse pair distances. Species-resolved SOAP and Cartesian chart residuals are library options and are off in the reported panels. Figure \ref{fig:active-residual} illustrates the conditioning on one coordinate, the background held fixed and both refits given the same new energy and gradient. Panel (a) matches the training gradients \(0.70\) and \(-0.45\); at the new site the background predicts energy \(0.17\) and the observation is \(0.72\). Panel (b), a free residual conditioned only on the new observation, reaches it and moves the training gradients by \(+0.27\) and \(-0.39\). Panel (c) is the Schur anchor at gradient noise \(0.004\): the new observation is still reached, and the training gradients move by \(+0.002\) and \(-0.001\). Section \ref{sec:si-active-residual} derives the complement. This drawing specifies the algebra of the anchor. Convergence and the saddle index are checked on measured forces.

\subsection{Chord leak and one all-image step}
\label{sec:orgc0baac2}

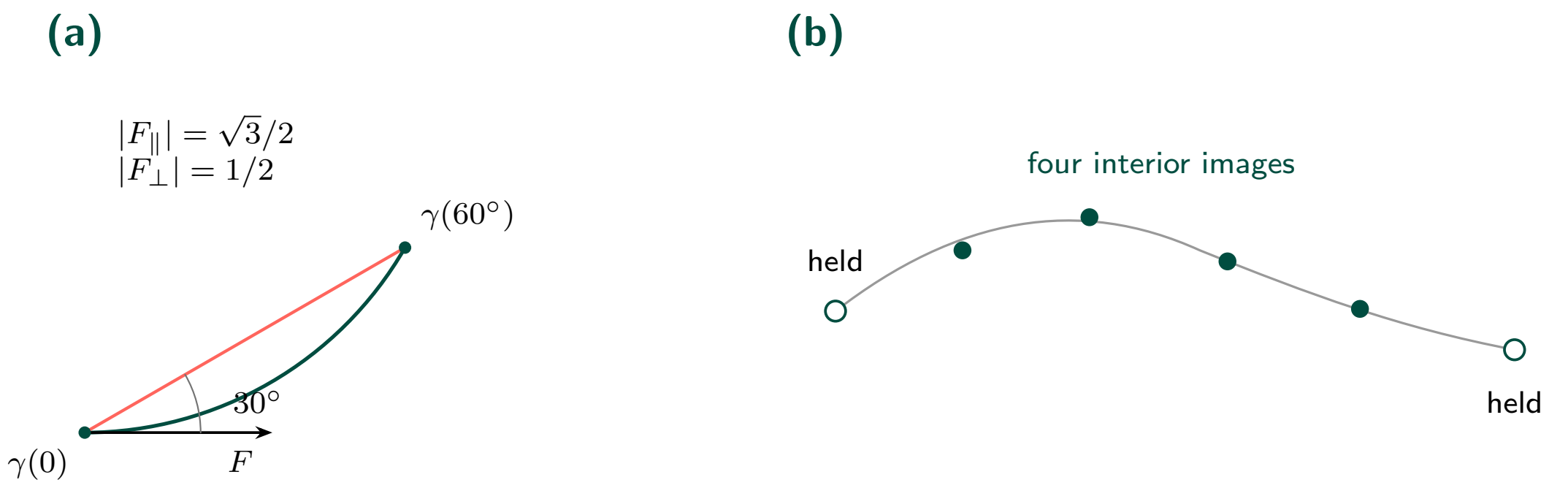
\begin{figure}[H]
\centering
\begin{tikzpicture}[>=Stealth, font=\sffamily\scriptsize,
                    line cap=round, line join=round]
  \def\R{3.35}
  \node[font=\sffamily\small\bfseries, text=RuhiTeal, anchor=west] at (-0.15,3.85) {(a)};
  \node[font=\sffamily\small\bfseries, text=RuhiTeal, anchor=west] at (6.55,3.85) {(b)};

  \begin{scope}[shift={(0.35,0.25)}]
    \draw[RuhiTeal, line width=0.95pt] (0,0) arc (-90:-30:\R);
    \draw[RuhiCoral, line width=0.8pt] (0,0) -- ({\R*sin(60)}, {\R*(1-cos(60))});
    \draw[-{Stealth[length=4.4pt]}, line width=0.7pt] (0,0) -- (1.70,0);
    \node[anchor=north west] at (1.15,-0.02) {$F$};
    \draw[black!55] (1.05,0) arc (0:30:1.05);
    \node[anchor=west] at (1.20,0.28) {$30^{\circ}$};
    \fill[RuhiTeal] (0,0) circle[radius=1.6pt];
    \fill[RuhiTeal] ({\R*sin(60)}, {\R*(1-cos(60))}) circle[radius=1.6pt];
    \node[anchor=north east] at (0,0) {$\gamma(0)$};
    \node[anchor=south west] at ({\R*sin(60)}, {\R*(1-cos(60))}) {$\gamma(60^{\circ})$};
    \node[anchor=west, align=left] at (0.15,2.55)
      {$|F_{\parallel}|=\sqrt{3}/2$\\
       $|F_{\perp}|=1/2$};
  \end{scope}

  \begin{scope}[shift={(7.15,1.35)}]
    \draw[black!40, line width=0.6pt] (0,0)
      .. controls (1.1,0.85) and (2.2,1.05) .. (3.3,0.55)
      .. controls (4.3,0.15) and (5.1,-0.15) .. (6.15,-0.35);
    \foreach \i/\x/\y in {0/0/0, 1/1.15/0.55, 2/2.30/0.85, 3/3.55/0.45, 4/4.75/0.02, 5/6.15/-0.35} {
      \ifnum\i=0
        \draw[RuhiTeal, line width=0.7pt, fill=white] (\x,\y) circle[radius=2.6pt];
      \else\ifnum\i=5
        \draw[RuhiTeal, line width=0.7pt, fill=white] (\x,\y) circle[radius=2.6pt];
      \else
        \fill[RuhiTeal] (\x,\y) circle[radius=2.3pt];
      \fi\fi
    }
    \node[anchor=south] at (0,0.22) {held};
    \node[anchor=north] at (6.15,-0.58) {held};
    \node[anchor=south, text=RuhiTeal] at (2.95,1.05) {four interior images};
  \end{scope}
\end{tikzpicture}
\caption{\textbf{Chord leak and one all-image step.}
(a) The arc $\gamma(\theta)=R(\sin\theta,\,1-\cos\theta)$ turns through
$\varphi=60^{\circ}$. The chord meets the initial tangent at
$\varphi/2=30^{\circ}$. A unit force along that tangent has along-chord
component $\sqrt{3}/2$ and across-chord component $1/2$.
(b) $N=6$ images. The all-image update measures the four interior
images and holds the two endpoints. The reported calculations use the
one-image schedule.}
\label{fig:chord-leak}
\end{figure}

The reported panels use the one-image schedule of the main text, in which each outer iteration adds one measured geometry chosen by the planner. GPRidge also implements an all-image schedule, which measures every interior image in one outer iteration; the reported runs do not use it. Figure \ref{fig:chord-leak} evaluates the chord-leak identity on a circular arc that turns through \(60^{\circ}\). The chord meets the initial tangent at \(30^{\circ}\). A unit force along the initial tangent has along-chord component \(\sqrt{3}/2\) and across-chord component \(1/2\), and \((\sqrt{3}/2)\tan 30^{\circ}=1/2\). Panel (b) is one all-image step on \(N=6\) images. The four interior images are measured, and the endpoints are held.

\subsection{Schematic index checks}
\label{sec:org13e3d14}

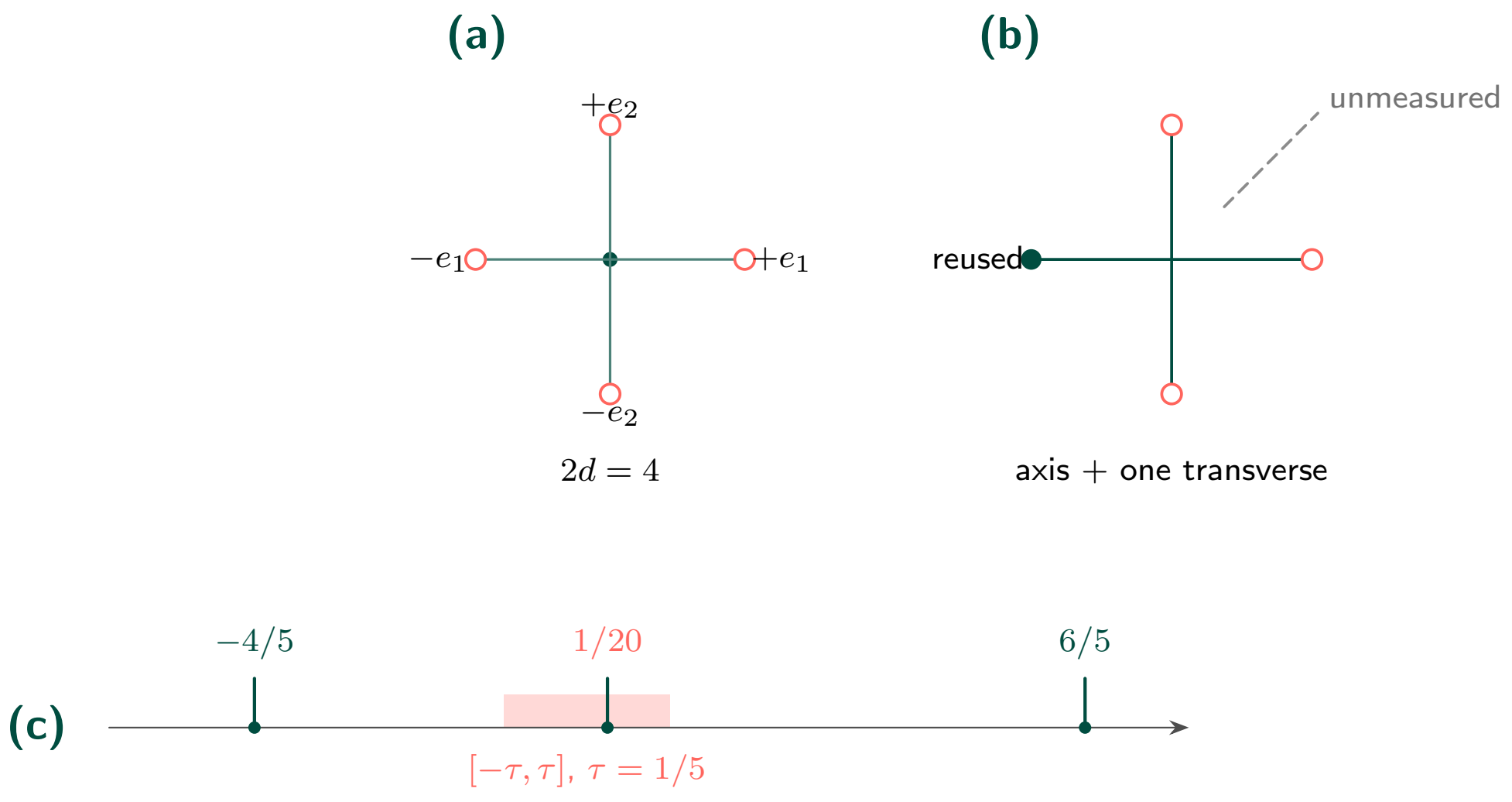
\begin{figure}[H]
\centering
\begin{tikzpicture}[>=Stealth, font=\sffamily\scriptsize,
                    line cap=round, line join=round]
  \node[font=\sffamily\small\bfseries, text=RuhiTeal, anchor=west] at (0,3.35) {(a)};
  \node[font=\sffamily\small\bfseries, text=RuhiTeal, anchor=west] at (4.55,3.35) {(b)};
  \node[font=\sffamily\small\bfseries, text=RuhiTeal, anchor=east] at (-2.95,-2.55) {(c)};

  \begin{scope}[shift={(1.55,1.45)}]
    \fill[RuhiTeal] (0,0) circle[radius=1.8pt];
    \foreach \x/\y/\lab/\anc in {
        1.15/0/$+e_1$/west,
        -1.15/0/$-e_1$/east,
        0/1.15/$+e_2$/south,
        0/-1.15/$-e_2$/north} {
      \draw[RuhiTeal!70, line width=0.6pt] (0,0) -- (\x,\y);
      \draw[RuhiCoral, fill=white, line width=0.7pt] (\x,\y) circle[radius=2.4pt];
      \node[anchor=\anc, inner sep=1.5pt] at (\x,\y) {\lab};
    }
    \node[anchor=north] at (0,-1.55) {$2d=4$};
  \end{scope}

  \begin{scope}[shift={(6.35,1.45)}]
    \draw[RuhiTeal, line width=0.7pt] (-1.20,0) -- (1.20,0);
    \draw[RuhiTeal, line width=0.7pt] (0,-1.15) -- (0,1.15);
    \fill[RuhiTeal] (-1.20,0) circle[radius=2.5pt];
    \draw[RuhiCoral, fill=white, line width=0.7pt] (1.20,0) circle[radius=2.4pt];
    \draw[RuhiCoral, fill=white, line width=0.7pt] (0,1.15) circle[radius=2.4pt];
    \draw[RuhiCoral, fill=white, line width=0.7pt] (0,-1.15) circle[radius=2.4pt];
    \draw[black!45, densely dashed, line width=0.7pt] (0.45,0.45) -- (1.25,1.25);
    \node[anchor=south west, text=black!55] at (1.20,1.15) {unmeasured};
    \node[anchor=east, inner sep=1.5pt] at (-1.20,0.02) {reused};
    \node[anchor=north] at (0,-1.55) {axis $+$ one transverse};
  \end{scope}

  \begin{scope}[shift={(1.35,-2.55)}, x=3.55cm]
    \fill[RuhiCoral!25] (-0.2,0) rectangle (0.2,0.28);
    \draw[black!70, ->] (-1.15,0) -- (1.45,0);
    \foreach \x/\lab in {-0.8/$-4/5$, 0.05/$1/20$, 1.2/$6/5$} {
      \draw[RuhiTeal, line width=0.8pt] (\x,0) -- (\x,0.42);
      \fill[RuhiTeal] (\x,0) circle[radius=1.5pt];
    }
    \node[anchor=south, text=RuhiTeal] at (-0.8,0.44) {$-4/5$};
    \node[anchor=south, text=RuhiCoral] at (0.05,0.44) {$1/20$};
    \node[anchor=south, text=RuhiTeal] at (1.2,0.44) {$6/5$};
    \node[anchor=north, text=RuhiCoral] at (0,-0.08) {$[-\tau,\tau]$, $\tau=1/5$};
  \end{scope}
\end{tikzpicture}
\caption{\textbf{Index checks.}
(a) The climbing-image check drawn at $d=2$. The four open circles are
the potential gradient at both signs of $e_1$ and of $e_2$, and $2d=4$.
In a search, $d=3N_{\mathrm{at}}-6$ for an isolated nonlinear molecule
and $d=3N_{\mathrm{at}}-5$ for an isolated linear molecule.
(b) The dimer measures the axis and one transverse direction, both
signs, and reuses one axis endpoint. The dashed direction is unmeasured.
(c) $\tau=1/5$. The eigenvalues $-4/5$ and $6/5$ lie outside
$[-\tau,\tau]$, and $1/20$ lies inside it, so this schematic spectrum
is unresolved. The table index is the full validator stencil, which is
a different count from chemical connectivity of the reactant and product.}
\label{fig:index-checks}
\end{figure}

Figure \ref{fig:index-checks} draws the two search checks and the unresolved-eigenvalue window of the article at a reduced dimension. Panel (a) is the climbing-image check at \(d=2\): the potential gradient at both signs of \(e_1\) and of \(e_2\), so \(2d=4\) probes. A search uses \(d=3N_{\mathrm{at}}-6\) for an isolated nonlinear molecule and \(d=3N_{\mathrm{at}}-5\) for an isolated linear molecule. Panel (b) measures the axis and one transverse direction, both signs, reuses one axis endpoint, and leaves a third direction unmeasured. Panel (c) takes \(\tau=1/5\). The eigenvalues \(-4/5\) and \(6/5\) lie outside \([-\tau,\tau]\), and \(1/20\) lies inside that interval, so this schematic spectrum is unresolved. These drawings are not the validator stencil. The index column in the tables is that full stencil. A count on the columns the search has measured decides only whether the search continues. Neither count tests whether the geometry connects the reactant to the product.

\subsection{Band tangent}
\label{sec:org1ded528}

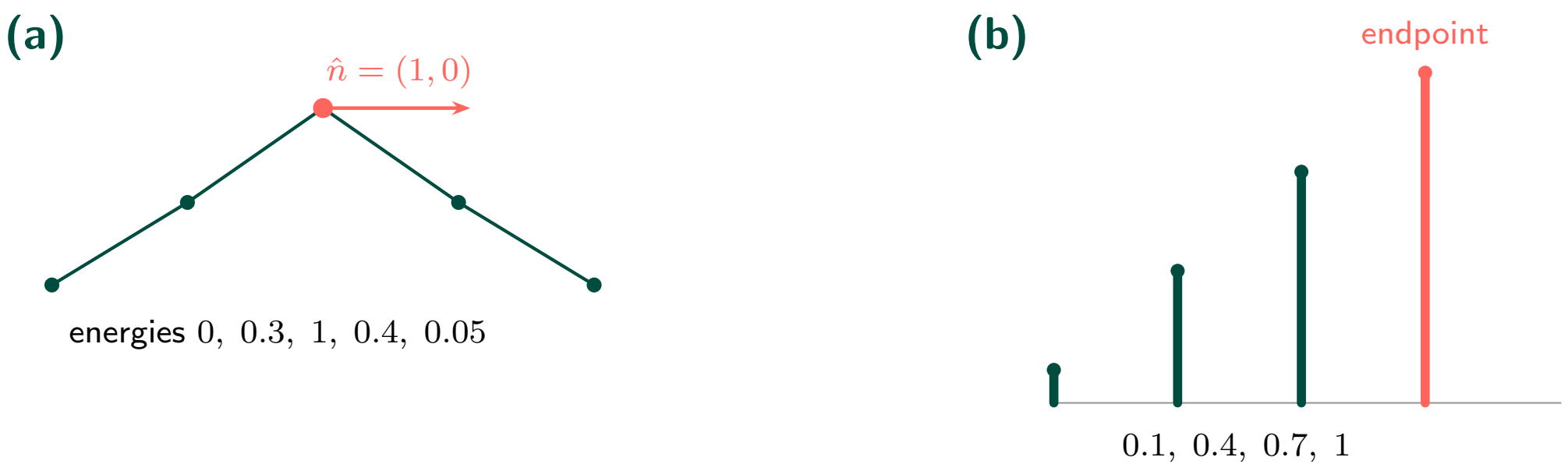
\begin{figure}[H]
\centering
\begin{tikzpicture}[>=Stealth, font=\sffamily\scriptsize,
                    line cap=round, line join=round]
  \node[font=\sffamily\small\bfseries, text=RuhiTeal, anchor=west] at (0,3.65) {(a)};
  \node[font=\sffamily\small\bfseries, text=RuhiTeal, anchor=west] at (8.15,3.65) {(b)};

  \begin{scope}[shift={(0.55,1.55)}]
    \draw[RuhiTeal, line width=0.8pt]
      (0,0) -- (1.15,0.70) -- (2.30,1.50) -- (3.45,0.70) -- (4.60,0);
    \foreach \x/\y in {0/0, 1.15/0.70, 3.45/0.70, 4.60/0} {
      \fill[RuhiTeal] (\x,\y) circle[radius=1.8pt];
    }
    \fill[RuhiCoral] (2.30,1.50) circle[radius=2.4pt];
    \draw[-{Stealth[length=4.4pt]}, RuhiCoral, line width=0.85pt]
      (2.30,1.50) -- (3.55,1.50);
    \node[anchor=south, text=RuhiCoral] at (2.95,1.52) {$\hat n=(1,0)$};
    \node[anchor=north west, align=left] at (0,-0.15)
      {energies $0,\;0.3,\;1,\;0.4,\;0.05$};
  \end{scope}

  \begin{scope}[shift={(9.05,0.55)}]
    \draw[black!40] (0,0) -- (4.3,0);
    \foreach \i/\h/\col in {0/0.28/RuhiTeal, 1/1.12/RuhiTeal, 2/1.96/RuhiTeal, 3/2.80/RuhiCoral} {
      \draw[\col, line width=2.2pt] ({\i*1.05},0) -- ({\i*1.05},\h);
      \fill[\col] ({\i*1.05},\h) circle[radius=1.7pt];
    }
    \node[anchor=south, text=RuhiCoral] at (3.15,2.85) {endpoint};
    \node[anchor=north] at (1.55,-0.12) {$0.1,\;0.4,\;0.7,\;1$};
  \end{scope}
\end{tikzpicture}
\caption{\textbf{Band tangent.}
(a) Five images, numbered $0$ through $4$, at $(0,0)$, $(1,0.15)$,
$(2,0.35)$, $(3,0.15)$ and $(4,0)$. The energies are $0$, $0.3$, $1$,
$0.4$ and $0.05$, so image $2$ is the highest interior image.
The centered neighbor difference is $\mathbf{R}_{3}-\mathbf{R}_{1}=(2,0)$,
and the unit tangent is $(1,0)$. The drawing enlarges the rise.
(b) Energies $0.1$, $0.4$, $0.7$ and $1.0$. The maximum lies on the
product endpoint, so the band has not resolved a barrier.}
\label{fig:band-tangent}
\end{figure}

Figure \ref{fig:band-tangent} is the band tangent of the article on five images, numbered \(0\) through \(4\), at \((0,0)\), \((1,0.15)\), \((2,0.35)\), \((3,0.15)\) and \((4,0)\). Image \(2\) is the highest interior image. The energies are \(0\), \(0.3\), \(1\), \(0.4\) and \(0.05\), the neighbor difference is \(\mathbf{R}_{3}-\mathbf{R}_{1}=(2,0)\), and the unit tangent is \((1,0)\). The drawing enlarges the rise, and the neighbors of the peak share a height, so the drawn arrow is that same tangent. Panel (b) has energies \(0.1\), \(0.4\), \(0.7\) and \(1.0\), and the maximum lies on the product endpoint. A band whose maximum falls on an endpoint has not resolved a barrier.

\subsection{Kernel support}
\label{sec:org7ec5e10}

\begin{figure}[H]
\centering
\begin{tikzpicture}[>=Stealth, font=\sffamily\scriptsize,
                    line cap=round, line join=round]
  \def\u{1.55}
  \fill[RuhiTeal!18] (-1*\u,-0.22) rectangle (1*\u,0.22);
  \fill[RuhiTeal!18] (1*\u,-0.22) rectangle (3*\u,0.22);
  \draw[black!70, ->] (-1.35*\u,0) -- (4.15*\u,0);
  \foreach \q/\name in {0/$0$, 2/$2$} {
    \fill[RuhiTeal] (\q*\u,0) circle[radius=2.3pt];
    \node[anchor=north, text=RuhiTeal] at (\q*\u,-0.28) {\name};
  }
  \fill[RuhiTeal] (0.5*\u,0) circle[radius=2.2pt];
  \node[anchor=south, text=RuhiTeal] at (0.5*\u,0.28) {$1/2$};
  \node[anchor=south, text=RuhiTeal, align=center] at (0.5*\u,0.72)
    {square $1/4$\\supported};
  \fill[RuhiCoral] (3.5*\u,0) circle[radius=2.4pt];
  \node[anchor=south, text=RuhiCoral] at (3.5*\u,0.28) {$7/2$};
  \node[anchor=south, text=RuhiCoral, align=center] at (3.5*\u,0.72)
    {square $9/4$\\joins the batch};
  \node[anchor=north, text=black!60] at (4.05*\u,-0.02) {$q$};
  \node[anchor=north west, text=black!60] at (-1.2*\u,-0.55)
    {shaded: within one length of a training row};
\end{tikzpicture}
\caption{\textbf{Kernel support.}
One feature, $\ell=1$, with training rows at $q=0$ and $q=2$. The
shaded intervals are the closed unit balls. The image at $1/2$ has
squared distance $1/4$ and is supported. The image at $7/2$ has squared
distance $9/4$ and is evaluated on the potential. A squared distance of $1$
meets the closed ball and stays supported.}
\label{fig:kernel-support}
\end{figure}
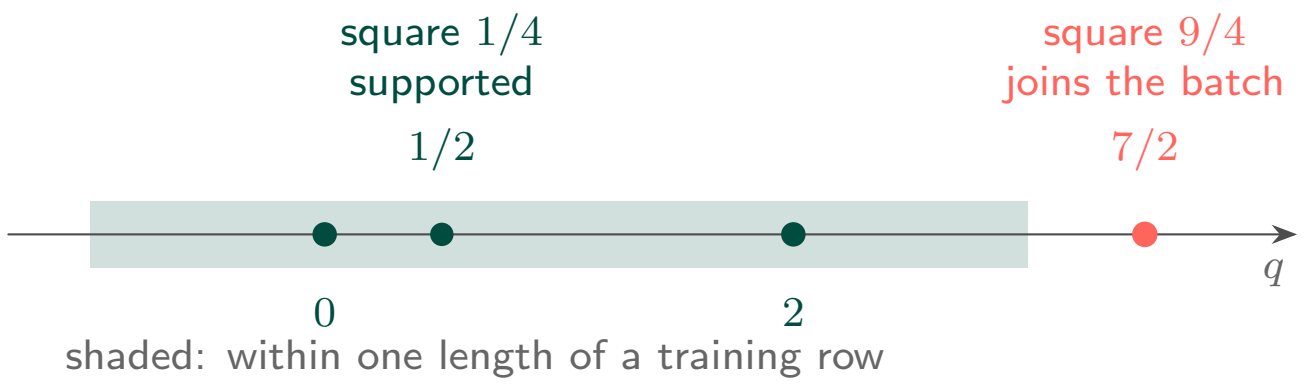

Figure \ref{fig:kernel-support} is the kernel-support test of the article for one feature with \(\ell=1\) and training rows at \(0\) and \(2\). The image at \(1/2\) has squared distance \(1/4\) to the nearer row and is supported. The image at \(7/2\) has squared distance \(9/4\) and joins the potential-evaluation batch. The shaded intervals are the closed unit balls around the training rows, so a squared distance of \(1\) stays supported. In the publication settings this batch is off for Baker--Chan and on for the Birkholz--Schlegel seed band.

\subsection{Where the elapsed time goes}
\label{sec:orge331c81}

The Supporting Information figure "Where the elapsed time goes" partitions one of the three passes of the GPRidge \texttt{v0.5.0} Grignard and Oxirane calculations on four MPI ranks with two CPU slots per rank. Their complete elapsed times are 263.173 and 84.373 s. Each calculation contains two complete Cartesian stencils: the earlier stencil reports unresolved weak modes, and the final stencil certifies the refined saddle. The earlier stencil is charged to the search and the final stencil to independent validation. GP mode extraction takes 0.363 s for Grignard and 0.023 s for Oxirane. The remaining dimer intervals are 12.6 and 29.7 s, outside the oracle, objective, conditioning, prediction, retry, and GP mode timers. The time outside the named components is 10.4 s and 7.6 s. The campaign timing figure reports medians and ranges across three complete repetitions; the accompanying table gives means and sample standard deviations. The component table in the Supporting Information gives the primary component values.

The main manuscript reports the complete-calculation timings and selected measured components; this figure shows their full partition for one of the three passes.
\subsection{Fit efficiency}
\label{sec:orgb82149d}

Figure~\ref{fig:pop-factors}, "Parallel efficiency of a GP fit", measures the complete hyperparameter fit replayed from the Grignard observations with GPRidge 0.5.0. The Tracking Application Live Performance (TALP) module of the Dynamic Load Balancing library measures MPI load balance and communication efficiency in separate profiled invocations. The TALP region covers the native replay from MPI initialization to finalization; the fit timer covers the hyperparameter fit within it. The unprofiled repetitions determine the timing comparisons. This fit isolates the GP algebra within the search.
\subsubsection{Strong and dense weak scaling of the fits}
\label{sec:si-retained-scaling}
\begin{figure}[H]
\centering
\includegraphics[width=\textwidth]{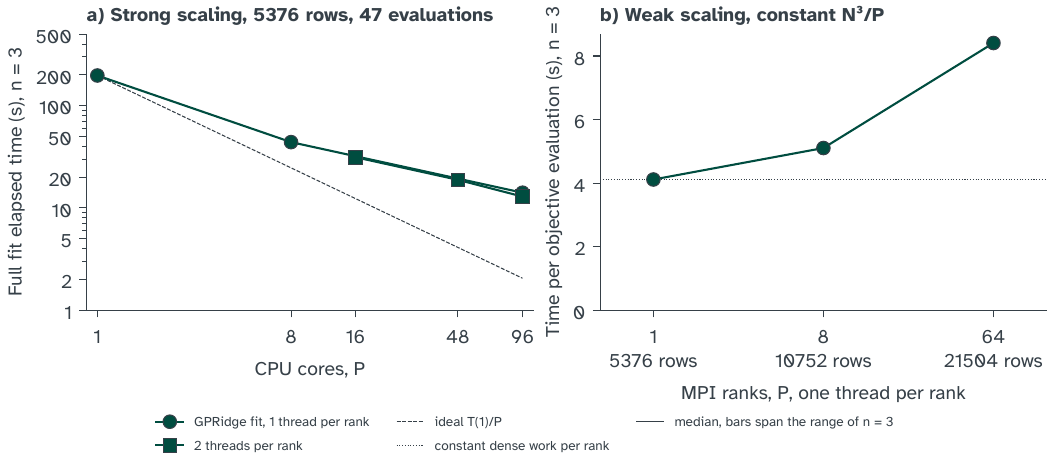}
\caption{\textbf{Strong and dense weak scaling of complete GP fits.}
Left: fixed 5376-row covariance and 47 objective evaluations with
reproducible arithmetic; the 96-core layouts span two nodes. Right:
covariance orders 5376, 10752 and 21504 on 1, 8 and 64 MPI processes,
respectively, with library arithmetic and one thread per process.
Points are medians of three unprofiled repetitions; error bars span their
minimum and maximum. The dashed line gives ideal strong scaling; the dotted line gives constant
time per objective evaluation under the dense weak-scaling work model. The fits make no potential evaluations.}
\label{fig:retained-strong-weak}
\end{figure}

The fit study includes covariance assembly, factorization and
objective derivatives. All layouts use the same observations and
initial parameters for a given matrix size. Each measurement has three
unprofiled repetitions; separate TALP runs measure the first and final
layouts of each plan. Figure \ref{fig:retained-strong-weak} shows medians and
ranges. Strong scaling uses reproducible arithmetic to preserve the fitted
parameters and objective trajectory across layouts. Dense weak scaling uses
library arithmetic, matching the arithmetic of the reaction calculations.
The two studies therefore have separate one-core reference measurements.

The strong-scaling calculation holds the 5376-row covariance, 47 objective
evaluations and 94 gradient evaluations for the two covariance components
fixed. Elapsed time falls from
197.12 s on one core to 12.91 s on 96 cores, with 48 MPI processes and two
threads per process across two nodes (Table \ref{tab:retained-strong}).
The speedup is 15.27, or 15.9 percent efficiency relative to the serial fit.
At the same 96-core allocation, 96 processes with one thread take 14.09 s.

\begin{table}[H]
\caption{\label{tab:retained-strong}Strong scaling of one Grignard fit with reproducible arithmetic. Values are medians of three repetitions; speedup uses the one-core median. The final two layouts span two nodes.}
\centering
\small
\begin{tabular}{rrrrrr}
Nodes & MPI processes & Threads & Cores & Elapsed (s) & Speedup\\
\hline
1 & 1 & 1 & 1 & 197.12 & 1.00\\
1 & 8 & 1 & 8 & 43.98 & 4.48\\
1 & 8 & 2 & 16 & 31.29 & 6.30\\
1 & 24 & 2 & 48 & 18.82 & 10.47\\
2 & 48 & 2 & 96 & 12.91 & 15.27\\
2 & 96 & 1 & 96 & 14.09 & 13.99\\
\end{tabular}
\end{table}

The dense weak-scaling calculation doubles the covariance order \(N\) when
the MPI process count \(P\) increases eightfold, keeping \(N^3/P\) constant.
This holds the leading dense-factorization work per process fixed;
memory per process and the work of other phases follow different laws.
The optimizer requires 47, 74 and 40 objective evaluations for the three
matrix sizes. Table \ref{tab:retained-weak} therefore reports both full-fit
time and time divided by the number of objective evaluations. The latter
gives weak-scaling efficiencies of 80.6 percent on eight processes and
48.9 percent on 64. Whole-fit times give 51.2 and 57.5 percent, respectively,
but include different numbers of optimizer evaluations. The separate
64-process TALP run gives MPI parallel efficiency of 0.67, load balance
of 0.92 and communication efficiency of 0.73.

\begin{table}[H]
\caption{\label{tab:retained-weak}Dense weak scaling with library arithmetic and one thread per MPI process. Times are medians of three repetitions. The 64-process layout spans two nodes. Efficiency compares elapsed time per objective evaluation with the serial value under the constant-\(N^3/P\) work model.}
\centering
\small
\begin{tabular}{rrrrrr}
Rows & Processes & Evaluations & Elapsed (s) & s/evaluation & Efficiency (\%)\\
\hline
5376 & 1 & 47 & 193.33 & 4.113 & 100.0\\
10752 & 8 & 74 & 377.71 & 5.104 & 80.6\\
21504 & 64 & 40 & 336.35 & 8.409 & 48.9\\
\end{tabular}
\end{table}

The timings include every repetition, numerical
validation, component timing and profiling result. These timings characterize the workloads on the measured layouts.

\subsubsection{Library and reproducible factor modes}
\label{sec:org957eb44}

A separate archived arithmetic study in Table \ref{tab:fit-modes} gives the fixed-work Grignard fit on one node with two 24-core Xeon Gold 6248R processors, rounded from the two repetitions in Table \ref{tab:fit-modes-si}. Each run seeds 120 observations, starts from row 20, and stops when the posterior refresh of the first MAP fit completes. It uses one thread per rank and a ScaLAPACK block size of 128. The library factor takes 10.0 s on 8 ranks and 8.3 s on 32, and the reproducible factor 16.0 s and 10.8 s. On this node and this fit, the reproducible factor takes 1.6 times as long at 8 ranks and 1.3 times as long at 32. Those are the measured ratios for this workload. The reproducible mode is the one whose bits do not depend on rank and thread count on one CPU microarchitecture.

\begin{table}[H]
\caption{\label{tab:fit-modes}Seconds to the first completed MAP fit of the fixed-work Grignard replay (first of two repetitions; the second differs by at most 0.25 s), by factorization mode and MPI ranks. This fit is not a publication search.}
\centering
\small
\begin{tabular}{lrrr}
Mode & 8 ranks & 16 ranks & 32 ranks\\
\hline
Library & 10.0 & 8.0 & 8.3\\
Reproducible & 16.0 & 11.1 & 10.8\\
\end{tabular}
\end{table}
\subsection{Factor timings that are not the publication searches}
\label{sec:orgb8ef501}

At \(n=50000\) on one node with two Xeon Gold 6248R processors, four ranks with 12 threads each, the library factor takes 57.0 s, the reproducible tile factor 53.8 s, Scheme II with INT8 residues through MKL 127.6 s, and Scheme II with FP64 residues 213.1 s. The two Scheme II engines return identical bits on that comparison. At \(n=12000\), the Scheme II factor is identical at 1, 2 and 4 ranks and 1 and 6 threads in five layouts, for both engines, with a relative residual of \(9.471\times10^{-16}\). On a node with two AMD EPYC 7742 processors and eight NVIDIA A100 GPUs (40 GB), Scheme II with INT8 tensor cores factors \(n=50000\) in 138.1, 91.0, 57.0 and 32.6 s on 1, 2, 4 and 8 GPUs, one rank per GPU. A repeat gives 139.4, 90.8, 57.3 and 33.3 s. At \(n=100000\) on 8 GPUs the factor takes 227.0 s, and the library factor on the CPUs of the same node takes 421.6 s. On an AMD Ryzen 9 9950X with an NVIDIA RTX 5070, the INT8 Scheme II product reaches 2.63 to 2.74 TFLOP/s of FP64-equivalent work at \(n=8192\) and 3.18 to 3.19 at \(n=16384\), and native cuBLAS DGEMM reaches 0.463 to 0.479. Section \ref{sec:si-predict-arithmetic} lists the same runs at one more digit. The \(n=12000\) Scheme II factor is also identical on the EPYC node for the FP64, built-in INT8 and CUDA engines. That is the only cross-processor comparison recorded here, and it covers Scheme II. It does not cover the reproducible or library modes, and it does not cover the 44 publication searches.
\section{Campaign options: derivations}
\label{sec:campaign-derivations}
The optional all-image-evaluated (AIE) loop uses one kernel, one posterior
solve, one lengthscale prior, one seed, one climbing-image
marker rule, and one oracle schedule. The reflected-force
formulation is given in the main text and in OCI-NEB
\cite{goswamiEnhancedClimbingImage2026}; this section focuses
on posterior marker selection and quantities specific to the
GP search.
Lean statements of the discrete identities sit in Section
\ref{sec:lean-identities}.
\subsection{What surrogate evidence establishes}
\label{sec:si-surrogate-validation}
Energy and Cartesian gradient observations constrain a single scalar
potential. They do not determine that potential between a finite set of
measured geometries. Consider the smooth potentials
\begin{equation}
 V_0(x,y)=\tfrac12(x^2+y^2),\qquad
 V_1(x,y)=V_0(x,y)+(x^2-1)^2.
\end{equation}
Their energies and both gradient components agree at $(\pm1,0)$.
Both have a stationary point at the origin, but their Hessians there are
$\operatorname{diag}(1,1)$ and $\operatorname{diag}(-3,1)$, respectively.
Thus identical joint training data admit both a minimum and a first-order
saddle at an unmeasured geometry. Posterior variance under a fitted kernel
is not a deterministic error bound for an arbitrary underlying potential.
This example does not rule out local convergence under additional
approximation and regularity assumptions; it shows why those assumptions
must be stated and checked separately.

\paragraph{Soft modes and local sensitivity.}
For a quadratic potential with nonsingular Hessian $H$, a constant gradient
perturbation $e$ displaces its critical point by $-H^{-1}e$. The displacement
is bounded by $\|e\|/\gamma$, with
$\gamma=\min_j|\lambda_j(H)|$ on the subspace without exact symmetry modes.
The most negative eigenvalue alone does not give this bound. For example,
\begin{equation}
 V(x,y)=-x^2+y^2/200,\qquad
 \mu(x,y)=V(x,y)-\epsilon y
\end{equation}
have the same index-one Hessian $\operatorname{diag}(-2,1/100)$.
The gradient error is $\epsilon>0$ and the stationary-point displacement
is $100\epsilon$, rather than the $\epsilon/2$ suggested by the negative
mode alone. Active length scales selected from the unstable direction
remain responsive to measured errors in stable directions.
\paragraph{Numerical curvature is a separate measurement.}
A small antisymmetric part of a force-difference Hessian measures a
consistency defect; it does not bound the symmetric truncation error.
For $V(x)=-\lambda x^2/2+a x^4/4$, with $\lambda,a>0$, the true curvature
at the origin is $-\lambda$, while the central gradient difference is
$\widehat H_h=-\lambda+a h^2$. The scalar stencil is symmetric for every
$h$, yet its sign is wrong when $a h^2>\lambda$. A stencil comparison or
an independent curvature calculation is needed to assess a soft mode.
Agreement across step sizes remains a numerical assessment, not a
rigorous bound without further control of derivative or oracle errors.

\paragraph{Consistent centered index stencils.}
Every column of the measured Hessian uses a centered force difference.
An available forward dimer image supplies one endpoint of the axis pair.
The need for a common stencil order can be seen from
\begin{equation}
 V(x,y,z)=-x^2+\tfrac12 k y^2+\tfrac12 z^2+c x^2y,\qquad k>0.
\end{equation}
At the origin, the Hessian is $\operatorname{diag}(-2,k,1)$.
Centered gradient differences recover this Hessian exactly for the cubic
polynomial. A forward difference in the $x$ direction instead adds
$(0,ch,0)^\top$ to its first column. Combining that column with centered
columns produces an antisymmetric-part Frobenius norm $|ch|/\sqrt{2}$,
despite an exact oracle. This error can obscure a positive soft mode.
Exact arithmetic and pointwise binary64 checks confirm these identities.

\paragraph{Cartesian motion through a nearly linear chart.}
For bond vectors $u$ and $v$, the expression
\begin{equation}
 \sin^2\theta=\frac{\|u\times v\|^2}{\|u\|^2\|v\|^2}
\end{equation}
avoids the cancellation in $1-\cos^2\theta$. Topology selection and
gradient evaluation use the same expression to identify the smooth linear
bend representation. The retraction retains Jacobian singular values
larger than $\sqrt{\epsilon_{\mathrm{mach}}}\,\sigma_{\max}$, bounding the
inverse norm by $1/(\sqrt{\epsilon_{\mathrm{mach}}}\,\sigma_{\max})$.
Corrections lie in the kept row space; the requested Cartesian
displacement keeps its component in the complementary space.
The local identities and a nearly linear substitution geometry
are checked in exact arithmetic.

A linearized valence-angle target can also leave its domain $[0,\pi]$.
For two bonds with relative transverse displacement $d$, a Cartesian
step reversing that displacement can produce the target
$\arctan(d)-2d/(1+d^2)<0$ for $0<d<1$; the reflected construction
produces a target above $\pi$. Neither target has an inverse in the
angle chart, although the Cartesian trial is a valid geometry. In this
case the retraction retains the bounded Cartesian trial, fixes frozen
atoms, and reports the chart residual as unresolved. Pair-separation and
measured-trust restrictions still apply. The domain inequality and
binary64 examples are checked in exact arithmetic.

\paragraph{Rotation residuals depend on the dimer separation.}
For a quadratic potential, the perpendicular gradient difference is
$r_F=h(Hv-(v^\top Hv)v)$, where $v$ is a unit orientation and $h$ is the
dimer separation. Decreasing $h$ can make $\|r_F\|$ smaller than a requested
force tolerance without improving the orientation. For
\begin{equation}
 H=\begin{pmatrix}-1/5&2/5\\2/5&2\end{pmatrix},\qquad v=(1,0)^\top,
\end{equation}
the directional curvature is $-1/5$ and $\|r_F\|=2h/5$, while the lowest
eigenvalue is $(9-\sqrt{137})/10$. At $h=1/100$, the residual is below
$1/20$, but the direction is not an eigenvector. Translational force
convergence and measured rotation convergence are therefore assessed
separately. A requested force tolerance is not an estimate of oracle noise.

\nolinkurl{scripts/derive_surrogate_validation_limits.py} in the software
repository checks these examples, the conditional variance identity, and the rank of
joint quadratic observations in exact SymPy arithmetic.
The Lean gradient-error and geometric-decay identities assume the
approximation bounds they propagate. They do not establish those bounds
for the implemented acquisition schedule.

\paragraph{Measured force-merit control.}
Let $m(x)=\max_a\|F_a(x)\|_2$ and $\phi(x)=m(x)^2$.
For an exact quadratic potential on a subspace with nonsingular Hessian,
the Newton displacement $s=-H^{-1}\nabla V(x)$ satisfies
\begin{equation}
 \nabla V(x+\alpha s)=(1-\alpha)\nabla V(x),\qquad
 \phi(x)-\phi(x+\alpha s)=\alpha(2-\alpha)\phi(x),
 \quad 0<\alpha\leq1.
\end{equation}
The identity holds for the maximum atomic norm because every atomic
force scales by the same factor. It supplies a local model for assessing
force reduction; an approximate minimum-mode step need not satisfy it.
The dimer includes a rejected measurement while restoring its accepted
center. A force-prediction error that causes rejection halves the permitted
displacement, down to a numerical floor. Accepted trials update the
radius from the measured reduction divided by its predicted reduction;
expansion additionally requires a step close to the current boundary.
All intermediate steps and retractions obey the same bound about the
accepted center. A measured escape from a resolved higher-index point
establishes a new center and force reference. The escape starts at the
first geometrically distinct downhill trial, which halving searches for,
and doubles the measured displacement while the measured energy keeps
falling; it stops at the first measured rise. A weak mode of
$-0.03~\mathrm{eV}\,\text{\AA}^{-2}$ lowers the energy by
$4\times10^{-5}$~eV over one $0.05$~\AA{} step, which a refit cannot
represent, so a single short step returns the dimer to the point it
left. For example, at the origin
of $V(x)=(-h_1x_1^2-h_2x_2^2+h_3x_3^2)/2$, with $h_i>0$, a
step $\delta e_2$ lowers the energy by $h_2\delta^2/2$ while increasing
the squared force from zero to $h_2^2\delta^2$. A force reference at the
refused stationary point would therefore reject a valid escape. If the
escape reaches the old ball's radius, projecting a further outward step
onto that ball returns the escape itself; centering the ball on the
measured recovery point permits continued motion. These statements do not
supply an escape direction when the additional curvature is unresolved.

The band has a separate empirical gradient-fidelity control. With
$f=\|\widehat g-g\|_2/\max(\|g\|_2,\epsilon_F)$ at the measured
frontier row, $f>3/4$ contracts the radius by $1/4$, while $f<1/4$
permits expansion by a factor of two. The exploration schedule's initial
radius is not an admissible lower bound for repeated contractions:
a persistently inaccurate local model can require a smaller radius.
Before the numerical floor is reached, repeated contractions
give $R_k=R_0/4^k$. Neither this recurrence nor the quadratic force-merit
identity proves global convergence on a nonlinear surface. Exact arithmetic verifies these identities.
\subsection{Dual HS-SVD and the stable basis}
\label{sec:si-hssvd}
The reported calculations leave dual HS-SVD off and factorize the noisy
Gram by Cholesky. This section derives the optional eigen-factorization.
It assumes the stated range and truncation hypotheses. It is not a
measurement of the Cholesky factor used for the 44 reactions.

A Gaussian process on energies and Cartesian gradients predicts the energy
at a query $x_*$ by
\begin{equation}
\mu(x_*)
  \;=\;
  \mathbf{k}_*^{\top}
  (K+\sigma^2 I)^{-1}\mathbf{y} ,
\label{eq:camp-dual-mean}
\end{equation}
where $\mathbf{y}$ stacks the observed energies and Cartesian
gradients, $K$ is the assembled signal kernel (noise stripped
from the diagonal), and $\mathbf{k}_*$ is the
cross-covariance between the query and the training set.
Equation \eqref{eq:camp-dual-mean} defines the dual predictor for
isotropic observation noise.

The gradient blocks of $K$ are nearly parallel on a band:
neighboring images differ by a fraction of a bond length,
so the rows of $K_{GG}$ are almost copies of one another.
Small signal eigenvalues can be sensitive to assembly roundoff.
An explicit feature representation motivates stable-basis methods
\cite{fasshauerStableEvaluationGaussian2012,fornbergStableComputationsGaussian2011}.

\paragraph{Mercer design.}
For the one-dimensional Gaussian
$k(x,z)=\exp(-\varepsilon^2(x-z)^2)$ and the Gaussian
weight $\rho(x)=(\alpha/\sqrt{\pi})\exp(-\alpha^2 x^2)$,
Mercer's theorem supplies eigenpairs
$(\lambda_n,\phi_n)$ with Hermite--Gaussian
$\phi_n$ and geometrically decaying $\lambda_n$,
\begin{equation}
\int k(x,z)\,\phi_n(z)\,\rho(z)\,dz
  \;=\;
  \lambda_n\,\phi_n(x) .
\label{eq:camp-mercer}
\end{equation}
Collect the values $\phi_n(x_i)$ into a matrix $\Phi$ and
set $A=\Phi\,\mathrm{diag}(\sqrt{\lambda})$. Then
\begin{equation}
K \;=\; A A^{\top} .
\label{eq:camp-AA}
\end{equation}
The physicists' Hermite polynomials satisfy $H_n'(u)=2nH_{n-1}(u)$.
A Hermite--Gaussian eigenfunction has the form
\(\phi_n(x)=c_n e^{-a x^2}H_n(bx)\), with constants \(a,b,c_n\).
Its derivative includes the Gaussian factor,
\[
 \phi_n'(x)=c_n e^{-a x^2}
 \left[2nbH_{n-1}(bx)-2axH_n(bx)\right].
\]
For $n=0$ the polynomial-derivative term is zero.
Under the regularity needed for termwise differentiation, each Mercer eigenvalue $\lambda_n$ remains and the feature factor is replaced by this full derivative. The recurrence alone does not
identify the joint Gram. Stacking energy and
gradient rows,
\begin{equation}
A_{\mathrm{full}}
  \;=\;
  \begin{pmatrix} \Phi \\ \Phi' \end{pmatrix}
  \mathrm{diag}(\sqrt{\lambda}) ,
\qquad
A_{\mathrm{full}} A_{\mathrm{full}}^{\top}
  \;=\;
  \begin{pmatrix}
    K_{EE} & K_{EG} \\
    K_{GE} & K_{GG}
  \end{pmatrix} ,
\label{eq:camp-Afull}
\end{equation}
reconstructs the joint Gram of the main text. The
$d$-dimensional separable kernel is the tensor product of
these one-dimensional factors.

\paragraph{Thin SVD and the stable basis.}
A thin SVD $A=USV^{\top}$ substituted into
Eq. \eqref{eq:camp-AA} gives
$K=US^{2}U^{\top}$. The stable basis at the training nodes
is therefore the left factor,
\begin{equation}
\Psi \;=\; U ,
\qquad
K \;=\; \Psi\,\mathrm{diag}(S^{2})\,\Psi^{\top} ,
\qquad
\Psi^{\top}\Psi \;=\; I .
\label{eq:camp-stable}
\end{equation}
Because the columns of $\Psi$ are orthonormal,
$\mathrm{cond}(\Psi)=1$ independently of
$\mathrm{cond}(K)$. The query basis is
$\Psi(x_*)=A(x_*)VS^{-1}$. Its inverse singular values still
amplify perturbations. Dropping modes changes the approximation and
requires an error assessment relative to the observation noise.

The number of separable Mercer modes grows
combinatorially with the number of inverse-distance
pairtypes, so the optional dual implementation does not materialize $A$. The
symmetric eigensolve of the assembled $K$ in
Eq. \eqref{eq:camp-stable} yields the corresponding invariant
subspaces and eigenvalues in exact arithmetic. It cannot undo
roundoff incurred while assembling the covariance.
That is the dual HS-SVD.

\paragraph{Noisy inverse.}
Write $P=\Psi\Psi^{\top}$ for the orthogonal projector
onto the exact nonzero range of $K$, and split
$\mathbf{y}=P\mathbf{y}+(I-P)\mathbf{y}$. On the range,
$K$ acts as $\Psi\,\mathrm{diag}(S^{2})\,\Psi^{\top}$, so
\begin{equation}
(K+\sigma^{2}I)\,
\Psi\,\mathrm{diag}\!\bigl(1/(S^{2}+\sigma^{2})\bigr)\,
\Psi^{\top}\mathbf{y}
  \;=\;
  P\mathbf{y} .
\label{eq:camp-range}
\end{equation}
On the complement $K$ vanishes, so
$(K+\sigma^{2}I)(\mathbf{y}-P\mathbf{y})
=\sigma^{2}(\mathbf{y}-P\mathbf{y})$ and
\begin{equation}
(K+\sigma^{2}I)^{-1}(\mathbf{y}-P\mathbf{y})
  \;=\;
  \sigma^{-2}(\mathbf{y}-P\mathbf{y}) .
\label{eq:camp-compl}
\end{equation}
Inverting Eq. \eqref{eq:camp-range} on $P\mathbf{y}$ and then
adding Eq. \eqref{eq:camp-compl} gives the noisy-eigen formula,\begin{equation}
(K+\sigma^{2}I)^{-1}\mathbf{y}
  \;=\;
  \Psi\,\mathrm{diag}\!\bigl(1/(S^{2}+\sigma^{2})\bigr)\,
  \Psi^{\top}\mathbf{y}
  \;+\;
  \sigma^{-2}\bigl(\mathbf{y}-\Psi\Psi^{\top}\mathbf{y}\bigr) .
\label{eq:camp-noisy}
\end{equation}
If $\mathbf{y}$ already lies in the range, the second
summand is zero. The dual weights are
$\boldsymbol{\alpha}=(K+\sigma^{2}I)^{-1}\mathbf{y}$ from
Eq. \eqref{eq:camp-noisy}, and the mean is
$\mu=\mathbf{k}_*^{\top}\boldsymbol{\alpha}$. The
tensor/affine reduced coefficient
$p_i=S_i^{2}(\Psi^{\top}\mathbf{y})_i/(S_i^{2}+\sigma^{2})$
is a different path: it is used only when $A$ is
materialized.

The Sexpat kernel on inverse-distance descriptors is
exactly the separable Mercer kernel of this construction
once $\varepsilon_p=1/\ell_p$. Cartesian gradients, not
descriptor gradients, are observed, so each gradient row of
$A_{\mathrm{full}}$ contracts $\partial_{q}\Phi$ with the
Jacobian $J=\partial q/\partial x$. That contraction puts
the chain-rule blocks $J^{\top}K_{qq}J'$ into
Eq. \eqref{eq:camp-Afull} without a finite-difference
curvature estimate.
\subsection{Lengthscale prior}
\label{sec:si-ls-prior}
Maximum-likelihood estimation of the lengthscale of a
noiseless stationary kernel is ill-posed
\cite{karvonenMaximumLikelihoodEstimation2023}: if the
observations differ from the prior mean by a constant
shift, the maximizer of the marginal likelihood is
$\ell=\infty$, and the predictive map fails to be Lipschitz
in the data. A band of molecular geometries is not a
noiseless interpolant, but the same mechanism appears as
two finite-$\ell$ failures.

Write $q^{(a)}$ for the inverse-distance descriptor of
training geometry $a$, $\rho_{0,p}$ for the standard
deviation of pairtype $p$ across the training set, and
$\ell_p$ for the pairtype lengthscale. Parameterize
$w_p=\log\ell_p$.

\paragraph{Collapse.}
If $\ell_p\ll\rho_{0,p}$, every pair of distinct geometries
is many lengthscales apart. The posterior variance at a
new point pins at the prior variance, and the acquisition
scores near-duplicates as informative. The active set then
grows by rows the model cannot tell apart.

\paragraph{Oversmooth.}
If $\ell_p\gg\rho_{0,p}$, every pair is a fraction of a
lengthscale apart. The posterior mean is a nearly flat
interpolant, the posterior variance collapses, and the
band walks to a spurious high saddle with a confident
wrong force.

\paragraph{Two-sided log-Gaussian.}
Both failures are movements of $w_p$ away from
$\log\rho_{0,p}$. The production prior is therefore
\begin{equation}
\log\ell_p
  \;\sim\;
  \mathcal{N}\bigl(\log\rho_{0,p},\,s^{2}\bigr) ,
\label{eq:camp-ls-prior}
\end{equation}
with a single shared width measured from the band,
\begin{equation}
s^{2}
  \;=\;
  \mathrm{Var}_{a<b}
  \bigl[\log\lVert q^{(a)}-q^{(b)}\rVert\bigr] .
\label{eq:camp-ls-width}
\end{equation}
Pairwise descriptor distances remain informative when each
coordinate of $q$ barely varies, because the set of
$\lVert q^{(a)}-q^{(b)}\rVert$ still spans the fine-step to
full-path ratio. The contribution to the MAP objective on
$w_p$ is
\begin{equation}
\mathcal{L}_{\ell}(w_p)
  \;=\;
  \frac{(w_p-\log\rho_{0,p})^{2}}{2s^{2}} ,
\qquad
\frac{\partial\mathcal{L}_{\ell}}{\partial w_p}
  \;=\;
  \frac{w_p-\log\rho_{0,p}}{s^{2}} .
\label{eq:camp-ls-obj}
\end{equation}
The prior is inactive when $s^{2}$ is not positive (a
single geometry, or every pairwise distance degenerate).
The amplitude $\sigma_f^{2}$ is centered on the robust
energy spread $(1.4826\,\mathrm{MAD}(V))^{2}$, not on the
current $\sigma_f^{2}$. That centre is a prior anchor in
energy units. A vanishing energy spread does not imply
small gradients, and it is not the force-convergence test.
Re-centring either mean on the
drifting SCG value removes the constraint: the lengthscale
ratchets upward and the variance either pins or blows up.

The reported band and dimer configurations enable this lengthscale prior.
\subsection{Sexpat on inverse distance}
\label{sec:si-sexpat}
A Cartesian squared-exponential kernel on atomic coordinates
is not invariant under translation or rotation, so the GP
would have to learn rigid motions that the energy does not
depend on. The production kernel acts on inverse
interatomic distances $q_p=1/r_p$, one lengthscale per
chemical pair type
\cite{koistinenMinimumModeSaddle2020}. Packed
log-parameters $\theta_0=\log\sigma_f^{2}$ and
$\theta_{1+p}=\log\ell_p$ give
\begin{equation}
k_{nm}
  \;=\;
  e^{\theta_0}
  \exp\Bigl(
    -\sum_p e^{-2\theta_{1+p}}(\Delta q_p)_{nm}^{2}
  \Bigr)
  \;=\;
  \sigma_f^{2}
  \exp\Bigl(
    -\sum_p (\Delta q_p)_{nm}^{2}/\ell_p^{2}
  \Bigr) .
\label{eq:camp-sexpat}
\end{equation}
At coincident geometries $\Delta q=0$ and $k=\sigma_f^{2}$.
The shape $\varepsilon_p=1/\ell_p$ is the Mercer shape of
Section~\ref{sec:si-hssvd}, so
Eq. \eqref{eq:camp-sexpat} and the separable expansion
agree term for term.
A Cartesian lengthscale is the wrong unit for this
descriptor kernel: $\ell_p$ is a scale on $q=1/r$,
not on $r$.
\subsection{No output normalization}
\label{sec:si-no-norm}
When output normalization is off, the output scale is $1$.
Energies and gradients are not divided by an empirical
energy standard deviation. Energy rows are still shifted
by one frozen residual-energy constant. With
\texttt{prior\_mean\_statistic=max} that constant is the maximum residual energy.
Gradients are not shifted. Predictions add the constant
back to the energy only, so the Cartesian force remains
the negative gradient of the predicted energy.

While hyperparameters are fitted with that scale fixed at
$1$, the objective frame divides the stacked residual by
one root-mean-square $s$ and divides the covariance by the
square of $s$. On exit it restores the log signal variance,
likelihood variance, jitter, and magnitude prior.
The root-mean-square mixes energy and gradient components,
which do not share units. For a stacked residual of length
$n$, the frame objective equals the physical negative log
marginal likelihood minus $n\log s$. Inside one fit $s$
does not depend on the hyperparameters, so the physical
minimizer is unchanged in exact arithmetic. This is a
numerical device, not a diagonal change of observation units.
The optional output-normalization switch divides centered
energies and Cartesian gradients by one positive energy
sample standard deviation.
Both blocks receive the same factor, so the standardized
gradient remains the Cartesian derivative of the
standardized energy. Gradient components are not given
unit variance. If the energy sample variance is zero, the
scale stays $1$. The robust energy median absolute
deviation remains an energy-variance prior centre. It does
not supply that scale, and it does not certify force
convergence.

Power-of-two covariance equilibration is a separate
transformation, given in the section on observation units,
conditioning, and stopping. It is neither this switch nor
the objective-frame root-mean-square.
\subsection{MAP objective and hyperparameter fitting}
\label{sec:si-map-scg}
Hyperparameters are a maximum \emph{a posteriori} point.
The reported rgmin L-BFGS fits minimize
\begin{equation}
\mathcal{L}(\boldsymbol{\theta})
  \;=\;
  \tfrac12\log\det C
  \;+\;
  \tfrac12\mathbf{y}^{\top}C^{-1}\mathbf{y}
  \;-\;
  \log p(\boldsymbol{\theta}) ,
\qquad
C \;=\; K+\sigma^{2}I ,
\label{eq:camp-map}
\end{equation}
the negative log posterior, not the negative log marginal
likelihood alone. The prior term is
Eq. \eqref{eq:camp-ls-obj} together with the half-Cauchy
amplitude contribution of the main-text SI.

$C$ is factorized by Cholesky, $C=LL^{\top}$. The
determinant of a triangular matrix is the product of its
diagonal, so
\begin{equation}
\log\det C
  \;=\;
  2\sum_{i}\log L_{ii} .
\label{eq:camp-logdet}
\end{equation}
The quadratic is a triangular solve
$\mathbf{b}=L^{-1}\mathbf{y}$,
\begin{equation}
\mathbf{y}^{\top}C^{-1}\mathbf{y}
  \;=\;
  \mathbf{b}^{\top}\mathbf{b} .
\label{eq:camp-quad}
\end{equation}
These reductions evaluate the likelihood terms; the configured priors supply the remaining MAP contribution.
The reported fits and predictions both use the Cholesky factor, with
\(\boldsymbol\alpha=L^{-\top}L^{-1}\mathbf y\).
The optional dual HS-SVD formula in Equation \eqref{eq:camp-noisy} describes another inverse representation.
Different factorizations of the same positive-definite covariance agree in exact arithmetic.
Their rounding errors can change an optimizer trajectory; this does not define a different mathematical MAP objective.
\subsection{The solver axis over a fixed objective}
\label{sec:si-solver-axis}
Equation \eqref{eq:camp-map} states a problem, not a
method. Any descent that reaches its stationary point is
admissible, so the minimizer is an axis of the
implementation rather than part of the model, and the
library exposes it as one: \texttt{scg},
\texttt{map\_lbfgs}, \texttt{inla}, and a family reached
through a C ABI as \texttt{rgmin:<method>}, with
\texttt{method} one of L-BFGS, BFGS, SR1, SR2, steepest
descent, Adam, Barzilai--Borwein, FIRE, FIRE2, dogleg, or a
nonlinear conjugate gradient rule.

What must not vary is the objective. Every entry point
builds the same \nolinkurl{MapObjectiveConfig}: the same
robust signal-variance anchor, the same barrier strength
capped at $1/2$, the same length-scale prior
$\rho_{0}$ of Eq. \eqref{eq:camp-ls-obj}, and the same
ceiling on $\log\sigma_{f}^{2}$. A comparison in which the
objective moved with the solver would report the
difference between two problems while appearing to report
the difference between two minimisers.

Two structural limits bound the axis. First, the generic
entry is gradient-only: it carries no curvature callback,
which is why the scaled conjugate gradient path stays a
separate entry point rather than one method among the
others, since it takes both SCG parameters and a curvature
callback. Second, Newton and rational function
optimization are excluded because they need an analytic
Hessian of the marginal likelihood, while the generic entry
that evaluates Eq. \eqref{eq:camp-map} forms gradients and
does not form that Hessian. Excluding them is a statement
about what that entry provides, not a ranking of the methods.
The hyperparameters carry a feasible set. Writing
$\boldsymbol{\theta}\in[\boldsymbol{\ell},\mathbf{u}]$ for
the admissible box, a step $\mathbf{p}$ is accepted at
\begin{equation}
\boldsymbol{\theta}^{+}
  \;=\;
  \Pi_{[\boldsymbol{\ell},\mathbf{u}]}
  \!\left(\boldsymbol{\theta}+\mathbf{p}\right),
\qquad
\Pi_{[\boldsymbol{\ell},\mathbf{u}]}(\mathbf{z})_{i}
  \;=\;
  \min\!\left(\max(z_{i},\ell_{i}),u_{i}\right),
\label{eq:camp-box}
\end{equation}
with either bound left empty meaning unbounded on that
side. The projection is what keeps a length scale inside
the range where the kernel is not numerically flat,
without the objective itself acquiring a penalty term that
would change the stationary point it defines.

A run that reaches the iteration cap is reported as having
exhausted its budget rather than as having converged. The
distinction matters when the axis is used as a
measurement: a method that stops at the cap has not found
the point of Eq. \eqref{eq:camp-map}, and averaging its
result with converged ones reports a number that belongs
to neither.
\subsection{Geodesic sequential IDPP}
\label{sec:si-sidpp}
A Cartesian linear path is $x(t)=(1-t)x_R+tx_P$.
Ordinary image-dependent pair-potential (IDPP) interpolation instead relaxes images against linearly interpolated pair distances,
\(r_p^{\rm IDPP}(t)=(1-t)r_p(x_R)+t r_p(x_P)\) \cite{smidstrupImprovedInitialPath2014}.
Sequential IDPP (SIDPP) inserts each image beside an already relaxed neighbor \cite{schmerwitzSequentialIdpp2024}.
The driver's \texttt{sidpp-geodesic} flag selects inverse-distance targets,
\begin{equation}
 q_p^{\rm target}(t)=\frac{1-t}{r_p(x_R)}+\frac{t}{r_p(x_P)},
 \qquad r_p^{\rm target}(t)=\frac{1}{q_p^{\rm target}(t)}.
\label{eq:camp-sidpp}
\end{equation}
The sequential image relaxation fits these targets in Cartesian coordinates.
The inverse-distance metric follows the motivation of geodesic interpolation \cite{zhuGeodesicInterpolationReaction2019}; Equation \eqref{eq:camp-sidpp} is a target interpolation, not a solution of a geodesic equation on the molecular configuration manifold.
The same inverse-distance metric is available to path reparameterization.
\subsection{Local stability of the reflected flow}
\label{sec:si-reflected-stability}
The reflected-force formulation cited in the main text has a
simple local stability consequence when its direction is the
exact lowest Hessian mode. For a symmetric Hessian, let
$A\mathbf{v}_1=\lambda_1\mathbf{v}_1$ with $\lambda_1<0$,
and let the remaining eigenvalues be $\mu_j>0$. The
linearized reflected-flow Jacobian has
eigenvalue $\lambda_1$ along $\mathbf{v}_1$ and $-\mu_j$ on
the perpendicular modes, so an index-one saddle is attracting.
This argument assumes exact mode alignment; an approximate
dimer axis or band tangent does not by itself establish those
eigenvalue signs.
\subsection{Posterior selection of the climbing image}
\label{sec:si-ci-posterior}
The highest-current-energy image defines the climbing-image
marker. Taking that image swaps the
marker whenever two maxima sit inside the energy
resolution, and the band never climbs the same barrier
twice. Writing the difference between a rival and the
incumbent as
\begin{equation}
\Delta \;\sim\; \mathcal{N}\!\left(\mu,\;\sigma^{2}\right),
\qquad
\mu = E_{r} - E_{i},
\qquad
\sigma^{2} = \sigma^{2}_{r} + \sigma^{2}_{i} ,
\label{eq:camp-ci-posterior}
\end{equation}
the marker moves when $\mu > \kappa\sigma$. At one standard
deviation the probability of a positive difference is about
$0.84$. A measured energy carries the oracle noise; a predicted one carries the posterior
variance. Using both for a measured energy inflates
$\sigma$, and the marker then refuses every handover. On
\texttt{25\_hcnh2} the higher-mean rule exhausts the
iteration budget; Eq. \eqref{eq:camp-ci-posterior}
converges and the marker never moves again; in the release
reaction \texttt{25\_hcnh2} takes 72 search calls in the archived GPRidge 0.5.0 record.
\subsection{Adaptive pruning of the training set}
\label{sec:si-prune}
Every energy and force the calculator returns is appended
to the training set. Adaptive pruning
\cite{goswamiAdaptivePruningIncreased2026} decides which
geometries the covariance $K$ carries. A pruned geometry remains in
the training set and can be re-selected by a per-image
neighborhood.

For the scalar kernel $k(s,s')=\sigma_f^2\exp[-(s-s')^2/\ell^2]$, the latent conditional variance after one noiseless observation, divided by $\sigma_f^2$, is
\[
 v_{\rm rel}(s)=1-\exp(-2s^2/\ell^2).
\]
For $0<\sigma_{\rm rel}<1$, equality with $\sigma_{\rm rel}^2$ occurs at
\[
 s_*=\frac{\ell}{\sqrt 2}\sqrt{-\log(1-\sigma_{\rm rel}^2)}
       =\frac{\ell\sigma_{\rm rel}}{\sqrt 2}+O(\sigma_{\rm rel}^3).
\]
The factor $\sqrt 2$ changes if the kernel is parameterized with $1/2$ in the exponent.
The implementation uses the numerical admission heuristic
\begin{equation}
 d_{\min}=\tfrac12\ell_{\min}\sigma_{\rm rel},\qquad
 \sigma_{\rm rel}=\sqrt{\sigma_n^2/\sigma_f^2}.
\label{eq:camp-dmin}
\end{equation}
This heuristic is not half of $s_*$ for the production kernel convention, and the scalar variance example does not prove a curvature-information cutoff.
The molecular kernel lengthscales act on inverse-distance features, whereas the midpoint filter compares Cartesian earth-mover distances. Its numerical threshold therefore depends on the fixed coordinate-unit convention of the run; it is not a unit-invariant Cartesian resolution theorem.
The dense dimer takes the larger of this numerical threshold and its geometric stencil threshold. Near proposals that predict a new best force or force convergence still receive calculator evaluations.
Pruning selects a subset for the factor while retaining every measured observation in the archive.
On a dense factor, an earth-mover-near observation whose force
residual exceeds the force tolerance replaces the nearest
representative, and the row count is unchanged. The count of
consecutive near probes resets only when the new force is lower.
An exact duplicate is unchanged. An HS-SVD factor appends that
observation. When a training-row ceiling is set, recent acquisitions and each
image's nearest row are kept. Among the rest, the row whose removal
changes the posterior mean least in the kernel norm is dropped, and
the norms are evaluated again before the next removal. The
reproducing property bounds the change at every geometry by
$\sqrt{k(x,x)}$ times that norm, so a distant row that still moves
the posterior is kept. With no fitted model the norm is undefined
and the rows farthest from the band are dropped. The dropped row
remains in the training set. The default ceiling is unset.
\subsection{All-image oracle}
\label{sec:si-aie}
After the inner band relaxes on $\mu$, the all-image loop
evaluates every intermediate on the true potential,
\begin{equation}
\mathcal{D}_{k+1}
  \;=\;
  \mathcal{D}_{k}
  \;\cup\;
  \bigl\{
    \bigl(x_i^{(k+1)},\,V(x_i^{(k+1)}),\,\nabla V(x_i^{(k+1)})\bigr)
  \bigr\}_{i=1}^{N-2} ,
\label{eq:camp-aie}
\end{equation}
and retrains. A one-image loop adds a single $x_{i^{\star}}$
chosen by posterior variance or IMSPE. If that $x_{i^{\star}}$
misses a ridge that exists only in $\mu$, the next inner
relaxation sits on the ridge and the single oracle never
contradicts it. Equation \eqref{eq:camp-aie} denies that
sit because every intermediate is measured.
\subsection{L-BFGS on the inner band}
\label{sec:si-lbfgs}
The inner stepper acts on the concatenated band
coordinates. The NEB force on an ordinary image is
\begin{equation}
F_i^{\mathrm{NEB}}
  \;=\;
  -\nabla V(x_i)
  \;+\;
  \bigl(\nabla V(x_i)\cdot\hat{\tau}_i\bigr)\hat{\tau}_i
  \;+\;
  F_i^{\mathrm{s}} .
\label{eq:camp-neb}
\end{equation}
That force is not $-\nabla$ of a scalar. Two limiting
cases fix the engine.

\paragraph{Potential gradient.}
When the force is $-\nabla V$, rgmin FIRE2 has a merit
function and is the eOn band default.

\paragraph{Band force.}
Equation \eqref{eq:camp-neb} includes a spring and a
tangent projection, so a line search on $V$ is not a
descent. The inner band steps with L-BFGS and clips the
step. Internal L-BFGS supplies a comparison implementation. rgmin FIRE is
opt-in. The eOn band engine is not constructed.
\subsection{Condition floor}
\label{sec:si-cond-floor}
The noisy condition number of the dual spectrum is
\begin{equation}
\kappa_{\mathrm{eff}}
  \;=\;
  \frac{S_{\max}^{2}+\sigma^{2}}{S_{\min}^{2}+\sigma^{2}} .
\label{eq:camp-kappa}
\end{equation}
Requiring $\kappa_{\mathrm{eff}}\le\kappa$ with
$\kappa=1/\sqrt{\varepsilon}$, where $\varepsilon$ is the unit roundoff,
and solving for the noise
gives the smallest lift that meets the bound,
\begin{equation}
\sigma^{2}_{\mathrm{floor}}
  \;=\;
  \frac{S_{\max}^{2}-\kappa\,S_{\min}^{2}}{\kappa-1} .
\label{eq:camp-floor}
\end{equation}
The effective noise is
$\max(\sigma^{2},\sigma^{2}_{\mathrm{floor}})$. If
$\sigma^{2}_{\mathrm{floor}}\le\sigma^{2}$, the fitted
nugget is left unchanged. The floor is a numerical
policy on the dual solve. It is not a kernel identity and
it is not a substitute for
Eq. \eqref{eq:camp-ls-prior}: a uniform lengthscale prior
would also move a healthy spectrum.
\section{The band as a Riemannian optimization problem}
\label{sec:si-riemannian}
Write an $N$-atom configuration as $x\in\mathbb{R}^{3N}$
and a band with $n$ movable images as
$X=(x_1,\dots,x_n)\in\mathcal{M}=\prod_{i=1}^{n}\mathcal{M}_i$.
In the classic Cartesian profile
$\mathcal{M}_i=\mathbb{R}^{3N}$. When the molecular
internal-coordinate chart is attached, $\mathcal{M}_i$ is
the image of the chart tangent basis
$N_i\in\mathbb{R}^{3N\times q}$ at the image anchor.
The tangent projector is $P_i=N_i N_i^{\top}$ when the
columns of $N_i$ are orthonormal. Otherwise
$P_i=N_i(N_i^{\top}N_i)^{-1}N_i^{\top}$.

The nudging step of NEB is itself a projection. With the
unit chord tangent $\tau_i$, the NEB perpendicular force is
$P_i^{\perp}F_i=(I-\tau_i\tau_i^{\top})F_i$. Composing the
two projectors gives the tangent space actually used by the
inner relaxation,
\begin{equation}
T_{X}\mathcal{M}
  \;=\;
  \bigoplus_i
  \operatorname{range}\bigl(P_i(I-\tau_i\tau_i^{\top})\bigr),
\qquad
\operatorname{proj}_X(V)
  \;=\;
  \bigl(P_i(I-\tau_i\tau_i^{\top})v_i\bigr)_{i}.
\label{eq:si-proj}
\end{equation}

\paragraph{Retraction.}
The retraction is the operation the string method already
performs after a step: move along the tangent vector and
re-space the images,
\begin{equation}
R_X(V)
  \;=\;
  \Pi_{\mathrm{eq}}(X+V),
\label{eq:si-retract}
\end{equation}
where $\Pi_{\mathrm{eq}}$ is the equal-arc-length
reparameterization. On the chart the addition is replaced by
the inverse-distance retraction already used for optimizer
steps. $R_X$ is a first-order retraction:
$R_X(0)=X$ and $\mathrm{d}R_X(0)=\mathrm{id}$ on
$T_X\mathcal{M}$. A normal step's arc-length correction is
second order when $P_i\tau_i=\tau_i$.

\paragraph{Merit and the climbing image.}
With the merit $f(X)=\sum_{i}E(x_i)$ the Riemannian
gradient is the projected true force with the sign of
physics,
$\operatorname{grad}f(X)=-\operatorname{proj}_X\bigl((F_i)_i\bigr)$.
A converged band, $\operatorname{grad}f=0$, is a
discretised minimum-energy path. The climbing image uses the
reflected-force expression cited in the main text. The
Riemannian analysis below concerns the projected band merit;
its local model uses the corresponding reflected quadratic.

\begin{theorem}[NEB projector is a Riemannian projector]\label{thm:si-neb-proj}
Let $P_i$ be an orthogonal projector and $\tau_i$ a unit vector with $P_i\tau_i=\tau_i$ (always so for $P_i=I$). Then $P_i(I-\tau_i\tau_i^{\mathsf T})$ is an orthogonal projector whose fixed space is $\operatorname{range}(P_i)\cap\tau_i^{\perp}$. Without $P_i\tau_i=\tau_i$ the product is a projector if and only if the two factors commute.
\end{theorem}

\begin{theorem}[Arc-length retraction is first order]\label{thm:si-retract}
$R_X(0)=X$ and $\mathrm{d}R_X(0)=\mathrm{id}$ on
$T_X\mathcal{M}$. This statement is not formalized: it rests on the
equal-arc-length reparameterization $\Pi_{\mathrm{eq}}$, which the Lean
library does not define.
\end{theorem}
\subsection{Riemannian trust region with truncated CG}
\label{sec:si-rtr}
Given $X_k$ and radius $\Delta_k$, minimize the model on
the tangent space,
\begin{equation}
m_k(\eta)
  \;=\;
  f(X_k)
  +\langle\operatorname{grad}f(X_k),\eta\rangle
  +\tfrac12\langle\operatorname{Hess}f(X_k)[\eta],\eta\rangle,
\qquad
\eta\in T_{X_k}\mathcal{M},
\quad
\lVert\eta\rVert\le\Delta_k,
\label{eq:si-model}
\end{equation}
with the Riemannian Hessian
$\operatorname{Hess}f(X)[\eta]
=\operatorname{proj}_X\bigl(\mathrm{D}\operatorname{grad}f(X)[\eta]\bigr)$.
Although each tangent space is linear, its projector \(P_X\) depends on the band.
Differentiating the projected gradient gives
\[
 \operatorname{Hess}f(X)[\eta]=P_XH\eta+P_X(\mathrm D P_X[\eta])\nabla f.
\]
The second term need not vanish. A frozen-projector approximation keeps only \(P_XH\eta\); it is not the exact Riemannian Hessian.
Finite differences of a projected band force also differentiate its changing tangent and spring terms, so they are not simply products with the block-diagonal energy Hessian.

\paragraph{Truncated CG.}
The subproblem is solved by truncated conjugate gradients
(Steihaug--Toint). Iterate CG on \eqref{eq:si-model}, and
stop at the trust boundary, at negative curvature
$\langle\operatorname{Hess}f[p],p\rangle\le 0$ (follow $p$
to the boundary), or at the inner tolerance
\[
\lVert r_j\rVert\le\lVert r_0\rVert\min(\lVert r_0\rVert^{\theta},\kappa).
\]
The candidate is $X^{+}=R_{X_k}(\eta_k)$ and the ratio
\begin{equation}
\rho_k
  \;=\;
  \frac{f(X_k)-f(X^{+})}{m_k(0)-m_k(\eta_k)}
\label{eq:si-rho}
\end{equation}
decides acceptance and the radius:
$\rho_k<\tfrac14\Rightarrow\Delta_{k+1}=\tfrac14\Delta_k$;
$\rho_k>\tfrac34$ and
$\lVert\eta_k\rVert=\Delta_k\Rightarrow
\Delta_{k+1}=\min(2\Delta_k,\bar\Delta)$;
otherwise $\Delta_{k+1}=\Delta_k$. The step is taken when
$\rho_k>\rho'$ with $\rho'\in(0,\tfrac14)$. These are the
constants of Absil, Baker and Gallivan
(Found.\ Comput.\ Math.\ 7, 303 (2007), Thm.~4.4).

The identification is what the paper's rho trust already
gestures at: it is the Euclidean special case
of~\eqref{eq:si-rho} with $\operatorname{proj}=I$ and
$R_X(V)=X+V$. The Riemannian statement makes the
chart-tangent and the nudging projection part of the
geometry instead of two separate post-processing steps.
The GP is the surrogate $f$; the true-PES oracle enters
only through the outer acquisition.
\section{A derived band-resolution certificate}
\label{sec:si-chord}
The climbing-image force criterion checks one image. A band that is kinked at the ridge can
satisfy it on a first-order saddle that is not on the
minimum-energy path. The optional band-resolution check uses the perpendicular force near the climbing image and the local path curvature to assess whether the image spacing resolves the ridge.

Let $\gamma(s)$ be the arc-length parametrised MEP, $C^3$
with curvature $\kappa$, and let three consecutive images
sit on it with spacing $h$. Write $c=\gamma(h)-\gamma(0)$
for the forward chord.

\paragraph{Chord misses the tangent.}
Taylor expansion of $\gamma$ about $s=0$ gives
\begin{equation}
\begin{aligned}
\angle\bigl(c,\gamma'(0)\bigr)
  &= \tfrac12\kappa h+\tfrac16\kappa' h^{2}+O(h^{3}), \\
\varphi
  &:= \angle\bigl(\gamma(h)-\gamma(0),\,\gamma(0)-\gamma(-h)\bigr)
   = \kappa h+O(h^{3}).
\end{aligned}
\label{eq:si-chord}
\end{equation}
The chord therefore misses the true tangent by half the
measured turning angle $\varphi$ up to $O(h^{2})$, and by
exactly $\varphi/2$ on a circular arc.

\paragraph{Force decomposition.}
An image on the MEP has a true force parallel to
$\gamma'$. For a circular arc, the chord tangent gives projections
$|F|\cos(\varphi/2)$ along the chord and
$|F|\sin(\varphi/2)$ across it. The chord-perpendicular
residual an on-path image must show is therefore
\begin{equation}
\lVert F_{\perp}^{\mathrm{chord}}\rVert
  \;=\;
  |F_{\parallel}^{\mathrm{chord}}|\,\tan(\varphi/2).
\label{eq:si-leak}
\end{equation}
The identity is written in terms of the along-chord
component so that an off-path image cannot enlarge its own
allowance.

\paragraph{Certificate.}
The optional band-resolution check requires both neighbors of the
climbing image to satisfy
\begin{equation}
\lVert F_{\perp}^{\mathrm{chord}}\rVert_{\infty}
  \;\le\;
  |F_{\parallel}^{\mathrm{chord}}|\,\tan(\varphi/2)
  +\varepsilon,
\label{eq:si-cert}
\end{equation}
with $\varepsilon$ the band's force tolerance. The circular-arc
identity motivates this check; for a general curve it does not
bound every higher-order discretization term. A violation flags
an image for further measurement or band refinement.

\paragraph{The leak can hide the violation.}
The leak in~\eqref{eq:si-leak} is itself the chord
discretisation error of the perpendicular force. A band
whose leak exceeds $\varepsilon$ cannot certify the
perpendicular force to $\varepsilon$ whatever residual it
shows, because the allowance in~\eqref{eq:si-cert} is then
set by the feature the certificate is meant to detect. At a
one-image spike the neighbors of the climbing image carry
a tangential force of several
$\mathrm{eV}\,\text{\AA}^{-1}$ and a turning angle near
$\pi/2$, and~\eqref{eq:si-cert} admits any perpendicular
force there. The stricter
certificate
\begin{equation}
|F_{\parallel}^{\mathrm{chord}}|\,\tan(\varphi/2)
  \;\le\;
  \varepsilon
\quad\text{and}\quad
\lVert F_{\perp}^{\mathrm{chord}}\rVert_{\infty}
  \;\le\;
  |F_{\parallel}^{\mathrm{chord}}|\,\tan(\varphi/2)
  +\varepsilon
\label{eq:si-cert2}
\end{equation}
at both neighbors is available but is not the default.
An eight-image reference band can carry the same neighbor
leak, so the bound asks for more resolution than that
reference, and refinements along a sharp ridge need not
move the barrier. The one-image spike the bound was written
for is refused by the midpoint condition of the next
paragraph instead.

\paragraph{Local order.}
Next to the saddle
$|F_{\parallel}|\simeq|\lambda_1|h$ and
$\varphi\simeq\kappa h$, so the leak is $O(h^{2})$ there.
One halving of the spacing restores the bound when it is
asked for.

\begin{theorem}[Chord misses the tangent by $\varphi/2$]\label{thm:si-chord-half}
For the circular arc $\gamma(\theta)=R(\sin\theta,1-\cos\theta)$ and $0<\varphi<\pi$, the angle between the chord $\gamma(\varphi)-\gamma(0)$ and $\gamma'(0)$ is $\varphi/2$.
\end{theorem}

\begin{theorem}[On-path leak is $|F_{\parallel}|\tan(\varphi/2)$]\label{thm:si-leak}
If $F\parallel\gamma'(0)$ and $0<\varphi<\pi$, then $\lVert F_{\perp}^{\mathrm{chord}}\rVert=|F_{\parallel}^{\mathrm{chord}}|\tan(\varphi/2)$, where the components are taken along and across the unit chord direction.
\end{theorem}
\subsection{Midpoint certificate}
\label{sec:si-midpoint}
A ridge can also hide between two images that both sit
below it, so a band whose highest image passes~\eqref{eq:si-cert}
need not bracket the barrier. The certificate therefore also
requires the climbing image to be at least as high as the
surrogate energy at the midpoint of every segment.

\paragraph{Corner cuts.}
The raw chord midpoint of two images on a curved path cuts
the corner of the valley and climbs its wall, so its
surrogate energy exceeds both images with no ridge between
them. The certificate therefore relaxes each midpoint on
the surrogate in the hyperplane perpendicular to the chord
(descent with an Armijo backtracking line search, no oracle
call, displacement bounded by one chord length) and
compares that valley-floor point with the climbing image.
A ridge the band straddles stays above the climbing image
after the relaxation; a corner cut drops to the floor. In the GPRidge 0.5.0 records the relaxed test holds without a refinement
on HCCH, CH\(_3\)O and H\(_2\)CNH: the highest relaxed midpoint lies
$0.013$, $0.062$ and $0.084~\mathrm{eV}$ below the climbing
image, and the searches take the same potential evaluations as without
the certificate (32, 31 and 33).

\paragraph{A refusal is measured.}
The relaxed midpoint of the worst segment costs one oracle
row: a true energy below the climbing image corrects the
surrogate and the certificate is re-run; a true energy
above it confirms the ridge and the refinement follows.
The same row verifies a hump the relaxation slid off. Both
cases are surrogate artefacts of unsampled space. At most
one row per segment.
\subsection{Equidistributing the leak}
\label{sec:si-equidistribution}
Refinement doubles the image count everywhere, but the leak
is a local error indicator, and the images can be moved
instead of multiplied. At image $j$ with mean adjacent
spacing $h_j$ write the leak as $\tfrac12 m_j h_j$ with the
monitor
\begin{equation}
m_j
  \;=\;
  \frac{2\,|F_{\parallel,j}^{\mathrm{chord}}|\,\tan(\varphi_j/2)}{h_j}
  \;\simeq\;
  |F_{\parallel,j}|\,\kappa_j,
\label{eq:si-monitor}
\end{equation}
the tangential force times the path curvature.

\paragraph{Equal leak per segment.}
Equidistribution asks every segment to carry the same leak,
$M_i\ell_i=\mathrm{const}$ with
$M_i=\max(m_i,m_{i+1})$, hence segment rest lengths
\begin{equation}
\ell_i
  \;=\;
  L\,\frac{1/M_i}{\sum_k 1/M_k},
\qquad
M_i\leftarrow\max\Bigl(M_i,\frac{2\varepsilon}{L}\Bigr),
\label{eq:si-rest}
\end{equation}
where $L$ is the band length and the floor is the monitor
below which a segment satisfies the leak bound even at the
whole length, so it is not resolvable further. The NEB
spring becomes
$k\bigl[(d_{i+1}-\ell_{i+1})-(d_i-\ell_i)\bigr]\tau_i$,
whose equilibrium is the rest lengths up to a common
offset. Every arc-length reparameterization places the
images at the cumulative fractions of $\ell$.

\paragraph{When it is applied.}
The equidistribution is applied once, on a converged band,
at the moment the certificate asks for more resolution.
The rest lengths of~\eqref{eq:si-rest} are measured on that
band, the image count is grown by the $O(h^{2})$ law,
$N\to\lceil N\sqrt{L_{\mathrm{leak},\max}/\varepsilon_F}\rceil$
with \(L_{\mathrm{leak},\max}\) the largest force leak and \(\varepsilon_F\) the force tolerance, both in eV/\AA{}.
The ratio is dimensionless. With at most a tripling, the band is resampled along its own
polyline at the cumulative rest fractions, and the refined
attempt holds that spacing through spring rest lengths.

Three other arrangements move the images for a structural
reason and are not the default. Refreshing the
equidistribution on every outer iteration moves the spring
targets during the relaxation, and the band does not settle.
Inserting images only next to the climbing image pins them
where the peak was while the ridge moves. Weighting the
springs by energy on the predicted profile crowds an
exothermic cell's images onto the reactant plateau, so a
plateau stationary point can be returned while the ridge
stays unresolved between two images. The monitor puts images
where the path turns under force. Where the path is straight
or the force vanishes the monitor is at its floor and the
images spread at equal arc length. Equal spacing can then
place most of the images on a flat approach and only a few
across the barrier.

\begin{theorem}[Monitor is leak per half spacing]\label{thm:si-monitor}
For spacing $h>0$,
\[
\frac{2|F_{\parallel}^{\mathrm{chord}}|\tan(\varphi/2)}{h}
=\frac{\lVert F_{\perp}^{\mathrm{chord}}\rVert}{h/2}.
\]
The monitor of Eq.~\eqref{eq:si-monitor} is the leak of
Eq.~\eqref{eq:si-leak} per half spacing.
\end{theorem}
\section{Index validation and the unmeasured complement}
\label{sec:si-index}
A stationary point of any index is a fixed point of the
climbing-image force, so force convergence and index validation are
separate. A nondegenerate index-one point requires one negative
physical Hessian eigenvalue and positive remaining eigenvalues.
The climbing-image check constructs an orthonormal Cartesian complement
$Q$ to the rigid motions and measures central differences of the oracle
gradient along every column of $Q$. Those columns, with the surrogate
Hessian as prior, enter a smoothed curvature spectrum. A column is
refined only when its own error hides the sign. A certified index
includes the stencil rows and leaves them unconditioned. A refused index
refits before the search continues. The dimer measures the axis and one
transverse direction.
The saved certificate Hessian is a prior only at the geometry it was
built for. A moved midpoint uses the surrogate spectrum at that
midpoint.

\paragraph{A negative Ritz value is a descent direction.}
The lowest Ritz value $\theta_1$ of any Krylov space is
the minimum of the Rayleigh quotient over that space,
hence $\lambda_1\le\theta_1$. A negative Ritz value
already proves a descent direction.

\paragraph{A measured negative direction leaves the complement open.}
Once a negative Ritz direction is known, residuals on its proper
Krylov subspace alone cannot rule out a second negative mode.
The unmeasured diagonal block can carry an additional negative
curvature while preserving the observed Hessian-vector products.
Thus acceptance requires information about that complement as well.
Every real complement curvature $c$ is compatible with the observed
products, including $c<0$, so the sign is measured before it counts.
The publication Baker calculations disable this inline climbing-image
check. Their reported index is determined by the separate
validator. The measured-span construction describes the optional
inline check and does not supply the publication Baker certificates.

\paragraph{Coverage and numerical resolution.}
The physical dimension is $d=3N-r$, where $r$ is the rank of the
rigid-motion basis: six for an isolated nonlinear molecule, five
for an isolated linear molecule. The climbing-image measurement
costs $2d$ potential evaluations. The dimer measurement is the axis and one
transverse direction. The decision resolution of each eigenvalue is
its posterior spread. Column information is the stencil's entrywise
error, including half the squared asymmetry of a doubly measured
entry. Here $A_h$ is the central-difference Hessian action at
separation $h$. Richardson extrapolation $(4A_{h/2}-A_h)/3$ cancels the
leading $O(h^2)$ error in the Hessian action on a refined column; the step difference
is an empirical truncation estimate. Neither estimate bounds all
symmetric truncation error.

The independent validator uses a mass-weighted Cartesian
force stencil and gives its empirical eigenvalue resolution
$\tau$. Acceptance requires one eigenvalue below $-\tau$ and all
other physical eigenvalues above $\tau$, together with the measured
maximum atomic force threshold. An unresolved mode prevents an
index-one claim regardless of its computed sign. Climbing-image
stencil rows are appended, and a certified index leaves them
unconditioned. Dimer stencil probes stay out of the training set.
An escape trial is a training row. The independent validation
stencil has a separate call count.

\paragraph{Escape.}
When the test fails on a resolved negative curvature, the image
leaves that point along the offending mode. On the band the image
takes one step toward the side the surrogate ranks lower. On the
dimer the escape is an order-one partitioned rational-function
search on the refit surrogate, and only the returned point is
measured. If that session is unavailable, the escape is a measured
walk from the first geometrically distinct downhill trial. The
displacement is doubled while the measured energy falls, and the
walk stops at the first rise. The measured recovery is the new trust center.
A weak second negative mode, refined below the stencil resolution or
within twice it, takes a measured partitioned rational-function search
on the oracle instead, starting from the Hessian of the complete
stencil. The mass-weighted model
maximizes along its lowest mode and minimizes along every other,
receives a Bofill update from each measured gradient, and has rigid
motion projected out at the current geometry. The Cartesian trust
radius is four dimer translation limits, halved when a step triples
the force. The search stops when the model has index one and the
measured force lies inside the tolerance. The probes stay out of the
training set; the end point is the next search's start and its only
new row.
An infinite measured upper bound skips the stationary-force veto.
A finite nonnegative bound keeps it. An unresolved sign supplies
neither an escape nor an index-one certificate. The band restart
is the fallback. The reflected force vanishes at a
stationary point of any index, which is why this certificate is
separate from force convergence.

\begin{theorem}[Lowest Ritz value bounds $\lambda_1$ from above]\label{thm:si-ritz-l1}
Let $T$ be a linear map on a finite-dimensional real inner product space, $\lambda_1$ the infimum of the Rayleigh quotient over nonzero vectors, and $\theta_1$ the infimum over the nonzero vectors of a nonzero subspace (a Krylov space for $PHP$). Then $\lambda_1\le\theta_1$. For symmetric $T$, $\lambda_1$ is an eigenvalue of $T$.
\end{theorem}

\begin{theorem}[Partial residuals do not exclude a second negative mode]\label{thm:si-no-partial-l2}
Let $A$ be symmetric and let $Q\in\mathbb R^{d\times m}$ with $m<d$ span a proper subspace. For every real $c$ there is a symmetric $B$ with $BQ=AQ$ (so $Q^{\top}BQ=Q^{\top}AQ$ and the residuals on the subspace coincide) and a unit vector $\mathbf v$ with $\mathbf v^{\top}B\mathbf v=c$. In particular \(c<0\) is possible, proving that an unmeasured negative direction cannot be excluded.
This establishes a negative Rayleigh quotient and at least one negative eigenvalue; it does not by itself establish a second negative eigenvalue in addition to a measured one.
\end{theorem}

If the measured subspace already contains a unit vector $u$ with $a=u^\top A u<0$, choose a unit $v$ orthogonal to that subspace.
The cross term $b=u^\top A v$ is fixed by the measured products.
A change in the unmeasured $vv^\top$ direction can set $c=v^\top Bv<b^2/a$ without changing $BQ=AQ$.
Then the restriction $\left(\begin{smallmatrix}a&b\\b&c\end{smallmatrix}\right)$ has negative first diagonal entry and positive determinant, hence two negative eigenvalues.
The min--max principle gives at least two negative eigenvalues of $B$.
Thus partial products cannot certify that the full Hessian has only one negative mode.

\begin{theorem}[Complete orthogonal reduction preserves index]\label{thm:si-index-exhaust}
Let $A\in\mathbb R^{d\times d}$ be symmetric and let
$U\in\mathbb R^{d\times d}$ satisfy $U^{\top}U=I$.
Then $T=U^{\top}AU$ has the same eigenvalues and inertia as $A$.
\end{theorem}
\begin{proof}
For every scalar $t$,
$\det(tI-T)=\det(U^{\top}(tI-A)U)=\det(tI-A)$,
since $\det(U)^2=1$. The characteristic polynomials agree.
\end{proof}
For $A=\operatorname{diag}(-2,-1,3)$,
Lanczos started at the first coordinate vector terminates with zero
residual after one direction and misses the second negative mode.
A complete basis recovers both negative eigenvalues. The
SymPy checks verify this example and an orthogonal change of basis.
They also exhibit a symmetric central-difference stencil whose
curvature sign differs from the exact Hessian, so a complete measured
basis still leaves finite-difference error. The example is an identity
of a complete orthogonal reduction. The climbing-image certificate
measures every physical column before the smoothed spectrum is formed.
The dimer certificate does not take the complement sign from that
spectrum, since every real complement curvature remains compatible
with the measured products.
\section{Band stepper ablation: ODE12r, the exponential metric and rigid-motion removal}
\label{sec:si-band-stepper}
The surrogate band of every reported molecular calculation is relaxed by L-BFGS on the posterior mean, with the per-atom step limit of \(0.2\) \AA{} and the eOn reset rules. Two other steppers and one coordinate policy were run on the same Baker and Birkholz settings, one repetition each, against a control at the same source revision, to decide whether any of them lowers the number of potential evaluations or total elapsed time at equal acceptance. The integrators and their control use source \texttt{2e7f80b17}; the rigid-motion arm and its control use \texttt{14c536a7e}. Both controls reproduce the 0.4.0 counts, 1041 Baker and 1858 Birkholz evaluations, rather than the headline 0.5.0 counts, 1215 and 1846.
\subsection{The arms}
\label{sec:orgd4c082c}

\begin{description}
\item[{ODE12r}] the adaptive first-order integrator of Makri, Ortner and Kermode \cite{makriPreconditioningSchemeMinimum2019} on the projected band force. A trial is one forward Euler step \(x' = x + h\,s\) with \(s = P^{-1}F\); the local error against the embedded second-order solution, \(e = \tfrac{h}{2}\lVert s - s'\rVert_\infty\), accepts the trial when the residual \(\lVert F\rVert_\infty\) falls by at least a factor \(1 - 10^{-2}h\), or stays within twice its value while \(e\) is below the tolerance \(0.1\), and sets the next \(h\) from the smaller of the error estimate \(\tfrac{h}{2}\sqrt{0.1/e}\) and the length at which the linear interpolant of the force is shortest in the \(P\)-norm, at most four times the last. A rejected trial keeps the band and shrinks \(h\) to between a tenth and a quarter. Every trial, kept or not, is one band evaluation on the surrogate. The integrator lives in the rgmin optimizer library; the band seam in rgsaddle keeps the evaluation at the start of a rejected step in a second cache slot so that the rejection costs one surrogate evaluation.
\item[{ODE12r-Exp}] ODE12r with \(P\) the exponential metric of Packwood and co-workers \cite{packwoodUniversalPreconditionerSimulating2016}, built per image from the pairs within twice the nearest-neighbour distance, with the exponent \(A = 3\), the diagonal stabilization \(0.1\) and the energy scale estimated once from the band force. The matrix is rebuilt when an atom has moved half a nearest-neighbour distance. In the form of the Makri paper the metric acts inside the band projection: the image is driven by \(-(P^{-1}g - t_P\,(t_P\cdot g))\) plus the spring along the tangent \(t_P\) normalized in the \(P\)-norm, the band relaxes to the minimum energy path of the metric's geometry, and convergence is measured on the residual \(\lVert g - P t_P (t_P\cdot g)\rVert_\infty\) of that path, on which the Euclidean perpendicular force does not vanish.
\item[{Rigid-motion removal}] L-BFGS as in production, with the molecular rigid-motion policy on: the band step of each image is projected off the six rigid generators and each trial image is aligned by the Kabsch rotation to its own previous geometry. The reported molecular searches leave this policy off; this arm measures its effect explicitly.
\end{description}

On a quadratic surface whose Hessian equals \(P\) the force interpolant vanishes at the Newton length, so once the error control admits it the integration takes Newton steps. The implementation reproduces Table II of the Makri paper on its 107-atom copper vacancy under a Morse potential with five images: 8.67 and 26.67 force evaluations per image to residuals of \(10^{-1}\) and \(10^{-3}\) eV/\AA{} without the metric against the published 8 and 27, and 7.33 and 19.33 with it against 8 and 19, at a spring constant of 1 eV/\AA{}\(^2\) read as the paper's dimensionless \(\kappa = k/\mu\). The integrator rows of the table below were measured with this form. An earlier form of the stepper, whose error estimate scaled the force difference by the coordinate magnitude and whose metric acted on the stepper as \(P^{-1}\) on the Euclidean projected force, costs 12.67 and 40.67 evaluations per image on the same copper vacancy and converged more of the panel cases (18 and 19 of 24 on Baker, 8 and 14 of 20 on Birkholz) at a comparable evaluation cost; neither form approaches the control.
\subsection{Outcome}
\label{sec:orgea0b233}

\begin{table}[tbp]
\caption{\label{tab:band-stepper-ablation}Band-stepper ablation on the Baker and Birkholz settings, one repetition each: cases that end on a first-order saddle, search evaluations of energy and force on the supplied potential summed over the cases, and total elapsed time including initialization, search and independent saddle validation. The controls reproduce one of the three passes of release 0.4.0. The integrator rows use source \texttt{2e7f80b17}; rigid-motion rows use \texttt{14c536a7e}. Each has its own matched control.}
\centering
\begin{tabular}{llrrr}
\toprule
Set & Band stepper & Passed & Calls & Wall (s)\\
\midrule
Baker & rg-lbfgs (production) & 24/24 & 1041 & 158\\
Baker & rgsaddle ODE12r & 14/24 & 10086 & 3337\\
Baker & rgsaddle ODE12r-Exp & 16/24 & 14180 & 7895\\
Baker & rg-lbfgs, rigid-motion removal & 23/24 & 998 & 157\\
Birkholz & rg-lbfgs (production) & 20/20 & 1858 & 555\\
Birkholz & rgsaddle ODE12r & 3/20 & 14150 & 16168\\
Birkholz & rgsaddle ODE12r-Exp & 8/20 & 17948 & 12155\\
Birkholz & rg-lbfgs, rigid-motion removal & 20/20 & 2735 & 963\\
\bottomrule
\end{tabular}
\end{table}

The calls are the search evaluations of energy and force on the supplied potential summed over the cases of a panel. Wall time sums the pipeline times present in the retained records, including initialization, search and independent validation. The native searches use the CPU implementation; allocation on a GPU node does not make these GPU measurements. A case passes when its converged image is a first-order saddle under the index test. The control reproduces one of the three passes of GPRidge 0.4.0, 1041 and 1858 evaluations, at both revisions. Neither integrator lowers the cost. The band relaxation is a surrogate step, so a stepper that needs many more surrogate steps for the same inner tolerance reaches the inner iteration limit more often, triggers more outer iterations and therefore more potential evaluations: on the Baker settings ODE12r needs 10086 evaluations for the 14 cases it converges, and ten cases exhaust their separate 1800 s job limits without completed cost records. Those missing costs are excluded from the displayed sum, so both its evaluation and wall-time sums are lower bounds; the metric recovers two of them at 14180 evaluations. On the Birkholz settings 17 of the 20 ODE12r searches and 12 of the 20 ODE12r-Exp searches run into the 900 s case budget, and the evaluations of the converged cases alone exceed the control's total. The metric's fixed point is also not the Euclidean path: the band relaxes to the minimum energy path of the metric's geometry, and the climbing image and the acquisition that follow it expect the Euclidean one. The rigid-motion arm saves 43 evaluations on the Baker panel but force-converges the HCONH\(_3^+\) cation on a point with two imaginary modes, and on the Birkholz panel it costs 877 evaluations more than the control, 1446 against 582 on the Grignard case. The production L-BFGS band is kept for every reported calculation, and the three arms stay available through \path{--neb-band-engine rg-saddle-ode12r}, \path{--neb-band-engine rg-saddle-ode12r-exp} and \path{--molecular-rigid-motion}. \section{Covariance algebra on an internal-coordinate chart}
\label{sec:si-operator-algebra}
The inverse-distance kernel of the main text is a squared
exponential on pair features $q_p=1/r_p$. The reported
calculations use that kernel on Cartesian inverse-distance features.
When an internal chart is attached, that
same squared exponential is written on the chart coordinates
and pulled back to Cartesian energy and force. This section
derives the chart, the joint energy--force tensor, the
hyperparameter derivatives that commute with the pull-back,
and the marginal-likelihood gradient the charted kernel is
fitted on. The identities assume a differentiable chart.
They are not a test of the inverse-distance program used for
the 44 reactions. Lean statements of the discrete identities
sit in Section \ref{sec:lean-identities}.
\subsection{Chart}
\label{sec:si-chart}
Let $\xi:\mathbb{R}^{3N}\to\mathbb{R}^{p}$ collect the
primitive internals of a topology: bond lengths, valence
angles, dihedrals, and the linear bends of
Section~\ref{sec:si-linear-bends}. Write
$B(x)=\partial\xi/\partial x$ for the Wilson $B$-matrix.
At an anchor $x_0$ the thin singular-value decomposition
$B(x_0)=U\Sigma V^{\mathsf T}$ has numerical rank $m$. The
chart is the $m$ leading left singular directions, scaled
by the kept singular values,
\begin{equation}
q(x)
  \;=\;
  \mathsf N\bigl(\xi(x)-\xi(x_0)\bigr),
\qquad
\mathsf N
  \;=\;
  \Sigma_m^{-1}U_m^{\mathsf T}
  \in\mathbb{R}^{m\times p}.
\label{eq:si-chart}
\end{equation}
Dihedral differences in $\xi(x)-\xi(x_0)$ are taken modulo
$2\pi$. Differentiation is local to a smooth branch of these differences,
with the primitive topology, anchor, kept rank and $\mathsf N$ fixed.
The chart Jacobian is $J(x)=\partial q/\partial x=\mathsf N B(x)$.
Changing a torsion branch or rebuilding the chart is a separate operation.

\paragraph{Anchor Jacobian.}
At $x_0$ one has $B(x_0)=U\Sigma V^{\mathsf T}$, with $U_m$, $\Sigma_m$ and $V_m$ the factors kept at rank $m$.
$U_m^{\mathsf T}$ drops every left singular vector past that block, so
\begin{equation}
J(x_0)
  \;=\;
  \Sigma_m^{-1}U_m^{\mathsf T}\,U\Sigma V^{\mathsf T}
  \;=\;
  V_m^{\mathsf T}.
\label{eq:si-anchor-J}
\end{equation}
The rows of $J(x_0)$ are orthonormal and span the internal
tangent space. Rigid translations and rotations lie in its
null space.

\paragraph{Rank.}
A non-linear molecule has $m=3N-6$ internal degrees of
freedom. A linear molecule has $m=3N-5$. Those are the
ranks the singular-value cut includes when the primitives
are complete.

\begin{theorem}[Anchor Jacobian is $V_m^{\mathsf T}$]\label{thm:si-anchor-J}
At the chart anchor, $J(x_0)=V_m^{\mathsf T}$.
\end{theorem}

\begin{theorem}[Chart rows are orthonormal at the anchor]\label{thm:si-anchor-ortho}
$J(x_0)J(x_0)^{\mathsf T}=I_m$.
\end{theorem}
\subsection{Pull-back and the joint energy--gradient tensor}
\label{sec:si-pullback}
The primitive covariance is the squared exponential on
chart coordinates, with $s^2>0$ and $\ell>0$,
\begin{equation}
k(q,q')
  \;=\;
  s^{2}\exp\!\Bigl(-\tfrac{|q-q'|^{2}}{2\ell^{2}}\Bigr),
\qquad
a\equiv\ell^{-2},
\quad
d\equiv q-q',
\quad
r^{2}\equiv|d|^{2}.
\label{eq:si-se}
\end{equation}
The pull-back operator $\mathcal P$ composes the kernel
with the chart,
$(\mathcal P k)(x,x')=k\bigl(q(x),q(x')\bigr)$. Energy and
Cartesian gradient are observed jointly. The observation
vector at $x$ is $[E(x);\nabla_x E(x)]$, so the covariance
is the image of $\mathcal P k$ under the linear
differential operator
$\mathcal D_x=[\,1;\,\nabla_x\,]$ on both arguments.
The chain rule gives $\nabla_x=J(x)^{\mathsf T}\nabla_q$ when
acting on a function composed with the fixed chart,
\begin{equation}
\mathcal D_x\,\mathcal D_{x'}^{\mathsf T}\,(\mathcal P k)
  \;=\;
  \begin{pmatrix}
    k &
    \bigl(J(x')^{\mathsf T}\nabla_{q'}k\bigr)^{\mathsf T}\\[2pt]
    J(x)^{\mathsf T}\nabla_{q}k &
    J(x)^{\mathsf T}\bigl(\nabla_q\nabla_{q'}^{\mathsf T}k\bigr)J(x')
  \end{pmatrix}.
\label{eq:si-joint}
\end{equation}

\paragraph{Squared-exponential derivatives.}
Differentiating~\eqref{eq:si-se} gives the closed forms
\begin{equation}
\nabla_q k=-k\,a\,d,
\qquad
\nabla_{q'}k=+k\,a\,d,
\qquad
\nabla_q\nabla_{q'}^{\mathsf T}k
  =k\bigl(a\,I-a^{2}\,d\,d^{\mathsf T}\bigr).
\label{eq:si-se-derivs}
\end{equation}
The energy--energy block of~\eqref{eq:si-joint} is $k$
itself. The displayed mixed blocks use gradients. For forces
$F=-\nabla_x E$, negate each energy--gradient block; the
gradient--gradient block keeps its sign because both arguments change
sign. Equivalently, the force Gram is $S K S^{\mathsf T}$ with
$S=\operatorname{diag}(1,-I_{3N})$ at each geometry. Joint
energy--gradient observations use every block of~\eqref{eq:si-joint}.
An energy-only assembly is never formed.

\paragraph{Linearity.}
Sum, projection, and Kronecker act on joint blocks by
linearity of $\mathcal D_x\mathcal D_{x'}^{\mathsf T}$.
A charted production run uses the pull-back alone.

\begin{theorem}[Energy-gradient block is the pull-back under D]\label{thm:si-joint}
If $q$ has Jacobian $J$ and $k(a,\cdot)$ has gradient $G(a,b)$ at $b$, then $x'\mapsto k(q(x),q(x'))$ has derivative $G(q(x),q(x'))\circ J(x')$, i.e.\ the energy--gradient block is $J(x')^{\mathsf T}\nabla_{q'}k$.
\end{theorem}

\begin{theorem}[Force-force block is the Jacobian sandwich]\label{thm:si-ff}
If $G(\cdot,b)$ has derivative $H(a,b)$ at $a$, the derivative in $x$ of $G(q(x),q(x'))\,J(x')u'$, evaluated on $u$, is
\[
H(q(x),q(x'))\bigl(J(x)u,J(x')u'\bigr).
\]
Thus the force--force block is
\[
J(x)^{\mathsf T}(\nabla_q\nabla_{q'}^{\mathsf T}k)J(x').
\]
\end{theorem}
\subsection{Hyperparameter derivatives}
\label{sec:si-dtheta}
Write $\theta=(\log s^{2},\log\ell)$. The operator
$\mathcal D_x\mathcal D_{x'}^{\mathsf T}$ does not depend
on $\theta$. The squared exponential and the fixed smooth chart
have continuous mixed derivatives, so observation and hyperparameter
differentiation commute,
\begin{equation}
\partial_{\theta}
\bigl(\mathcal D_x\mathcal D_{x'}^{\mathsf T}\mathcal P k\bigr)
  \;=\;
  \mathcal D_x\mathcal D_{x'}^{\mathsf T}\mathcal P\,
  (\partial_{\theta}k).
\label{eq:si-commute}
\end{equation}

\paragraph{Magnitude.}
The block is linear in $s^{2}$. Therefore
$\partial_{\log s^{2}}$ of the joint tensor is the tensor
itself.

\paragraph{Length scale.}
The identities $\partial_{\log\ell}k=k\,a\,r^{2}$ and
$\partial_{\log\ell}a=-2a$ give
\begin{align}
\partial_{\log\ell}\,k
  &= k\,a\,r^{2},
\label{eq:si-dlogl-k}\\
\partial_{\log\ell}\,\nabla_q k
  &= -\,k\,a\,d\,(a r^{2}-2),
\label{eq:si-dlogl-g}\\
\partial_{\log\ell}\,\nabla_q\nabla_{q'}^{\mathsf T}k
  &= k\Bigl[
       a\,I\,(a r^{2}-2)
       -a^{2}\,d\,d^{\mathsf T}(a r^{2}-4)
     \Bigr].
\label{eq:si-dlogl-h}
\end{align}
The joint blocks of $\partial_{\log\ell}K$ are the same
Jacobian sandwiches as in~\eqref{eq:si-joint}, with
$k$ replaced by~\eqref{eq:si-dlogl-k}--\eqref{eq:si-dlogl-h}.

\begin{theorem}[Hyperparameter derivatives commute with $\mathcal D$]\label{thm:si-commute}
Eq. \eqref{eq:si-commute} holds for each component of
$\theta$.
\end{theorem}

\begin{theorem}[Log-magnitude derivative is the tensor]\label{thm:si-dlogs}
$\partial_{\log s^{2}}$ of~\eqref{eq:si-joint} equals
\eqref{eq:si-joint}.
\end{theorem}
\subsection{Marginal-likelihood gradient}
\label{sec:si-nlml}
Let $C(\theta)=K(\theta)+\Sigma_{\mathrm{noise}}$ be the
positive definite Gram matrix of the joint residual observations $y$,
after subtracting a fixed prior mean, and write $\alpha=C^{-1}y$.
The residual and observation operators are held fixed during
hyperparameter differentiation. The negative log marginal likelihood is
\begin{equation}
\mathcal L
  \;=\;
  \tfrac12 y^{\mathsf T}C^{-1}y
  +\tfrac12\log|C|
  +\tfrac n2\log 2\pi.
\label{eq:si-nlml}
\end{equation}
Differentiating with respect to a packed parameter
$\theta_j$ and using
$\partial_{\theta}C^{-1}=-C^{-1}(\partial_{\theta}C)C^{-1}$
gives
\begin{equation}
\frac{\partial\mathcal L}{\partial\theta_j}
  \;=\;
  \tfrac12\operatorname{tr}\!\bigl(C^{-1}\partial_{\theta_j}C\bigr)
  -\tfrac12\,\alpha^{\mathsf T}(\partial_{\theta_j}C)\,\alpha.
\label{eq:si-nlml-grad}
\end{equation}

\paragraph{Charted fit.}
Under a chart the stable-basis (HS-SVD) representation is
not used, so~\eqref{eq:si-nlml-grad} is evaluated on the
dense whole-rank factor with the derivative tensors of the
previous subsection. Leave-one-out objectives read the
pair-distance geometry of the ambient kernel and are not
defined for a composite kernel. When one is configured, the
fit uses~\eqref{eq:si-nlml-grad} for the duration of the
call and restores the caller's objective.
The fitting objective adds $-\log\pi(\theta)$ for the configured
hyperparameter prior $\pi$; its gradient therefore also includes
$-\nabla_\theta\log\pi$. The likelihood formula does not select the
optimizer: SCG and the explicitly configured alternatives act on
the chosen objective.

\begin{theorem}[NLML gradient]\label{thm:si-nlml-grad}
Eq. \eqref{eq:si-nlml-grad} is the derivative of
\eqref{eq:si-nlml}. This statement is not formalized: the Lean library
does not contain the derivative of the matrix inverse and of
$\log\det$ along a hyperparameter path.
\end{theorem}
\subsection{Second-order tensors and the exact directional curvature}
\label{sec:si-nlml-curv}
The scaled-conjugate-gradient driver takes the curvature of
$\mathcal L$ along a search direction
$p=(p_u,p_v)$ in $(\log s^{2},\log\ell)$ exactly when the
kernel can supply it. Write $z=a r^{2}$, distinct from the signal scale $s$. The log-magnitude
second derivative of every block is the block itself, the
mixed derivative is the log-length first derivative, and
the second log-length derivatives are
\begin{align}
\partial^{2}_{\log\ell}\,k
  &= k\,z\,(z-2),
\label{eq:si-d2logl-k}\\
\partial^{2}_{\log\ell}\,\nabla_q k
  &= -\,k\,a\,d\,(z^{2}-6z+4),
\label{eq:si-d2logl-g}\\
\partial^{2}_{\log\ell}\,\nabla_q\nabla_{q'}^{\mathsf T}k
  &= k\Bigl[
       a\,I\,(z^{2}-6z+4)
       -a^{2}\,d\,d^{\mathsf T}(z^{2}-10z+16)
     \Bigr].
\label{eq:si-d2logl-h}
\end{align}
With fixed likelihood noise, $C_p=p_u K+p_v\partial_{\log\ell}K$ and
$C_{pp}=p_u^{2}K+2p_up_v\,\partial_{\log\ell}K
+p_v^{2}\,\partial^{2}_{\log\ell}K$,
\begin{equation}
\frac{d^{2}\mathcal L}{dp^{2}}
  \;=\;
  \tfrac12\operatorname{tr}(C^{-1}C_{pp})
  -\tfrac12\operatorname{tr}(C^{-1}C_p C^{-1}C_p)
  -\tfrac12\,\alpha^{\mathsf T}C_{pp}\alpha
  +\alpha^{\mathsf T}C_p C^{-1}C_p\alpha.
\label{eq:si-nlml-curv}
\end{equation}
Equation~\eqref{eq:si-nlml-curv} is the likelihood contribution.
The posterior adds the prior curvature $-p^{\mathsf T}\nabla^2_\theta\log\pi(\theta)p$. Its objective, gradient and curvature must use the same prior.

\begin{theorem}[Directional curvature of the NLML]\label{thm:si-nlml-curv}
Eq. \eqref{eq:si-nlml-curv} is the second derivative of
\eqref{eq:si-nlml} along $p$. This statement is not formalized, for the
reason given for Eq. \eqref{eq:si-nlml-grad}.
\end{theorem}
\section{Atom types and identifiable length scales}
\label{sec:si-types}
The Sexpat kernel carries one length scale per unordered
pair of atom types. With the types equal to the elements, a
transferring hydrogen shares its C--H length scale with
every spectator C--H bond, whose inverse distance barely
moves along the path. One scale then serves both regimes.

\paragraph{Scales are fixed by structure.}
The number of scales is settled before the first fit, from
structure, not from the optimizer.

\paragraph{Bond-graph split.}
The covalent bond graphs of reactant and product (Cordero
radii, minimum image, the conventional
$1.25\,(R_a+R_b)$ tolerance) mark every atom whose bond set
changes as reactive. Each such atom takes its element's
reactive type. The Birkholz--Schlegel type files use fixed labels
generated from the reactant and product with the combined
normalized k-nearest-neighbour radial and SOAP descriptors.
The radial block uses a $4~\text{\AA}$ cutoff and four neighbors
per species. SOAP uses a Gaussian density width of
$0.3~\text{\AA}$, $n_{\max}=5$, and $l_{\max}=4$. The generator
clusters descriptor distances by the largest gap within each
element and reaction-role group. Each resulting type starts
tied to its bond-graph base type. The reported fits release no
finer length scale. The descriptor sets the fixed
labels offline. The GP covariance uses inverse pair distances;
SOAP contributes only to the preprocessing of the labels.

\paragraph{Stationarity of the fitted objective.}
The refit is skipped when the fitted objective is stationary at
the current $\theta$. That test reads the gradient of the objective
that was fitted: the leave-one-out gradient when the fit is on the
leave-one-out objective, not the marginal-likelihood gradient.
\section{Noise floor and truncation of the stable basis}
\label{sec:si-chop}
The stable basis diagonalises the noise-free Gram matrix,
$K=\Phi\,\mathrm{diag}(s_i^{2})\,\Phi^{\mathsf T}$ with
orthonormal $\Phi$. Every posterior formula uses
\begin{equation}
(K+\sigma^{2}I)^{-1}
  \;=\;
  \Phi\,\mathrm{diag}\Bigl(\tfrac{1}{s_i^{2}+\sigma^{2}}\Bigr)
  \Phi^{\mathsf T}
  +\frac{1}{\sigma^{2}}\bigl(I-\Phi\Phi^{\mathsf T}\bigr).
\label{eq:si-dual-inverse}
\end{equation}

\paragraph{Noise floor.}
The kept eigenvalues carry the backward error of a
symmetric eigensolver,
$|\delta s_i^{2}|\le c\,n\varepsilon\,\|K\|_2$, so
$s_i^{2}+\sigma^{2}$ is resolvable only when $\sigma^{2}$
(or $s_i^{2}$) exceeds $n\varepsilon\,s^{2}_{\max}$. The
backend therefore evaluates~\eqref{eq:si-dual-inverse}
with $\sigma^{2}_{\mathrm{eff}}=\max(\sigma^{2},f)$, where
$f$ is the floor at which the squared spectrum has
condition $1/(n\varepsilon)$,
\begin{equation}
\frac{s^{2}_{\max}+f}{s^{2}_{\min}+f}
  \;\le\;
  \frac{1}{n\varepsilon},
\qquad
f\approx n\varepsilon\,s^{2}_{\max}.
\label{eq:si-floor}
\end{equation}
A singular-value bound $1/\sqrt{\varepsilon}$ on the
eigenvalues floors $\sigma^{2}$ at
$\sqrt{\varepsilon}\,s^{2}_{\max}\approx
1.5\times 10^{-8}\,s^{2}_{\max}$ whenever the kernel has a
near-null direction.

\paragraph{Truncation.}
Dropping component $i$ moves its direction into the
complement term of~\eqref{eq:si-dual-inverse}. Its exact
weight $1/(s_i^{2}+\sigma^{2})$ lies in
$[1/(s^{2}_{\mathrm{drop}}+\sigma^{2}),\,1/\sigma^{2}]$
with $s^{2}_{\mathrm{drop}}$ the largest discarded
eigenvalue. Assigning a discarded direction to the complement uses the upper end, \(1/\sigma^2\).
Relative to its exact inverse weight, the increase is \(s_i^2/\sigma^2\), bounded by \(s_{\mathrm{drop}}^2/\sigma^2\). The plateau
test that also truncates the spectrum inspects the shape of
the $s_i$ sequence alone, so it is allowed to drop a
component only when $s_i^{2}\le\tau\sigma^{2}$ for the
target tolerance $\tau$. The complement is formed from
residual vectors $r=k_*-\Psi\Psi^{\mathsf T}k_*$, never
from $|k_*|^{2}-|\Psi^{\mathsf T}k_*|^{2}$: when $k_*$
lies in the kept range that difference is rounding, and
multiplied by $1/\sigma^{2}$ it produced negative local
posterior variances.

\begin{theorem}[Noisy inverse splits range and complement]\label{thm:si-noisy-inv}
Let $K=\Psi\,\mathrm{diag}(s_i^2)\Psi^{\mathsf T}$ with $\Psi^{\mathsf T}\Psi=I$, $\sigma^2\neq0$ and $s_i^2+\sigma^2\neq0$. Then $(K+\sigma^2I)\bigl[\Psi\,\mathrm{diag}\bigl((s_i^2+\sigma^2)^{-1}\bigr)\Psi^{\mathsf T}+\sigma^{-2}(I-\Psi\Psi^{\mathsf T})\bigr]=I$; $\Psi$ need not be square.
\end{theorem}
\section{Linear-bend chart for collinear triples}
\label{sec:si-linear-bends}
The valence angle $\theta$ of a triple $a$--$b$--$c$ with
$u=x_a-x_b$ and $v=x_c-x_b$ has
$\partial\theta/\partial x\propto 1/\sin\theta$. At
collinearity the angle is a cone, not a smooth coordinate,
and its Jacobian row is undefined.

\paragraph{Smooth pair.}
The same two-dimensional bending manifold is charted by the
pair of linear bends
\begin{equation}
\beta_n(x)
  \;=\;
  n\cdot(\hat u+\hat v),
\qquad
n\in\{n_1,n_2\},
\quad
n_1\perp n_2\perp\hat u,
\label{eq:si-linbend}
\end{equation}
with the reference direction \(n\) held fixed, the exact gradient is
$\partial\beta_n/\partial x_a=(I-\hat u\hat u^{\mathsf T})n/|u|$,
$\partial\beta_n/\partial x_c=(I-\hat v\hat v^{\mathsf T})n/|v|$
and
$\partial\beta_n/\partial x_b=-(\partial_a+\partial_c)\beta_n$,
smooth for every non-degenerate leg. The references are
$n_1=\hat u\times e_{\min}$ (normalized, $e_{\min}$ the
Cartesian axis least aligned with $\hat u$) and
$n_2=\hat u\times n_1$.

\paragraph{Which angles are collinear.}
The chart already cuts the $B$-matrix at
$\sqrt{\varepsilon}$ relative. The angle row carries
$1/\sin\theta$, so the angle is a resolvable coordinate if
and only if $1/\sin\theta\le 1/\sqrt{\varepsilon}$, i.e.\
$\sin^{2}\theta\ge\varepsilon$. Angles that fail this inequality are replaced by
their bend pair. Dihedrals whose inner triple is collinear
(undefined plane) are dropped. A path topology resolves
collinearity across every geometry of the path, because the
charts built from it are evaluated at all of them.

\paragraph{One-sided limit.}
An angle that stays bent on the path can still be met
exactly collinear at a query. The topology gives the
bending-plane normal $n_0$ of the first bent geometry, and
the Jacobian at the cone point is the finite one-sided
limit in that plane,
$\partial\theta/\partial x_a=-p/|u|$,
$\partial\theta/\partial x_c=-p/|v|$ with
$p=n_0\times\hat u$, continuous with the analytic gradient
on the bent side. For linear HCCH the resulting chart has
$3N-5=7$ tangent dimensions.

\begin{theorem}[Linear bend is smooth off a degenerate leg]\label{thm:si-linbend}
For a reference direction $n$, the coordinate $\beta_n(u,v)=\langle n,\hat u+\hat v\rangle$ is $C^k$ for every $k$ at every pair of nonzero legs $u,v$.
\end{theorem}
\section{Minimum-image proximity graph}
\label{sec:si-mic-graph}
The chart topology comes from a cutoff-free proximity
graph, Gabriel (edge $ab$ iff no third atom lies in the
open ball with diameter $ab$) or relative neighborhood
(no third atom $c$ with $\max(|c-a|,|c-b|)<|a-b|$). Under
periodic boundaries every displacement, including the ball
centre and the lune test, is the minimum-image displacement
of the cell.

\paragraph{Prefix blockers.}
Let $d$ be the minimum-image displacement from $a$ to $b$
and $m=x_a+d/2$. Any atom $c$ that blocks $ab$ satisfies
$|c-a|<|d|$: for Gabriel,
$|c-a|\le|c-m|+|m-a|<|d|/2+|d|/2$; for the relative
neighborhood graph the condition contains $|c-a|<|d|$
directly. Hence with the neighbors of $a$ sorted by
minimum-image distance, the blockers of the candidate at
rank $r$ lie in the strictly closer prefix, and the
scan terminates at the first candidate at distance
$\ge|d|$.

No neighbor count or radius enters. A cutoff pair list or
a $k$-nearest walk would only bound the candidate set by a
$k$ or $r_c$ that the exact graph does not admit.

\begin{theorem}[Gabriel blockers lie in the closer prefix]\label{thm:si-gabriel}
If $c$ lies in the open ball with diameter $ab$ then $|c-a|<|b-a|$; if $\max(|c-a|,|c-b|)<|a-b|$ then also $|c-a|<|a-b|$.
\end{theorem}
\section{Factored derivative geometry}
\label{sec:si-factored-geometry}
The joint covariance is assembled from blocks indexed by
pairs of coordinate dimensions $(n,m)$. Each block needs,
per pair type, the derivatives of the pair-type distance
\begin{equation}
d_{\mathrm{pt}}(i,j)
  \;=\;
  \sum_{f\in\mathrm{pt}}\bigl(q_f(i)-q_f(j)\bigr)^{2}
\label{eq:si-dpt}
\end{equation}
between configurations $i$ and $j$. Cached as three
$n_{\mathrm{conf}}\times n_{\mathrm{conf}}$ fields per pair
type for every one of the $(1+D)(2+D)/2$ blocks, that
footprint is of order
$n_{\mathrm{conf}}^{2}D^{2}n_{\mathrm{pt}}$: $811~\mathrm{MB}$
at $52$ rows and $15~\mathrm{GB}$ at $225$ rows for
$\mathrm{H_2PO_4^-}$ (seven atoms and $D=21$ Cartesian derivative dimensions).

\paragraph{Separability.}
With $J_{f,n}=\partial q_f/\partial x_n$ the blocks
factor over the configuration pair,
\begin{align}
\partial_{x_n(i)} d_{\mathrm{pt}}
  &= 2c_n(i)-2\sum_f J_{f,n}(i)\,q_f(j),
\qquad
c_n=\textstyle\sum_f J_{f,n}q_f,
\label{eq:si-fact-i}\\
\partial_{x_m(j)} d_{\mathrm{pt}}
  &= 2c_m(j)-2\sum_f q_f(i)\,J_{f,m}(j),
\label{eq:si-fact-j}\\
\partial_{x_n(i)}\partial_{x_m(j)} d_{\mathrm{pt}}
  &= -2\sum_f J_{f,n}(i)\,J_{f,m}(j).
\label{eq:si-fact-ij}
\end{align}
Each factor is a matrix of rank at most $|\mathrm{pt}|+1$
over the configuration pair, so a block is three small
matrix products of $Q$
($n_{\mathrm{conf}}\times d_{\mathrm{feat}}$) and $J_n$.
The geometry is kept in this factored form whenever the
dense cache would not fit half of the available memory, at
$O(n_{\mathrm{conf}}\,d_{\mathrm{feat}}\,D)$ storage. The
factored and walked blocks agree to $10^{-10}$ in the unit
test.

\begin{theorem}[Pair-type mixed derivative is a Gram of $J$]\label{thm:si-factored}
Eq. \eqref{eq:si-fact-ij} is the mixed derivative of
\eqref{eq:si-dpt}.
\end{theorem}
\section{Arithmetic of the posterior contraction}
\label{sec:si-predict-arithmetic}
For the posterior coefficients \(\alpha_i\) and a computed covariance
column \(k_i(x)\), the prediction contains the scalar contraction
\begin{equation}
 p(x)=\sum_{i=1}^{n}\alpha_i k_i(x).
 \label{eq:predict-product}
\end{equation}
Cancellation in this sum can amplify rounding error even when every
operand is finite. The arithmetic controls described here change this
contraction. Matrix entries and posterior coefficients remain binary64.
The covariance, factorization, likelihood derivatives, and optimizer
use their own arithmetic. In particular, selecting a prediction
product cannot change the meaning of the sum and correction fields used
by the likelihood-gradient reductions.

The executables \texttt{gprn} and \texttt{gprd} accept \texttt{-{}-{}predict-sum} with values
\texttt{compensated}, \texttt{scheme1}, \texttt{scheme2}, and \texttt{kernel\_float}. Leaving it unset
selects \texttt{compensated}. The pair-kernel mean uses compensated summation of
rounded binary64 products in this mode. The general training product
also accumulates the multiplication residual from a fused multiply-add.
These are distinct recurrences: compensation of addition alone cannot
recover bits discarded by multiplication. Each prediction accumulator
captures its selected mode when constructed.
\subsection{Exact slice and residue products}
\label{sec:org0ad0004}

The \texttt{scheme1} product splits each operand into signed integer slices
with separate powers of two, following the error-free decomposition
principle of Ozaki et al. \cite{ozakiGeneralizationErrorFree2013}.
For a dot product of length \(n\leq 2^{31}-1\), the slice width is
\begin{equation}
 t=\left\lfloor\frac{52-\lceil\log_2\max(1,n)\rceil}{2}\right\rfloor.
\end{equation}
Each slice has integer coefficients of magnitude less than \(2^t\).
Consequently, the sum of the absolute slice products is less than
\(n2^{2t}\leq2^{52}\). Binary64 multiplication and addition represent each
integer slice contraction exactly. The powers of two remain separate
from these integer products, avoiding intermediate floating-point
underflow and overflow. A signed integer accumulator combines all slice
pairs and rounds only the final result to binary64.

The \texttt{scheme2} product uses integer residues and the Chinese remainder
theorem, the construction underlying Ozaki Scheme II
\cite{ozakiSchemeIIModular2026}. A finite binary64 operand is represented
as a signed integer significand times a power of two. Blocks contain at
most 128 products, and each exponent window spans at most 120 binary
positions. The magnitude of a reconstructed window coefficient \(C\)
therefore satisfies
\begin{equation}
 |C|\leq128(2^{53}-1)^2 2^{120}<2^{233}.
 \label{eq:predict-window-bound}
\end{equation}
The eight moduli are 998244353, 985661441, 975175681, 469762049,
167772161, 1004535809, 2013265921, and 1811939329. They are pairwise
coprime, and their product \(M\) exceeds \(2^{234}\). Thus the residues
identify a unique signed coefficient in \((-M/2,M/2)\), which contains
the interval in Equation \eqref{eq:predict-window-bound}.

Every window and block contributes to the same signed accumulator.
Its 70 words of 64 bits represent an integer times \(2^{-2372}\).
The absolute sum of at most \(2^{31}-1\) finite binary64 products is
less than \(2^{2079}\), below the accumulator's upper limit \(2^{2108}\).
The lower limit accommodates significand alignment within a window.
Storage for this accumulator is independent of the product length.
Rounding a window or a block separately would lose cancellation across
that boundary; the accumulator retains those bits through the full sum.

For accumulator magnitude \(A\neq0\), let \(b\) be the index of its highest
set bit and \(e\) its power-of-two scale. The final rounding discards
\begin{equation}
 q=\max(b-52,-1074-e,0)
\end{equation}
low bits. The rounding bit, the disjunction of all lower bits, and the
retained least significant bit implement rounding to nearest with ties
to even. Including the subnormal spacing \(2^{-1074}\) in this single
rounding avoids double rounding near underflow. Exact cancellation
returns zero; a final value outside the finite binary64 range overflows.
This construction concerns the product of the supplied binary64
operands. It does not recover error in a computed covariance entry or
in the fitted coefficients, and rank-local rounding followed by an MPI
reduction is not a single exact global contraction.
\subsection{Certifying the rounded product}
\label{sec:org90609ee}

The CPU slice and residue products first test whether a compensated
calculation determines the correctly rounded answer. For nonzero
operands in \([2^{-400},2^{400}]\) in magnitude, every product is a multiple
of \(2^{-904}\). Products and their fused multiply-add residuals are normal
or zero, and the partial sums remain finite for \(n\leq2^{31}-1\).
Wider exponent ranges use the full slice or residue calculation.

Each product is represented by its rounded value and exact residual.
The summation applies an exact two-sum transform to those terms
\cite{ogitaAccurateSumDot2005}, retaining a leading sum \(s\) and computed
correction \(c\). Let \(r_j\) be the residuals from the two-sum transforms,
\(T=\operatorname{fl}(\sum_j |r_j|)\), \(u=2^{-53}\), and
\(\gamma_k=ku/(1-ku)\). There are at most \(m=2n\) transformed additions.
The error in the accumulated correction satisfies
\begin{equation}
 \left|c-\sum_j r_j\right|
 \leq \frac{\gamma_{m-1}}{1-\gamma_{m-1}}T
 <4nuT,\qquad 1\leq n\leq2^{31}-1.
 \label{eq:predict-certificate-bound}
\end{equation}
The denominator accounts for rounding in the positive accumulation
that produces \(T\). The implementation rounds the bound outward. If all
residuals vanish, the leading sum already represents the exact product.

For \(z=\operatorname{fl}(s+c)\), a further two-sum transform recovers
\(\tau=(s+c)-z\) exactly. The interval \([\tau-B,\tau+B]\), with \(B\) the
outward bound from Equation \eqref{eq:predict-certificate-bound}, encloses
the exact product minus \(z\). The result is certified only when this
interval lies strictly between the midpoints to the two neighboring
binary64 numbers. A midpoint or overlapping interval invokes the full
exact calculation. The certificate changes the cost of deciding the
rounded result; it does not replace that result with a tolerance-based
approximation. The compensated default and device residue product keep
their respective arithmetic.
\subsection{Device arithmetic and verification}
\label{sec:org42e0678}

The \texttt{kernel\_float} option executes the residue product on a CUDA device
using vectors from the kernel\textsubscript{float} library. The product retains integer
residues and one accumulator across all 128-term chunks of a product.
This implementation
does not replace covariance storage with half or bfloat16 values.
A separate eight-term demonstration returned 17 in binary64, half, and
bfloat16 arithmetic; equality on that input establishes neither a
rounding-error bound nor a production speedup.

The GPU algorithm in Ozaki Scheme II uses symmetric residues modulo
integers no larger than 256, forms their products with INT8 Tensor Core
matrix multiplications, and reconstructs the scaled matrix product. Its exact integer
accumulation requires the contraction length to be below \(2^{17}\), or
the contraction to be blocked. Our eight larger moduli and exponent
windows instead implement an exact dot product of the stored operands.
The device kernel assigns one product to each thread block. Its first 32 threads form four partial sums for
each of the eight residue channels. For vector width \(w\), group
\(g\in\{0,1,2,3\}\) processes indices \(w(g+4q)+\ell\), where \(q\geq0\) and
\(0\leq\ell<w\), restricted to the current chunk. These sets partition
the chunk. For modulus \(m\), the channel result is
\begin{equation}
 r=\left(\sum_{g=0}^{3}r_g\right)\bmod m.
\end{equation}
Each partial residue lies in \([0,m)\), and reduction after each addition
keeps the intermediate below \(2m<2^{32}\). The partition therefore
preserves the integer residue. One thread reconstructs each window into
the product's accumulator, with one binary64 rounding after all chunks.

The energy and pair contributions for a query share a kernel launch.
Their operands and offsets occupy one upload buffer. Each caller thread
uses pinned host storage, device storage and a private CUDA stream.
The host call waits for that stream before publishing the results or
reusing the buffers. A device change releases storage on its owning
device before allocating on the selected device. The operands are
uploaded for each batch; that storage avoids repeated allocation.
This kernel uses integer vector arithmetic, without the Tensor Core
matrix products of the cited implementation. Transfers and launch costs
contribute to device time. Published throughput for large matrix products
therefore does not predict the cost of this posterior contraction.

\nolinkurl{proofs/lean/GprOptim/ProductAccumulator.lean} proves coprimality, the
window bound, signed reconstruction uniqueness, and accumulator
capacity over integers. It also proves the integer inequality underlying
Equation \eqref{eq:predict-certificate-bound}. \nolinkurl{scripts/derive_product_accumulator.py} checks
the constants and cancellation identities with SymPy.
\nolinkurl{scripts/verify_product_accumulator.sollya} checks rounding ties,
subnormal cases, and a product residual. These statements do not
constitute a formal verification of the compiled C++ or CUDA program.

The executable oracle, \nolinkurl{scripts/check_predict_products.py} with
\nolinkurl{proofs/predict_product_probe.cpp} as its shared-library interface, compares each of the three alternate products
against an exact rational sum rounded to binary64 on 42 individual products and a batch of 44 products per
mode. They include 129- and 257-term sums, cancellation across blocks,
exponent spreads of 1000, and products that overflow individually but
cancel to a subnormal result. The identities
\begin{align}
 2^{54}+1-2^{54}&=1,\\
 (1+2^{-52})(1-2^{-52})-1&=-2^{-104}
\end{align}
exercise distinct losses: rounding a partial sum and rounding a product.
A fitted pair-kernel test also exercises the selected contraction through
the GP prediction interface. The recorded campaign timings in the main text
belong to their source and runtime. They provide no measured
speedup for these alternate products.

The certificate was checked against 143 exact rational sums, including
midpoint and cancellation cases. It accepted 113; both exact backends
matched the rational result in all 143 cases. The device checks also
ran four concurrent batches. CUDA race checking reported no shared-memory
hazards for this test set.

The warp partition passes the rational comparisons at vector
widths 1, 2 and 4. The executable
\nolinkurl{scripts/check_cuda_product_workspace.py} checks a further
850 products per width with changing operands at fixed addresses,
growing and shrinking batches, empty products and four concurrent caller
threads. It also verifies that invalid requests leave output untouched.
CUDA memory and race checking report no errors for width 4. These checks
establish arithmetic and storage behavior; the saddle timings below
use their stated source revisions.

Table \ref{tab:predict-batch-latency} measures the host call cost of separate
and batched residue products on an RTX 5070. Each product has 128 terms.
The measurements include allocation, transfer, execution, and result
retrieval. Each method is warmed up, followed by seven repetitions in
alternating order within one allocation. The entries are the mean and
sample standard deviation. This measures the effect of
grouping products, and does not measure a complete saddle search.

\begin{table}[htbp]
\caption{\label{tab:predict-batch-latency}Host call time for exact residue products, in milliseconds.}
\centering
\begin{tabular}{rll}
Products & Separate calls (ms) & One batch (ms)\\
\hline
8 & \(3.960\pm0.006\) & \(0.510\pm0.004\)\\
32 & \(15.824\pm0.012\) & \(0.527\pm0.003\)\\
256 & \(126.507\pm0.070\) & \(0.631\pm0.011\)\\
\end{tabular}
\end{table}

\subsection{Production comparisons}
\label{sec:org9ec9d40}

This ancillary prediction-arithmetic experiment uses GPRidge 0.6.0, source \texttt{00b92df42}; it is separate from the GPRidge 0.5.0 reaction results. Each cell uses four MPI ranks and
two CPU slots per rank. Table \ref{tab:predict-production} contains one run per cell;
these entries have no estimated sampling uncertainty. Pipeline time
includes the search and its validation. Strict acceptance requires one
imaginary mode and no unresolved vibrational modes, together with the
force criterion. Reference-energy differences are diagnostics. Method calls
refer to the band search for Baker and the dimer search for Birkholz.
Pipeline time includes the seed band and validation for both sets. The
CUDA residue kernels appear in the repeated ethane comparison below.

\begin{table}[htbp]
\caption{\label{tab:predict-production}Single production trials for the three CPU prediction products. Every reaction finishes inside the 900 s Grignard deadline.}
\centering
\begin{tabular}{lllrr}
Case & Product & Outcome & Method calls & Pipeline (s)\\
\hline
hcch & compensated & pass & 29 & 2.55\\
hcch & scheme1 & pass & 29 & 2.41\\
hcch & scheme2 & pass & 29 & 2.53\\
ethane & compensated & pass & 92 & 8.58\\
ethane & scheme1 & pass & 118 & 11.81\\
ethane & scheme2 & pass & 118 & 12.34\\
Grignard & compensated & pass & 558 & 299.45\\
Grignard & scheme1 & pass & 568 & 305.35\\
Grignard & scheme2 & pass & 568 & 304.79\\
\end{tabular}
\end{table}

Three repetitions of the ethane cell compare the reusable workspace, the warp partition and compensated arithmetic. All nine cells run through the same reproducer in one
allocation on a Ryzen 9 9950X with an RTX 5070, using four MPI ranks and
two CPU slots per rank. The order reverses on alternate repetitions.
The band deadline is 30 s and the complete-cell deadline is 60 s.
Table \ref{tab:predict-workspace-production} reports the mean and sample
standard deviation of complete pipeline times. A timed-out search is a
reaction that did not finish, and its observed evaluation counts are lower bounds.

\begin{table}[htbp]
\caption{\label{tab:predict-workspace-production}Repeated ethane searches with compensated arithmetic and two CUDA residue kernels. Calls are search calls; pipeline time includes independent saddle validation. Each completed search has one imaginary mode and no unresolved modes and satisfies the force criterion.}
\centering
\begin{tabular}{llrl}
Product and implementation & Passes & Search calls & Pipeline (s)\\
\hline
Compensated & 3/3 & 83 & \(8.45\pm0.16\)\\
CUDA residues, reusable workspace & 0/3 & 78, 80, 80 & timeout\\
CUDA residues, warp partition & 3/3 & 140 & \(29.78\pm0.14\)\\
\end{tabular}
\end{table}

The warp partition completes the three searches within the deadline,
but takes more calls and wall time than compensated arithmetic. Its
correctly rounded products do not reduce the search cost on this cell.
These workstation measurements are separate from the single CPU trials in
Table \ref{tab:predict-production}, from the A100 trials below and from the
panels in the main text.

Two matched A100 allocations repeat hcch and ethane three times per
implementation, with four MPI ranks and two CPU slots per rank. One
allocation compares the initial device kernel with reusable storage; the
other compares reusable storage with the warp partition. Each allocation
includes a compensated control. The production device work uses one GPU,
shared by the ranks. Table \ref{tab:predict-a100-repeated} reports complete
pipeline means and sample standard deviations. The band deadline remains
30 s. All hcch searches pass with 32 search calls; the compensated ethane
searches pass with 123 calls. Every device ethane search times out.

\begin{table}[htbp]
\caption{\label{tab:predict-a100-repeated}Repeated A100 production comparisons. Each row represents three hcch and three ethane trials. A timeout denotes zero of three completed ethane searches within the band deadline.}
\centering
\small
\begin{tabular}{p{0.14\linewidth}p{0.35\linewidth}rr}
Comparison & Product and implementation & hcch time (s) & Ethane time (s)\\
\hline
Storage & Initial CUDA residues & \(12.46\pm0.32\) & timeout\\
Storage & CUDA residues, reusable workspace & \(9.21\pm0.03\) & timeout\\
Storage & Compensated & \(4.26\pm0.08\) & \(18.08\pm0.09\)\\
Warp & CUDA residues, reusable workspace & \(9.10\pm0.09\) & timeout\\
Warp & CUDA residues, warp partition & \(8.49\pm0.07\) & timeout\\
Warp & Compensated & \(4.27\pm0.06\) & \(17.98\pm0.06\)\\
\end{tabular}
\end{table}

Both device changes reduce the hcch wall time within their matched
comparisons. The device contraction remains slower than compensated
arithmetic, and the ethane failures prevent a production speedup claim.

The compensated campaigns pass 24 of 24 Baker cells and 20 of 20 Birkholz
cells, with 1193 band-search calls and 2082 dimer-search calls respectively.
The Birkholz seed bands require a separate 760 calls. Their slowest
pipelines take 19.82 and 447.35 s. Exact prediction products increase the
ethane call count. Two exact-arithmetic Grignard trials complete in about 305 s, while the timed-out alternative is identified in the table. The recommended production arithmetic remains compensated.
Correct rounding of a local contraction does not constrain the sequence
of decisions made by the nonlinear search.

These single trials
are separate from the repeated production measurements in the main text.

The reference-energy diagnostic exceeds its positive tolerance of
0.05 eV for four compensated cells:
HCN + H\(_2\) (Baker, 0.07885 eV), DACP2 (0.17080 eV), Grignard
(0.14825 eV), and oxirane (0.15202 eV). The supplied reference geometries were also audited with the same
campaign oracle. Their maximum atomic forces are 0.28390, 0.14893,
0.19796, and 0.14864 eV/Angstrom, respectively; all exceed the campaign
threshold of 0.0514221 eV/Angstrom. The Grignard reference
has two unresolved vibrational modes. Energy offsets from these
geometries therefore do not quantify error against independently
converged saddles.
\subsection{Scaling a GP fit}
\label{sec:orga4c5c93}

A fixed training set separates GP fitting from oracle cost. Thirty fits
reuse 40 archived Grignard band observations and the same initial kernel,
with zero potential evaluations. Each implementation runs three times at 1, 2, 4,
8 and 16 MPI ranks, with two CPU slots per rank on one Xeon Platinum 8358
node. Rank and implementation order reverse in the middle repetition.
All runs use the same ScaLAPACK runtime and OpenBLAS dispatch.

The distributed column-vector gather uses one sum
reduction. Each component has one owner, with zero contributions from
all other ranks. The result is replicated with an allreduce when all
ranks need it, or returned by a reduction to the root. This removes the
root buffer containing one full vector per rank. Overlapping partial
sums use their ordered reduction. The comparison uses
the root gather and broadcast.

\begin{table}[htbp]
\caption{\label{tab:fit-disjoint-scaling}Warm fits of 40 Grignard observations. Entries are means and sample standard deviations in seconds over three repetitions. All fits return successfully with zero potential evaluations.}
\centering
\begin{tabular}{rll}
MPI ranks & Root gather and broadcast (s) & Disjoint reduction (s)\\
\hline
1 & \(4.08\pm0.68\) & \(3.71\pm0.01\)\\
2 & \(3.14\pm0.20\) & \(2.94\pm0.04\)\\
4 & \(2.70\pm0.03\) & \(2.57\pm0.13\)\\
8 & \(2.70\pm0.06\) & \(2.62\pm0.08\)\\
16 & \(2.82\pm0.04\) & \(2.75\pm0.02\)\\
\end{tabular}
\end{table}

The first one-rank control takes 4.864 s; its other two repetitions take
3.694 and 3.691 s. All three contribute to the reported mean. At each
rank count, the paired fitted parameters and both training residuals
match exactly. Across rank counts, the largest change in a fitted log
parameter is \(5.51\times10^{-9}\) and in the force-component RMSE is
\(4.24\times10^{-11}\). The disjoint reduction reaches a mean speedup of
1.44 at four ranks relative to one rank, falling to 1.35 at sixteen.
The fit therefore does not sustain strong scaling across this range.
These measurements concern this fit and do not establish a
speedup of the full saddle search. Performance traces use TALP and Nsight.
\subsection{Exact factorization timings}
\label{sec:orgbc087bc}

The run-time choice \texttt{-{}-{}linalg-mode reproducible|library} selects a tile-ordered Cholesky with exact accumulators or the ScaLAPACK and SLATE routines. Table \ref{tab:factor-cpu} lists the factor of a synthetic symmetric positive definite matrix on one node with two Xeon Gold 6248R processors, four ranks with 12 threads each. The relative residual is not computed at \(n=50000\).

\begin{table}[htbp]
\caption{\label{tab:factor-cpu}Factor time (s) on one Xeon Gold 6248R node, 4 ranks with 12 threads each.}
\centering
\begin{tabular}{lrr}
Mode & \(n=12500\) & \(n=50000\)\\
\hline
Library & 3.227 & 57.036\\
Reproducible & 1.257 & 53.796\\
Scheme II, FP64 residues & 4.611 & 213.069\\
Scheme II, INT8 residues (MKL) & 3.657 & 127.598\\
\end{tabular}
\end{table}

At \(n=12000\) the Scheme II factor is identical for 1 rank with 1 thread, 1 rank with 6 threads, 2 ranks with 6 threads, 4 ranks with 6 threads, and 4 ranks with 1 thread, for the FP64 and the INT8 (MKL) engines, and the relative residual is \(9.471\times10^{-16}\) in all ten runs. The times of the FP64 engine are 80.251, 16.114, 8.919, 5.118 and 21.981 s in that order, and those of the INT8 engine 61.417, 12.037, 7.203, 4.181 and 17.935 s.

On the GPU node described next, the \(n=12000\) Scheme II factor is identical for the FP64 engine, the built-in INT8 engine at 4 ranks with 6 threads, and the CUDA engine at 1, 2, 4 and 8 ranks with 4 threads, with a relative residual of \(9.461\times10^{-16}\) (outer width 2048). The residual differs from the \(9.471\times10^{-16}\) of the Xeon runs for the same factor.

Table \ref{tab:factor-gpu} lists the CUDA engine on a node with two AMD EPYC 7742 processors and eight NVIDIA A100 SXM4 GPUs (40 GB), one rank per GPU with 16 threads. The CPU rows use 8 ranks with 16 threads. The \(n=50000\) Scheme II factor is identical for the FP64 residue engine on the CPUs and for the CUDA engine on every GPU count.

\begin{table}[htbp]
\caption{\label{tab:factor-gpu}Factor time (s) on the A100 node. The first value of the GPU rows is the first run, the second the repeat.}
\centering
\begin{tabular}{lrllr}
Engine & Ranks & \(n=12500\) & \(n=50000\) & \(n=100000\)\\
\hline
Library, CPU & 8 & 9.329 & 57.609 & 421.644\\
Reproducible, CPU & 8 & 1.594 & 79.576 & \\
Scheme II FP64, CPU & 8 & 5.898 & 303.324 & \\
Scheme II INT8, CUDA & 1 & 3.063, 3.211 & 138.104, 139.428 & \\
Scheme II INT8, CUDA & 2 & 2.243, 2.427 & 90.970, 90.761 & \\
Scheme II INT8, CUDA & 4 & 1.743, 2.080 & 57.004, 57.254 & \\
Scheme II INT8, CUDA & 8 & 1.279, 1.885 & 32.598, 33.334 & 227.037\\
\end{tabular}
\end{table}

On an AMD Ryzen 9 9950X with an NVIDIA RTX 5070, the INT8 CUDA Scheme II product has a relative difference from a binary64 product of \(8.8\times10^{-15}\) at \(n=8192\) and \(1.4\times10^{-14}\) at \(n=16384\). It runs at 2628.3, 2737.9 and 2705.8 GFLOP/s of FP64-equivalent work at \(n=8192\) and at 3176.0, 3185.0 and 3194.1 GFLOP/s at \(n=16384\) (host to host). The native cuBLAS DGEMM runs at 463.2, 478.9 and 478.9 GFLOP/s at \(n=8192\) and at 477.9, 479.1 and 478.9 at \(n=16384\).

The separate fixed-work Grignard fit replays 120 observations from row 20 with ScaLAPACK block size 128 and one thread per rank on a node with two 24-core Xeon Gold 6248R processors, and stops at the posterior refresh of the first MAP fit. Table \ref{tab:fit-modes-si} lists both repetitions.

\begin{table}[htbp]
\caption{\label{tab:fit-modes-si}Seconds to the first completed MAP fit, one of the three passes and repetition 2.}
\centering
\begin{tabular}{llll}
Mode & 8 ranks & 16 ranks & 32 ranks\\
\hline
Library & 10.047, 10.055 & 8.041, 8.042 & 8.294, 8.297\\
Reproducible & 16.073, 15.833 & 11.057, 11.055 & 10.801, 10.812\\
\end{tabular}
\end{table}
\subsection{Contraction before a variance solve}
\label{sec:org8a2dcc4}

A scalar linear functional of joint energy and force observations has
weights \(w\), query covariance \(K_{qq}\), and training--query covariance
\(K_{tq}\). For a fixed training factor \(LL^{\mathsf T}\) its latent variance is
\begin{equation}
 v_w=w^{\mathsf T}K_{qq}w-
       \left\|L^{-1}(K_{tq}w)\right\|_2^2.
\end{equation}
The identity \(L^{-1}(K_{tq}w)=(L^{-1}K_{tq})w\) allows the contraction to
precede the triangular solve. One right-hand side suffices, while the
energy--force and cross-point correlations remain in \(K_{tq}w\). The
configured observation subset, its noise, and its factor are unchanged.
The implementation contracts in extended precision before the binary64
solve. This order also avoids cancellation between separately rounded
solutions for nearby stencil points.

\nolinkurl{proofs/lean/GprOptim/VarianceSubset.lean} proves the
real-arithmetic identity by matrix associativity, and
\nolinkurl{scripts/derive_conditional_cholesky.py} verifies the corresponding
quadratic form symbolically. The finite-precision example in
\nolinkurl{scripts/verify_functional_contraction.sollya} uses a signed
two-point stencil for which contracting first retains the exact result
and contracting separately rounded solutions loses a term. The example quantifies cancellation for this factor and stencil. The functional-variance and active-posterior checks pass in serial
and on four ranks with ScaLAPACK and SLATE.
\subsection{Reusing background covariance work}
\label{sec:org4b639e9}

The marginal variance of a query \(q\) and its cross covariance with active
observations \(a\) share a training--query covariance block. For a subset
\(S\) with factor \(LL^{\mathsf T}\), define \(W=L^{-1}K_{Sq}\). The two
posterior moments are
\begin{align}
 \operatorname{diag} C_{qq}
   &=\operatorname{diag} K_{qq}-\operatorname{diag}(W^{\mathsf T}W),\\
 C_{qa}&=K_{qa}-(L^{-\mathsf T}W)^{\mathsf T}K_{Sa}.
\end{align}
When the query has fewer channels than the active observations, both
moments use the same assembled \(K_{Sq}\) and forward solution \(W\).
This removes a repeated covariance assembly and triangular solve.
The backward solve and rectangular contraction retain their order.
The subset, regularization and output scaling remain the same.

The subset factor depends on the fitted background and selected
observations. Conditioning or fitting the independent active correction
leaves that background unchanged, so its factor can be reused.
Changing the selected observations or rebuilding the background requires
a new subset factor. The rectangular identity is proved in
\nolinkurl{proofs/lean/GprOptim/VarianceSubset.lean};
\nolinkurl{scripts/derive_conditional_cholesky.py} also checks the
quadratic and rectangular forms with symbolic arithmetic.

The 58 affected GP checks pass in serial and on four ranks with both
ScaLAPACK and SLATE. They compare marginal variances with the joint
covariance and verify factor reuse across active updates, together with
invalidation on a changed subset, likelihood noise or kernel.

For Grignard, the number of subset factorizations decreases
from 30 to 24, and the cumulative variance timer decreases from 43.31
to 34.57 s. The pipeline decreases from 446.63 to 438.82 s in this single
comparison. The paired sources \texttt{0fe73fe7d} and \texttt{b8dd0ed6d} both use 547 dimer calls; this is a separate variance-reuse experiment. The archived 0.5.0 reaction uses 545 dimer calls and 40 seed-band calls. Reported peak memory
is 1257.55 and 1256.68 MB, respectively. These trials establish preserved
search results and less repeated factorization work; they do not
establish a distribution of walltime improvements.

\section{Implementation details}
\label{sec:convergence}
The surrogate loop alternates relaxation on the GP posterior mean with
true energy-and-gradient evaluations. The observations define a
training set; acquisition policies choose which proposed geometries to
measure, and hyperparameter fitting updates the statistical model. True
forces and independent curvature measurements assess the returned geometry.
The distinctions between model uncertainty, force convergence, and saddle
index are stated in Section \ref{sec:si-surrogate-validation}.

At fixed kernel and noise parameters, conditioning on derivative data
cannot increase the exact posterior variance (Section \ref{sec:derivatives-detail}).
An inverse-distance representation supplies rigid-motion invariance
(Section \ref{sec:si-sexpat}). Neither property establishes contraction of
the implemented saddle search.
\subsection{How derivative observations accelerate learning}
\label{sec:derivatives-detail}
A DFT evaluation returns the energy \(V(\mathbf{x})\) together with the
atomic force \(-\nabla V(\mathbf{x})\), a vector of \(3N\) components.
Those \(3N\) extra scalars do not enter the GP as an independent field.
They couple to the energy through the joint energy-gradient
covariance of Rasmussen and Williams (Sec. 9.4).
\texttt{Distance::dist\_at\_vec} at unit lengthscale stores
\(-2(\Delta q_p)^2\) with \(q_p=1/r_p\), and
\nolinkurl{SexpatCF::buildSharedKernel} flips the sign to
\(Q_p=2(\Delta q_p)^2\). Packed log-parameters
\(\theta_0=\log\sigma_f^2\) and \(\theta_{1+p}=\log\ell_p\) give
\begin{equation}
k(\theta)_{nm}
  = e^{\theta_0}
    \exp\Bigl(-\tfrac12\sum_p e^{-2\theta_{1+p}}\,Q_{p,nm}\Bigr)
  = \sigma_f^2\exp\Bigl(-\sum_p (\Delta q_p)^2/\ell_p^2\Bigr).
\label{eq:sexpat-detail}
\end{equation}
The energy and derivative paths use the same exponent.
The Lean \texttt{se\_kernel} is the scalar textbook
form \(\sigma_f^2\exp(-d^2/(2\ell^2))\) on \(\mathbb{R}\); the two
agree after \(\ell_{\mathrm{Lean}}=\ell/\sqrt{2}\). The
supplementary proofs give symmetry, the diagonal,
nonnegativity, the bound, and stationarity of that scalar \(k\).
They do not formalize the inverse-distance packing.

The production joint Gram is the blocked Solak assembly in
\texttt{CovarianceAssembly},
\begin{equation}
K_{\mathrm{joint}} = \begin{pmatrix} K_{EE} & K_{EG} \\ K_{GE} & K_{GG} \end{pmatrix},
\label{si-eq:block_kernel}
\end{equation}
With \(K_{EE}=k(\theta)\) untruncated, \(K_{EG}\) and \(K_{GG}\) are gradient blocks, \(K_{EG}\propto-\tfrac12 d_1\) and \(K_{GG}\propto-\tfrac12 d_{12}+\tfrac14 d_1\odot d_2\), where \(d_1\) and \(d_2\) are the first derivatives of the quadratic in the exponent of Eq. \ref{eq:sexpat-detail} and \(d_{12}\) is the mixed second derivative. Eq. \ref{si-eq:block_kernel} carries no further minus: the observation is the Cartesian gradient, and the force is \(-\nabla V\).
The assembly is never written as the matrix product \(L k L^\top\).
A gradient observation constrains the slope of the energy surface.

Let \(C_{EE}=K_{EE}+\Sigma_E\) and \(C_{GG}=K_{GG}+\Sigma_G\) be the
observation covariance blocks, including independent energy and derivative
noise. At fixed kernel and noise parameters, define
\begin{align}
 S &= C_{GG}-K_{GE}C_{EE}^{-1}K_{EG},\\
 r_G(\mathbf{x}) &= k_{G,*}-K_{GE}C_{EE}^{-1}k_{E,*}.
\end{align}
\begin{equation}\label{eq:schur}
 P_\mathrm{joint}^2(\mathbf{x})
 =P_\mathrm{energy}^2(\mathbf{x})-r_G(\mathbf{x})^\top S^{-1}r_G(\mathbf{x}).
\end{equation}
Here \(r_G\) is the covariance between the query and derivative observations
conditional on the energy observations. Substituting the marginal derivative
cross-covariance \(k_{G,*}\) into this expression is incorrect. The block
inverse identity gives Eq. \ref{eq:schur}; the supplementary Lean comparison
assumes a nonnegative variance reduction and propagates that inequality.

\begin{theorem}[Derivative observations reduce posterior variance]
\label{thm:deriv}
For a positive definite joint observation covariance at fixed kernel and
noise parameters, Eq. \eqref{eq:schur} gives
$P_\mathrm{joint}^2(\mathbf{x})\le P_\mathrm{energy}^2(\mathbf{x})$.
The inequality is strict exactly when $r_G(\mathbf{x})\ne0$.
\end{theorem}

This comparison concerns exact conditioning under one statistical model.
Changing hyperparameters, pruning observations, or truncating a numerical
factorization changes that comparison. An orthonormal eigenbasis has
condition number one, while inversion still depends on the spectrum of the
observation covariance. Section \ref{sec:si-hssvd} derives an optional eigen-factorization. The reported calculations factorize the noisy Gram by Cholesky. Basis orthogonality is an algebraic property of that optional factor. The theorem assumes exact conditioning at fixed hyperparameters, and it is not a test of the floating-point search.

A joint observation supplies one energy and up to \(3N\) Cartesian derivative
values. These values can constrain reaction directions efficiently, but
they need not supply independent constraints. For example, at the three
points \((0,0,0)\), \((1,0,0)\), and \((0,1,0)\), the polynomial \(z^2\) and its
gradient vanish. Three joint observations therefore give a design matrix
of rank nine for the ten coefficients of a general three-dimensional
quadratic. They do not determine the quadratic despite supplying twelve
scalar values. The exact block identity and this rank calculation are
checked by \nolinkurl{scripts/derive_surrogate_validation_limits.py} in the software
repository. Reductions in true evaluations must be measured with matched
search configurations, rather than inferred from the number of force
components.
\subsection{Active-set management for extended substrates}
\label{sec:active_set}
Frozen substrate atoms provide the local environment for the GP. Moving
atoms always enter the representation; selected frozen atoms contribute
distances to moving atoms. Pairs of frozen atoms are excluded. Selection
changes the surrogate representation, while the physical potential uses
the full substrate.

Radius selection includes frozen atoms within \(d_{\mathrm{act}}\) of a moving
atom. The monotonic rule includes every activated atom; the windowed rule
removes atoms that leave the radius. Nearest-neighbour selection includes
\(k\) frozen atoms, ordered by their minimum distance to the moving atoms.
The initial radius can determine \(k\), or the user can specify it.
Distances between moving atoms do not change under a rigid motion of the
moving set, so the frozen anchors must fix it: with two anchors the
rotation about the line through them changes no pair feature, and the GP
predicts no force along it. The library's nearest-neighbour selector
therefore extends \(k\) to the nearest anchors that span a plane. The
reported Pt(111) calculations use release 0.5.0 and seed \(k\) from the
activation radius, whose substrate atoms already span a plane. Their
selection satisfies this condition without the extension. Periodic
systems require minimum-image distances for each rule.

For \(M\) training geometries, each supplying one energy and
\(3n_{\mathrm{mov}}\) gradient components, the joint covariance has
\(M(1+3n_{\mathrm{mov}})\) rows. Frozen atoms add pair features. Holding \(k\)
fixed bounds the number of features between moving and frozen atoms,
while the covariance order grows with the number of training geometries.
Several neighbors can change simultaneously. When membership changes,
the pair features are recomputed for every training geometry and the GP is
refitted. The measured energies and forces are reused
\cite{koistinenMinimumModeSaddle2020}.

A comparison of selection rules must use the same potential, moving
atoms, initial geometries and convergence criteria. It must measure both
potential evaluations and elapsed time, because fewer pair features can
reduce model cost without reducing the number of observations. All atoms move in the molecular
calculations reported in the main text; substrate selection applies to
extended systems with fixed atoms.
\subsubsection{Uncertainty quantification and the trust region}
\label{sec:laplace}
The optional dual HS-SVD posterior evaluates predictions in the basis
of Section \ref{sec:si-hssvd}; hyperparameter fitting uses the
noisy-covariance Cholesky reduction. The reported molecular calculations
disable dual HS-SVD and use Cholesky for fitting, prediction and variance.
We introduce Laplace, CCD, slice, PSIS, SVGD, BMA, and RFF as
extensions of the optional solve. The extensions calibrate when an outer
iteration may skip the true
potential and, where selected, they supply a
hyperparameter-marginalised predictor. They do not replace the MAP
mean or the stable basis.

Each dimer outer iteration probes
\(\sigma^2(\mathbf{x}_\text{mid})\) at the current midpoint and, if the
variance lies below \(\tau_\text{accept}\), substitutes the GP mean for
the true-potential call. Near the saddle the training set is dense and
the optimizer takes many surrogate steps. Far from the data the
variance forces a true evaluation. The Laplace-mixture variance
(Section \ref{sec:laplace}) includes hyperparameter uncertainty, so
\(\tau_\text{accept}\) has units of eV\(^2\).

Laplace-slice marginalisation replaces the central-composite-design (CCD)
quadrature of Section \ref{sec:laplace} with Neal 2003 doubling slice
sampling on the negative log marginal likelihood
target. \cite{murraySliceSamplingCovariance2010} Cheap-mean dispatch
sends the dimer's mean predicts through MAP \texttt{calculatePotential} and
keeps slice-marginalised variance for the acceptance test, so the
per-iter cost is dominated by the same SCG-bounded hyperparameter
optimization as the MAP arm. Three choices bound the cost of that arm: (i) cheap-mean dispatch (mean from
MAP, variance from the slice cache), the 2008 Carlin--Louis
empirical-Bayes split \cite{carlinBayesAndEmpirical2008}; (ii) a bounded integrated nested
Laplace approximation (INLA) trust update for SCG, capping the
per-iter negative log marginal likelihood (NLML) budget at three
evaluations after the first warm-started fit (Rasmussen and Williams
\S5.5 \cite{rasmussenGaussianProcessesMachine2006}); (iii) the axial-only
CCD grid of Section \ref{sec:laplace} (\(K = 1 + 2p\) instead of
\(1 + 2p + 2^p\)).

Slice-sampling Laplace marginalisation pays a per-call
\(K \times \mathcal{O}(N^3)\) Cholesky burst for \(K\) posterior samples.
Replacing the sampling loop with the analytical
INLA Gaussian-mode (delta method; Rue 2009
\cite{rueApproximateBayesianInference2009} sec. 3.2, GPstuff INLA mode 0
\cite{vanhataloGPstuffBayesianModeling2013}) drops the per-call cost to
\(\mathcal{O}(p N n_q)\) via a closed-form between-sample variance:

\begin{equation}
\label{eq-inla-delta}
\mathrm{Var}_\text{INLA-Delta}(\mathbf{x}^*) =
  \mathrm{Var}_\text{MAP}(\mathbf{x}^*) +
  \mathbf{g}(\mathbf{x}^*)^\top \mathbf{H}^{-1} \mathbf{g}(\mathbf{x}^*),
\end{equation}

where \(\mathbf{H}\) is the Hessian of \(-\log p(\boldsymbol{\theta} | \mathbf{y})\)
at \(\boldsymbol{\theta}_{\mathrm{MAP}}\) and
\[
 g_j=\partial_{\theta_j}m_*+
 (\partial_{\theta_j}\mathbf k_*)^\top\alpha+
 \mathbf k_*^\top\partial_{\theta_j}\alpha,\qquad
 \partial_{\theta_j}\alpha=C^{-1}
 \left(\partial_{\theta_j}r-(\partial_{\theta_j}C)\alpha\right),
\]
where \(r=y-\mathsf{Obs}\,m\), \(C=K+N+\delta I\), and \(\alpha=C^{-1}r\).
For a hyperparameter-independent mean and residual, the first and residual-derivative terms vanish.
The derivative of the cross-covariance remains. The training solves are cached at
\(\boldsymbol{\theta}_{\mathrm{MAP}}\) via the cached training Cholesky.
The MAP term is the usual posterior variance. The second term is the
delta-method contribution from the hyperparameter Hessian. The random-importance-sampling
Pareto-smoothed-importance-sampling (PSIS) sampler's result depends on
rank topology because the importance weights change under any change
in reduction order.

A third, optional layer applies Random Fourier Features
(RFF) \cite{rahimiRandomFeaturesLargeScale2007} to the per-call mean
predict path. The squared-exponential kernel
\(k(\mathbf{x}, \mathbf{y}) = \sigma^2 \exp(-\|\mathbf{x} -
\mathbf{y}\|^2 / (2 \ell^2))\) is approximated as \(\phi(\mathbf{x})^\top
\phi(\mathbf{y})\) via \(D_{rff}\) random cosine features
\(\phi_i(\mathbf{x}) = c \cos(\mathbf{w}_i^\top \mathbf{x} + b_i)\),
with \(c = \sigma\sqrt{2/D_{rff}}\),
\(\mathbf{w}_i \sim \mathcal{N}(0, \ell^{-2} I)\) and \(b_i \sim U(0, 2\pi)\).
The regression appends one constant column for the kernel offset, with
the energies centered on \(E_0 = \mathrm{train.E}[0]\) before the fit, so
the feature space has dimension \(D_{rff}+1\). The per-call
cross-covariance product becomes one feature evaluation and one inner
product in that feature space, and the cost of that evaluation does not
grow with \(N\). Equation \ref{eq-inla-delta} is unchanged, and RFF replaces
only the cheap-mean predict path. Earth mover's-distance (EMD) pruning
bounds the training set at \(n_{\mathrm{obs}} \le 30\), and INLA-Delta
supplies the calibrated variance for the acceptance test. RFF caps the
per-predict cost at \(\mathcal{O}(D_{rff} D)\).

The feature count \(D_{rff}\) trades coefficient noise against fit cost:
too few features leave random-feature coefficient noise in the dimer's
curvature step. The Cholesky on the \((D_{rff}+1)\)-dimensional Gram
matrix costs \(\mathcal{O}(D_{rff}^3)\).

Empirical Bayes (EB) and Stein variational gradient descent
(SVGD) \cite{liuSteinVariationalGradient2016} are the fallbacks when
the LAPLACE--INLA mixture mean wanders along passenger
hyperparameter directions. EB uses the MAP mean (skipping the
second-order TCOV correction) and routes the variance through the
LAPLACE mixture. SVGD replaces the deterministic CCD grid with \(K\)
particles that adapt to the posterior via the Liu--Wang update
\(\boldsymbol{\theta}_k^{t+1} = \boldsymbol{\theta}_k^t + \epsilon
\phi^*(\boldsymbol{\theta}_k^t)\) with \(\phi^*\) in the RKHS of the RBF
kernel. Particles concentrate in informative directions and spread in
passenger directions.

When the training set is small, hyperparameters remain uncertain and a
single GP can underestimate the predictive spread. Bayesian model
averaging (BMA) over a grid of hyperparameter
settings \cite{larsenMachinelearningenabledOptimizationAtomic2023}
computes

\begin{equation}
\mathrm{Var}_\text{BMA}(\mathbf{x}) = \sum_i w_i \sigma_i^2(\mathbf{x}) + \sum_i w_i (\mu_i(\mathbf{x}) - \mu_\text{BMA})^2
\end{equation}

where \(w_i \propto \exp(-\mathcal{L}_i)\) are posterior model weights.
The second term, the variance of the model means, is non-negative (law
of total variance), so
\(\mathrm{Var}_\text{BMA} \geq \sum_i w_i \sigma_i^2\). This bounds the weighted mean of member variances; it need not exceed the largest member variance or give a more conservative trust region than every individual model.
As the training set grows and the weights concentrate on the best
model, BMA reduces to single-model GP.

The Student-t process (TP) places an inverse-gamma hyperprior on the
signal variance. \cite{shahStudenttProcessesAlternatives} For a \emph{fixed}
kernel matrix the TP and GP posterior means coincide
(\(\mathbf{k}_*^\top K^{-1}\mathbf{y}\)). In practice the TP and GP
optimise different objectives. The TP marginal likelihood
includes the hyperprior, so the two processes
select different hyperparameters, produce different posterior means,
and can produce different saddle estimates. For the inverse-gamma parameterization, let $a$ and $b$ be its shape and scale, let $n$ be the number of observations, and let $S=r^\top K^{-1}r$ for a fixed positive-definite base covariance $K$.
The posterior scale factor is $(2b+S)/(2a+n-2)$ when $2a+n>2$.
If $a,b$ stay fixed and $S/n$ has a finite limit, the scale factor approaches that limit; this does not imply identical fitted hyperparameters or saddle estimates.
The derivative of $n/(n+2a-2)$ with respect to $n$ is $(2a-2)/(n+2a-2)^2$.
It is positive for $a>1$, zero for $a=1$, and negative for $a<1$ on the domain where the denominator is positive.

An eigendecomposition-based factorized likelihood decouples noise from
kernel hyperparameters by eigendecomposing the kernel matrix once and
sweeping noise over an exponential grid, with the signal variance
determined analytically (GP) or via the hyperprior (TP) at each grid
point.
The Student-\(t\) process and factorized likelihood follow the DTU CatLearn
implementation; this is distinct from the SUNCAT CatLearn ML-NEB search
baseline \cite{garridoTorresLowScalingAlgorithm2019}.
\subsection{MAP priors and optional GP-NEB extensions}
\label{sec:production-stack}
The fixed molecular protocols use rgmin L-BFGS for MAP fitting and
Cholesky for the noisy training covariance (the Supporting Information).
The alternatives below combine weakly-informative priors on the kernel
hyperparameters during the SCG MAP solve (\texttt{gpr/ml/SCG.cpp}), a deterministic axial CCD
quadrature for the optional posterior over \(\boldsymbol{\theta}\)
(\texttt{LaplaceINLA::buildGrid}), and a Phase 1 / Phase 2 GP-NEB outer loop
with an OCI-NEB minimum-mode-following (MMF) dimer burst on the
per-image GP signal (\texttt{gpr/neb/GPNEB.cpp},
the optional implementation). The MAP + HS-SVD posterior is
optional. The priors regularise the SCG MAP target, and CCD--INLA
extends its uncertainty calculation.
\subsubsection{Weakly-informative priors on the SCG MAP target}
\label{sec:bayes-priors}
The SCG MAP solve minimizes \(-\log p(\mathbf{y}\mid\boldsymbol{\theta})
-\log p(\boldsymbol{\theta})\), where the prior \(p(\boldsymbol{\theta})\)
is the product of a half-Cauchy prior on the kernel signal standard
deviation and an independent half-Student-t prior on each lengthscale.
Both shapes follow the weakly-informative
prescription. \cite{gelmanPriorDistributionsVariance2006,polsonHalfCauchyPriorGlobal2012}
The kernel's own \texttt{PriorGaussian} seed on \(\sigma_f^2\) and
\texttt{PriorSqrtt} seed on \(\boldsymbol{\ell}\) are zeroed in the SCG
objective so the half-Cauchy and half-Student-t are not
double-counted.

The signal scale carries a half-Cauchy\((0, A)\) prior with
\(A = \sqrt{2}\). Working in \(w = \log\sigma_m^2\) and dropping
constants, the contribution to the NLML objective and its gradient
are

\begin{equation}
\label{eq:half-cauchy-prior}
\mathcal{L}_{\mathrm{HC}}(w)
   = \log\!\left(1 + \frac{\sigma_m^2}{A^2}\right) - \tfrac{1}{2}\,w,
\qquad
\frac{\partial \mathcal{L}_{\mathrm{HC}}}{\partial w}
   = \frac{\sigma_m^2}{A^2 + \sigma_m^2} - \tfrac{1}{2}.
\end{equation}

Each lengthscale carries a half-Student-t\((\nu = 4, \sigma = 1)\)
prior, the GP-dimer / GP-NEB shape of the Koistinen
papers. \cite{koistinenNudgedElasticBand2017,koistinenNudgedElasticBandInverse2019}
On \(w_i = \log\ell_i\):

\begin{equation}
\label{eq:half-student-t-prior}
\mathcal{L}_{\mathrm{HST}}(w_i)
   = \tfrac{5}{2}\,\log\!\left(1 + \frac{\ell_i^2}{4}\right) - w_i,
\qquad
\frac{\partial \mathcal{L}_{\mathrm{HST}}}{\partial w_i}
   = \frac{5\,\ell_i^2}{4 + \ell_i^2} - 1.
\end{equation}

Both contributions are scaled by an \(N_{\text{eff}}\) schedule
matching the previous interior-point barrier:
\(\beta(n_{\mathrm{obs}}) = \min\!\bigl(10^{-4} + 10^{-3}\, n_{\mathrm{obs}},\;
0.5\bigr)\). The schedule keeps the prior essentially inert on
few-observation sets, where the kernel seeds already regularise SCG,
and ramps to the half-prior strength as evidence accumulates.
A constant-strength prior over-shrinks
\(\boldsymbol{\theta}_{\mathrm{MAP}}\) on few-observation sets; the ramp
avoids this while still pulling hard cases through the smooth tail
when \(n_{\mathrm{obs}}\) grows.
\subsubsection{Exact distributed posterior cross-covariance and IMSPE acquisition}
\label{sec:imspe-acquisition}
Integrated mean squared prediction error (IMSPE) is an optional
variance-reduction acquisition \cite{sacksDesignAnalysisComputer1989,rasmussenGaussianProcessesMachine2006}.
The publication driver uses the tree-structured Parzen estimator \cite{bergstraAlgorithmsHyperparameter2011} described in
Section \ref{sec:si-oie-schedule}. IMSPE does not enter those calculations.

Let \(C\) be the regularized joint training covariance and \(k_{Di}\) the
covariance between its training energy--gradient observations and an energy
query at image \(i\). The latent posterior energy covariance is
\begin{equation}
\label{eq-kpost}
 K_{\mathrm{post}}(x_i,x_j)
 =k(x_i,x_j)-k_{Di}^{\mathsf T}C^{-1}k_{Dj}.
\end{equation}
For an additional scalar energy observation with independent noise
variance \(\sigma_n^2\), the reduction in summed energy variance over the
interior images is
\begin{equation}
\label{eq-imspe-score}
 \Delta_i=\sum_{j=1}^{n-2}
 \frac{K_{\mathrm{post}}(x_j,x_i)^2}
 {K_{\mathrm{post}}(x_i,x_i)+\sigma_n^2}.
\end{equation}
This scalar formula illustrates the Schur update. A potential evaluation
also returns all Cartesian derivatives. The implemented joint-observation
score conditions on that complete block. If \(Z_i\) denotes its energy and
gradient channels, \(F\) all target force channels and \(N_i\) the observation
noise of \(Z_i\), its reduction in summed force variance is
\begin{equation}
 \Delta_i^F=\operatorname{tr}\!\left[
 C_{FZ_i}(C_{Z_iZ_i}+N_i)^{-1}C_{Z_iF}\right].
\end{equation}
Here each \(C_{AB}\) is a block of the current latent posterior covariance.
The optional IMSPE policy selects the largest finite nonnegative
\(\Delta_i^F\) among eligible images. Its ranking depends on cross-covariances
as well as the marginal variances. MAP fitting does not make the
training-conditional shift identical across queries and does not make
this ranking equivalent to maximum variance or a force-based score.

The distributed implementation reuses the training Cholesky factor.
A triangular solve forms \(Y=L^{-1}K_{DQ}\) for the chain query block,
and the posterior is \(K_{QQ}-Y^{\mathsf T}Y\). The query block includes
the energy and Cartesian-derivative channels required by the selected
score. Its cost grows with both the number of observations and query
channels.

\begin{algorithm}[htbp]
\caption{Double-ended band search used by the reported calculations. Baker--Chan returns the climbing image at 0.0514221 eV \AA$^{-1}$. The Birkholz--Schlegel seed band stops at 0.1 eV \AA$^{-1}$ and hands its maximum to the single-ended dimer.}\label{alg:gpneb-outer}
\begin{algorithmic}[1]
\Require Endpoints, 18 intermediate images, potential $\mathcal O$, force tolerance
\Ensure Candidate geometry, measured observations and search status
\State Interpolate a geodesic SIDPP band; evaluate its endpoints and interior images
\For{each outer iteration within the calculation budget}
    \State Condition the joint energy--gradient GP on the training observations
    \State Refit log hyperparameters by MAP L-BFGS when the kernel-hold test requires it
    \State Relax the band on the GP mean with the external L-BFGS band engine
    \State End relaxation on its force, trust, cycle or progress criterion
    \State Select required measurements for current or proposed geometries
    \State Apply the tree-structured Parzen estimator \cite{bergstraAlgorithmsHyperparameter2011} to replaceable acquisitions
    \State Add unsupported images to the batch when support batching is enabled
    \State Evaluate the selected geometries with $\mathcal O$ and store their observations
    \State Resolve a proposed restore or continuation using the measured-force reward
    \If{the current climbing image meets the measured atomic-force criterion}
        \State \Return candidate geometry and observations (path-only plan)
        \State or hand the maximum, tangent, fitted GP and training set to the dimer (path-to-saddle plan)
    \EndIf
\EndFor
\State \Return unfinished search and its observations
\end{algorithmic}
\end{algorithm}

\subsubsection{OCI-NEB dimer burst on the climbing image's per-image GP}
\label{sec:ocineb-burst}
When both outer-loop signals of Section \ref{sec:production-stack} close
(maximum chain variance below \texttt{unc\_convergence} and worst-image NEB
force below \texttt{mmf\_force\_threshold}, default \(0.1\)), the OCI-NEB
controller takes the climbing image. The dimer attaches to the
climbing image's per-image GP \(\mathcal{G}_i\), or to the global
\(\mathcal{G}\) if the local set is not yet ready. \(\mathcal{G}_i\) is
the local surrogate trained on observations clustered around image
\(i\) under a hierarchical-Bayes \(\tau^2\) prior on
\(\boldsymbol{\theta}_i - \boldsymbol{\theta}_{\text{global}}\). That
construction supplies the optional OCI-NEB burst's per-image surrogate
(Section \ref{sec:alternative-paths}). A re-fit on the burst cluster would
displace \(\boldsymbol{\theta}\) from the NEB-tuned values the next outer
prediction needs. Algorithm \ref{alg:ocineb-burst} snapshots
\(\boldsymbol{\theta}\) before the burst and restores it after the dual
update of both training sets.

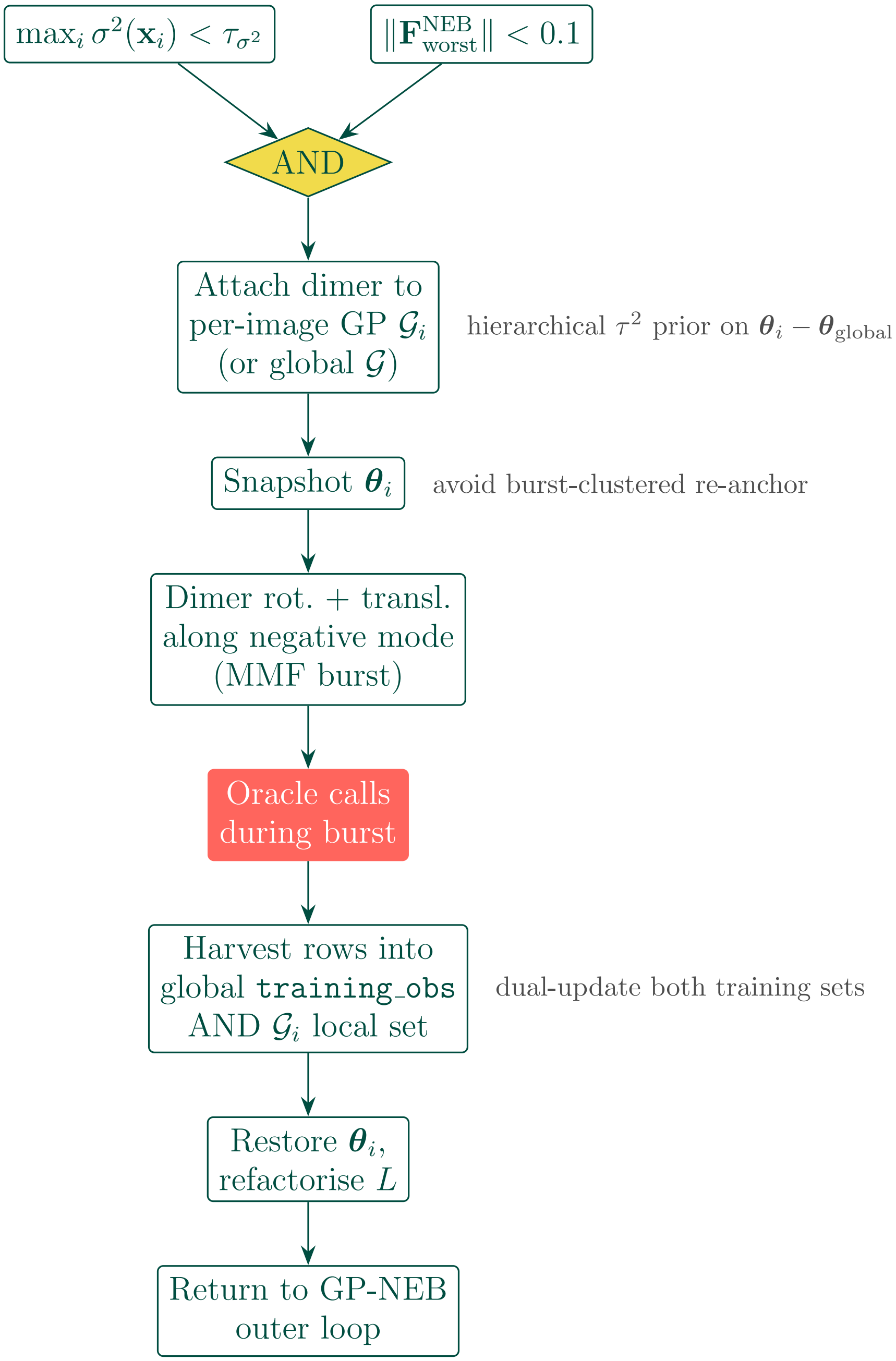
\begin{figure}[htbp]
\centering
\begin{tikzpicture}[
    font=\fontfamily{qhv}\selectfont,
    node distance=8mm and 12mm,
    every node/.append style={font=\fontsize{12}{14}\selectfont},
    every edge/.append style={font=\fontsize{11}{13}\selectfont},
    annote/.style={font=\fontsize{10}{12}\selectfont, text=black!70},
    block/.style={draw=RuhiTeal, line width=0.6pt, fill=white,
                  rounded corners=2pt, inner sep=4pt, align=center,
                  text=RuhiTeal},
    oracle/.style={draw=RuhiCoral, line width=0.6pt, fill=RuhiCoral,
                   rounded corners=2pt, inner sep=4pt, align=center,
                   text=white},
    decide/.style={draw=RuhiTeal, line width=0.6pt, fill=RuhiYellow,
                   diamond, aspect=2.4, inner sep=2pt, align=center,
                   text=RuhiTeal},
    flow/.style={->, draw=RuhiTeal, line width=0.6pt,
                 >={Stealth[length=2.5mm]}}
]
\node[block] (var) {$\max_i \sigma^2(\mathbf{x}_i) < \tau_{\sigma^2}$};
\node[block, right=of var] (force) {$\|\mathbf{F}^{\mathrm{NEB}}_{\text{worst}}\| < 0.1$};
\node[decide, below=of force, xshift=-22mm] (and) {AND};
\node[block, below=of and] (attach) {Attach dimer to\\per-image GP $\mathcal{G}_i$\\(or global $\mathcal{G}$)};
\node[block, below=of attach] (snap) {Snapshot $\boldsymbol{\theta}_i$};
\node[block, below=of snap] (dim) {Dimer rot.\ + transl.\\along negative mode\\(MMF burst)};
\node[oracle, below=of dim] (eval) {Oracle calls\\during burst};
\node[block, below=of eval] (harv) {Harvest rows into\\global \texttt{training\_obs}\\AND $\mathcal{G}_i$ local set};
\node[block, below=of harv] (rest) {Restore $\boldsymbol{\theta}_i$,\\refactorise $L$};
\node[block, below=of rest] (ret) {Return to GP-NEB\\outer loop};
\draw[flow] (var) to (and);
\draw[flow] (force) to (and);
\draw[flow] (and) to (attach);
\draw[flow] (attach) to (snap);
\draw[flow] (snap) to (dim);
\draw[flow] (dim) to (eval);
\draw[flow] (eval) to (harv);
\draw[flow] (harv) to (rest);
\draw[flow] (rest) to (ret);
\node[annote, right=2mm of attach] {hierarchical $\tau^2$ prior on $\boldsymbol{\theta}_i - \boldsymbol{\theta}_{\text{global}}$};
\node[annote, right=2mm of snap] {avoid burst-clustered re-anchor};
\node[annote, right=2mm of harv] {dual-update both training sets};
\end{tikzpicture}
\caption{\textit{OCI-NEB dimer burst on the climbing image's per-image
GP.} The figure shows the two-signal conjunction (chain variance
and worst-image NEB force), the dimer attachment to the local
surrogate, the harvest of oracle rows into both training sets, and
the restore of the snapshotted hyperparameters.}
\label{fig:ocineb-burst}
\end{figure}

\begin{algorithm}[htbp]
\caption{OCI-NEB dimer burst (implementation)}\label{alg:ocineb-burst}
\begin{algorithmic}[1]
\Require Chain $path$, climbing image index $i$, session $\mathcal{S}$
\Ensure Refined CI position; both training sets updated
\State $\mathcal{G} \gets \mathcal{S}.\text{image\_gps}[i]$ if per-image set ready, else $\mathcal{S}.\text{gp}$
\State $\boldsymbol{\theta}_{\text{snap}} \gets \mathcal{G}.\text{packInlaHyperparameters}()$
\State construct dimer with tight rotation tolerances and translation cap
\State \Call{Dimer.attachExternalGP}{$\mathcal{G}$};\quad \Call{Dimer.initialize}{seed, mode}
\State \Call{Dimer.execute}{oracle} \Comment{rotation + translation burst}
\For{each new oracle row $k$ since burst start}
    \State \Call{Session.appendObservation}{$i$, $\mathbf{x}_k, e_k, \mathbf{g}_k$}
    \Comment{dual-update: global $+$ per-image}
\EndFor
\State $\mathcal{G}.\text{setParameters}(\boldsymbol{\theta}_{\text{snap}})$
\State \Call{$\mathcal{G}$.updateModelWithFullData}{$\mathcal{S}.\text{image\_training\_obs}[i]$ or $\mathcal{S}.\text{training\_obs}$}
\State classify outcome by final dimer curvature $\lambda$ and the cosine to the chain tangent
\end{algorithmic}
\end{algorithm}
\subsection{All-image-evaluated GP-NEB}
\label{sec:gp-aie}
A band is \(N\) images \(x_0,\ldots,x_{N-1}\), with \(x_0\) and \(x_{N-1}\) the reactant and product. The reported panels use the one-image schedule described in the main manuscript. The schedule in this section is the all-image alternative. It is implemented, and it was not selected for the Baker--Chan or Birkholz--Schlegel calculations.

The inner stepper is L-BFGS on the concatenated interior coordinates. The NEB force is generally not the gradient of the physical energy, the step is clipped, and a physical-energy line search is not used to judge the nudged relaxation. During path relaxation the force condition is
\begin{equation}
\max_{i\in\mathcal I}\max_a \bigl\| F_{ia}^{\mathrm{NEB}} \bigr\|_2 < \varepsilon
\label{eq:aie-conv}
\end{equation}
where \(a\) indexes atoms, \(\mathcal I\) contains the interior images, and \(\varepsilon\) is the configured force tolerance. With climbing-image-only convergence, the climbing phase instead checks the physical gradient at \(i^\star\) against the measured-force tolerance, and the primary criterion is the largest Euclidean norm of an atomic force. The climbing-image reflection preserves the norm of the complete force vector, and it need not preserve the largest atomic norm. Additional path or inner-relaxation criteria are optional.

The geodesic SIDPP seed \cite{smidstrupImprovedInitialPath2014,schmerwitzSequentialIdpp2024,zhuGeodesicInterpolationReaction2019} redistributes the band in \(q=1/r\), and with the geodesic flag on the same inverse-distance metric measures later reparameterizations of the band, with band steps taking a trust per internal coordinate from the image spacing.

Each outer iteration \(k\) updates the GP on \(\mathcal{D}_k=\{(x_j,V_j,\nabla V_j)\}\). A MAP fit minimizes the negative log posterior with the selected optimizer. The reported molecular runs use rgmin L-BFGS; scaled conjugate gradients and integrated hyperparameter fits are optional. The first trial of a fit without a warm start moves the log hyperparameters by at most one unit. The L-BFGS line search takes the midpoint of its zoom bracket when the value at the far end diverges, or when it differs from the value at the near end by more than \(10^{3}\bigl(1+\lvert\varphi(\ell)\rvert+\lvert\varphi'(\ell)\rvert(h-\ell)\bigr)\) for bracket \([\ell,h]\), and it interpolates otherwise. The fitted state supplies the starting point for an update, and the inner band steps on the posterior mean. The all-image outer iteration then evaluates every intermediate image on the potential at the relaxed band and retrains,
\begin{equation}
\mathcal{D}_{k+1}
  = \mathcal{D}_k
    \cup
    \bigl\{
      \bigl(x_i^{(k+1)},\, V(x_i^{(k+1)}),\, \nabla V(x_i^{(k+1)})\bigr)
    \bigr\}_{i=1}^{N-2} .
\label{eq:aie-oracle}
\end{equation}
Evaluating every intermediate image supplies more measured constraints per outer iteration, and it does not exclude an unmeasured ridge between images or establish the climbing image's Hessian index. Both acquisition schedules can use a minimum-mode dimer for off-path refinement, and that use is reported separately from the band convergence criterion. Measured-gradient disagreement can contract the permitted travel between energy and force evaluations, and the dimer assesses trial progress through the measured maximum atomic force and includes rejected observations for refitting. The Supporting Information derives the local force-merit identity and reports the measured-control validation cases.

An image on the minimum-energy path carries a force parallel to the true tangent, and the chord tangent at turning angle \(\varphi\) leaks a perpendicular residual
\begin{equation}
\lVert F_{\perp}^{\mathrm{chord}}\rVert
  \;=\;
  |F_{\parallel}^{\mathrm{chord}}|\,\tan(\varphi/2).
\label{eq:chord-leak}
\end{equation}

Optional path-resolution checks compare the forces on the climbing image's neighbors with this leak, and they compare segment-midpoint energies with the climbing-image energy. These checks are separate from the primary climbing-image-only force criterion. The leak is \(O(h^{2})\) next to the saddle, and refinement halves the two climbing-image segments when a segment midpoint sits above the climbing image (the Supporting Information section on the band-resolution certificate). When enabled, these checks add measured stencil rows around the climbing image, and they remain separate from the chosen convergence criterion.
\subsubsection{Alternative paths}
\label{sec:alternative-paths}
The reported dimer refinement uses MAP hyperparameters fitted with rgmin
L-BFGS and a Cholesky posterior. SCG is an alternative hyperparameter
optimizer. Laplace slice, axial CCD, BMA, and SVGD extend the uncertainty
calculation. In the optional per-image GP-NEB path, an
\texttt{ActiveLearningSession} carries one local GP \(\mathcal{G}_i\) per image.
The OCI-NEB dimer attaches to \(\mathcal{G}_i\). Algorithm \ref{alg:ocineb-burst} snapshots \(\boldsymbol{\theta}\) and restores it after the dual training-set update, so the burst does not move the climbing-image hyperparameters.
\subsection{Architecture and public API}
\label{sec:implementation}
\label{sec:org35d1722}

\begin{figure}[htbp]
\centering
\begin{tikzpicture}[
  every node/.style={font=\small\sffamily},
  module/.style={draw=RuhiTeal, fill=white, rounded corners=3pt,
                 minimum width=2.8cm, minimum height=0.7cm, align=center,
                 line width=0.8pt},
  newmodule/.style={module, fill=RuhiCoral!15, draw=RuhiCoral},
  backend/.style={module, fill=RuhiTeal!8},
  arrow/.style={-{Stealth[length=5pt]}, RuhiTeal, line width=0.6pt},
  dasharrow/.style={arrow, dashed},
  label/.style={font=\scriptsize\sffamily, RuhiTeal},
]

\node[backend] (dist) at (0.0, 0) {Distance\\+ cache};
\node[backend] (cov) at (3.8, 0) {Sexpat kernel\\(inverse distance)};
\node[backend] (linalg) at (7.6, 0) {Linear algebra\\(Eigen / ScaLAPACK)};

\node[module] (gpr) at (3.8, 1.7) {GP regression};
\node[module] (scg) at (0.0, 1.7) {MAP fitting\\L-BFGS\\SCG (optional)};
\node[module] (hssvd) at (7.6, 1.7) {Cholesky posterior\\optional dual HS-SVD};

\node[module] (activegp) at (0.0, 3.4) {ActiveGP\\surface};

\node[module] (dimer) at (0.0, 5.1) {Atomic dimer};
\node[module] (neb) at (3.8, 5.1) {NEB\\rgmin L-BFGS};
\node[module] (minim) at (7.6, 5.1) {GP minimizer};

\node[newmodule, text width=2.9cm, font=\scriptsize\sffamily] (rg) at (0.0, 7.3) {rg sessions\\SellaMin\\SellaSaddle\\MinMode\\RgsaddleBandEngine};
\node[module, fill=RuhiYellow!20, draw=RuhiYellow!80!black] (python) at (3.8, 7.3) {Python / ASE\\nanobind};
\node[module, fill=RuhiYellow!20, draw=RuhiYellow!80!black] (eon) at (7.6, 7.3) {eOn .con\\CI-NEB / MMF};

\draw[arrow] (python) -- (rg);
\draw[arrow] (python) -- (neb);
\draw[dasharrow] (python) -- (eon);
\draw[arrow] (rg.west) -- ++(-0.4,0) |- (activegp.west);
\draw[arrow] (activegp.east) -- ++(0.55,0) |- (gpr.west);
\draw[arrow] (dimer.east) -- ++(0.55,0) |- (gpr.north west);
\draw[arrow] (neb) -- (gpr);
\draw[arrow] (minim.south) -- ++(0,-0.35) -| (gpr.east);
\draw[dasharrow] (eon) -- (neb);
\draw[arrow] (gpr) -- (scg);
\draw[arrow] (gpr) -- (hssvd);
\draw[arrow] (gpr) -- (cov);
\draw[arrow] (gpr) -- (linalg);
\draw[arrow] (cov) -- (dist);

\end{tikzpicture}
\caption{\textbf{Architecture of \texttt{gpr\_optim}.} The GP
regression core sits on shared covariance functions, distance
computation and linear algebra backends. The reported molecular
calculations use L-BFGS fitting and Cholesky; SCG and dual HS-SVD
are optional library paths. ActiveGP is the
potential rg sessions drive (SellaMin, SellaSaddle,
MinMode, RgsaddleBandEngine). rg sessions are coral.
The production dimer, NEB and minimizer consume the core
directly. The inner band engine is selected by
\texttt{--neb-band-engine}; rgmin L-BFGS is the default when
the rg library is linked, and rgmin FIRE is opt-in. Python (yellow) passes an ASE calculator. The eOn engine is opt-in.}
\label{fig:architecture}
\end{figure}
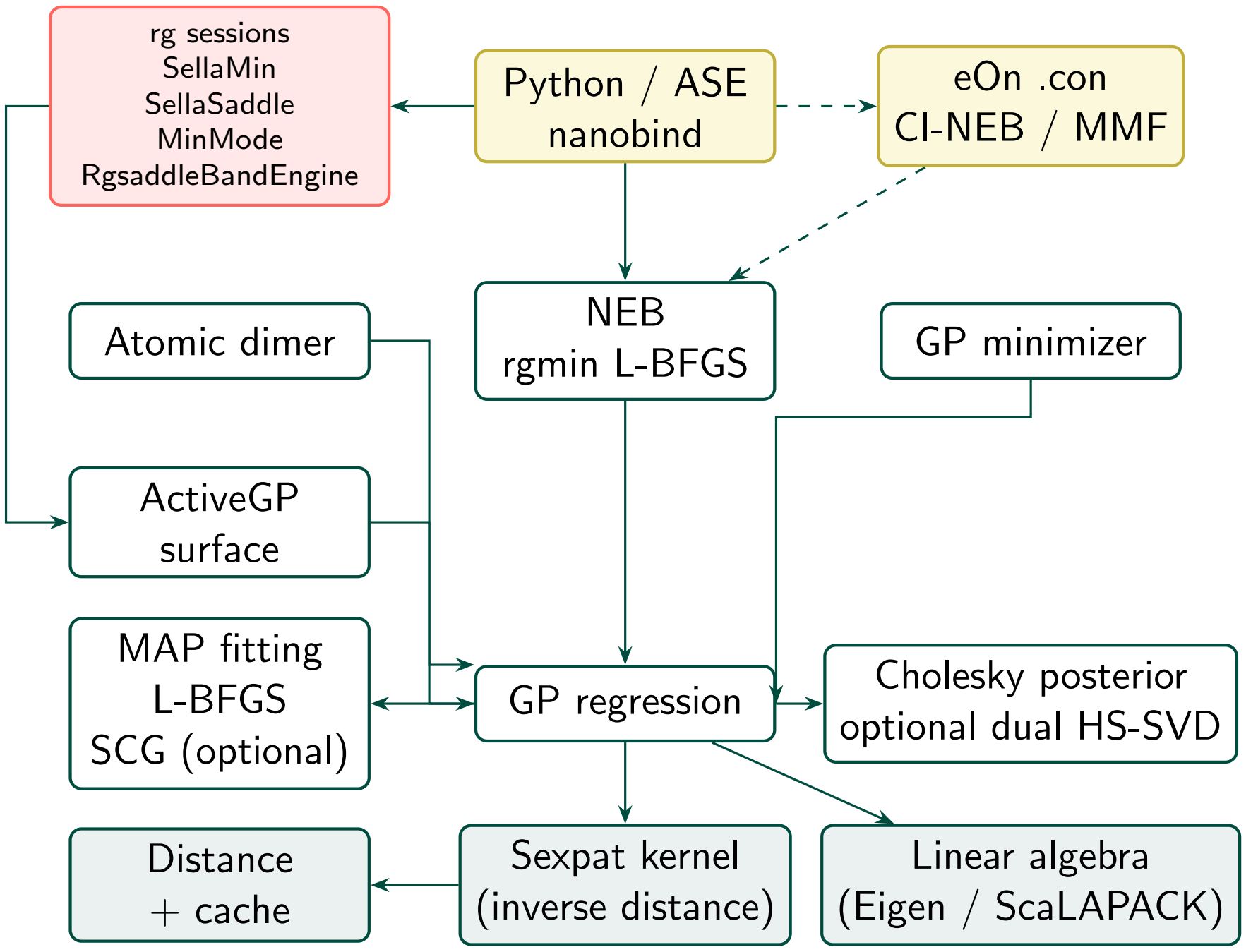

We split \texttt{gpr\_optim} into five core modules that implement the
surrogate loop and three application modules that consume it
(Figure \ref{fig:architecture}).

\begin{itemize}
\item \textbf{GP regression} (\texttt{gpr/ml/}): training, prediction, variance
estimation, and hyperparameter optimization. The reported molecular
runs use rgmin L-BFGS; SCG and INLA mode solves are optional. The
optional dual HS-SVD backend, selected by \texttt{-{}-{}gp-hssvd},
evaluates the posterior mean and variance using the stable basis of
Section \ref{sec:si-hssvd}. It factorizes the assembled signal kernel
\(K = \Psi\,\mathrm{diag}(S^2)\,\Psi^\top\).
Both production panels disable dual HS-SVD and use a Cholesky
factorization of \(C = K + \sigma^2 I\). Cholesky also supplies the
factorization inside the SCG objective (Section \ref{sec:si-hssvd}). The
\texttt{StudentTProcessRegression} subclass overrides the likelihood and
gradient computations and shares the rest of the infrastructure
through virtual dispatch.
\item \textbf{Covariance functions} (implementation): squared
exponential kernel with per-pairtype ARD on the inverse
interatomic distance
representation. \cite{koistinenMinimumModeSaddle2020}
\item \textbf{Distance computation} (\texttt{gpr/auxiliary/}): pairwise inverse-distance
evaluation with frozen atoms, multiple atom types, and the
EMD/RMSD permutation-invariant
metrics. \cite{goswamiBayesianOptimizationGaussian2026}
\item \textbf{Linear algebra backends} (\texttt{gpr/linalg/}): compile-time dispatch to
Eigen (default), cuSOLVER (runtime-loaded CUDA), \texttt{Kokkos}, the
C++17 parallel STL, or ScaLAPACK (MPI-distributed).
\item \textbf{Dimer, NEB, minimizer}: three application modules on the same
surrogate loop. The NEB module implements the improved tangent
estimate \cite{henkelmanImprovedTangentEstimate2000} and the climbing
image \cite{henkelmanClimbingImageNudged2000}. Energy-weighted spring
constants and event-driven geodesic SIDPP reparameterization
(library and CLI default off) use the
inverse-distance metric of geodesic reaction-path
interpolation \cite{smidstrupImprovedInitialPath2014,schmerwitzSequentialIdpp2024,zhuGeodesicInterpolationReaction2019}.
The reported one-image schedule uses required-measurement priorities
and a tree-structured Parzen estimator \cite{bergstraAlgorithmsHyperparameter2011} for eligible acquisitions. Maximum variance,
IMSPE and expected improvement are selectable alternatives. Section
\ref{sec:si-oie-schedule} gives the production NEB picker and the
geodesic default. The reported searches relax the band by L-BFGS. A FIRE2 band optimizer is also available, because the NEB force is not a potential gradient.
\item \textbf{ActiveGP / rgsaddle}: \texttt{ActiveGP} is the potential rg
sessions drive. \texttt{SellaMin}, \texttt{SellaSaddle}, \texttt{MinMode} and
\texttt{RgsaddleBandEngine} are the drivers. Section
\ref{sec:rg-activegp} gives the \texttt{dlopen} client, the
\texttt{ForceGate} and \texttt{HessUpdate} enums, and the internals /
log-cell charts.
\end{itemize}

The \texttt{gprn} chain driver pairs with the \texttt{gprd} dimer binary. It
accepts reactant and product \texttt{.con} frames and runs the outer
evaluation loop on the same layered chain (MAP / Laplace-CCD, with
PSIS/Delta/RFF arms available). The loop is all-image-evaluated
(AIE: every intermediate image queried per outer iteration), its
GP-accelerated variant (GP-AIE), or OIE. It writes the relaxed
multi-frame chain through the readcon spec-v2 format, so the same
Python plotting helpers consume eOn climbing-image NEB /
minimum-mode-following reference runs and the GP-accelerated runs
without a sidecar \texttt{.dat}. Parsed readcon fixed-atom flags become the
moving-atom mask: IDPP/SIDPP auxiliary relaxations project updates
onto moving coordinates, and the inverse-distance path metric keeps
moving--moving plus moving--fixed rows while dropping fixed--fixed
rows. Surface-sensitive distances remain; the fixed substrate atoms
do not move.

The GP-OIE branch updates its GP from the accumulated training set, relaxes the band on the GP mean, and normally evaluates one selected image per outer iteration: the
maximum-variance image on the layered-chain modes, the
maximum-NEB-force image on MAP. Convergence requires that the
surrogate reach a fixed point of the inner relaxation and that the
freshly oracle-evaluated NEB force at the worst image drop below
the user tolerance, so the reported barrier is an oracle energy
rather than a GP-mean energy.

Downstream sources include headers such as \nolinkurl{gpr_optim/AtomicDimer.h} through a flat namespace. The cuSOLVER backend loads
\texttt{libcudart}, \texttt{libcusolver}, and \texttt{libcublas} at runtime via
\texttt{dlopen}, as in the vesin neighbor-list library. Build time does
not require a CUDA toolkit; the same binary runs on machines with
and without a GPU. We cache the Cholesky factor on the GPU in a
persistent \texttt{DeviceState} singleton between consecutive operations
on the cuSOLVER Cholesky path, which removes three host-to-device
transfers per optimization iteration.
\subsubsection{rg sessions, ActiveGP and eOn}
\label{sec:rg-activegp}
\texttt{ActiveGP} answers as a potential.
\texttt{force()} returns the GP when the posterior is resolved
and the oracle when it is not, so rgsaddle, ASE or eOn share
one potential. A two-sided test in multiples of the
model's own likelihood noise decides the path: below
\(\sigma_{\mathrm{lo}}\) the GP answers; between
\(\sigma_{\mathrm{lo}}\) and \(\sigma_{\mathrm{hi}}\) the
oracle is called and the row is learned; above
\(\sigma_{\mathrm{hi}}\) the oracle still answers but the
row is not incorporated. Acquisition and incorporation
are split, so a quasi-Newton history sees a constant
potential between \texttt{modelState()} epochs.

The rg sessions drive that potential.
\texttt{RgsaddleClient} loads \texttt{librgsaddle} with \texttt{dlopen}
from a path or \texttt{RGSADDLE\_LIBRARY} and refuses any ABI other
than major 1. \texttt{SellaMin} is the order-0 settle (product-side
roll-down after IRC). \texttt{SellaSaddle} is the order-1 P-RFO
search and the order-1 validity check exposed through the
library. \texttt{MinMode} is the dimer or Lanczos lowest-mode invert
step. \texttt{RgsaddleBandEngine} is rgsaddle-owned band assembly
(FIRE or L-BFGS) on the same \texttt{activeGpRgsaddleSurface} callback.
The rg gradient is \(\partial E/\partial x\); \texttt{ActiveGP} force
is \(-\nabla E\).

The reported searches relax the band by L-BFGS. A FIRE2 band optimizer is also available, because the NEB force is not a potential gradient.

Force and Hessian-update knobs are rgsaddle \texttt{IntEnum} values,
not implied scalars. \texttt{ForceGate} is \texttt{L2\_NORM} (0),
\texttt{LINF\_NORM} (1) or \texttt{MAX\_FORCE\_ON\_ATOM} (2); the host
wrappers default to per-atom max force (2). \texttt{HessUpdate}
is \texttt{BFGS} (0) or \texttt{TS\_BFGS} (1); rgsaddle \texttt{CartesianPes}
defaults to mass-weighted BFGS, and TS-BFGS leaves a
negative mode negative. \texttt{InternalPes}
is the rg internals chart: rgsaddle loads a vocn auto-find
primitive set and does not generate bond / angle /
dihedral topology. \texttt{CellInternalPes} packs
\([q_{\mathrm{int}};\,\mathrm{cell}]\); the log-cell chart
is \texttt{CellChart::LogDeform}. Mask-false cell degrees of
freedom drop out of the packed chart.

eOn and the GP search share the same energy-and-force call. The NEB driver reads the reactant and product geometries and writes the relaxed chain, so the same plots cover the eOn climbing-image and minimum-mode searches and the GP-accelerated runs.
\subsubsection{Distributed linear algebra}
\label{sec:org8818eb0}

The ScaLAPACK backend \cite{blackfordScaLAPACKUsersGuide1997} compiles
in with \texttt{-Dlinalg\_backend=scalapack}. Application sources call the
same \texttt{gpr::linalg} interface under every backend. The seven
Cholesky entry points (\texttt{cholesky}, \texttt{forward\_solve},
\texttt{forward\_solve\_mat}, \texttt{back\_solve}, \texttt{cholesky\_inverse},
\texttt{trace\_product}, \texttt{log\_det\_cholesky}) dispatch to \texttt{pdpotrf},
\texttt{pdtrsm}, and \texttt{pdpotri} over a 2D block-cyclic BLACS process grid.
A \texttt{ScalapackState} singleton stores the grid context, the block
size (default 64), and a cached Cholesky factor that persists
across consecutive triangular solves. Scatter and gather operations
redistribute between Eigen column-major storage and the
block-cyclic layout. For \(n_\text{procs} = 1\), the BLACS grid contains
one process. ScaLAPACK calls in library arithmetic use this grid. Their
operation order can differ from Eigen and StdPar, producing different
bit patterns. A separate reproducible path uses
canonical tile kernels.

Those seven functions provide noisy-covariance Cholesky solves,
including the SCG objective. The optional dual posterior factorizes the
assembled signal kernel \(K\) (Section \ref{sec:si-hssvd}),
\(K = \Psi\,\mathrm{diag}(S^2)\,\Psi^\top\); it does not form the
explicit Mercer design \(A\). On more than one MPI rank the
dual-driver ladder (Section \ref{sec:si-solver-axis}) takes a
distributed symmetric eigensolve (\texttt{pdsyev}) of that assembled \(K\).
At small assembled-kernel orders the BLACS grid has coarser blocking
than the matrix and the panel broadcasts of a distributed \texttt{pdsyev}
can dominate the arithmetic, so no strong scaling is assumed for
\texttt{pdsyev} in this regime.

The training-covariance Cholesky reuses the singleton
\texttt{ScalapackState} \texttt{cached\_L()} across consecutive triangular solves.
Auxiliary factorizations that do not outlive the GP cannot share
that cache: the \((D_{rff} + 1) \times (D_{rff} + 1)\)
Bayesian-linear-regression Gram matrix the RFF mean-predict path
solves at every retrain, and the Williams-Rasmussen polarisation
block on the Laplace probe loop. Sharing would overwrite the GP's
\(L\) between the two factorizations and fault at the next
\texttt{Cpdgemr2d}. We route them through the \texttt{KernelInverse} operator
hierarchy (\texttt{DenseLLTInverse} for the Eigen backend,
\texttt{ScalapackPDPotrfInverse} for the ScaLAPACK backend), where each
instance owns its own factored state: the Eigen implementation
owns an \texttt{Eigen::LLT}; the ScaLAPACK implementation owns its own
distributed \(L\) slab plus descriptor and runs \texttt{pdpotrf} / \texttt{pdtrsm}
against that slab. The same ownership distributes the
per-rank-local \(Z^\top \mathrm{diag}(\rho)\, Z\) Gram assembly
across MPI ranks: each rank assembles a contiguous slab of the
design matrix, partial sums into a \((D_{rff} + 1)^2\) accumulator,
and a single \texttt{MPI\_Allreduce} sums the accumulator instead of
scattering \(Z\). A \texttt{ScalapackPDPotrfInverse} then factorizes that
Gram.
\subsection{Performance}
\label{sec:org2e8e849}

The surrogate saddle search repeats three linear-algebra costs on a
growing derivative-enhanced training set. Every outer dimer or NEB
iteration appends one observation tile (energy plus \(D\) derivative
channels) and must assemble and factor the training covariance.
The inner relaxation and dimer rotation then evaluate the
predict-mean at many geometries against that factor. Between outer
iterations the self-consistent hyperparameter walk re-fits
\(\boldsymbol{\theta}\) at fixed training inputs, five to seven
trials per outer iteration, and rebuilds the same
lengthscale-invariant pair geometry at each new
\(\boldsymbol{\theta}\). Three complementary reductions cut those
costs: a 2D block-cyclic covariance assembly plus a rank-block
Cholesky append on the training factor; an early contraction of the
predict-mean against the training weights \(\boldsymbol{\alpha}\);
and a geometry handle that caches pair geometry across the inner
sweep. Distance and kernel evaluation stay SIMD-vectorized. The optional scaled
conjugate gradient (SCG) optimizer for the negative log marginal likelihood
uses an exact directional curvature, so the NLML path does not pay
a probe-point factorization.

The production panels use Cholesky of \(C = K + \sigma^2 I\)
for training, prediction and variance. The optional dual HS-SVD
posterior solve is described in Section \ref{sec:si-hssvd}. The SCG
objective also uses Cholesky to evaluate the scalar NLML and its gradient.

Distance computation and kernel evaluation are vectorized with
Google Highway (AVX2 / AVX-512 / NEON selected at runtime). Two
fused kernels avoid intermediate allocations:
\texttt{accumulate\_invr\_diff\_sq} for the lengthscale-weighted
inverse-distance sum, and \texttt{se\_kernel\_exp\_truncate\_sq} for the
kernel evaluation directly on squared distances.

Atomic positions stay fixed during hyperparameter optimization.
The raw \(1/r_{ij}\) values are kept after the first evaluation;
subsequent SCG iterations recompute only the
lengthscale-weighted sum. FPS-EMD
\cite{goswamiAdaptivePruningIncreased2026} is a hyperparameter
subset: farthest-point sampling on the earth-mover distance
selects the points used to fit \(\boldsymbol{\theta}\), while the
full training matrix \(X\) stays for prediction. The
configuration-pair loop is parallelized with OpenMP and combines
with SIMD for two-level parallelism.

The library default sets \texttt{OPENBLAS\_NUM\_THREADS} and
\texttt{OMP\_NUM\_THREADS} to 1 (one BLAS thread per MPI rank). Thread fan-out plus spin-barrier overhead dominates the arithmetic at the small matrix sizes this library forms, and threaded reductions produce rank-dependent floating-point order. An explicit environment value
overrides the default.

OpenMP and the ScaLAPACK backend are mutually exclusive at the
project level. Tile-based assembly already saturates the processor
via MPI ranks, and the threaded BLAS used by \texttt{pdpotrf} / \texttt{pdtrsm}
contends with user-space OpenMP regions. Enabling the OpenMP
runtime against the ScaLAPACK build adds team-setup overhead at small
distance-loop call sites that exceeds any parallel benefit. The build
system rejects the combination and points either to ScaLAPACK MPI
rank scaling or to Eigen with OpenMP for shared-memory thread
scaling. On the Eigen + OpenMP path with BLAS pinned to a single
thread, the parallel squared-exponential gradient kernel and the
ARD grid-search / BMA hyperparameter evaluation use the OpenMP
threads.

The negative log marginal likelihood is not defined everywhere in
hyperparameter space. When the lengthscale is too large relative
to the interatomic distances, the kernel matrix approaches a
rank-1 matrix (all entries equal to \(\sigma^2\)) and the Cholesky
decomposition fails. For small training sets (\(n < 10\)) these
singular regions occupy enough of the domain that SCG routinely
encounters trial points with no gradient information.

SCG \cite{mollerScaledConjugateGradient1993} handles those points by
construction. At each iteration it computes a search direction
\(\mathbf{d}_k\) conjugate to previous directions, evaluates the
directional curvature along \(\mathbf{d}_k\) from the current
Cholesky factor, directional covariance derivatives, and
solve-based trace identities, and sets the step size \(\alpha_k\)
from that curvature. For the NLML objective this curvature is the
second directional derivative \(\mathbf{d}_k^\top H \mathbf{d}_k\), obtained by differentiating the covariance likelihood gradient (Rasmussen--Williams Equation 5.9) once more, so the optimizer skips the
extra probe-point factorization that the classical
finite-difference SCG variant would require. \texttt{LOO} and \texttt{LOCAL\_LOO}
refuse that Hessian: the exact-Hessian assembly returns false and
SCG uses the finite-difference curvature probe instead. The
MPI-distributed ScaLAPACK backend maps the exact curvature onto
distributed triangular solves and scalar reductions.  When the NLL evaluation at the
trial point \(\boldsymbol{\theta}_k + \alpha_k \mathbf{d}_k\)
returns NaN (Cholesky failure), SCG increases the regularization
parameter \(\lambda\) and reduces the step, backing away from the
singular region.

That backtracking succeeds because the set of hyperparameters
where the kernel matrix is positive definite is \emph{open}:
eigenvalues are continuous functions of matrix entries, so any
feasible point has a neighborhood of feasible points. The retry
grows a Levenberg--Marquardt damping term by a fixed factor per
attempt, which shrinks the effective step toward the feasible
ball. The damping is bounded; when a non-finite objective
persists at the configured damping limit the optimizer reports
failure and the caller restores the last feasible hyperparameters
rather than accepting the step.

When the Cholesky decomposition fails from floating-point rounding
rather than a non-PD matrix, the library retries once with a
diagonal shift computed from the standard backward-error bound
\(\gamma_n = n\varepsilon/(1-n\varepsilon)\) scaled by the largest
diagonal entry. If that single rescue also fails, the
decomposition is reported as failed rather than retried further.
Adding jitter \(\delta\) shifts all eigenvalues by \(\delta\), so any
matrix whose minimum eigenvalue exceeds \(-\delta\) becomes
positive definite after the shift.
\subsubsection{Derivation of the exact directional curvature}
\label{sec:org0dc1cec}

The curvature formula was derived by symbolic differentiation and
checked by identity before translation into production C++. The
GP objective is
\(f(\boldsymbol{\theta}) = \tfrac{1}{2}(\log\!\det C + \mathbf{y}^\top
C^{-1}\mathbf{y})\). Perturbing along a unit direction
\(\mathbf{d}\) with step \(t\) and expanding
\(C(t) = C_0 + tD + \tfrac{1}{2}t^2 E\), where
\(D = \partial_t C\) and \(E = \partial_t^2 C\) at \(t{=}0\), gives the
closed form
\begin{equation}
  \mathbf{d}^\top H \mathbf{d} \;=\; \tfrac{1}{2}\big(
    -\mathrm{tr}(A D A D) + \mathrm{tr}(A E)
    + 2 \alpha^\top D A D \alpha - \alpha^\top E \alpha\big)
    \,+\, \kappa_{\text{prior}}.
\end{equation}
with \(A = C^{-1}\), \(\alpha = A\mathbf{y}\), and
\(\kappa_{\text{prior}}\) the second directional derivative of the
log-prior. Symbolic verification in the repository confirms the
identity: automatic differentiation of \(f(C(t))\) is simplified and
compared to the closed form, both in \(A\)-form (above) and in the
equivalent \(C\)-form that uses only the covariance matrix and its
directional derivatives. Both residuals simplify identically to
zero.

The production evaluation avoids forming \(A\) explicitly. With \(C=LL^{\top}\) and \(A=(LL^{\top})^{-1}\), \(\mathrm{tr}(A D A D)=\|L^{-1} D L^{-\top}\|_F^2\), by a forward solve and a back solve on \(D\) and a Frobenius product. With \(v=D\alpha\), \(\alpha^{\top} D A D \alpha=v^{\top}(LL^{\top})^{-1}v\), two triangular solves. The \(\mathrm{tr}(A E)\) term uses the existing trace
identity \(\mathrm{tr}(A E) = \sum_{ij} A_{ij} E_{ij}\) with \(A\)
obtained from the Cholesky inverse. The precomputed inverse costs
\(O(n^3)\) once plus \(O(n^2)\) per hyperparameter, versus
\(O(k n^3)\) for solve-based traces where \(k\) is the number of
hyperparameters. The prior curvature \(\kappa_{\text{prior}}\) is
evaluated exactly using a second-order dual-number system that
threads second derivatives through \(\log\), \(\sqrt{\cdot}\), and
the Student-\(t\) prior without a symbolic runtime dependency. For
the ScaLAPACK backend the same identity maps onto distributed
triangular solves (\texttt{pdtrsm\_}) and a scalar \texttt{MPI\_Allreduce},
avoiding the probe-point Cholesky (\texttt{pdpotrf\_}) that the
finite-difference variant would otherwise require every SCG
iteration.

\subsubsection{Distributed covariance assembly}
\label{sec:mpi-cov}
The covariance matrix \(K\) for derivative observations has a block
structure determined by the spatial dimensions of the atomic
system. For \(N\) observation configurations with \(D\) derivative
dimensions (the unique nonzero entries in the derivative-dimension
index vector), \(K\) is \(N(1+D) \times N(1+D)\). Each block
corresponds to a pair of derivative dimensions \((d_1, d_2)\) and is
computed independently:

\begin{align}
K_{00}(i,j) &= \sigma_f^2 \exp\!\Bigl(-\sum_p \frac{(q_p(x_i) - q_p(x_j))^2}{\ell_{t(p)}^2}\Bigr),\quad q_p(x)=1/r_p(x), \label{eq:kff}\\
K_{d,0}(i,j) &= \frac{\partial K_{00}(i,j)}{\partial x_{i,d}}\quad\text{(\texttt{ginput4})}, \label{eq:kdf}\\
K_{d,d}(i,j) &= \frac{\partial^2 K_{00}(i,j)}{\partial x_{i,d}\partial x_{j,d}}\quad\text{(\texttt{ginput2})}, \label{eq:kdd}\\
K_{d_1,d_2}(i,j) &= \frac{\partial^2 K_{00}(i,j)}{\partial x_{i,d_1}\partial x_{j,d_2}}\quad\text{(\texttt{ginput3})}. \label{eq:kd1d2}
\end{align}

The ScaLAPACK backend assembles \(K\) directly into its 2D
block-cyclic distributed layout, with no replicated full-matrix
intermediate and no final reduction. Algorithm \ref{alg:mpi-cov}
selects exactly one rank per tile by the BLACS ownership
predicate on the tile-corner global index. Owned tiles are
evaluated as the sum of the squared-exponential and constant
covariance blocks (with the likelihood diagonal added on
\(i = j\)) and written into the rank's local block-cyclic buffer.
The assembled distributed \(K\) is consumed in place by \texttt{pdpotrf}
without a gather or broadcast.

\begin{algorithm}[htbp]
\caption{Distributed covariance assembly via 2D block-cyclic ownership}\label{alg:mpi-cov}
\begin{algorithmic}[1]
\Require Observations $\mathbf{x}$, tile size $s$, observation tile count $N_{\text{obs}}=N(1+D)/s$, rank $r$
\Ensure Distributed covariance buffer $K_{\text{dist}}$ in 2D block-cyclic layout
\State $K_{\text{dist}} \gets \mathbf{0}$ \Comment{rank-local block-cyclic slice only}
\For{$i = 0,\ldots,N_{\text{obs}}-1$}
    \For{$j = 0,\ldots,i$} \Comment{lower triangle}
        \If{$\texttt{owns\_global\_entry}(i\cdot s, j\cdot s)$}
            \State $T \gets$ \Call{TileCovariance}{$\mathbf{x}_{[i\cdot s\,:\,(i+1)s]}, \mathbf{x}_{[j\cdot s\,:\,(j+1)s]}$}
            \If{$i = j$} \State $T \gets T + \Sigma_{\text{lik}}$ \EndIf
            \State write $T$ into $K_{\text{dist}}$ at owned local block
        \EndIf
    \EndFor
\EndFor
\State \Call{pdpotrf}{$K_{\text{dist}}$} \Comment{distributed Cholesky in place; no Allreduce}
\end{algorithmic}
\end{algorithm}

Per-rank work scales as \(O(N^2(1+D)^2 / P)\) for the assembly and
\(O(N^3(1+D)^3 / P)\) for the factorization. No collective fires
during assembly; the only MPI traffic is the panel-by-panel
synchronization inside the distributed Cholesky. The same
tile-by-tile ownership pattern is reused on the cross-covariance
assembly for the prediction and variance paths.

Callers that need the full dense kernel on every rank (the
geometry-handle objective sweep and the cross-covariance
consumers) use a
round-robin assembly instead: each block and its transpose
partner is computed by exactly one rank via assignment, so every
entry of the reduced buffer has at most one nonzero contributor
and the rest hold exact zeros. IEEE-754 addition returns the
unique contributor unchanged under any reduction tree
(\(x + 0 = x\) exactly), which makes a plain \(O(\log P)\)
\texttt{MPI\_Allreduce} identical to any rank-ordered reduction for these
buffers. The deterministic rank-ordered reduction is therefore
reserved for overlapping partial sums (per-rank trace terms,
training-slice means, scalar likelihood contributions), and the
\(N^{2}\)-sized assembly payloads never pay its \(O(P)\) root traffic.
A bitwise parity test pins the contract: the distributed
assembly must reproduce the every-rank-serial assembly exactly,
at every rank count.

The same ownership rule is used for the first-derivative NLL
trace term. For each hyperparameter, the GP gradient needs
\(\operatorname{tr}(K^{-1}\partial K/\partial\theta_i)\). The
ScaLAPACK path computes one temporary lower-triangle inverse of
the current distributed Cholesky factor with \texttt{pdpotri\_}, then
contracts that inverse directly with the owned lower-triangle
entries of \(\partial K/\partial\theta_i\) as the derivative
blocks are generated. The off-diagonal entries enter with
coefficient two, so for symmetric \(A=K^{-1}\) and symmetric
\(D_i=\partial K/\partial\theta_i\),
\begin{equation}
\operatorname{tr}(A D_i)
  = \sum_p A_{pp}(D_i)_{pp}
    + 2\sum_{p>q} A_{pq}(D_i)_{pq}.
\end{equation}
The contraction avoids storing one full distributed derivative
matrix per hyperparameter on the fast path. If the temporary
inverse cannot be formed, the backend falls back to the
solve-based trace against the cached Cholesky factor. Symbolic
verification in the repository confirms the dense trace,
Cholesky-solve trace, lower-triangle contraction, and blockwise
quadratic-form identities; the C++ backend test compares the
distributed lower-triangle contraction against the solve-based
trace.
\subsubsection{Early alpha contraction of the predict-mean kernel-vector product}
\label{sec:org40271b4}

The training-side optimizations above run once per outer dimer
iteration. The predict-mean call at the dimer middle point runs
many times per outer iteration (once per inner relaxation step
and once per dimer rotation), so a constant-factor reduction in
its per-call cost compounds over the entire saddle search.
The squared-exponential atomic-pairtype (\(\mathrm{S}_\text{expAt}\))
kernel dominates that call.

The naive predict-mean writes the energy + gradient prediction as
\(\boldsymbol{\mu}_* = K_{*X}^\top \boldsymbol{\alpha}\), with
the cross-covariance \(K_{*X}\) blocked into a function-function
(\(K_{ff}\)), function-gradient (\(K_{df}\)), and gradient-gradient
(\(K_{dd}\)) tile per derivative-dimension pair \((d_1, d_2)\). Per
test point the dense implementation iterates \(D \times D\) tiles,
each requiring a fresh pairtype-distance derivative pass.
Following
\cite{schmitzAlgorithmicDifferentiationAutomated2022} we introduce
a contraction of the operator tensor product with
\(\boldsymbol{\alpha}\) \emph{before} assembling the kernel matrix.
Define the per-training-point scalar
\begin{equation}
\varphi_i(\mathbf{x}_*) \;=\; \alpha_i^{E}\,k(\mathbf{x}_i,\mathbf{x}_*) \;+\; \sum_d \alpha_i^{G,d}\,\frac{\partial k}{\partial x_i^{d}}(\mathbf{x}_i,\mathbf{x}_*),
\label{eq:phi-scalar}
\end{equation}
so that the predict-mean reduces to \(\mu(\mathbf{x}_*) = \sum_i
\varphi_i(\mathbf{x}_*)\) and the predict-mean gradient to
\(\boldsymbol{g}(\mathbf{x}_*) = \sum_i \nabla_{\mathbf{x}_*}
\varphi_i(\mathbf{x}_*)\). The trade-off is structural: the dense
form materializes a \((D\!+\!1)\times(D\!+\!1)\) block per
\((\mathbf{x}_i, \mathbf{x}_*)\) pair and contracts with
\(\boldsymbol{\alpha}\) only at the end; the contracted form
evaluates a single scalar per \((\mathbf{x}_i, \mathbf{x}_*)\)
pair and differentiates that scalar with respect to
\(\mathbf{x}_*\) at the end.

For the inverse-interatomic-distance representation the alpha
fold collapses still further. Each atom pair \(p\) contributes
\(u_p(\mathbf{x}) = 1/\lVert \mathbf{x}_{a_p} - \mathbf{x}_{b_p} \rVert\)
to the pairtype distance, and \(\partial u_p / \partial \mathbf{x}\)
touches only the six atomic-coordinate slots of the two atoms in
pair \(p\). The per-pair scalars are
\begin{align}
c_{i,p}      &= \sum_{\text{6 slots of }p} \alpha_i^{G,d}\,\bigl(\partial u_p/\partial \mathbf{x}_i\bigr)_{d}, \label{eq:cip}\\
s_{i,m}      &= \alpha_i^E - \sum_p \frac{2}{\ell_{t(p)}^{2}} (u_p(\mathbf{x}_i) - v_p(\mathbf{x}_m))\, c_{i,p}, \label{eq:sim}\\
A_{p,m}      &= \sum_i k(\mathbf{x}_i,\mathbf{x}_m)\,\bigl[s_{i,m}\,(u_p(\mathbf{x}_i) - v_p(\mathbf{x}_m)) + c_{i,p}\bigr], \label{eq:apm}
\end{align}
the predict-mean energy and gradient at test point \(\mathbf{x}_m\)
reduce to
\begin{align}
\mu(\mathbf{x}_m) &= \sum_i k(\mathbf{x}_i,\mathbf{x}_m)\, s_{i,m}, \\
g_\beta(\mathbf{x}_m) &= \sum_{p \ni \beta} \frac{2}{\ell_{t(p)}^{2}} A_{p,m}\,
\bigl(\partial v_p/\partial \mathbf{x}_m\bigr)_{\beta},
\end{align}
where the sum over \(p\) runs over the (at most \(\sim\) pairs-per-atom) pairs
that touch atomic-coordinate \(\beta\). The rank-1 outer-product
factorization of the mixed second derivative
\(\partial^2 d^2_p / \partial \mathbf{x}_i^a \partial
\mathbf{x}_m^b = -2 (\partial u_p / \partial \mathbf{x}_i)_a\,
(\partial v_p / \partial \mathbf{x}_m)_b\) (verified symbolically
in the repository) is what removes the \(D \times D\) inner
iteration: the alpha fold consumes the rank-1 structure and the
per-pair work drops from \(\mathcal{O}(D^2)\) to
\(\mathcal{O}(\text{pairs} \cdot 6)\).

The per-pair contraction applies across every production
mean-predict surface. The maximum a posteriori mean uses it
directly. The Laplace integrated nested approximation re-uses it
per factor of the central composite design grid, so the \(1 + 2p\)
samples that build the calibrated mean inherit the per-call
factor reduction. The slice mixture and the Pareto-smoothed
importance-sampling variant use it per \(\boldsymbol\theta_k\)
sample in the mean accumulator, in place of the dense
block-by-block walk against the materialised
\(\mathbf{K}(\boldsymbol\theta_k)\); the within-sample variance
pass includes the materialised cross-covariance because the
triangular solve against the cached training Cholesky needs it.
The empirical Bayes mean dispatch routes through the same
maximum-a-posteriori predictor and inherits the contraction at
zero additional cost.
\subsubsection{Reduced-coefficient prediction}
\label{sec:stable_predict}
The optional dual HS-SVD predictor forms \(\boldsymbol\alpha\) from
Eq. \ref{eq:camp-noisy} and evaluates
\(\mu(\mathbf{x}_*) = \mathbf{k}_*^\top\boldsymbol\alpha\), or AlphaK of
the same \(\boldsymbol\alpha\). The orthogonal eigenbasis preserves the
regularized spectrum and does not guarantee accurate prediction.
An alternative explicit-feature representation uses a reduced
coefficient \(\mathbf{p}\) to evaluate the mean.
With \(A_\text{train} = U S V^\top\) and \(\Psi = U\), the
weight-times-design product collapses,

\begin{equation}\label{eq:reduced_coeff}
A_\text{train}^\top (K+\sigma^2 I)^{-1}\mathbf{y} = V\,\mathrm{diag}\!\left(\tfrac{S}{S^2+\sigma^2}\right)(\Psi^\top \mathbf{y}) = V\,\mathrm{diag}(S^{-1})\,\mathbf{p},
\end{equation}

with the reduced rank-\(r\) coefficient \(p_i = \tfrac{S_i^2}{S_i^2 +
\sigma^2}(\Psi^\top \mathbf{y})_i\). The factor
\(S_i^2/(S_i^2+\sigma^2) \in [0,1]\) bounds
\(\lVert\mathbf{p}\rVert_2\) by \(\lVert\mathbf{y}\rVert_2\). In this tensor/affine predictor the posterior mean is the energy-feature row times \(V\,\mathrm{diag}(S^{-1})\,\mathbf{p}\) from Eq. \ref{eq:reduced_coeff}, without forming \(\boldsymbol\alpha\). The Cartesian gradient is \(\partial\mu/\partial x=\sum_j(\partial\mu/\partial\tilde q_j)\,J_{j,:}\) with \(J=\partial\tilde{\mathbf{q}}/\partial\mathbf{x}\) on the lengthscale-normalized descriptor, and the force is \(-\partial\mu/\partial x\).
Eq. \ref{eq:reduced_coeff} is the tensor/affine coefficient identity.
It avoids forming large dual weights but does not preclude cancellation
in feature evaluation or the final contraction.
When dual HS-SVD is selected, the posterior of Section \ref{sec:si-hssvd}
forms an explicit \(\boldsymbol\alpha\) from that eigen-factorization
(as \(\mathbf{k}_*^\top\boldsymbol\alpha\) or AlphaK). The reported
calculations do not select it. They factorize \(C=K+\sigma^2 I\) by
Cholesky. The sensitivity of the optional eigen-factorization is governed
by the regularized covariance, including the noise and spectral truncation
policy.

\begin{figure}[htbp]
\centering
\maxsizebox{\textwidth}{0.55\textheight}{\begin{tikzpicture}[
  font=\sffamily\scriptsize,
  every node/.style={align=center},
  box/.style={draw=RuhiTeal, line width=0.65pt, rounded corners=2pt,
              fill=white, text=black, inner sep=3.5pt,
              text width=5.15cm, minimum height=0.78cm},
  boxprod/.style={box, fill=RuhiTeal!10},
  boxderiv/.style={box, fill=RuhiYellow!22, draw=RuhiYellow!80!black,
                   dashed, line width=0.6pt},
  boxobj/.style={box, fill=RuhiCoral!12, draw=RuhiCoral},
  boxout/.style={box, fill=RuhiTeal!4, text width=2.35cm, minimum height=0.72cm},
  arr/.style={-{Stealth[length=4pt]}, RuhiTeal, line width=0.6pt},
  arrdash/.style={arr, dashed, RuhiTeal!65},
  arrcoral/.style={-{Stealth[length=4pt]}, RuhiCoral, line width=0.6pt},
  tit/.style={font=\sffamily\footnotesize\bfseries, RuhiTeal},
  sub/.style={font=\sffamily\tiny, RuhiTeal!70!black},
  panel/.style={draw=RuhiTeal!30, line width=0.45pt, rounded corners=3pt,
                fill=RuhiTeal!3, inner sep=0pt},
]

\node[panel, minimum width=5.7cm, minimum height=5.55cm] at (2.85, 6.85) {};
\node[panel, minimum width=5.7cm, minimum height=5.55cm] at (9.25, 6.85) {};
\node[panel, minimum width=5.7cm, minimum height=3.55cm] at (2.85, 2.15) {};
\node[panel, minimum width=5.7cm, minimum height=3.55cm] at (9.25, 2.15) {};

\node[tit] at (2.85, 9.25) {A.\ Tensor Mercer};
\node[sub] at (2.85, 8.92) {theory path; not the default posterior};
\node[boxderiv] (mercer) at (2.85, 8.25)
  {Hermite--Gaussian pairs $(\lambda_n,\phi_n)$\\[-1pt]
   $H_n'=2n\,H_{n-1}$ builds the force rows};
\node[boxderiv] (Afull) at (2.85, 7.20)
  {$A_{\mathrm{full}}=[\Phi;\ \Phi'J]\,\mathrm{diag}(\sqrt{\lambda})$\\[-1pt]
   $K_{\mathrm{joint}}=A_{\mathrm{full}}A_{\mathrm{full}}^\top$};
\node[boxderiv] (svdA) at (2.85, 6.15)
  {thin SVD $A=USV^\top$\\[-1pt]
   $\Psi(x)=A(x)VS^{-1}$, $\Psi(X)=U$};
\node[boxderiv] (ptens) at (2.85, 5.10)
  {$p_i=\frac{S_i^2}{S_i^2+\sigma^2}(\Psi^\top y)_i$\\[-1pt]
   $\mu=A_E(x_\ast)Wp$ (no $\alpha$)};
\draw[arrdash] (mercer) -- (Afull);
\draw[arrdash] (Afull) -- (svdA);
\draw[arrdash] (svdA) -- (ptens);

\node[tit] at (9.25, 9.25) {B.\ Production dual HS-SVD};
\node[sub] at (9.25, 8.92) {orthogonal basis; regularized eigenvalues};
\node[boxprod] (assembleK) at (9.25, 8.25)
  {Assemble signal $K$ (Sexpat on $q$)\\[-1pt]
   strip $\sigma^2$ from the diagonal};
\node[boxprod] (eigK) at (9.25, 7.20)
  {$K=\Psi\,\mathrm{diag}(S^2)\,\Psi^\top$\\[-1pt]
   exact-arithmetic relation to A; no tensor $A$};
\node[boxprod] (solveN) at (9.25, 6.15)
  {$\alpha=(K+\sigma^2 I)^{-1}y$\\[-1pt]
   inverse sensitivity set by noisy spectrum};
\node[boxprod] (predD) at (9.25, 5.10)
  {mean $K_\ast^\top\alpha$ or AlphaK$(x_\ast;\alpha)$\\[-1pt]
   variance: same solve on $k_\ast$};
\draw[arr] (assembleK) -- (eigK);
\draw[arr] (eigK) -- (solveN);
\draw[arr] (solveN) -- (predD);

\node[font=\sffamily\tiny, RuhiTeal!80!black] at (6.05, 7.20) {$K=AA^\top$};

\node[tit, RuhiCoral!85!black] at (2.85, 3.55) {C.\ SCG objective};
\node[sub, RuhiCoral!70!black] at (2.85, 3.22) {trace from the eigenbasis; scalars from $L$};
\node[boxobj] (Cmat) at (2.85, 2.50)
  {$C=K+\sigma^2 I=Q\,\mathrm{diag}(\mu)\,Q^\top$\\[-1pt]
   $K$ from panel B};
\node[boxobj] (cholC) at (2.85, 1.40)
  {$\log\det C=2\sum\log L_{ii}$\\[-1pt]
   $\alpha=L^{-\top}L^{-1}y$};
\draw[arrcoral] (Cmat) -- (cholC);

\node[tit] at (9.25, 3.55) {D.\ Dimer and NEB};
\node[sub] at (9.25, 3.22) {shared posterior; analytic surrogate HVP};
\node[boxout] (mean) at (8.00, 2.50) {GP mean $\mu$\\[-1pt]forces $-\nabla\mu$};
\node[boxout] (var)  at (10.50, 2.50) {posterior\\[-1pt]variance};
\node[boxout] (hvp)  at (8.00, 1.40) {analytic HVP\\[-1pt]$H(x)v$};
\node[boxout] (acq)  at (10.50, 1.40) {acquisition\\[-1pt]on the band};
\draw[arr] (predD.south) -- ++(0,-0.18) -| (mean.north);
\draw[arr] (predD.south) -- ++(0,-0.18) -| (var.north);
\draw[arr] (mean.south) -- (hvp.north);
\draw[arr] (var.south) -- (acq.north);

\node[boxderiv, text width=0.7cm, minimum height=0.32cm] at (1.1, -0.15) {};
\node[right, font=\sffamily\tiny] at (1.55, -0.15) {A theory};
\node[boxprod, text width=0.7cm, minimum height=0.32cm] at (4.3, -0.15) {};
\node[right, font=\sffamily\tiny] at (4.75, -0.15) {B production};
\node[boxobj, text width=0.7cm, minimum height=0.32cm] at (8.0, -0.15) {};
\node[right, font=\sffamily\tiny] at (8.45, -0.15) {C SCG};

\end{tikzpicture}}
\caption{Tensor Mercer derivation (A), optional dual HS-SVD (B),
SCG objective (C) and dimer/NEB consumers (D). The reported
calculations use Cholesky of $C=K+\sigma^2 I$, as in the objective
of panel C, and leave panel B off. $K=AA^\top$ identifies the two
factorizations under an exact feature model. The SCG objective takes
$K$ from the assembled Gram and evaluates $\log\det$ and $\alpha$ from
Cholesky of $C$.}
\label{fig:hssvd-four-panel}
\end{figure}
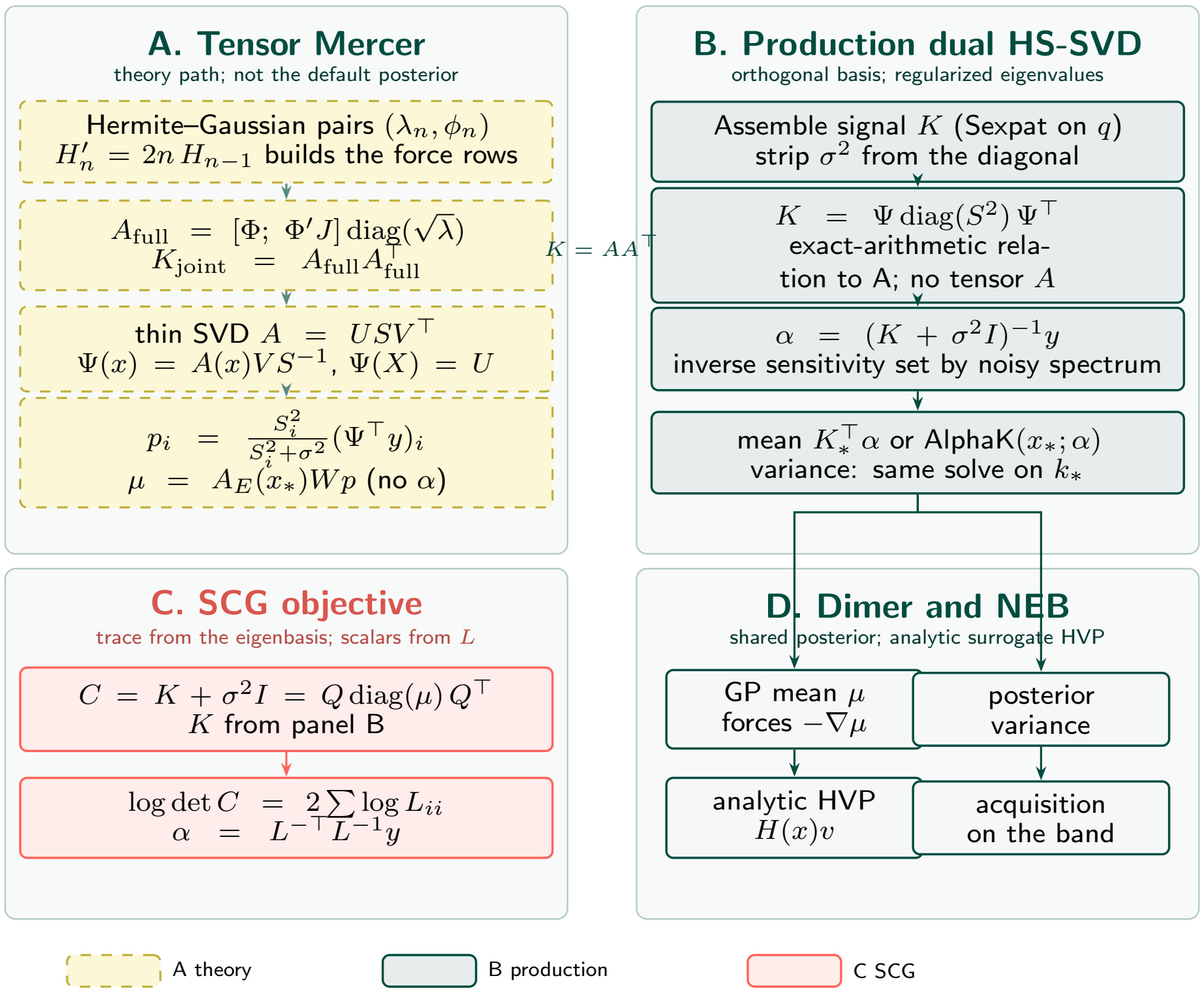

Output normalization is configurable and changes the numerical scaling
of the fit. The dual path runs without it, which does not establish that
normalization is unnecessary for every kernel or observation set, and
it neither confers a general accuracy advantage over Cholesky nor
removes the conditioning of prediction or of the hyperparameter
objective.

An optional dimer rotation path uses analytic Hessian-vector products of the GP mean. The reported molecular driver instead uses the center and displaced-image force difference with \(\Delta=0.01\) \AA{} described in the article. The minimum-mode solve is a symmetric
eigenproblem over the GP Hessian action, so Davidson and locally
optimal block preconditioned conjugate-gradient ideas
\cite{davidsonIterativeCalculationFew1975,knyazevTowardOptimalPreconditioned2001}
apply without finite differencing. The action is analytic and reuses
the descriptor chain rule of the mean. With the lengthscale-normalized
inverse-distance descriptor \(\tilde{\mathbf q}(\mathbf{x})\), its
Jacobian \(J = \partial\tilde{\mathbf q}/\partial\mathbf{x}\), and the
squared-exponential kernel
\(k_i = \sigma_f^2 \exp(-\lVert\mathbf{r}_i\rVert^2)\) with
\(\mathbf{r}_i = \tilde{\mathbf q}(\mathbf{x}) - \tilde{\mathbf q}_i\),
the dual-mean descriptor gradient and its directional curvature along
\(\mathbf{v}_{\tilde q} = J\mathbf{v}\) accumulate in descriptor space,

\begin{align}
\mathbf{g}_{\tilde q} &= \sum_i k_i\Big[w^E_i\,(-2\mathbf{r}_i) + \text{(force-row weights)}\Big],\label{eq:hvp_descgrad}\\
\mathbf{h}_{\tilde q} &= \sum_i k_i\Big[w^E_i\,\big(-2\mathbf{v}_{\tilde q} + 4(\mathbf{r}_i\!\cdot\!\mathbf{v}_{\tilde q})\,\mathbf{r}_i\big) + \text{(force-row terms)}\Big],\label{eq:hvp_deschv}
\end{align}

and map to Cartesian by the second-order chain rule

\begin{equation}\label{eq:hvp}
H(\mathbf{x})\,\mathbf{v} = J^\top \mathbf{h}_{\tilde q} + \big(\partial_{\mathbf{x}}J^\top\!\cdot\!\mathbf{v}\big)\,\mathbf{g}_{\tilde q},
\end{equation}

the descriptor-Hessian action \(J^\top\mathbf{h}_{\tilde q}\) plus the
curvature of the descriptor map itself, the directional descriptor
Hessian \(\partial_{\mathbf{x}}J\!\cdot\!\mathbf{v}\) contracted with the
descriptor gradient. The dual weights \(\boldsymbol\alpha\) split into an energy part \(w^E\) and gradient-row coefficients \(\alpha^G\) for the derivative observations. Eq. \ref{eq:phi-scalar} multiplies those coefficients by \(\partial k/\partial x_i^d\) and has no force minus. The predicted force is \(-\nabla\mu\). That optional analytic rotation path uses these weights and a prepared
Hessian-vector operator, so each such rotation step contracts against
the same \(\boldsymbol\alpha\) with no second Mercer tensor. These
weights serve both posterior refresh and dimer curvature evaluation.
When dual HS-SVD is selected, its posterior is refreshed directly after
full-data appends; that path avoids a redundant ScaLAPACK
training-covariance assembly and Cholesky factorization.

\Needspace{15\baselineskip}
\subsection{Language bindings}
\label{sec:bindings}
Python users can pass an ASE calculator to the same search library used by
\texttt{gprn} and \texttt{gprd}. This dimer example starts from a user-supplied geometry
near a candidate saddle:

\begin{lstlisting}[language=Python]
from ase.calculators.emt import EMT
from ase.io import read
from gpr_optim import GaussianProcess
from gpr_optim.dimer import GPDimer

atoms = read("near_saddle.con")
gp = GaussianProcess(uncertainty="map")
dimer = GPDimer(atoms, oracle=EMT(), gp=gp)
converged = dimer.run(fmax=0.05, steps=200)
\end{lstlisting}

Replace \texttt{EMT()} with the ASE calculator for the system and potential of
interest. The GP includes each measured energy and force, and the ASE bridge
converts force signs to the gradient convention used internally. For path-to-saddle refinement, pass the same GP and the band maximum to
\texttt{GPDimer}; its training set includes the GP observations. The C callback API and Fortran module expose
the same energy-and-force contract to compiled applications. The ASE dimer tutorial and the Python and Fortran binding guides are included with the code.
\subsection{Build systems and reproducibility}
\label{sec:build-repro}
Meson and CMake both build the library and its tests. CMake supports
\texttt{FetchContent} for downstream consumers. A companion environment file
pulls the developer-local tools (meson, ninja, gcovr).

Hyperparameter tuning (lengthscale ranges, noise grid
resolution, pruning thresholds) uses Optuna across chemical
systems. Those settings depend on the chemical system. The
library provides presets (\texttt{default}, \texttt{fast}, \texttt{accurate}) as
starting points.

Reduction order is selected by a single build flag,
\texttt{deterministic\_reductions}. Its default, \texttt{true}, replaces every
overlapping cross-rank floating-point sum with a
\(\textsc{Gather} \to\) rank-0 long-double accumulate \(\to\)
\(\textsc{Bcast}\) pattern. Rank ordering controls the reduction order at a fixed process count,
at a cost of \(\mathcal{O}(p)\) latency per reduction. Agreement across
process counts additionally depends on the arithmetic within each rank. The dense assembly
payloads stay out of this pattern entirely: their
one-owner-per-block support makes the plain \(\textsc{Allreduce}\)
exact (Section \ref{sec:mpi-cov}). The deterministic path
carries only the scalar and small-vector sums (trace terms,
slice means, likelihood contributions). The disjoint-support routing removes that payload class
from the deterministic tree by construction. Overlapping cross-rank sums accumulate in rank order by default and one-owner assembly reductions are exact; the reported
molecular calculations use that default with compensated binary64 posterior products and library factorization, and per-reaction search counts are identical across the three repetitions. Setting the build flag to
\texttt{false} uses native cross-rank reductions for throughput.

A Lean 4 development of the
labelled theorems of the supplementary material gives the
scalar squared-exponential identities, the Schur comparison
under an assumed decomposition (Eq. \ref{eq:schur}), the
local stability result for an exactly aligned reflected mode,
the BMA variance inequality, and
the gradient error at a surrogate critical point
\cite{moura2021lean4}. Those proofs are not required to run the
engine. They do not formalize Sexpat packing
(Eq. \ref{eq:sexpat-detail}), the blocked \([E;G]\) assembly, or the
HS-SVD factorization of assembled \(K\). The supplementary material
includes the labelled theorem statements and Lean source excerpts,
rendered with \url{https://github.com/VIDA-NYU/Lea}. Parallel reduction
order is model-checked separately in
\nolinkurl{docs/verification/parallel_reduction.tla}
\cite{konnovIntroducingTechniquesProofs2019}.

\section{Scope of the algebraic checks}
\label{sec:lean-identities}
The accompanying \texttt{lean/} sources contain statements about exact real arithmetic, orthogonal projectors and covariance identities.
Their conclusions depend on the hypotheses stated in each declaration.
An implication that assumes the desired bound does not establish that bound for the search, and a scalar component identity does not verify a multidimensional algorithm.
These artifacts do not certify the C++ implementation, floating-point error, numerical rank selection, spectral completeness, or convergence of a refitted surrogate search.

The physical acceptance rule is independent of these algebraic checks: the molecular calculations require a measured force and a complete independent Hessian with resolved index one.
An inner relaxation can stop on a surrogate-force criterion; that inner stop does not accept the geometry.
The covariance, Householder reflection and variance-mixture identities are derived in Sections \ref{sec:si-operator-algebra}, \ref{sec:si-hssvd} and \ref{sec:laplace} under their stated assumptions.
\section{Scale-free trust radius}
\label{sec:scalefree-tr}
A radius fixed in absolute distance does not track image spacing, which changes from reaction to reaction.
The trust radius used by the optional scale-free arm is the median pairwise distance in the training set times a dimensionless factor \(\tau\),
\begin{equation}
r = \tau \cdot \operatorname*{median}_{i<j}\lVert x_i - x_j \rVert,
\label{eq:scalefree-radius}
\end{equation}
with \(\tau=0.5\).
The radius grows when the ratio of actual to predicted decrease exceeds \(0.75\) and shrinks when that ratio is below \(0.25\).
A surrogate step is admitted only when the predictive standard deviation is at most a constant times the largest force component.
The reported Baker, Birkholz, and Pt(111) searches use the fixed radius; this arm is the dimensionless alternative.

\bibliography{rg_main}
\end{document}